\pdfoutput=1
\documentclass[iop]{emulateapj}
\begin{document}
\title{Uncloaking globular clusters in the inner Galaxy\footnote{Based
    partly on observations with the NASA/ESA {\it Hubble Space
      Telescope}, obtained at the Space Telescope Science Institute,
    which is operated by the Association of Universities for Research
    in Astronomy, Inc., under NASA contract NAS 5-26555. This paper
    also includes data gathered with the 6.5 meter Magellan Telescopes
    located at Las Campanas Observatory, Chile. }}
\author{Javier Alonso-Garc\'{i}a}
\affil{Departamento de Astronom\'{i}a y Astrof\'{i}sica, Pontificia Universidad Cat\'{o}lica de Chile, 782-0436 Macul, Santiago, Chile} 
\affil{Department of Astronomy, University of Michigan, Ann Arbor, MI 48109-1090} 
\email{jalonso@astro.puc.cl}
\author{Mario Mateo}
\affil{Department of Astronomy, University of Michigan, Ann Arbor, MI 48109-1090} 
\email{mmateo@umich.edu}
\author{Bodhisattva Sen}
\affil{Department of Statistics, Columbia University, New York, NY 10027} 
\email{bodhi@stat.columbia.edu}
\author{Moulinath Banerjee}
\affil{Department of Statistics, University of Michigan, Ann Arbor, MI 48109-1107}
\email{moulib@umich.edu}
\author{M\'{a}rcio Catelan}
\affil{Departamento de Astronom\'{i}a y Astrof\'{i}sica, Pontificia Universidad Cat\'{o}lica de Chile, 782-0436 Macul, Santiago, Chile} 
\email{mcatelan@astro.puc.cl}
\author{Dante Minniti}
\affil{Departamento de Astronom\'{i}a y Astrof\'{i}sica, Pontificia Universidad Cat\'{o}lica de Chile, 782-0436 Macul, Santiago, Chile} 
\affil{Vatican Observatory, Vatican City State V-00120, Italy}
\affil{Department of Astrophysical Sciences, Princeton University, Princeton NJ 08544-1001, USA}
\email{dante@astro.puc.cl}
\author{Kaspar von Braun}
\affil{NASA Exoplanet Science Institute, California Institute of Technology, Pasadena, CA 91125-2200}
\email{kaspar@ipac.caltech.edu}

\begin{abstract}
Extensive photometric studies of the globular clusters located towards
the center of the Milky Way have been historically neglected. The
presence of patchy differential reddening in front of these clusters
has proven to be a significant obstacle to their detailed study.  We
present here a well-defined and reasonably homogeneous photometric
database for 25 of the brightest Galactic globular clusters located in
the direction of the inner Galaxy.  These data were obtained in the
$B$, $V$, and $I$ bands using the Magellan 6.5m telescope and the {\it
  Hubble Space Telescope}. A new technique is extensively used in this
paper to map the differential reddening in the individual cluster
fields, and to produce cleaner, dereddened color-magnitude diagrams
for all the clusters in the database. Subsequent papers will detail
the astrophysical analysis of the cluster populations, and the
properties of the obscuring material along the clusters' lines of
sight.
\end{abstract}


\section{Introduction}
The globular cluster system of the Milky Way has long been used to
learn about the evolution of the Galaxy. As the Galactic globular
clusters (GGCs) constitute some of the oldest systems in the Milky
Way, the study of the stellar populations of these {\it fossils} can
give us important clues of the early stages of the Galaxy's formation.
The tool most extensively used in this task has been the analysis of
the color-magnitude diagrams (CMDs) (e.g., \citet{an05,ma09a}).  However, the
presence of significant differential extinction in low-latitude
fields, particularly near the Galactic center, greatly complicates
traditional CMD analyses.  As a result, the study of many GGCs located
towards the inner Galaxy has been historically neglected.

Various recent studies have tried to overcome the difficulties
associated with the study of inner GGCs to better exploit them as
probes of the stellar populations near the Galactic Center.  One
obvious approach is to use near-infrared photometry to study these clusters
(e.g.,\citet{da00,va07}) to take advantage of the smaller extinction in these
bands.  But to extract precise information about the ages of the GGCs,
an accurate location of the main sequence turn-off (MSTO) point is
crucial in most methods of cluster dating \citep{st96,sa97,gr03}.
Infrared photometry sufficiently deep to reach the main sequence (MS)
with enough precision to accurately locate the turn-off (TO) point is
difficult to achieve, and only now are we starting to see the first
results after the careful application of new adaptive optics
techniques on big telescopes \citep{mo09}. Deep photometry for
non-highly reddened GGCs is much more easily obtained in the optical,
but at the cost of increased extinction. A number of techniques have
been developed to produce extinction maps for individual clusters as a
means of dealing with this issue (e.g., \citet{me04,vo01}). In general,
the resolution and accuracy of optical extinction maps are not
adequate to fully eliminate the effects of differential reddening at a
level of precision to produce deep optical CMDs of similar high
quality now typical for high-latitude clusters with little or no
differential extinction (e.g., \citet{ro00a,pi02}).

In this paper we present a new optical photometric database consisting
of a sample of 25 GGCs located towards the Galactic Center. We use a
new dereddening technique (\citet{al11}, from now on referred to as
Paper I) to map the differential extinction along their fields, and
produce new, cleaner, differentially dereddened CMDs of the clusters.
In section 2, we give the criteria used to define our sample of
GGCs. Section 3 summarizes the steps that we followed to obtain
precise astrometry and optical photometry of the stars in these
clusters suitable for our analyses. In section 4 we apply the
dereddening technique described in Paper I, and provide an overview of
the differentially dereddened CMDs of the clusters in our sample,
along with extinction maps along their fields. We also describe the
characteristic features of the environments in which they are located,
and briefly summarize previous optical and infrared photometric
studies where appropriate. Finally, in section 5, we provide a
summary of our results.

\section{Selection of our sample members}
The cluster sample presented in this study consists of 25 GGCs 
in the direction of the inner Galaxy, all located within $30 \deg$ of
the Galactic Center. Because the precision of our dereddening
technique (Paper I) requires good sampling and photometry down to a
few magnitudes below the MSTO, and also depends on the density of
stars and the spatial variations in the extinction, the clusters in
our sample were chosen to also satisfy the following criteria:

\begin{itemize}
\item Exhibit moderate mean extinction, implying that they may suffer
  from extinction variations. We explicitly restricted our
  sample to clusters with a mean reddening of $E(B-V) \ge 0.07$ mag.
\item Be sufficiently luminous to possess a well-defined MS. Our
  analysis requires a significant number of stars (at least a few
  hundreds) to calculate the extinction in a region. We therefore
  chose clusters with a luminosity satisfying $M_{\rm V} \le -6$.
\item Be relatively nearby.  With our dereddening technique, the stars
  that provide most of the information about the differential
  extinction are those located in the CMD sequences most nearly
  orthogonal to the reddening vector (subgiant branch (SGB) and upper
  MS). Since one of the goals of this project is to accurately
  calculate the relative ages of these clusters, we must also reach the
  MS with good photometric precision at the TO in order to carry out a reliable
  age/metallicity analysis. Hence, we chose clusters with an apparent
  distance modulus of $(m-M)_{\rm V} \le 16.6$.
\item Be sufficiently extended so that we can define maps that cover a
  significant solid angle around the clusters. We therefore chose
  clusters with a tidal radius $r_t \ge 7.5$ arcmin.
\end{itemize}

\begin{figure}
\plotone{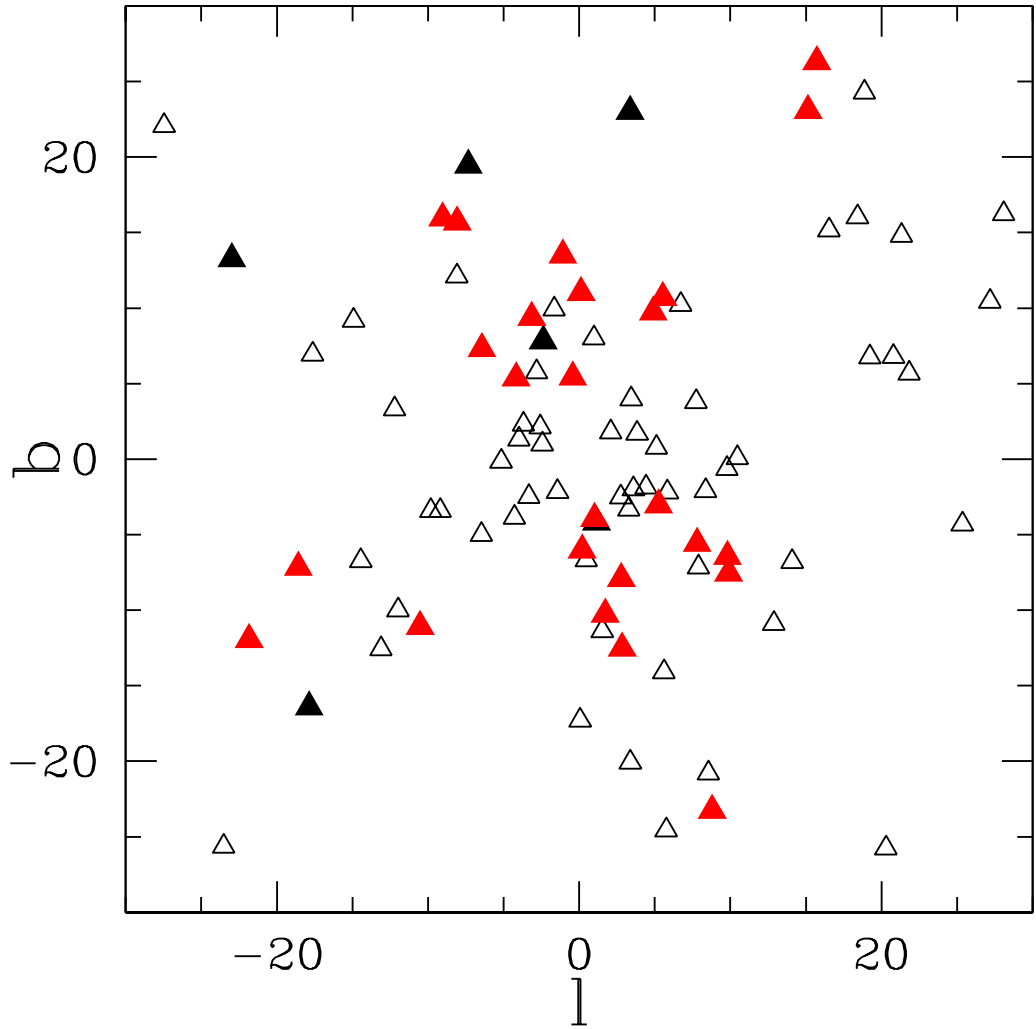}
\caption{Location, in galactic coordinates, of the clusters that
  follow the criteria to belong to our sample (solid triangles) and
  the rest of the inner GGCs (open
  triangles). The position of the 25 GGCs that we were able to observe
  and comprise our final sample is in red.}
\label{figpos}
\end{figure}

We were able to observe 25 of the 31 GGCs that fulfill these
requirements, according to the 2003 version of the \citet{ha96}
catalog. Their positions are shown in Figure \ref{figpos} along with
the rest of the clusters located in the inner Galaxy.  Their
characteristics, according to the most recent (December 2010)
version\footnote{According to this updated version of the catalog, two
  of the clusters in our sample, NGC 6287 and NGC 6355, do not fulfill
  anymore the apparent distance modulus requirement to belong to our
  sample. However, we have still kept these clusters in our studied
  sample, since previous studies of these objects have been very
  scarce, and their published parameters are accordingly quite
  uncertain.} of the Harris catalog, are given in Table \ref{tabgc}.

\section{Observations and data reduction}
We obtained optical photometric data for our sample of GGCs using the
Magellan 6.5m Baade Telescope located at the Las Campanas Observatory
(LCO) in Chile, and the {\it Hubble Space Telescope} ({\it HST}). In
this section we provide a complete summary of these observations and
explain in detail the steps followed for the reduction and calibration
of the data. The final photometry has an absolute precision with respect to the calibrating stars of
${\sigma}\sim0.02-0.03$ magnitudes in most cases. The internal
precision among the stars observed in each field is around ${\sigma}\sim0.01-0.02$ magnitudes, sufficient to
achieve the precise extinction maps that will allow us to produce
clean CMDs. From these we will aim to derive the cluster
parameters with an accuracy similar to those derived from
high-Galactic latitude clusters CMDs that suffer little or no differential
extinction.

\subsection{Observations}
The GGCs in our sample were observed over four nights, May, 30th to
June, 2nd 2005, with the LCO 6.5 m Magellan Baade Telescope, using the
Inamori Magellan Areal Camera and Spectrograph (IMACS) in imaging
mode. We used the f/4 camera to image a field of view (FOV) of $15.46'
\times 15.46'$, with a pixel size of $0.11''$.  All fields were
observed using standard Johnson-Cousins $B$,$V$, and $I$ filters
\citep{be79}. Two sets of observations with different exposure times
(short and long) were taken in the $B$ and $V$ filters, and three
(extra-short, short, and long) in $I$, for every cluster. Although the
nights during our observing run were not all completely photometric,
the seeing conditions were very good, with average values of $\sim
0.6''$ in $V$ for the whole run. Table \ref{tabobsmag} lists the
details of these ground-based photometry observations.

In order to obtain useful photometry of the inner regions of the more
centrally crowded clusters, we supplemented our Magellan images with
images taken with the Advanced Camera for Surveys (ACS) aboard {\it
  HST}. These data were obtained in by our group's Snapshot program
10573. The ACS has a FOV of $3.37' \times 3.37'$, with a pixel size of
$0.05''$.  Five clusters of our sample were observed using the
$f435w(B_{435})$, $f555w(V_{555})$, and $f814w(I_{814})$
filters. Table \ref{tabobsacs} lists the details of these new {\it
  HST} observations.

To better calibrate our photometry, we also used images available
through the {\it HST} data archive for all the clusters in our sample.
The data taken from the {\it HST} data archive are comprised of
$f439w(B_{439})$, $f555w(V_{555})$, $f606w(V_{606})$, and
$f814w(I_{814})$ images obtained with the Wide Field Planetary Camera
2 (WFC2), and of $f435w(B_{435})$, $f606w(V_{606})$, and $f814w(I_{814})$ images taken with the ACS. Table \ref{tabobsarc}
lists the different {\it HST} programs that we use data from.

\subsection{Data reduction and photometry derivation}
For the ground-based data, the initial processing of the raw CCD
images was done with the routines in the {\it ccdred} package of the
Image Reduction and Analysis Facility (IRAF). The images were first
corrected for bias, then flatfielded using a combination of dome and
twilight flats. Afterward, since different frames for every cluster
were taken with different exposure times in the $B$, $V$, and $I$
filters (see Table \ref{tabobsmag}), and they had small spatial
offsets between them, the frames of every individual cluster were
aligned, using the IRAF task {\it imalign}, and average-combined for
every exposure time and filter, with the IRAF task {\it imcombine}.

Stellar photometry was carried out on the ground-based processed
images using an updated version of DoPHOT \citep{sc93}. This version
works on any platform and accepts images consisting of real data
values. It also provides better aperture corrections than previous
versions, allowing for variations in aperture corrections as a
function of field position and stellar magnitude (see Appendix \ref{ap1}).

We also carried out an astrometric analysis of the cluster fields in
our sample. We derived coordinates (Right Ascension $\alpha$ and
Declination $\delta$) for all stars detected in our photometric study
by comparison with bright stars obtained in each field from the Two
Micron All Sky Survey (2MASS) catalog stars available through the
Infrared Processing and Analysis Center (IPAC) website. In practice,
more than 100 stars were available as astrometric references within
the fields of every chip of our CCD camera. A third order polynomial
fit, done with the IRAF task {\it mscpeak}, produced dispersions of
${\sigma}\sim0.25''$, consistent with the catalog precision. Using the
astrometric information of the images we could also calculate if there
were variations in the pixel area coverage accross the images, which
could lead to miscalculations in the measured fluxes. But the pixel
area changed by less than $1 \% $ across any of the chips of the
camera, and therefore we did not need to apply any corrections to the
measured photometry.

To calibrate our ground-based data, we first transformed the
photometry from each of the individual eight CCDs in the camera in
IMACS to a common zero point, and then transformed this system to a
standard system--in our case, the Johnson-Cousins system
\citep{be79}. We arbitrarily chose one of the CCDs, chip 2, as the one
defining the instrumental system and brought the photometry from the
other chips to this system (see Appendix \ref{ap2}). After this was
done, we placed our photometry in the Johnson-Cousins photometric
system by observing \citet{la92} fields over a range of airmasses
during the nights of our 2005 observing run (see Appendix
\ref{ap3}). Since these nights were not completely photometric we used
\citet{st00} photometric standard stars and the photometry from the
clusters obtained with the {\it HST} to fine-tune the calibration of
our ground-based data (see Appendix \ref{ap3}).

The photometry of the data from the {\it HST} observations was
obtained using the programs HSTPHOT for the WFPC2 data \citep{do00}
and DolPHOT for the ACS data \citep{do00}. These programs have been
specifically tailored to analyze data from these cameras. They take
the images for every cluster retrieved from the the {\it HST} in the
different filters, and analyze them all at the same time. The programs
align the images, combine them and provide the final combined-frame
and individual-frame photometry, both in the {\it HST} filter system and in
the Johnson-Cousins system. We follow the author's prescriptions to
give the input parameters for the analysis to both programs.
Photometric goodness-of-fit parameters were employed to select only
objects with high-quality measurements. Only measurements for which an
object was classified as stellar (HSTPHOT and DolPHOT types 1 and 2),
and for which $-0.1 \leq sharpness \leq 0.1$, $crowding \leq 0.5$, and
errors in all measured magnitudes ${\sigma}\leq 0.1$ were retained.

\section{Dereddened color-magnitude diagrams and extinction maps}

In this section we present the results of applying the dereddening
method described in Paper I to all the clusters in our sample of inner
GGCs. We provide the CMDs of these clusters before and after
correcting for differential extinction and describe any
improvements. We also provide the extinction maps that the dereddening
technique generates, and compare them with the extinction maps
supplied by \citet{sc98}, from now on referred as the SFD extinction
maps. For each cluster, we provide references to its more recent --
but not necessarily all-- optical and infrared photometric studies.
We also report studies that explore the presence of different stellar
populations in the observed GGCs, and speculate on the possibility of
several stellar populations in some of the sampled clusters from the
presence of a color-extended HB, as suggested in some recent studies
\citep{da08b,gr10,ca11}.

A detailed explanation of the dereddening method is provided in Paper
I, but we summarize here the different steps that we follow to
apply the method, along with the results from every step, to obtain the
final extinction maps and dereddened CMDs. The technique
is applied independently to the two available colors ($B-V$ and $V-I$), and it
is only at the end of the process that the extinction maps obtained
from both analyses are averaged.

\begin{figure}[htbp]
\epsscale{1.25}
\plotone{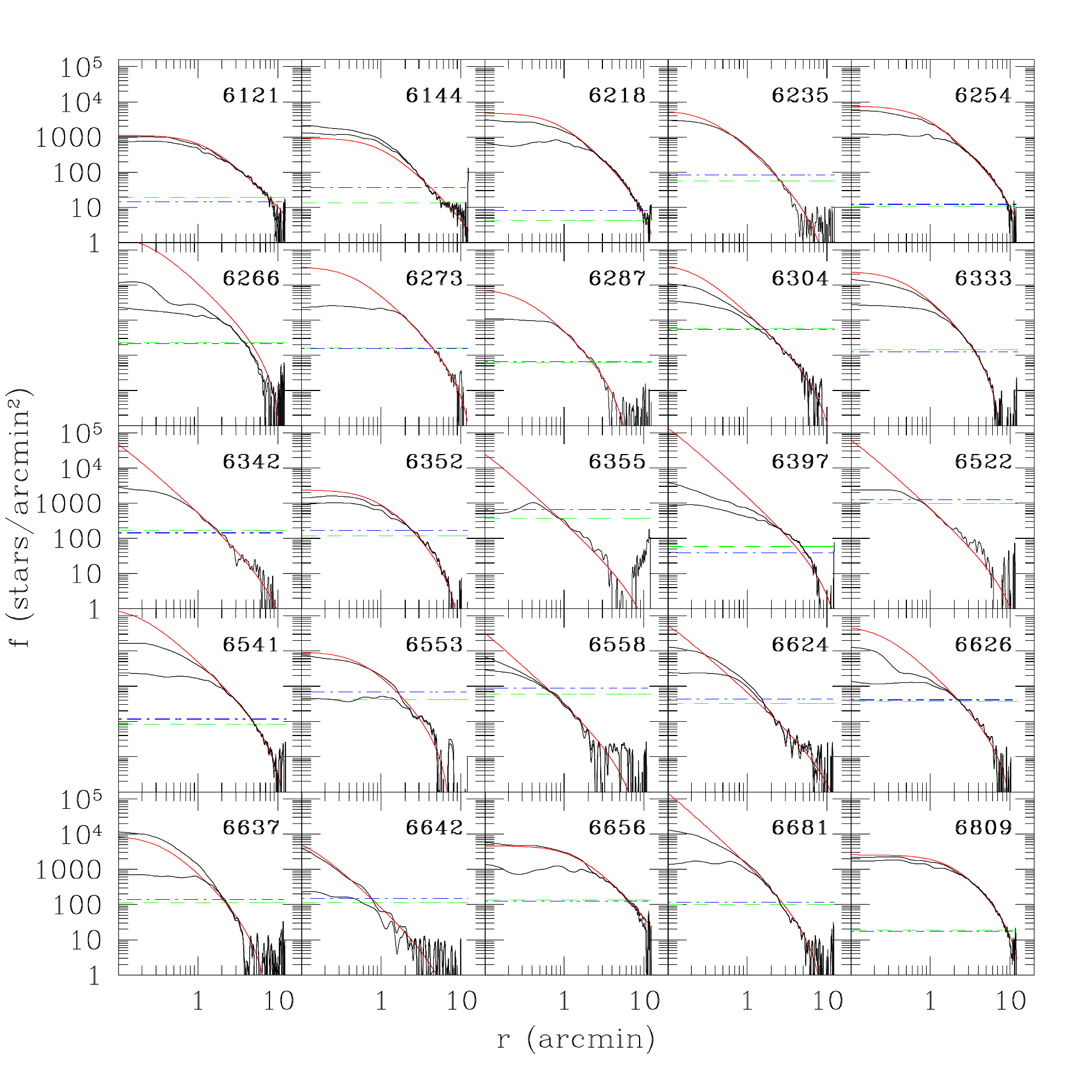}
\caption{Radial density profile of the stars in the cluster from the
  observations (black) and from fitting a King profile (red), once the
  constant non-member field component has been subtracted. The
  straight green dashed line shows the non-member field component
  according to the fit. The straight blue dot-dashed line, showing the
  non-member field component according to the Besan\c{c}on model, is
  plotted for comparison. The general disagreement in the inner regions
  between observations and models is due to the lower completeness of
  these regions, due to crowding effects. Whenever there are two black
  density profiles, one belongs to the space+ground observations, while
  the other corresponds to ground-only observations, and generally
  shows a lower completeness.} 
%NGC 6144, NGC 6637 and NGC 6809 show a
%  higher number of stars observed than predicted from the models,
%  which make us suspect that maybe there is a mistake with the values
%  of $r_c$ and $r_t$ in the literature for this clusters.}
\label{figking1}
\end{figure}

\begin{figure}[htbp]
\epsscale{1.25}
\plotone{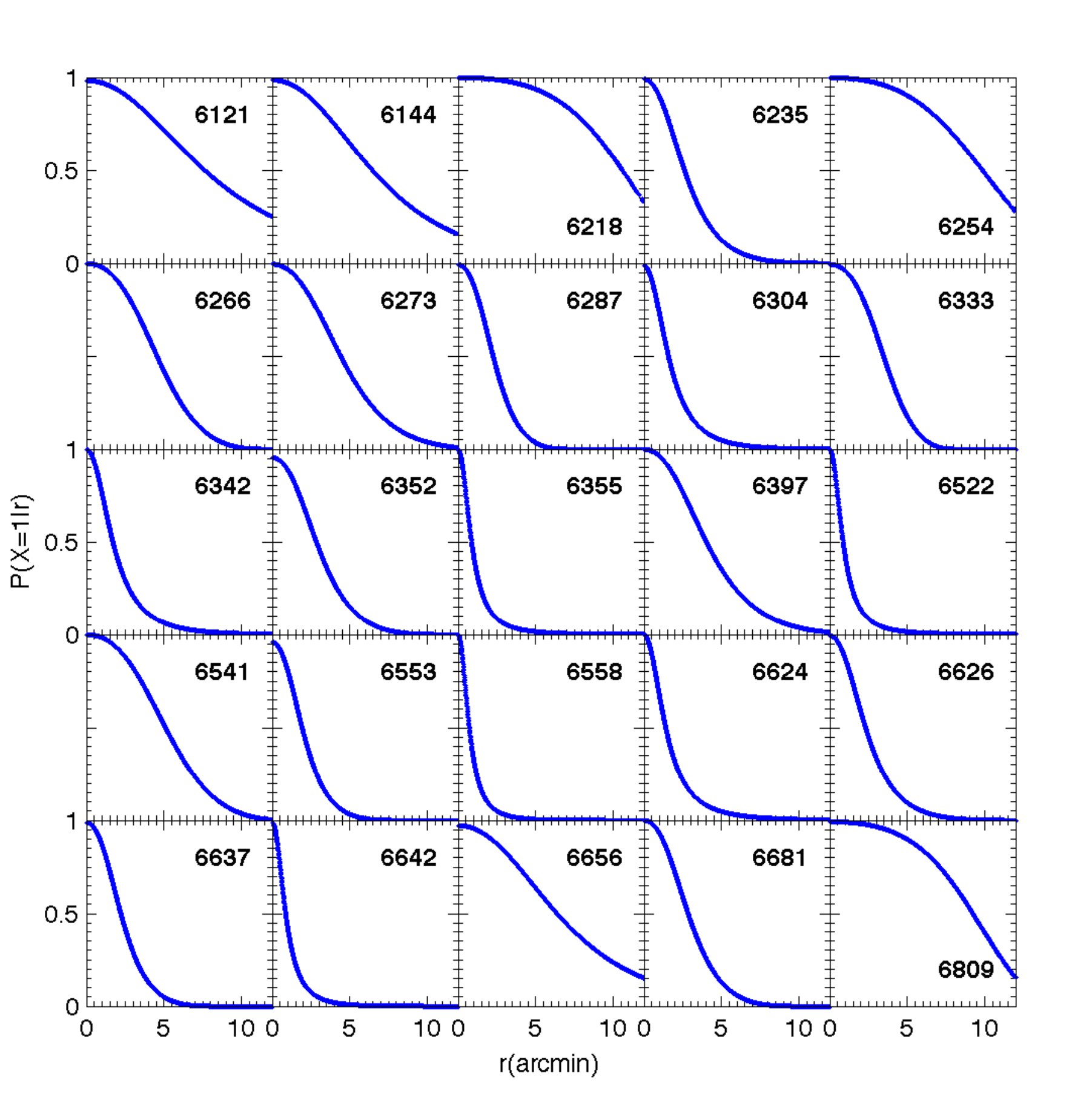}
\caption{Probability to belong to the cluster as a function of
  distance to the center of the cluster. We should note, however, that
  both for the central regions where the crowding is more significant
  (as mentioned in Paper I), as for regions where extinction variations
  across the field are significant, the detection limit changes, in
  the case of the high crowding by being lower, and in the case of
  significative extinction by, although being nominally the same,
  corresponding to different intrinsic luminosities in different parts
  of the field. This is going to produce deviations from the model as
  we have seen in Figure \ref{figking1}. In our analysis, we assume that we are
  missing an equal percentage of stars from the cluster and from the
  field due to these facts, and hence we use the values provided in
  this figure.}
\label{figking2}
\end{figure}

\begin{figure}
\epsscale{1.1}
\plotone{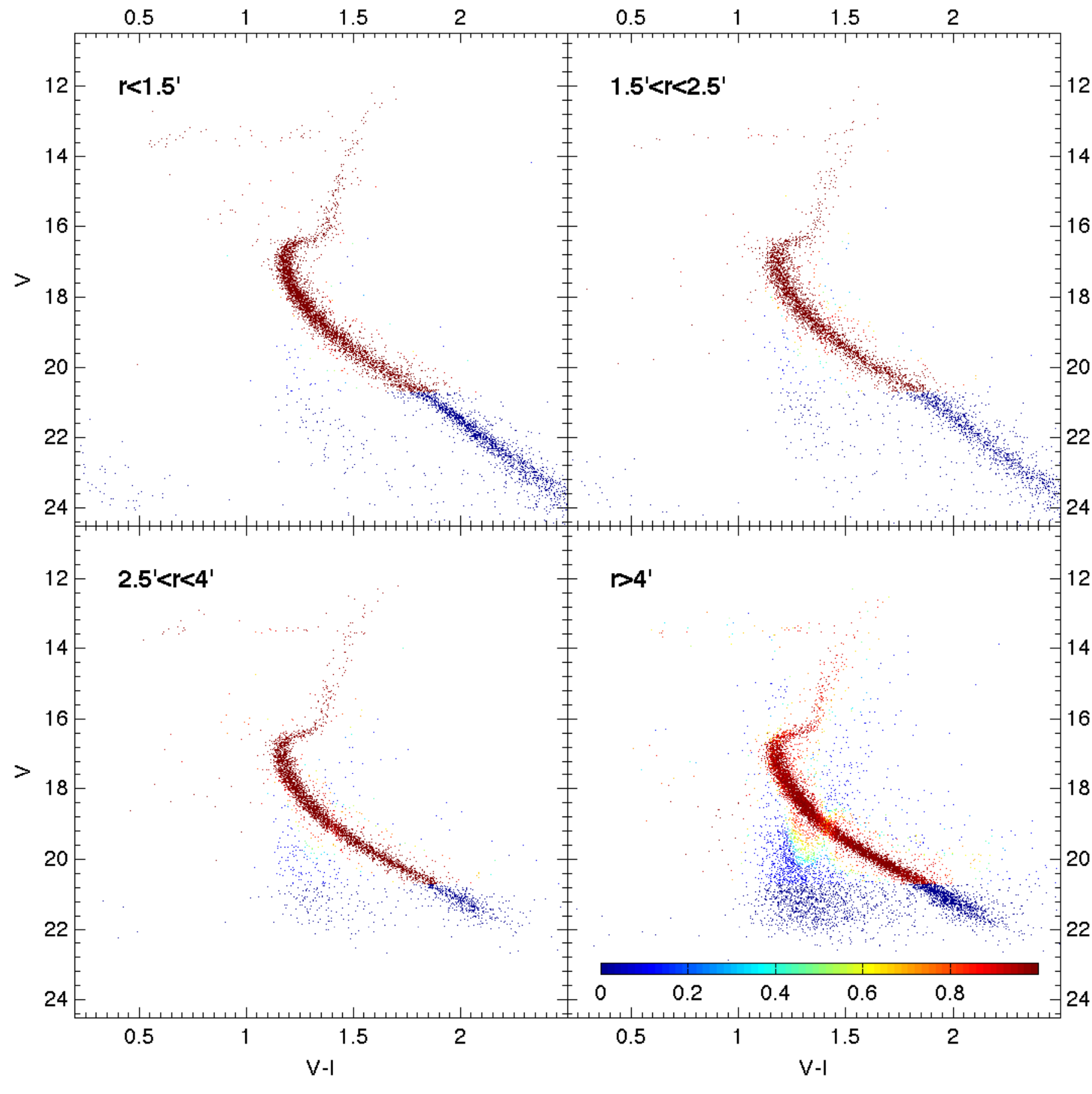}
\caption{Observed dereddened CMD for M4 at four different distances from
  the cluster center, with the probability of every star to belong to
  the cluster as a function of position in the sky, color and
  magnitude, $P(X=1|r,c,m)$, represented by the different colors of
  the stars as indicated in the color bar. Notice that we restrict our
  analysis to stars brighter than the completeness limit shown in
  Table \ref{tabclimit}.\label{figprobm4}}
\end{figure}

The first step of our dereddening method is to calculate the
probabilities for the observed stars to belong to the cluster. In
order to do that, we first calculate the surface density of the stars
as functions of distance $r$ to the cluster center, as explained in
Paper I, and by a fit with an empirical \citet{ki62} profile plus
constant field model (see Figure \ref{figking1} and Table
\ref{tabking}) obtain, using Equation 12 from Paper I, the conditional
probabilities of the stars being members of the cluster as a function
of position in the sky $P(X=1|r)$ (see Figure \ref{figking2}), where
$X$ indicates membership to the cluster ($X=1$ indicates observation
of a member, and $X=0$ indicates observation of a non-member). Then we
calculate, following Paper I, the probability densities of the
observed stars as a function of $r$, color $c$, and magnitude $m$, and
the probability density of being field members as function of $c$ and
$m$ by comparison with a Besan\c{c}on model \citep{ro03} (see table
\ref{tabbesancon}).  Finally, applying Equation 20 from Paper I, we
are able to find $P(X=1|r,c,m)$, the conditional probabilities of the
stars being members of the cluster as a function of position in the
sky and in the CMD (see Figure \ref{figprobm4} in this paper, and
Figure 4 from Paper I as examples of the method). We restrict our
analysis to stars down to the completeness limit of our observations,
and areas in which the probability $P(X=1|r)>0.1$ (see Table
\ref{tabclimit}), as calculated in Equation 13 in Paper I. There are a
few cases in which we do not calculate $P(X=1|r,c,m)$ for one of the
colors, since the observed CMD of the cluster is not calibrated in
that color (NGC 6558 in $B-V$, and NGC 6235, NGC 6342 and NGC 6355 in
$V-I$), and a comparison with the Besan\c{c}on models can lead to
mistakes. In those cases, we equal the total probability to just
$P(X=1|r)$ (see Equation 13 in Paper I), and proceed with the next
steps of the analysis, which are not affected by any constant offset
in the color calibration.

\begin{figure}[htbp]
\epsscale{1.25}
\plotone{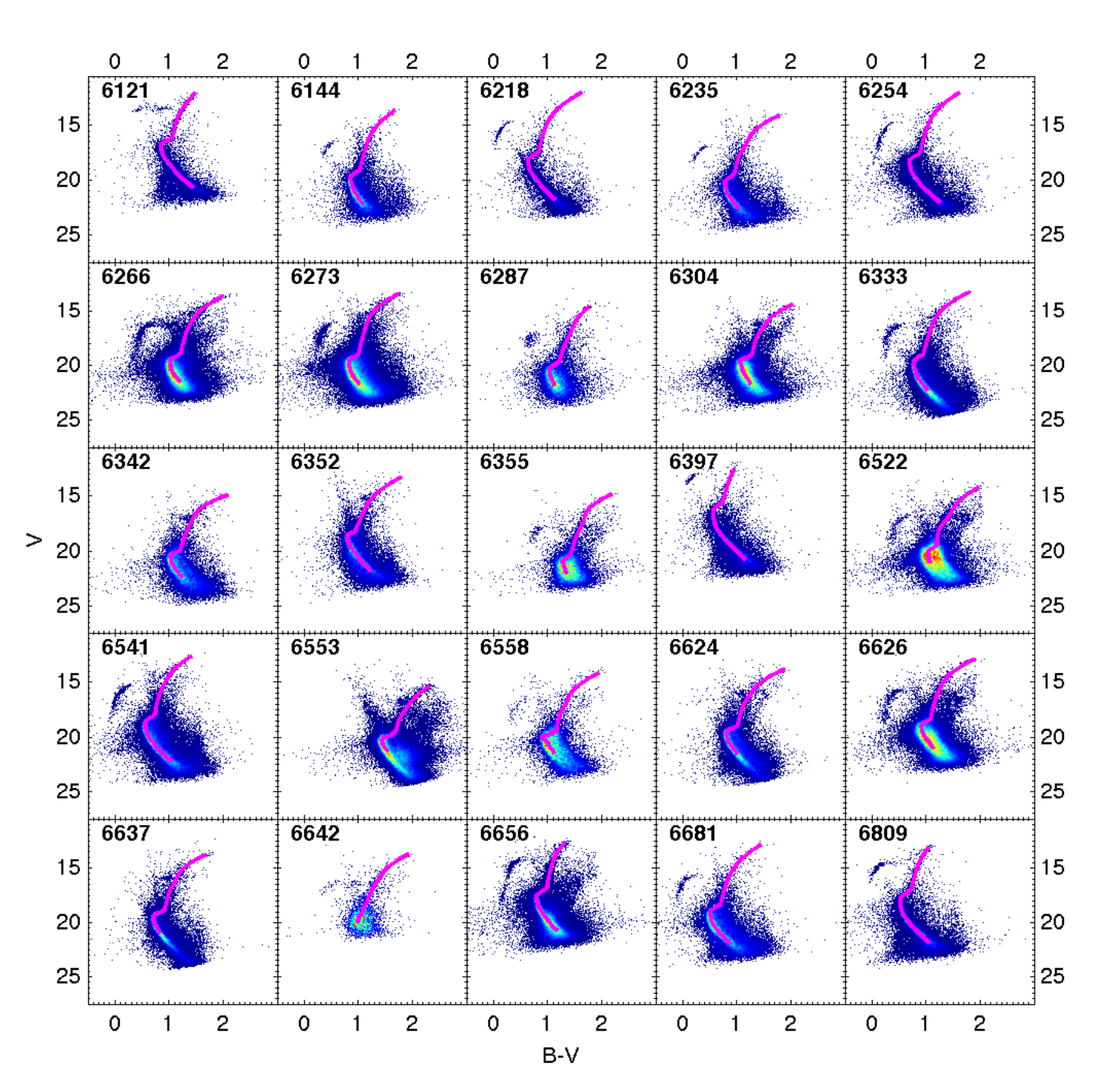}
\caption{Ridgelines in the $V$ vs. $(B-V)$ CMDs for the clusters in
  the sample, down to the completeness limit.}
\label{figridgebv}
\end{figure}

\begin{figure}[htbp]
\epsscale{1.25}
\plotone{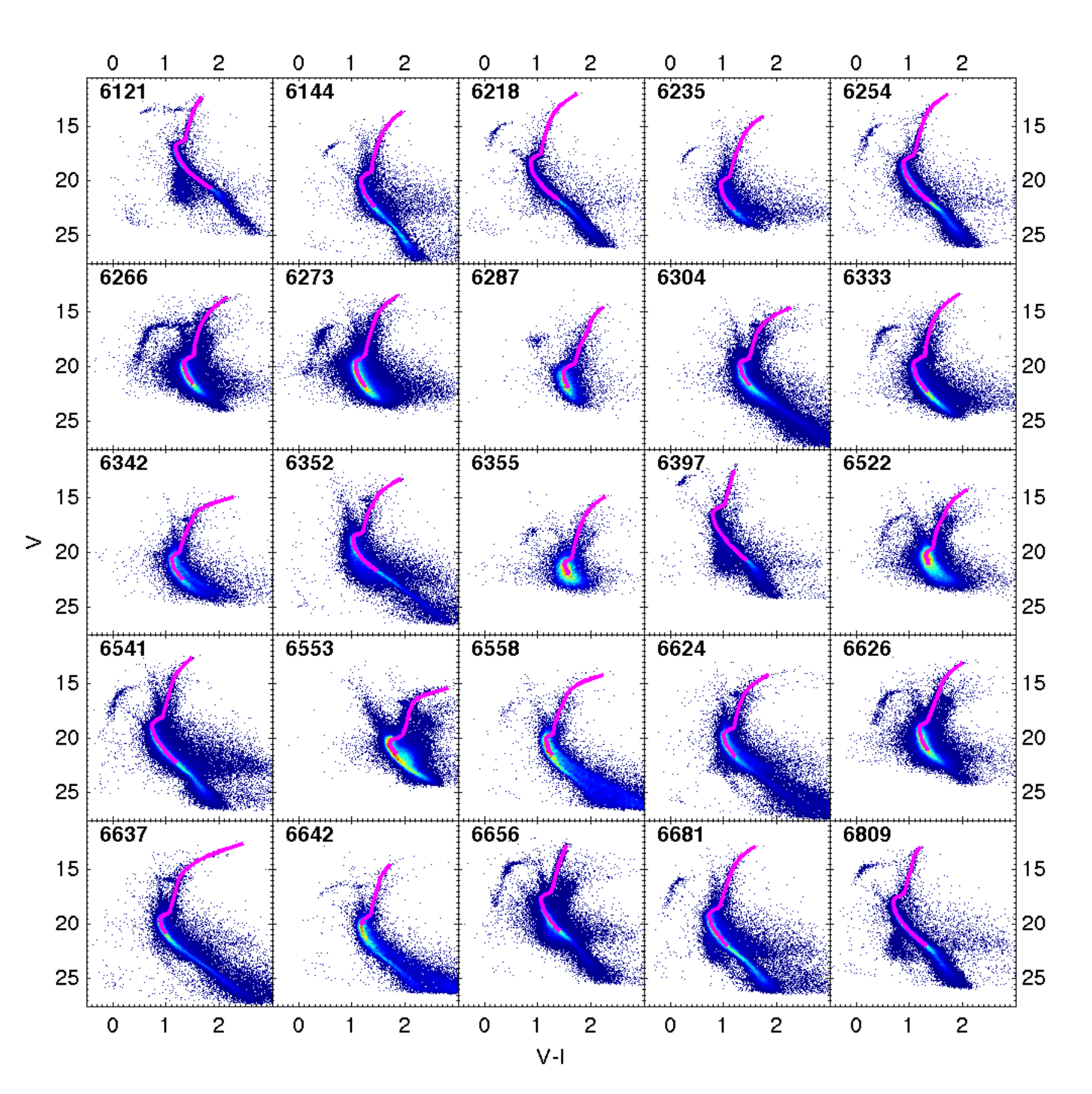}
\caption{Ridgelines in the $V$ vs. $(V-I)$ CMDs for the clusters in
  the sample, down to the completeness limit.}
\label{figridgevi}
\end{figure}

Once the cluster membership probabilities have been calculated, we
need to build the ridgeline for the CMDs of every cluster in our
sample. We follow the three-step recipe described in Paper I to obtain
the ridgelines\footnote{For some clusters the ridgelines were better
  calculated if we used only stars where the membership probability as
  a function of distance was $P(X=1|r)>0.5$, instead of $P(X=1|r)>0.1$
  as in Paper I. For homegeneity on the process through the sample, we
  apply this more severe cut to calculate the ridgelines of all the
  observed clusters.} for both available colors (see Figures
\ref{figridgebv} and \ref{figridgevi}).

After that, we move the stars along the reddening vector \citep{sc98}
until they intersect the ridgeline, and smooth the resulting
individual color excesses by building a bivariate non-parametric
regression of the extinction as a function of the spatial coordinates
right ascension and declination, which allows us to generate an
extinction map\footnote{For some clusters the reddening maps were
  better obtained when, in the iterative process described in section
  2.4 of Paper I, we use stars with extinctions at most $2\sigma$ away
  from the extinction value calculated in the previous iteration, as
  opposed to the $3\sigma$ threshold mentioned in Paper I. Again for
  homegeneity on the process through the sample, we apply this more
  severe cut to calculate the extinction maps of all the observed
  clusters.}.  From this map, we take the relative extinction that
corresponds to every star observed. We then plot the CMD of the stars
in our observation after having been corrected for the differential
reddening and use it as the input for the next iteration.

\begin{figure}[htbp]
%\epsscale{0.77}
\plotone{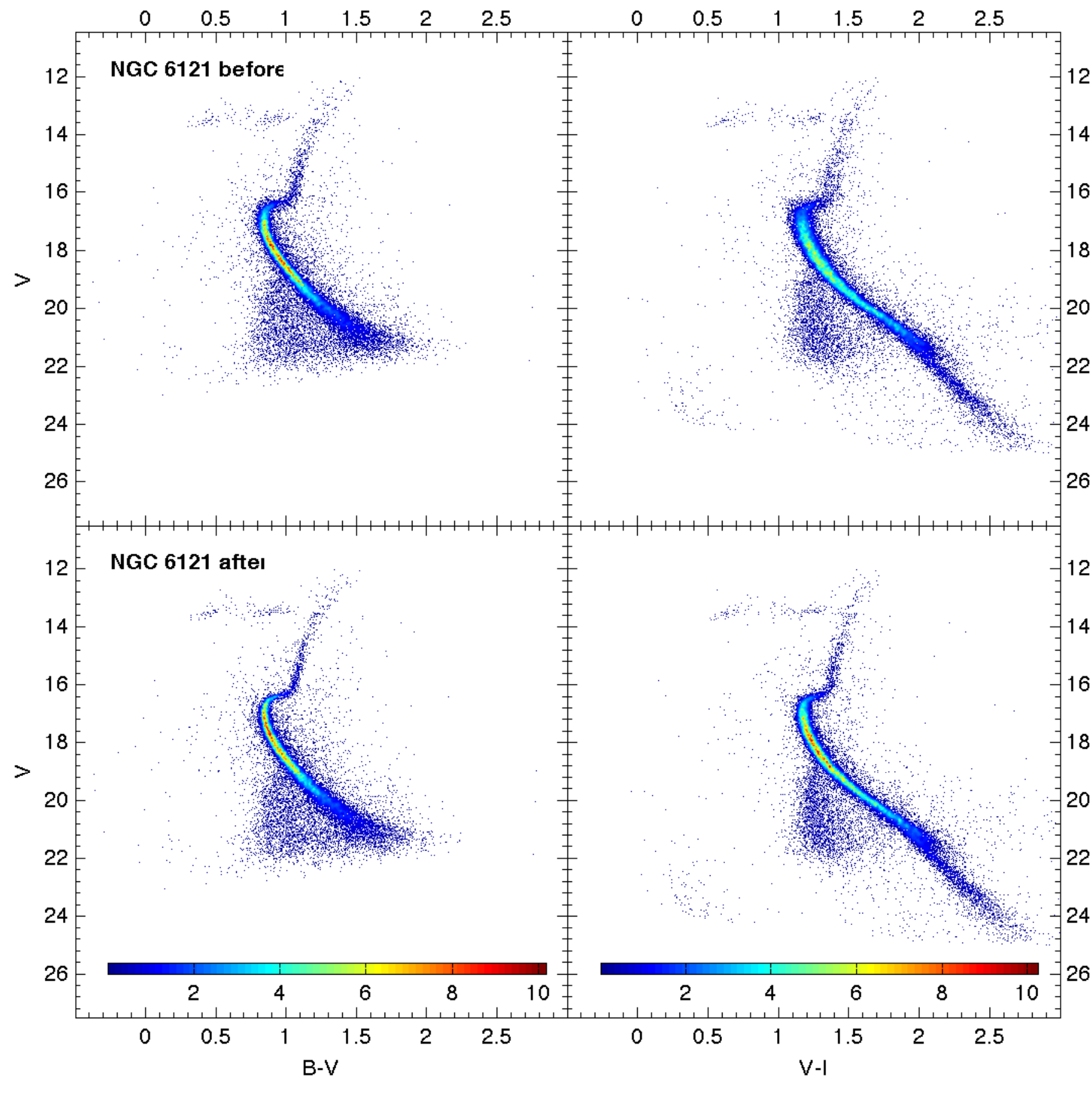}
\begin{tabular}{ccc}
\includegraphics[scale=0.28]{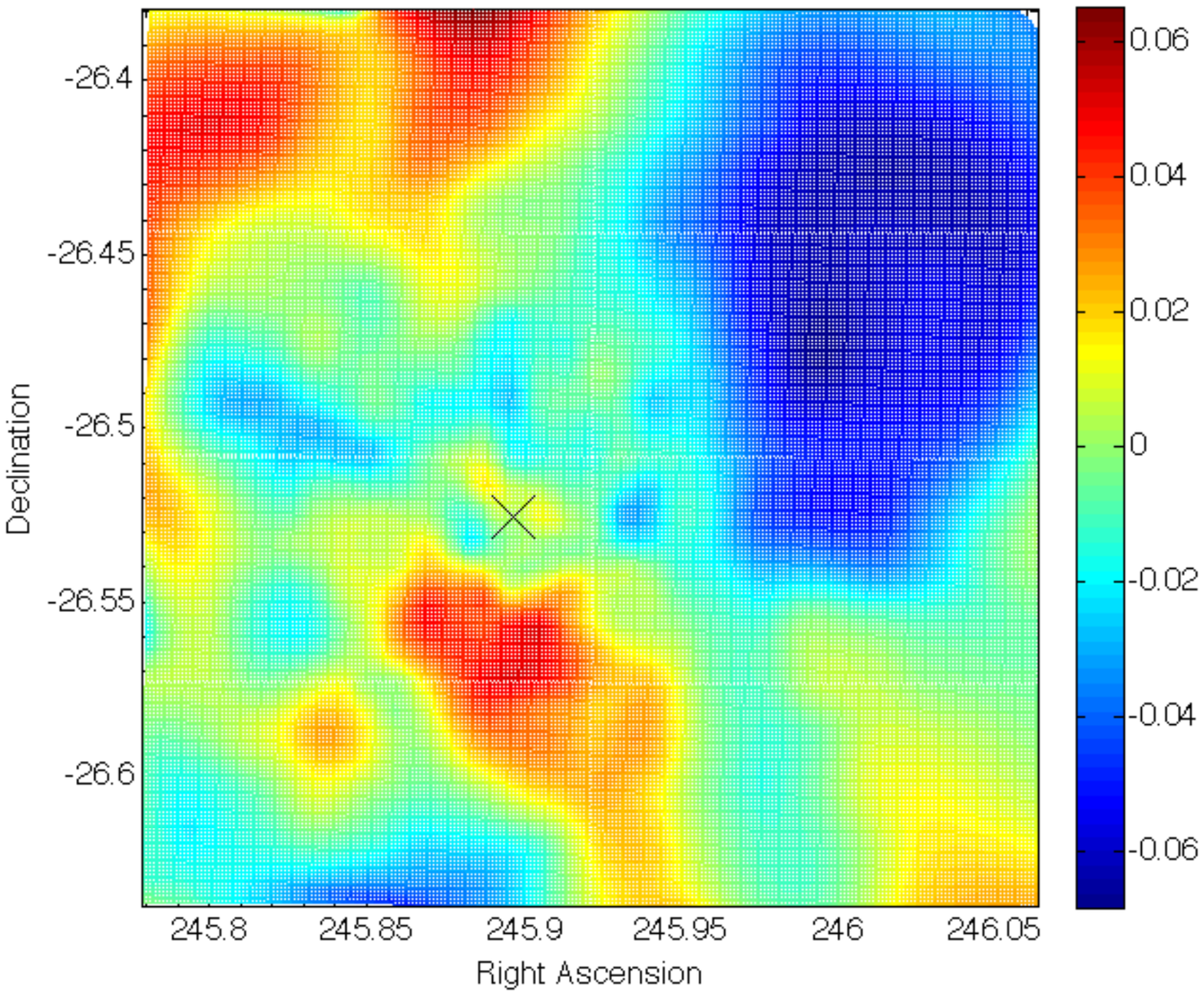} &
\includegraphics[scale=0.28]{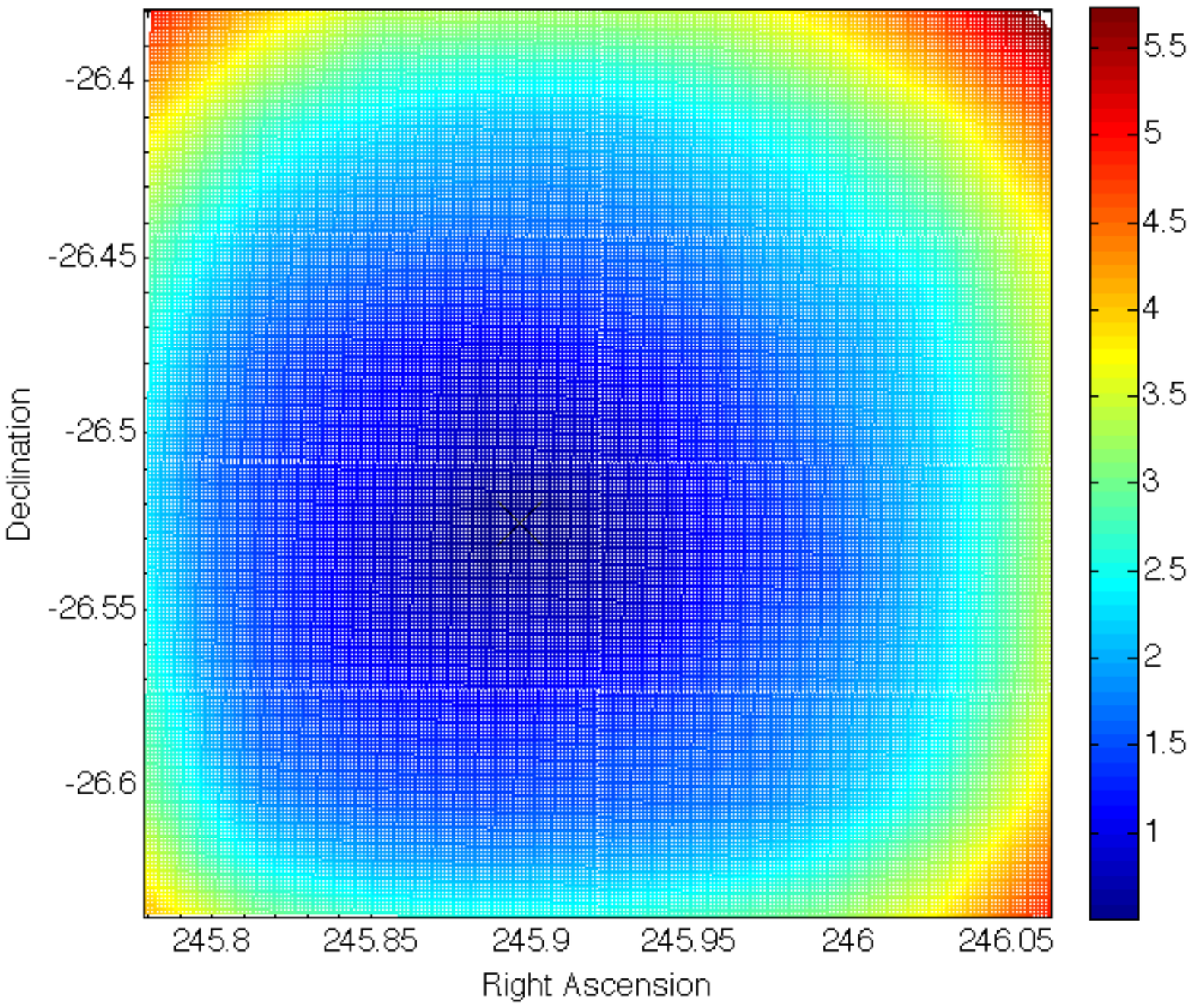} &
\includegraphics[scale=0.28]{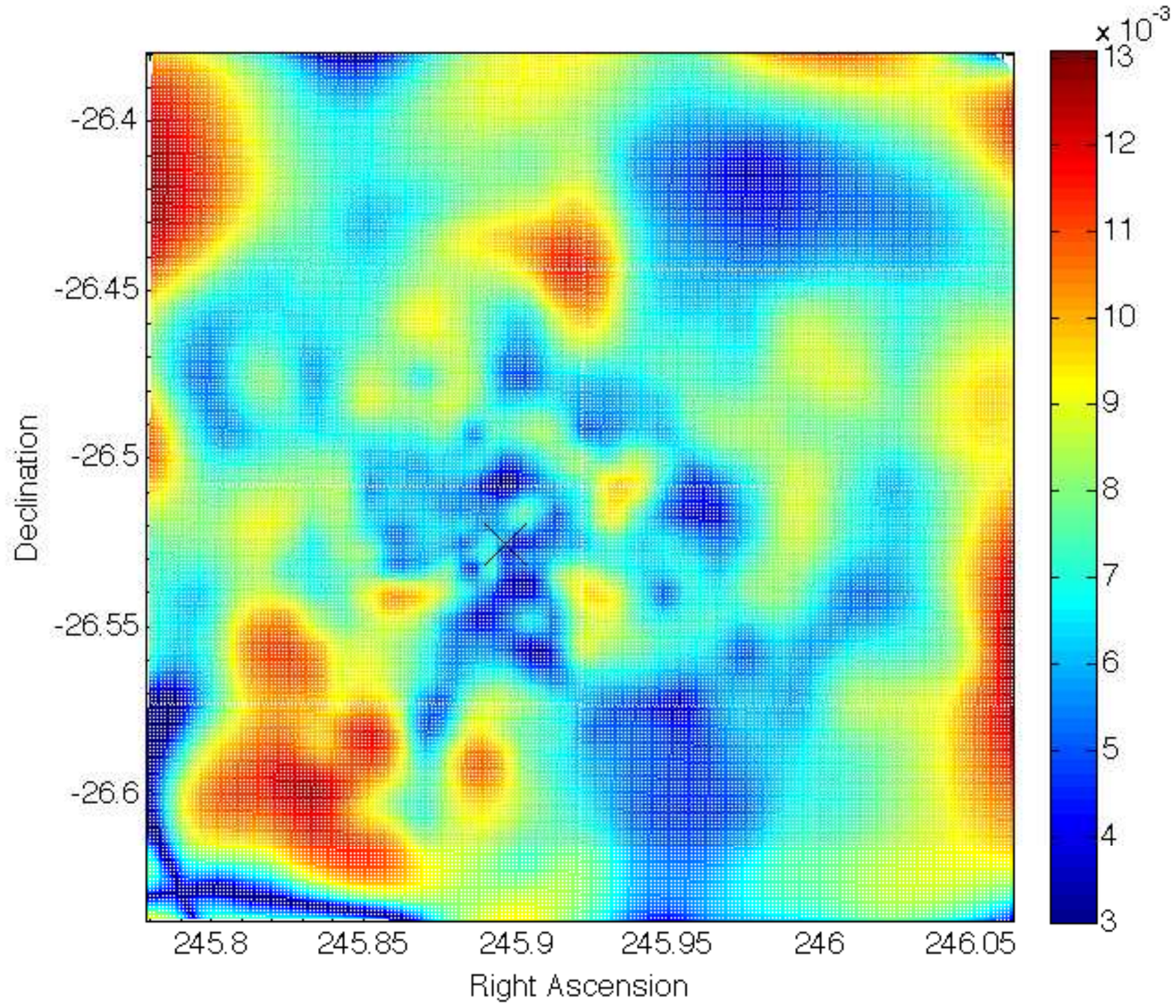} \\
\end{tabular}
\caption{\footnotesize On top, CMDs of the cluster NGC 6121 - M 4,
  before and after being differentially dereddened. Only our Magellan
  photometry was used to build the $B-V$ vs. $V$ CMD. ACS photometry
  (from project 10775) and Magellan photometry were used to build the
  $V-I$ vs. $V$ CMD. Color bars show the range in the densities of
  stars in the CMD ($\times 10^4$ stars per square magnitude). On the
  bottom, extinction map (left) for the cluster field, along with its
  resolution (middle) and its precision (right), as provided by our
  technique. The x marks the position of the cluster center. The color
  code gives, respectively, the color excesses $E(B-V)$ for the
  extinction map, the bandwidths used in the resolution map, and the
  standard deviation $\sigma$ of the color excesses in the precision
  maps.}
\label{figngc6121}
\end{figure}

\begin{figure}[htbp]
%\epsscale{0.77}
\plotone{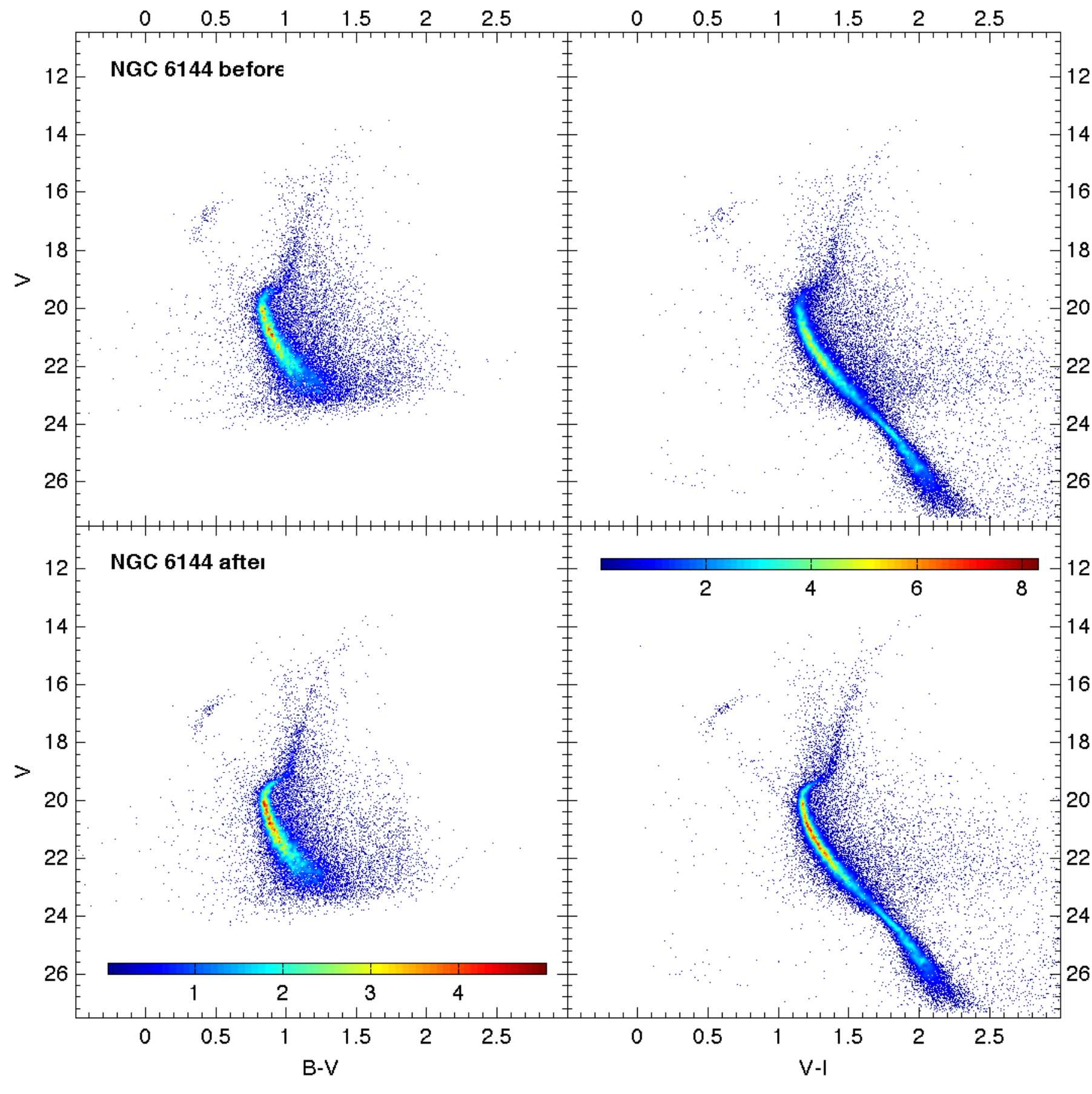}
\begin{tabular}{ccc}
\includegraphics[scale=0.28]{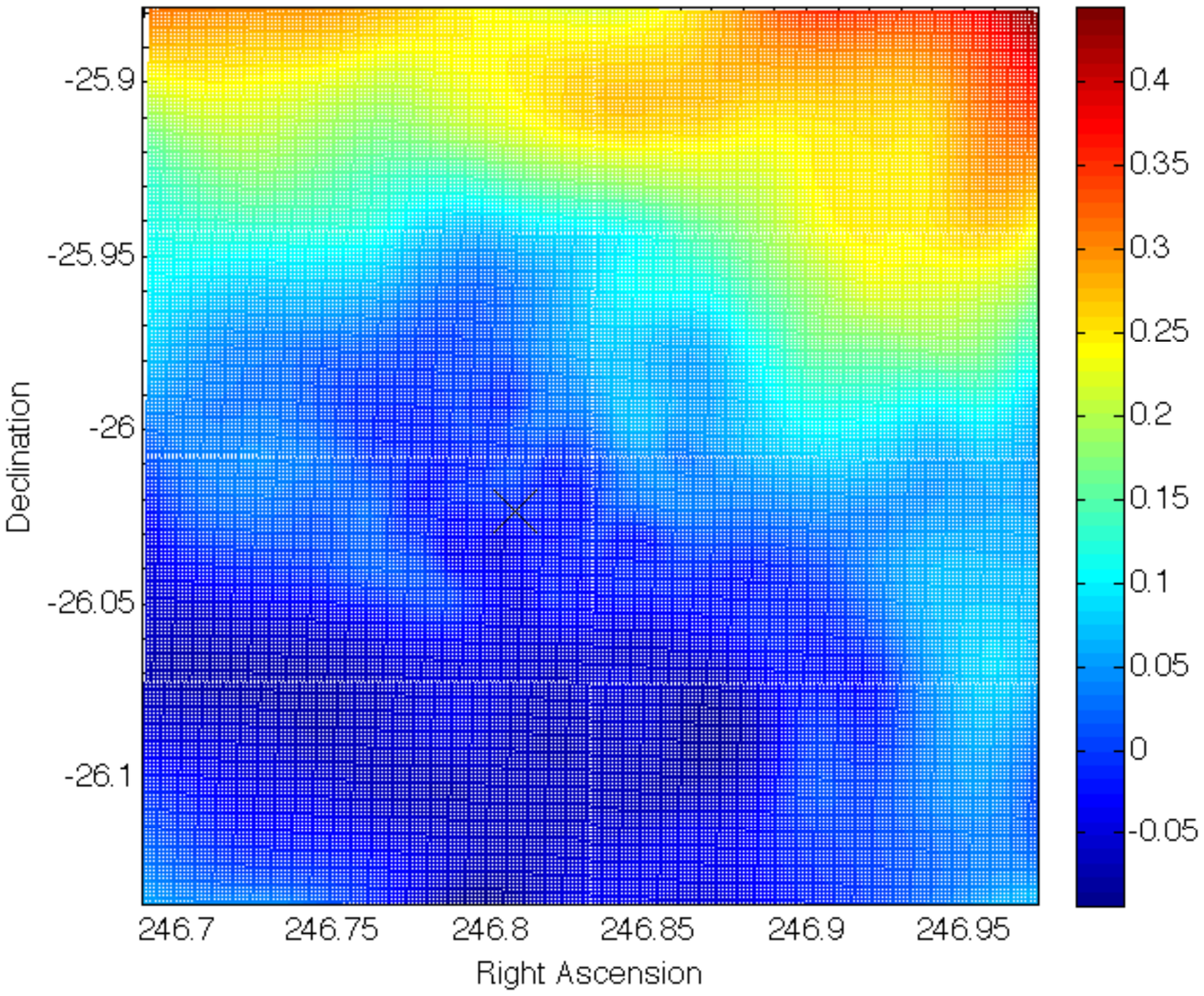}  &
\includegraphics[scale=0.28]{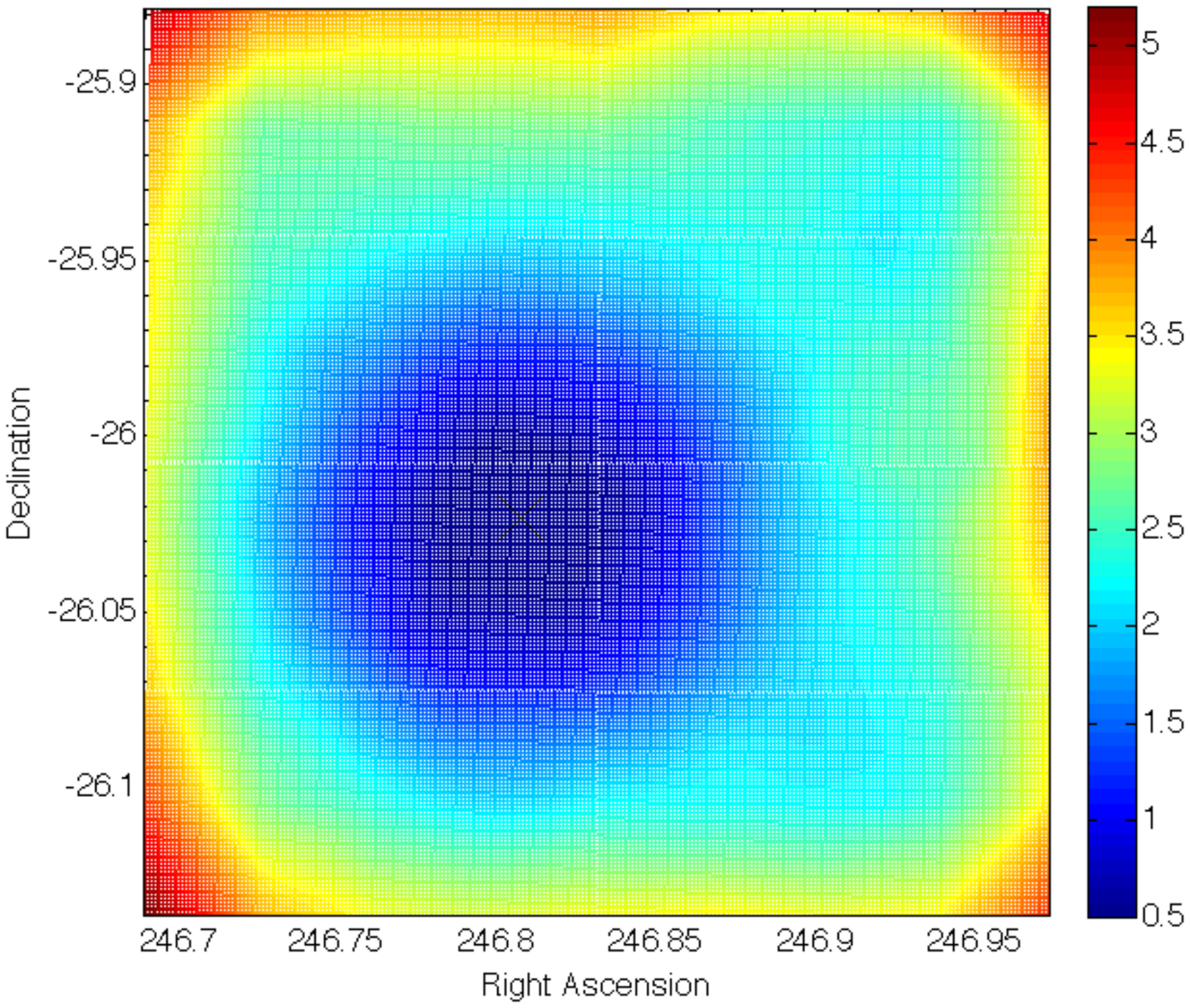}  &
\includegraphics[scale=0.28]{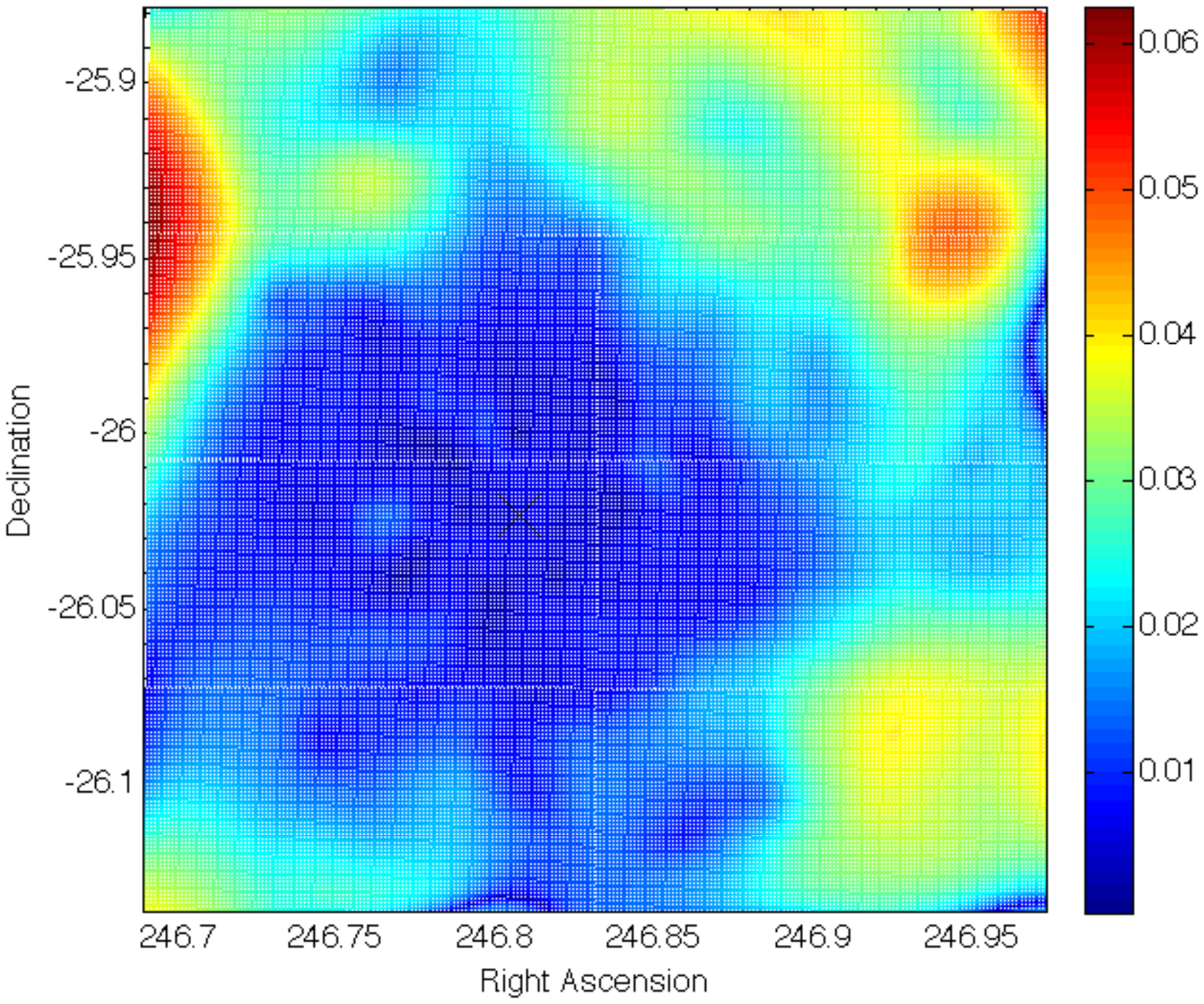} \\
\end{tabular}
\caption{\footnotesize As in Figure \ref{figngc6121}, but for the
  cluster NGC 6144. Only our Magellan photometry was used to build the
  $B-V$ vs. $V$ CMD. ACS photometry (from project 10775) and Magellan
  photometry were used to build the $V-I$ vs. $V$ CMD. Notice that the
  $B-V$ vs. $V$ CMD could not be correctly calibrated in color using
  the method described in the text because of the lack of calibrating
  data in the $B$ filter.}
\label{figngc6144}
\end{figure}

\begin{figure}[htbp]
%\epsscale{0.77}
\plotone{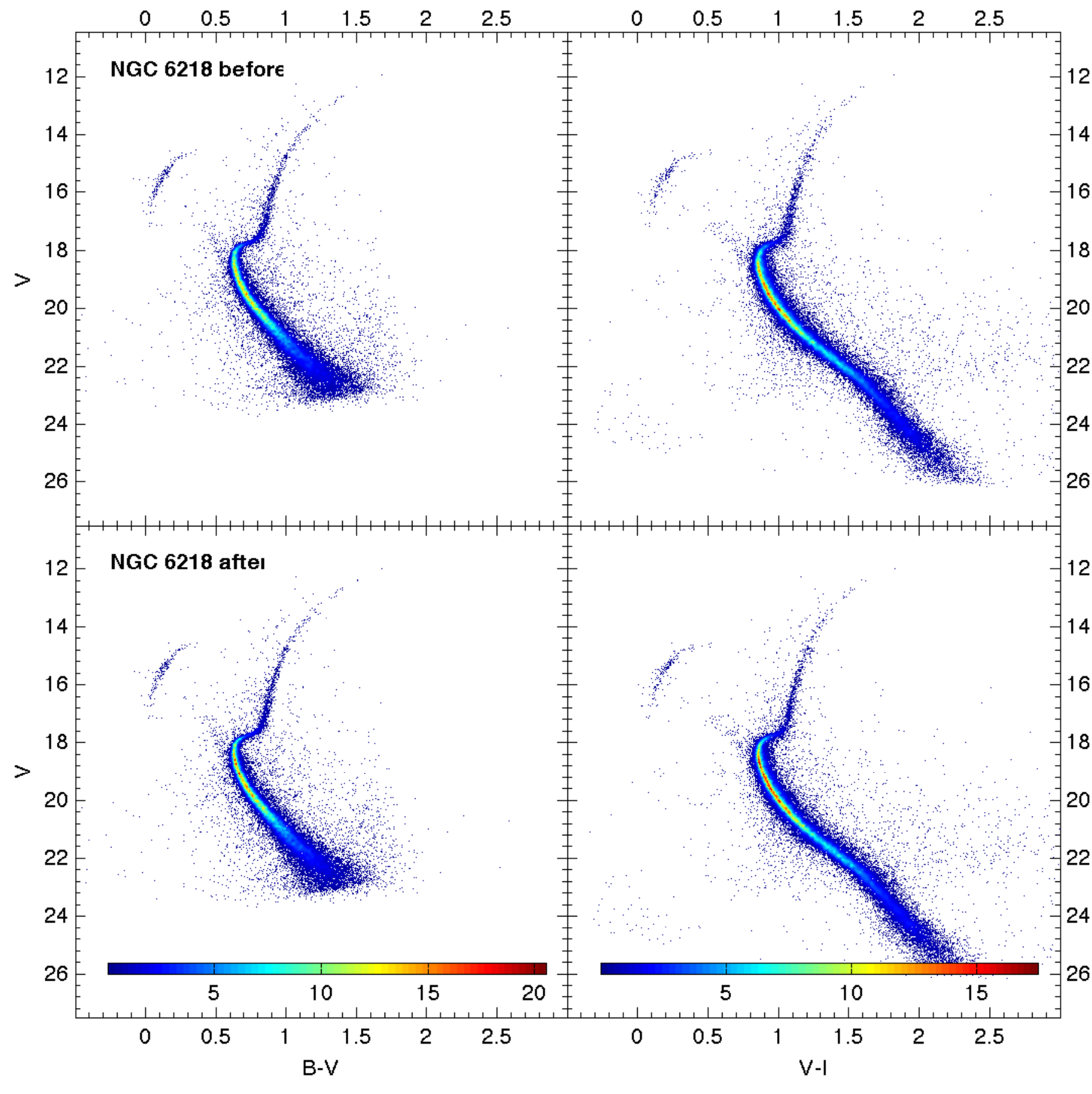}
\begin{tabular}{ccc}
\includegraphics[scale=0.28]{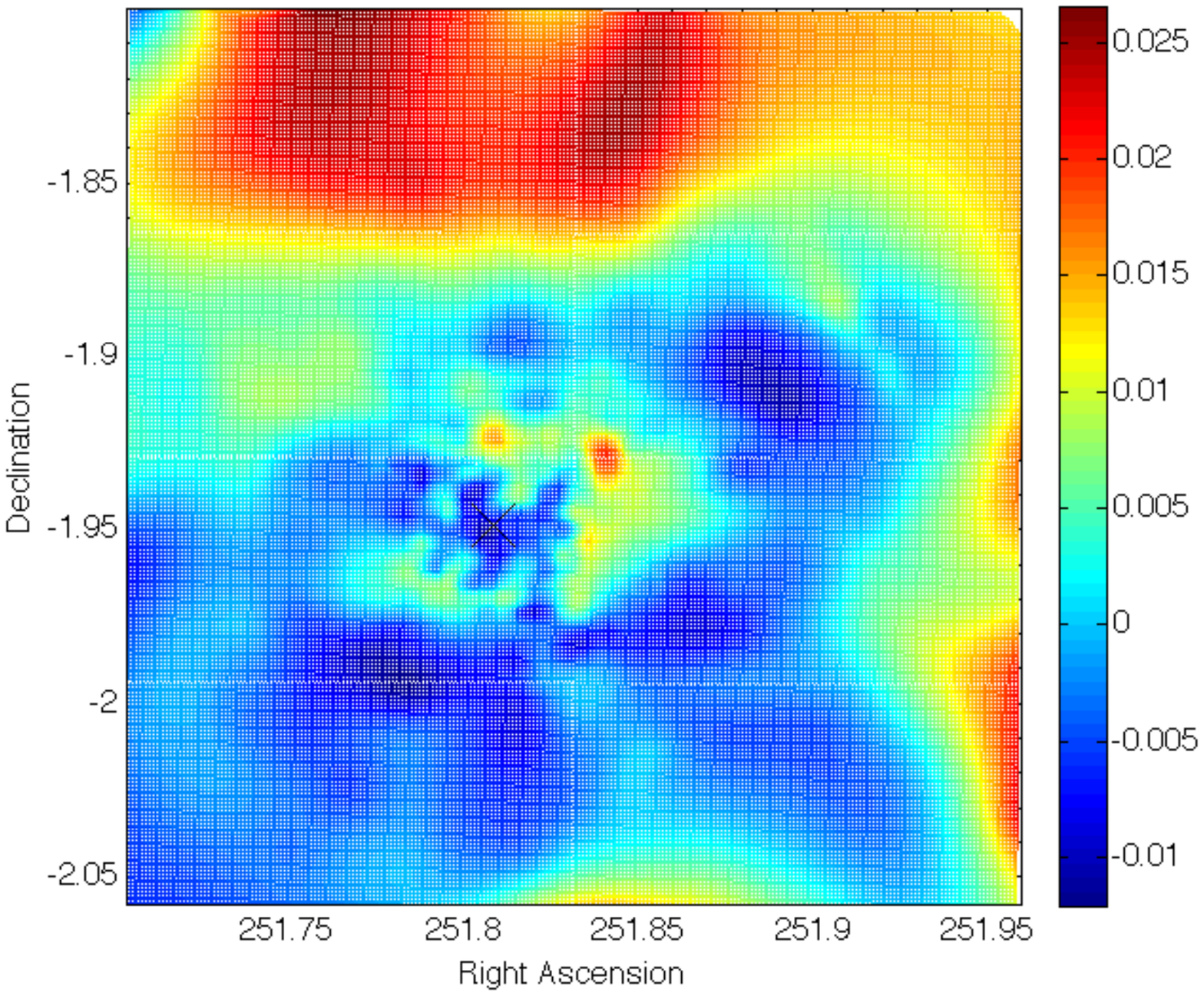}  &
\includegraphics[scale=0.28]{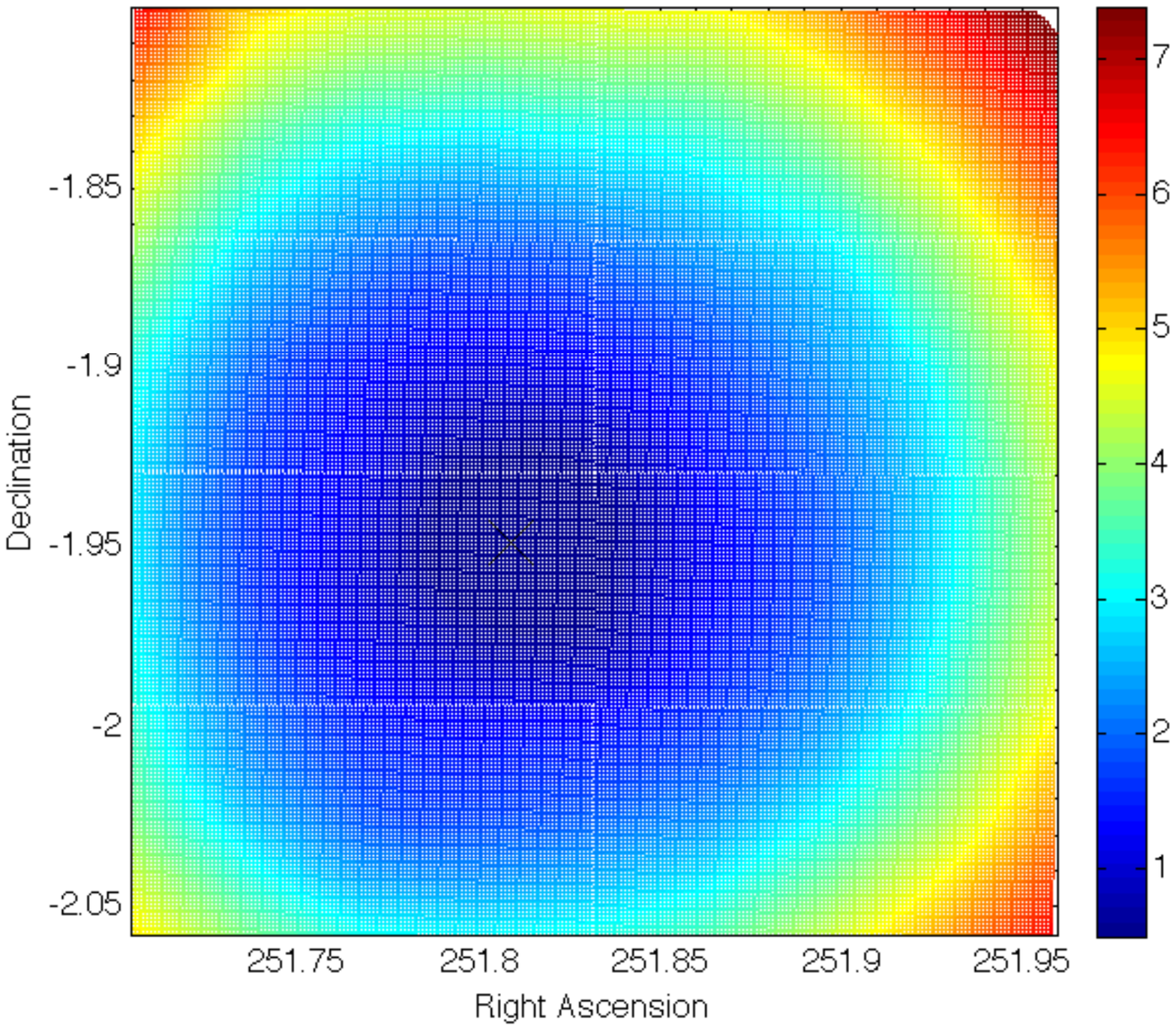}  &
\includegraphics[scale=0.28]{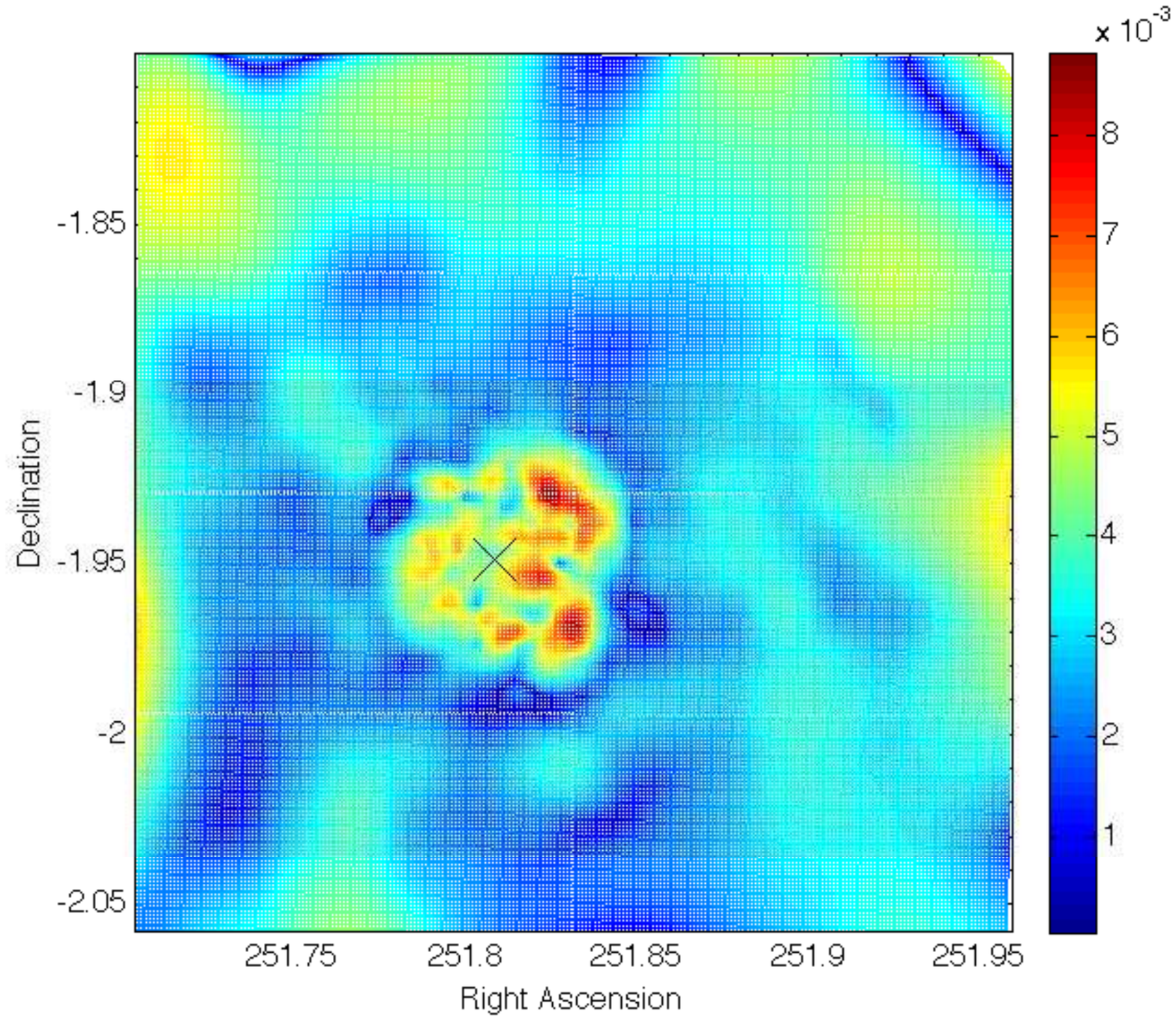} \\
\end{tabular}
\caption{\footnotesize As in Figure \ref{figngc6121}, but for the
  cluster NGC 6218 - M 12. ACS photometry (from our
  project 10573) and Magellan photometry were used to build the $B-V$
  vs. $V$ CMD. ACS photometry (from project 10775) and Magellan
  photometry were used to build the $V-I$ vs. $V$ CMD.}
\label{figngc6218}
\end{figure}

\begin{figure}[htbp]
%\epsscale{0.77}
\plotone{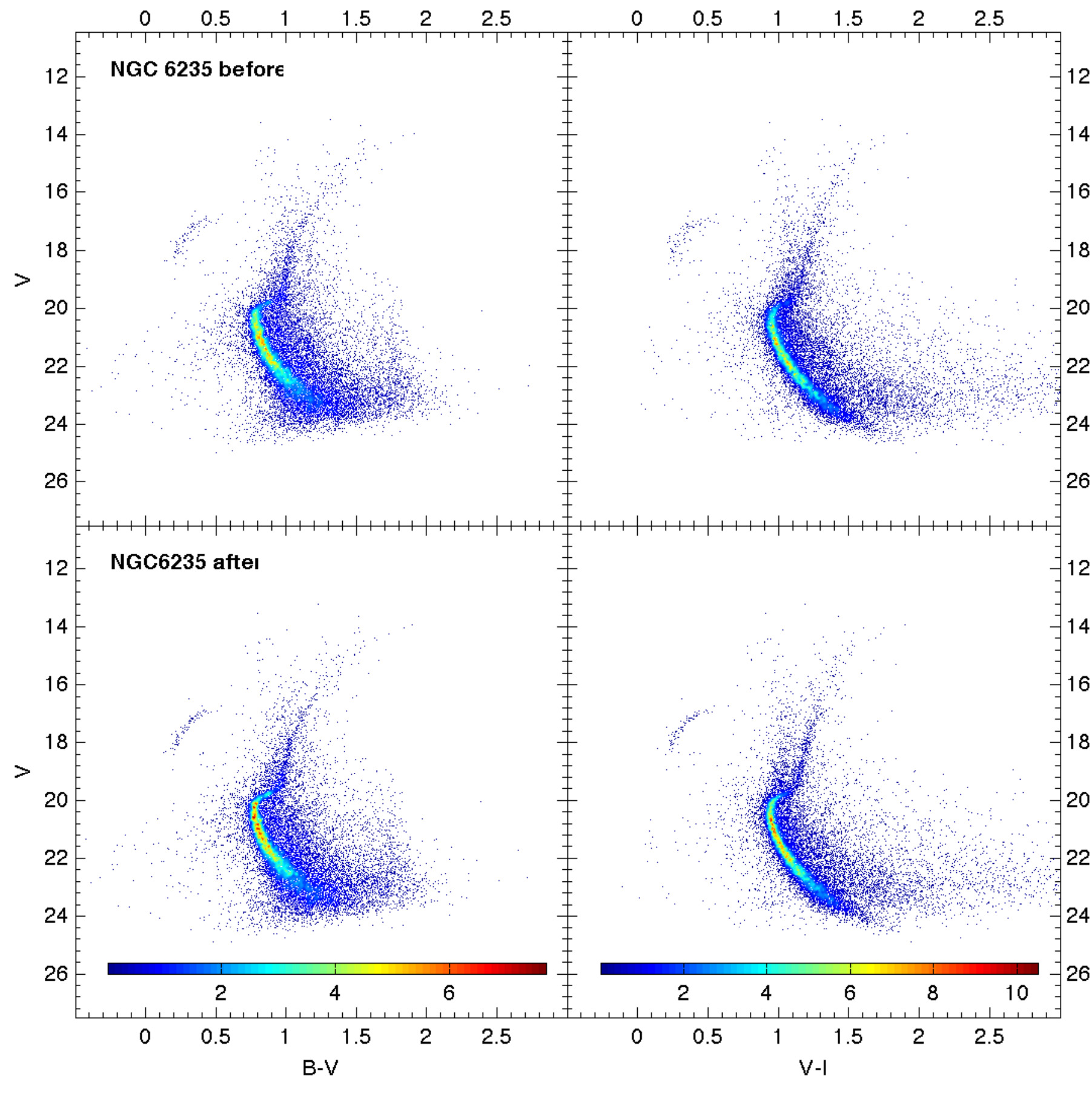}
\begin{tabular}{ccc}
\includegraphics[scale=0.28]{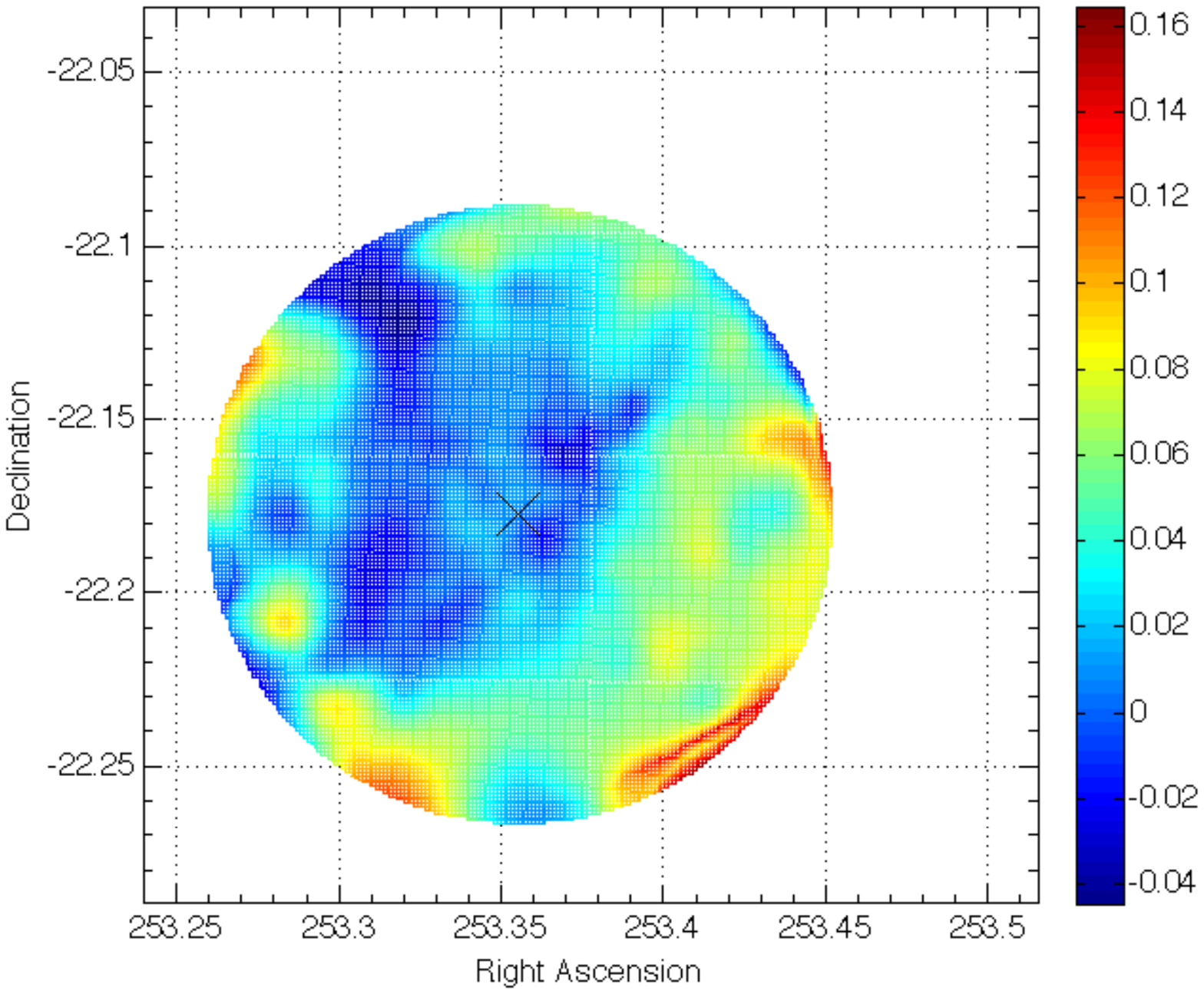}  &
\includegraphics[scale=0.28]{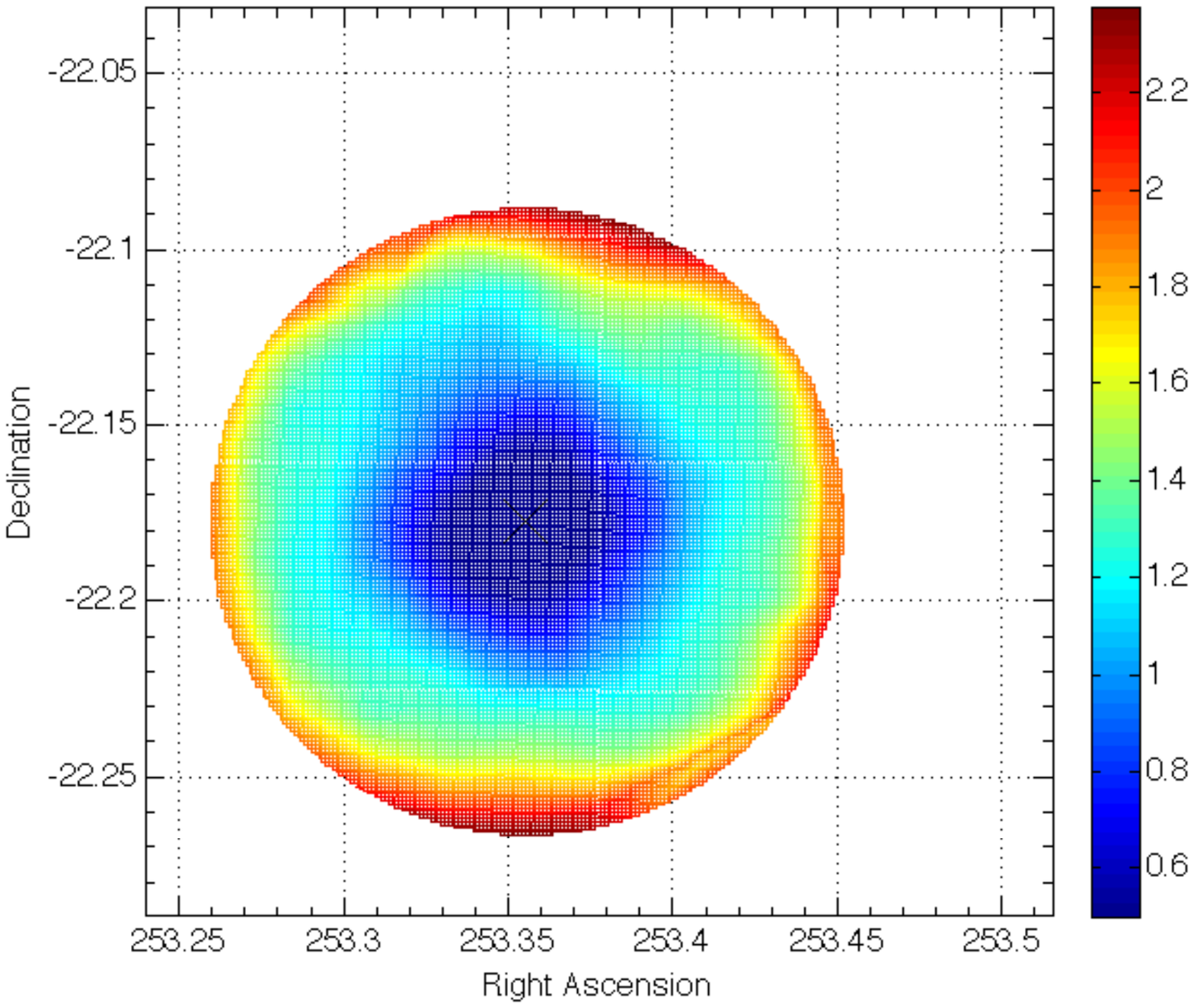}  &
\includegraphics[scale=0.28]{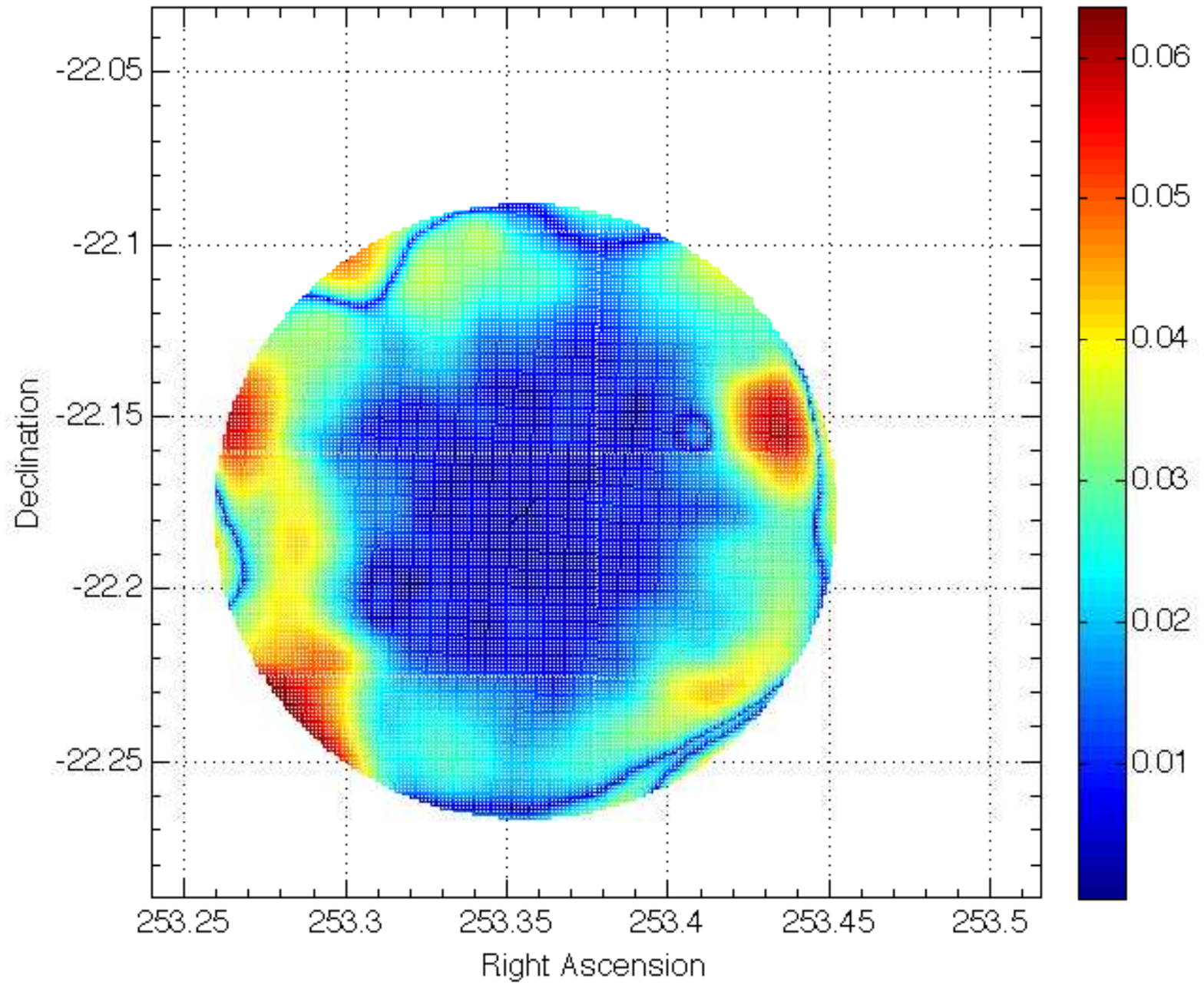} \\
\end{tabular}
\caption{\footnotesize As in Figure \ref{figngc6121}, but for the
  cluster NGC 6235. Only our Magellan photometry was used to build the
  CMDs in both colors. Notice that the $V-I$ vs. $V$ CMD could not be
  correctly calibrated in color using the method described in the text
  because of the lack of calibrating data in the $I$ filter.}
\label{figngc6235}
\end{figure}

\begin{figure}[htbp]
%\epsscale{0.77}
\plotone{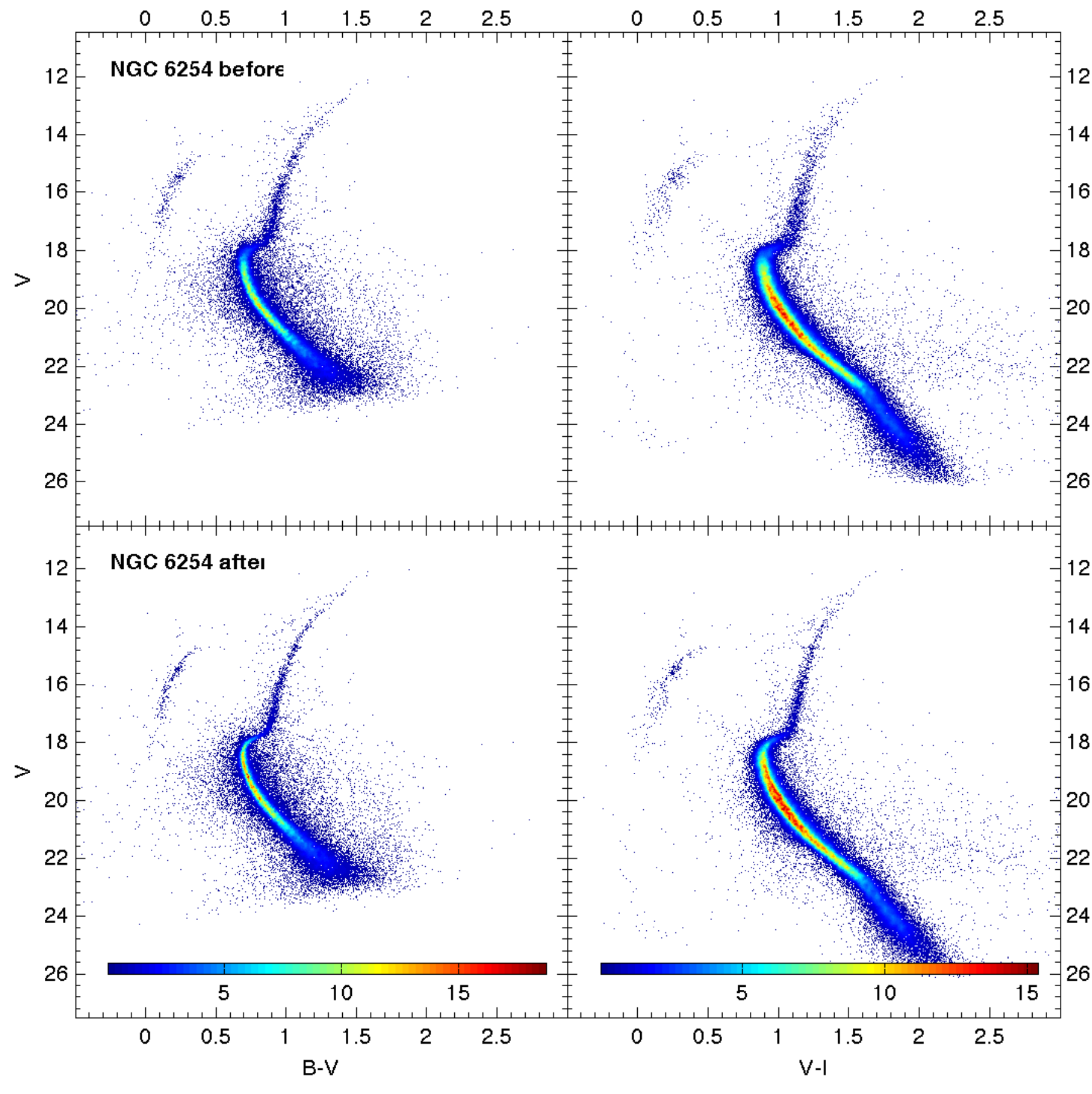}
\begin{tabular}{ccc}
\includegraphics[scale=0.28]{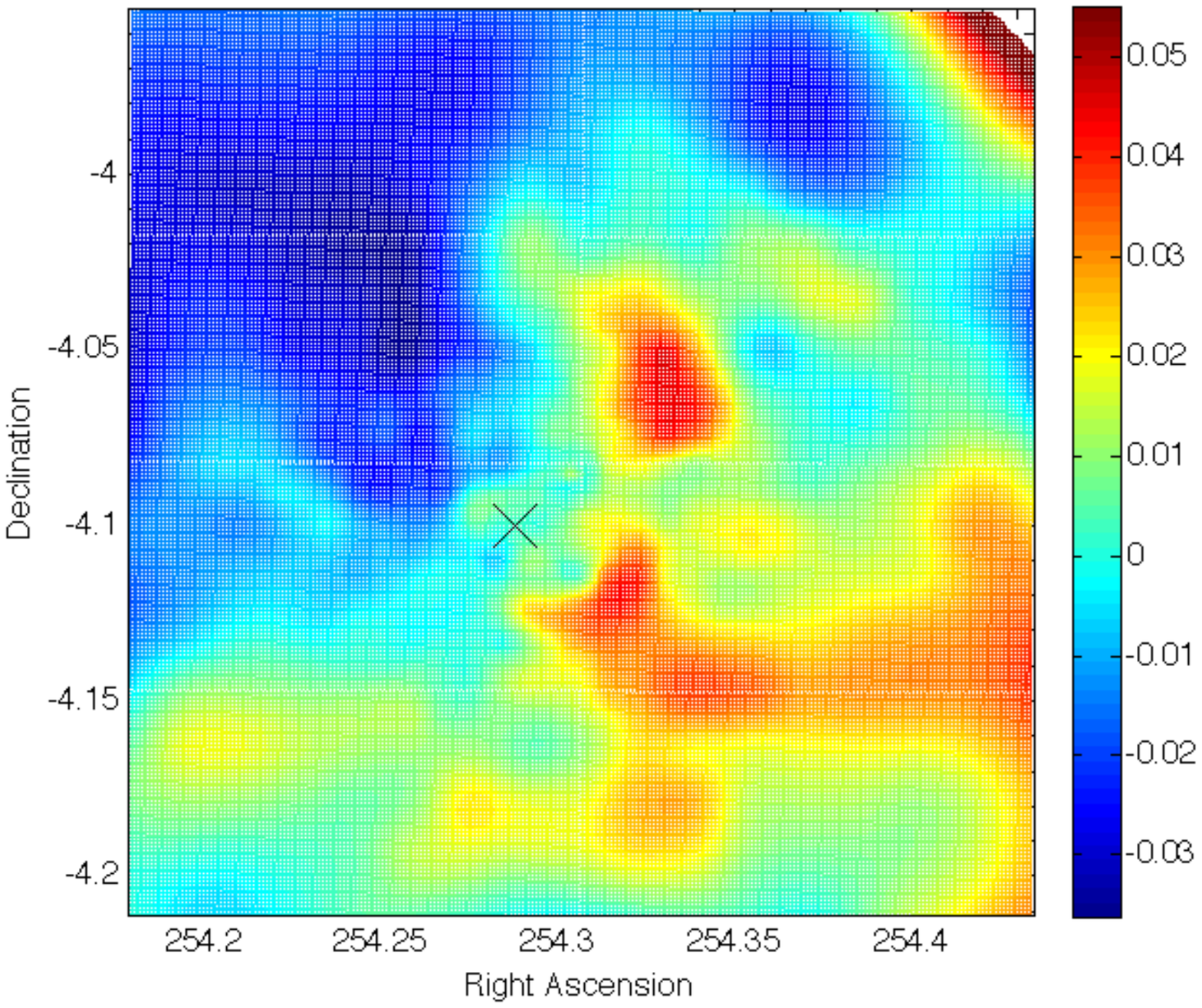}  &
\includegraphics[scale=0.28]{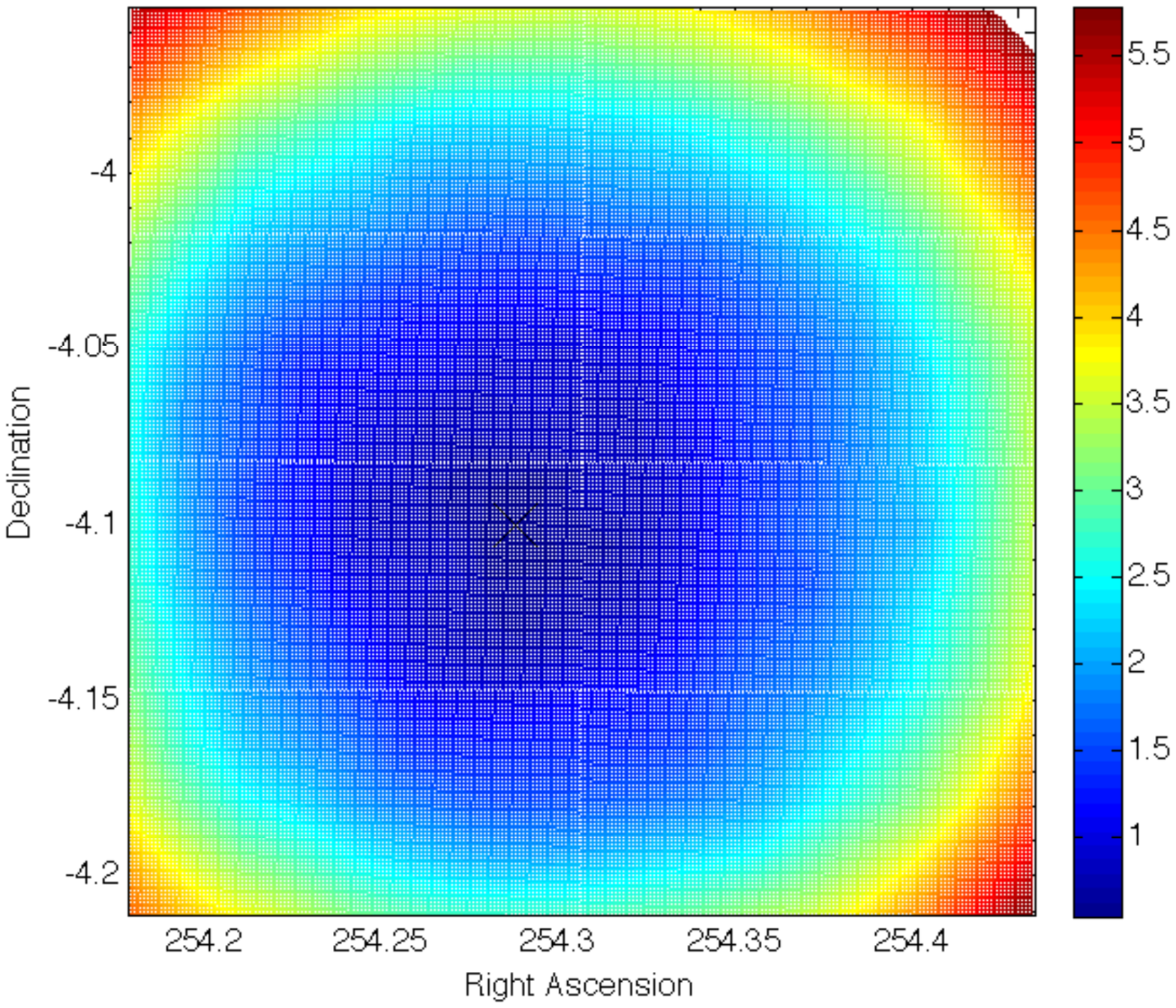}  &
\includegraphics[scale=0.28]{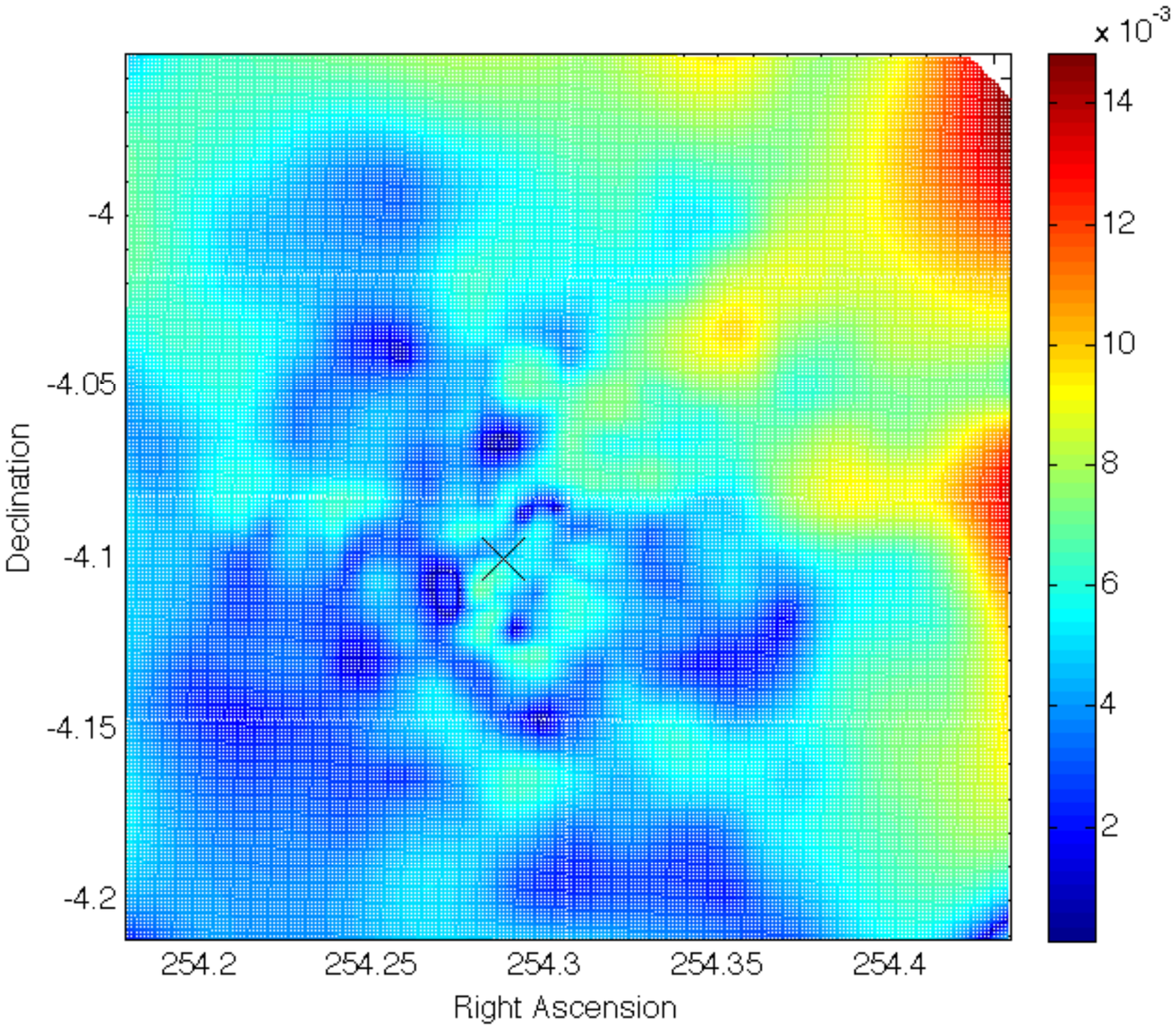} \\
\end{tabular}
\caption{\footnotesize As in Figure \ref{figngc6121}, but for the NGC
  6254 - M 10. Only our Magellan photometry was used to build the
  $B-V$ vs. $V$ CMD. ACS photometry (from project 10775) and Magellan
  photometry were used to build the $V-I$ vs. $V$ CMD.}
\label{figngc6254}
\end{figure}

\begin{figure}[htbp]
%\epsscale{0.77}
\plotone{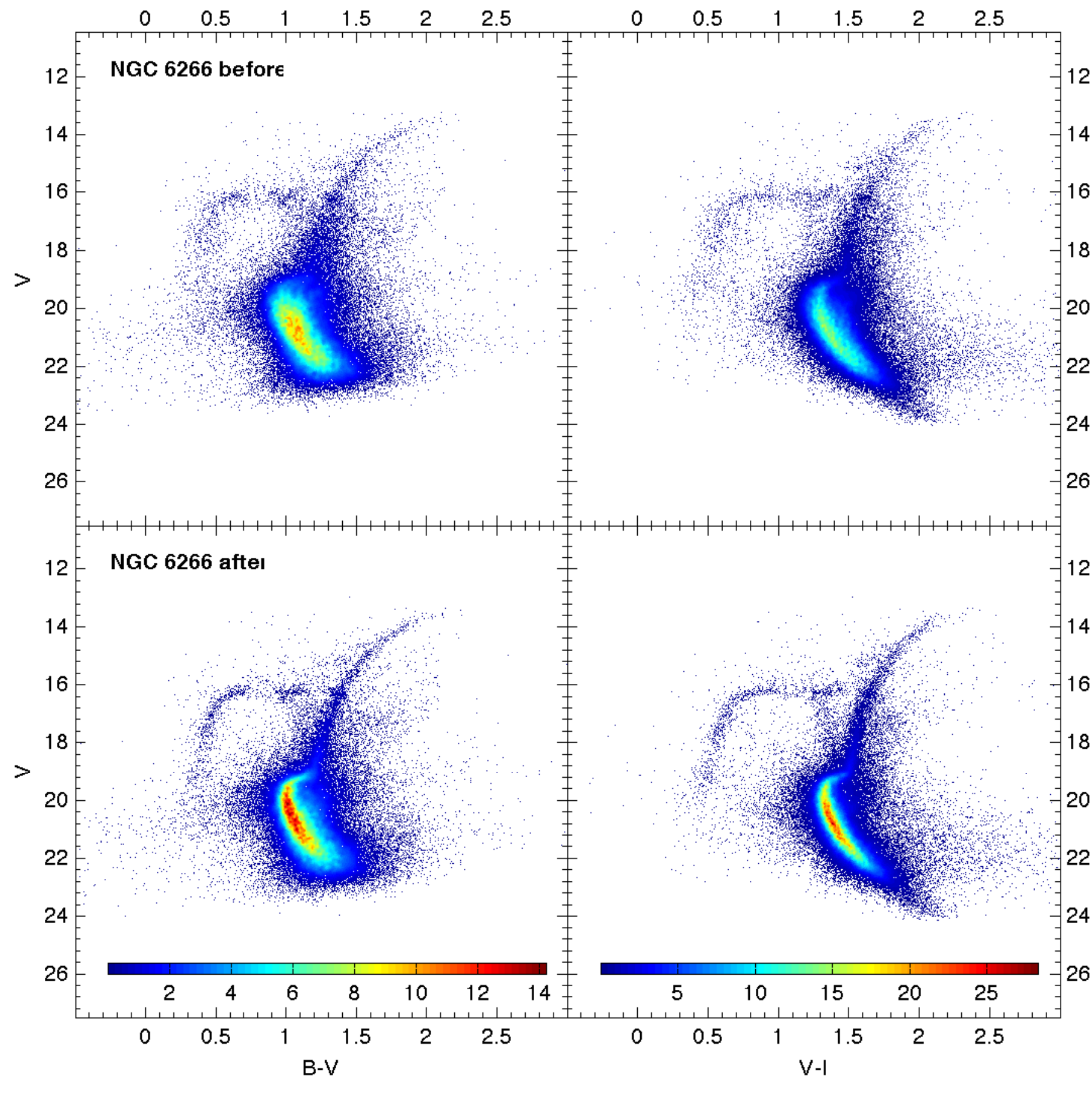}
\begin{tabular}{ccc}
\includegraphics[scale=0.28]{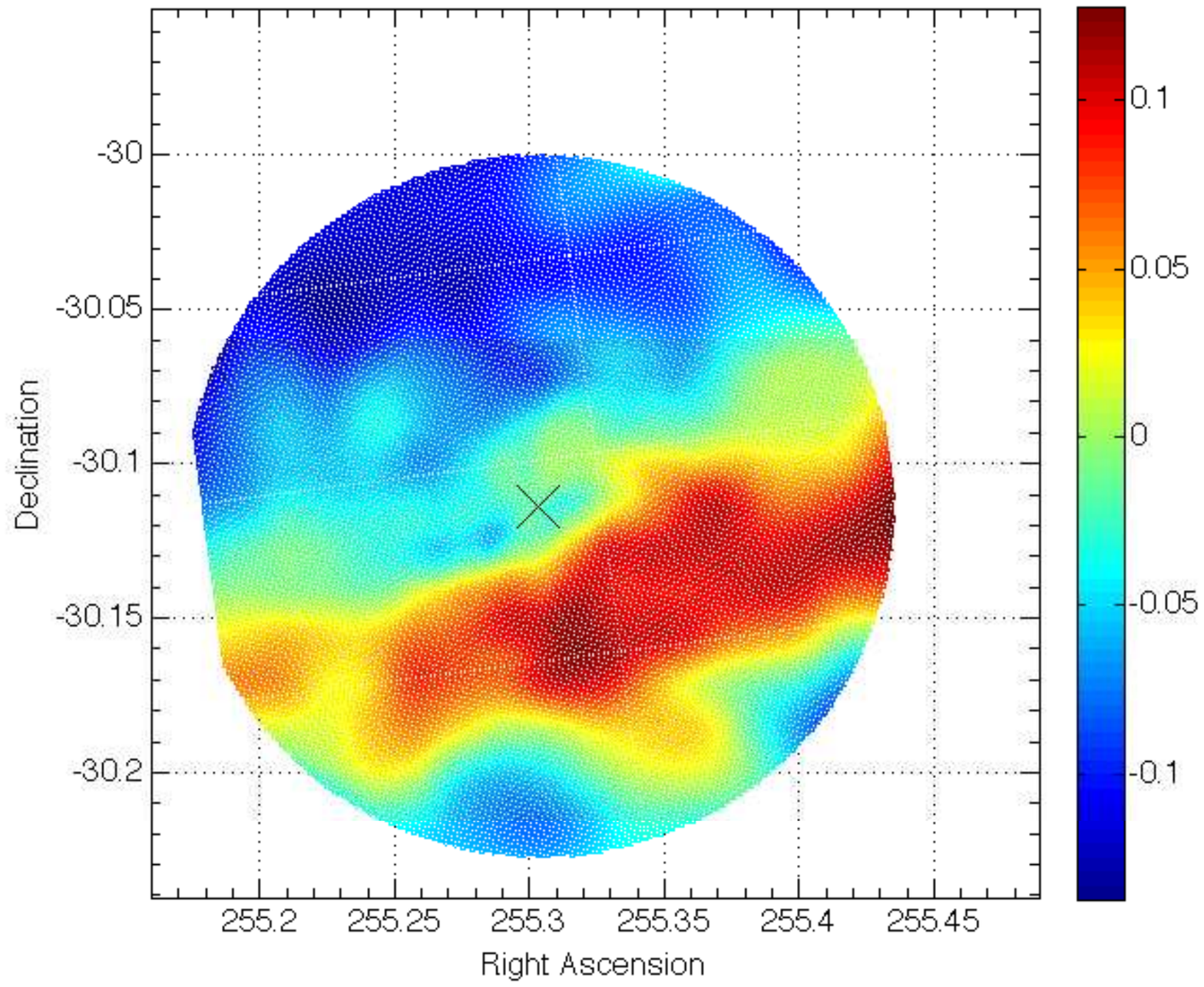}  &
\includegraphics[scale=0.28]{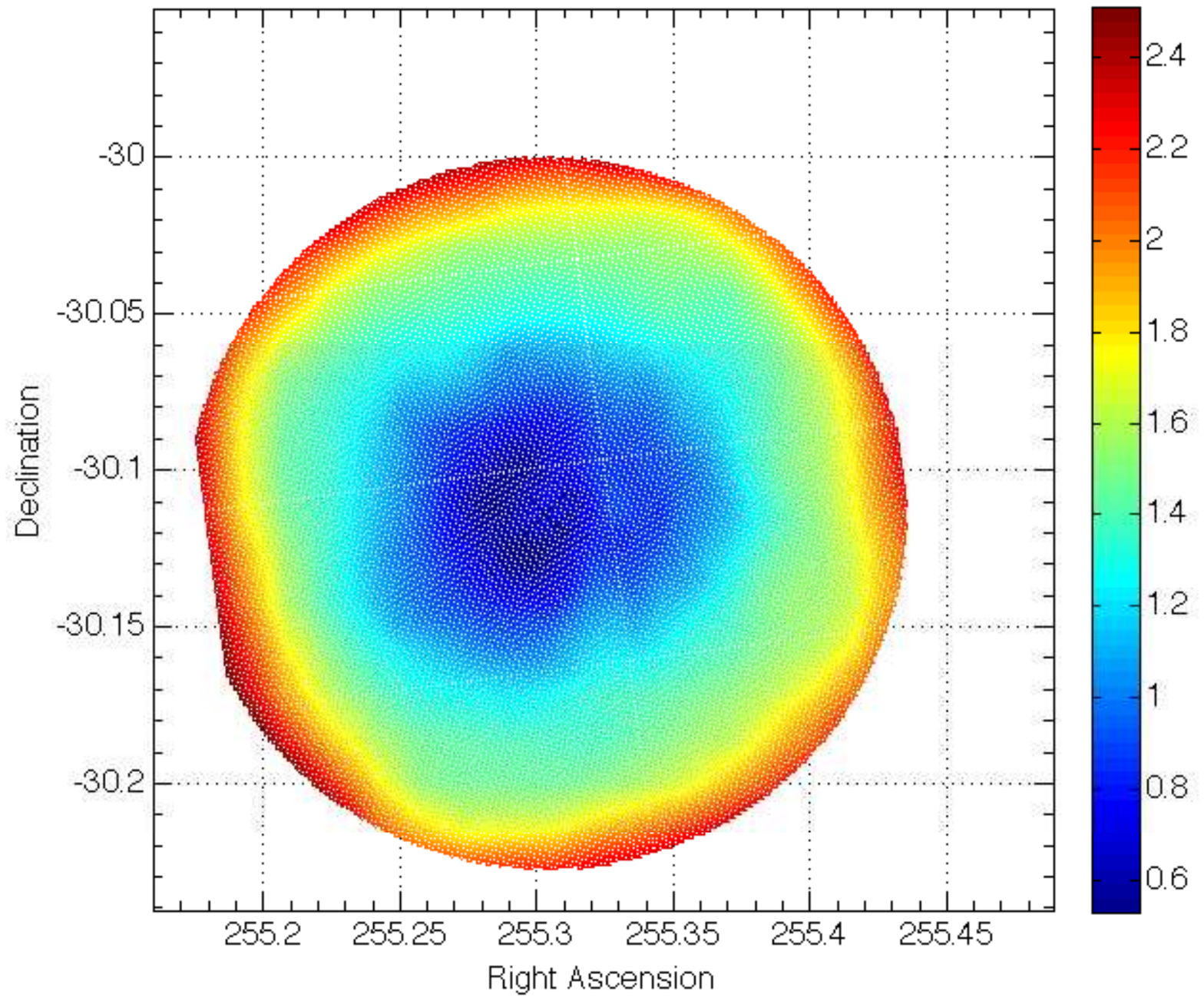}  &
\includegraphics[scale=0.28]{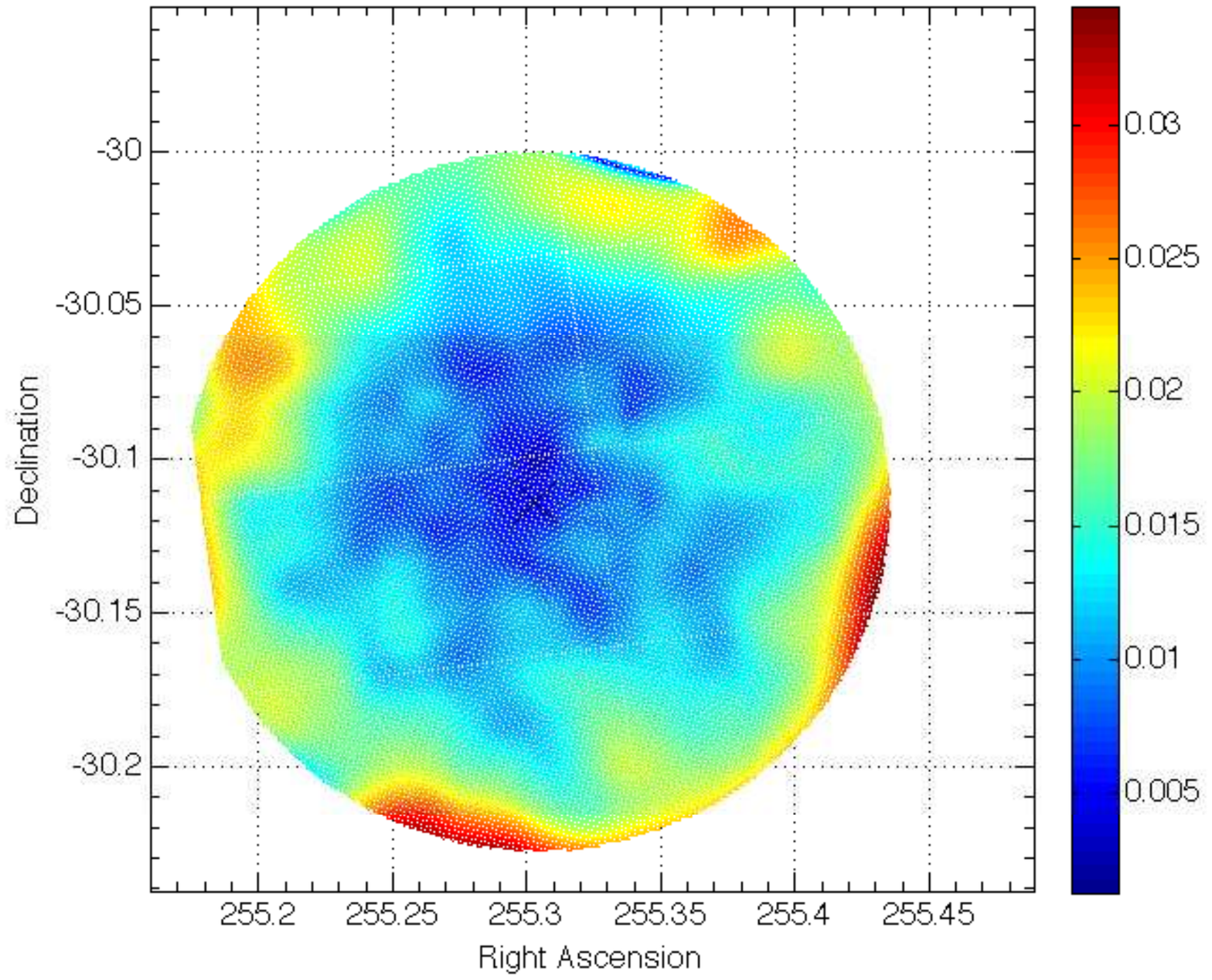} \\
\end{tabular}
\caption{\footnotesize As in Figure \ref{figngc6121}, but for the
  cluster NGC 6266 - M 62. Only our Magellan photometry was used to
  build the $B-V$ vs. $V$ CMD. WFPC2 photometry (from project 8709)
  and Magellan photometry were used to build the $V-I$ vs. $V$ CMD.}
\label{figngc6266}
\end{figure}

\begin{figure}[htbp]
%\epsscale{0.77}
\plotone{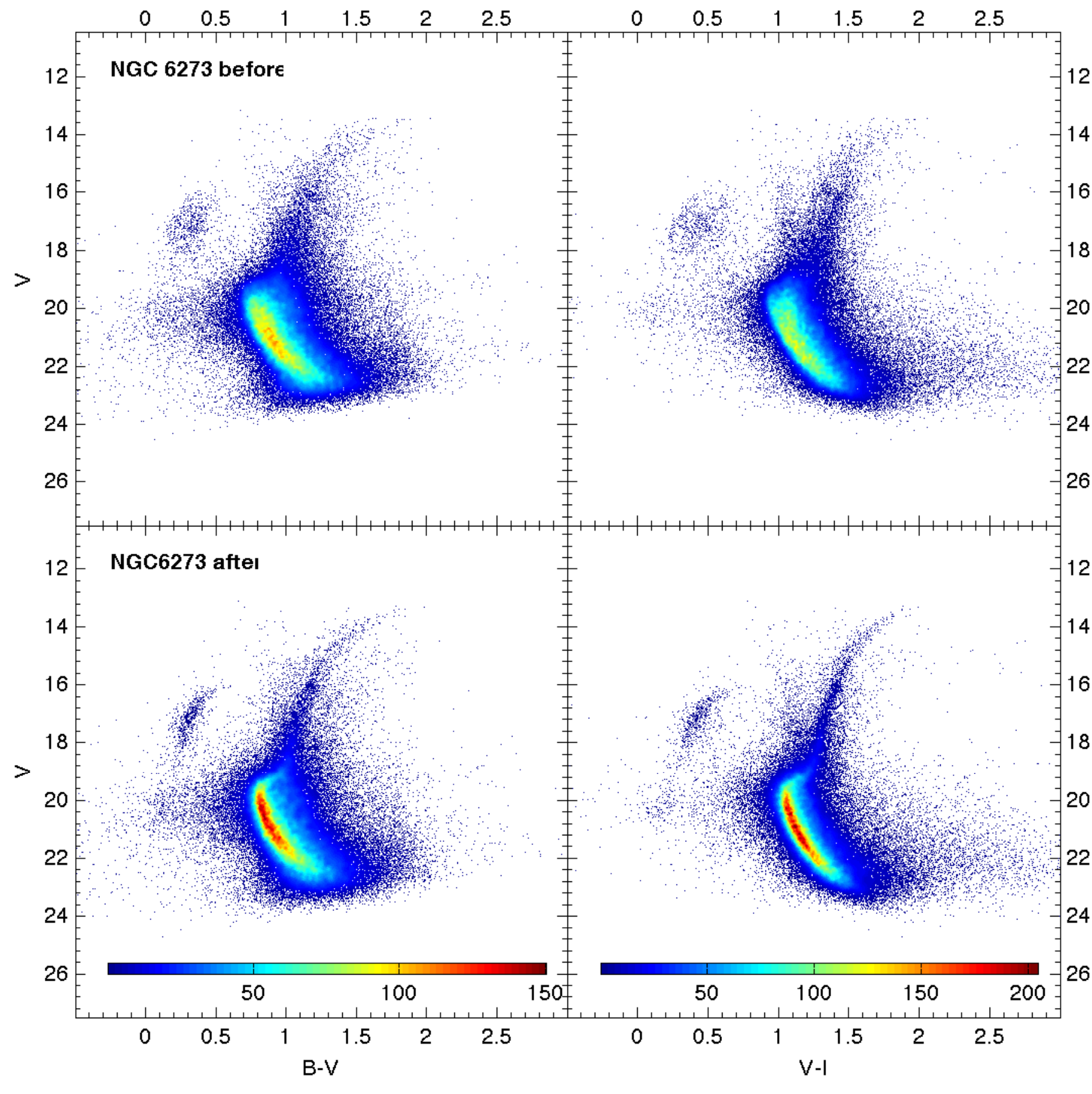}
\begin{tabular}{ccc}
\includegraphics[scale=0.28]{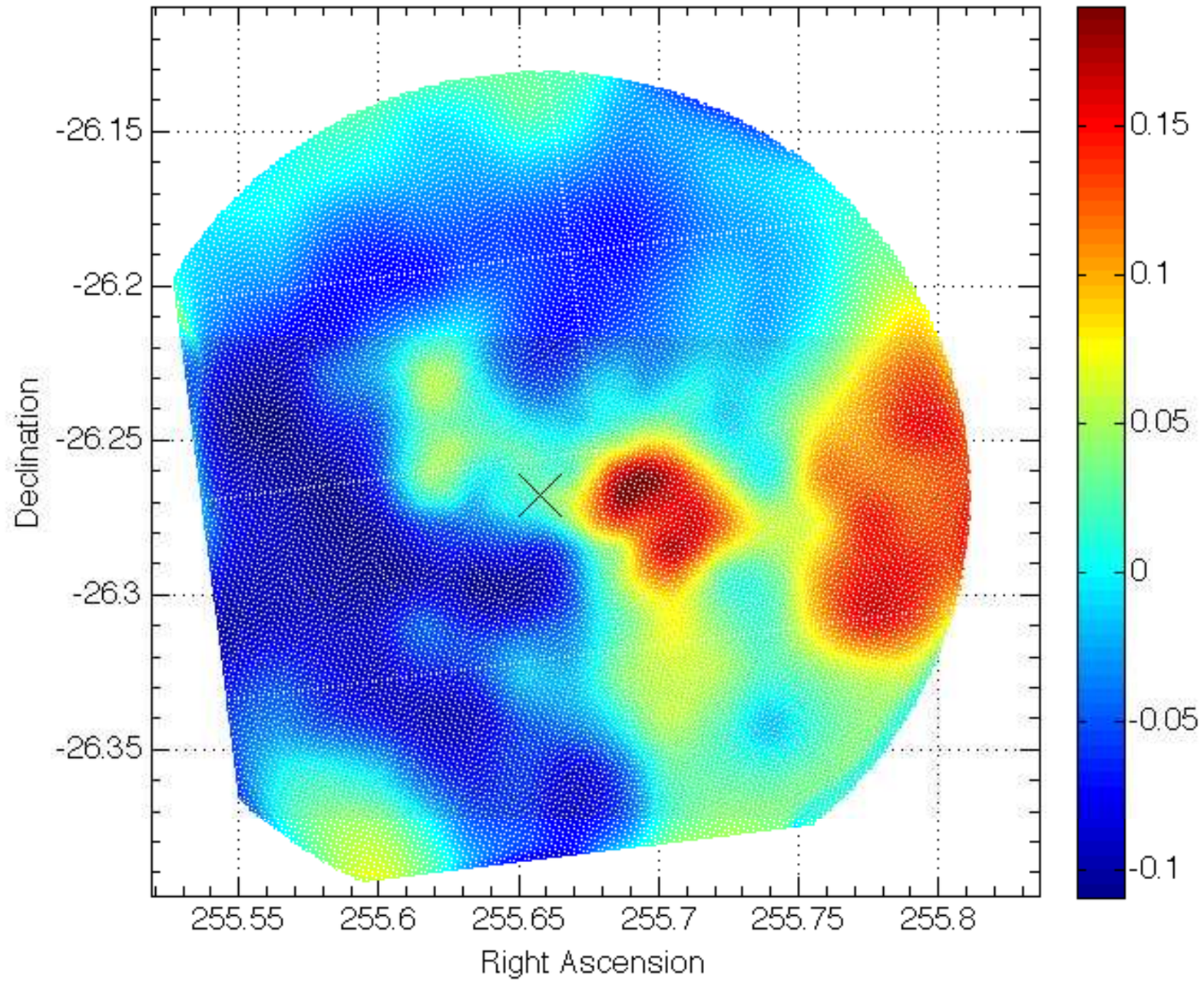}  &
\includegraphics[scale=0.28]{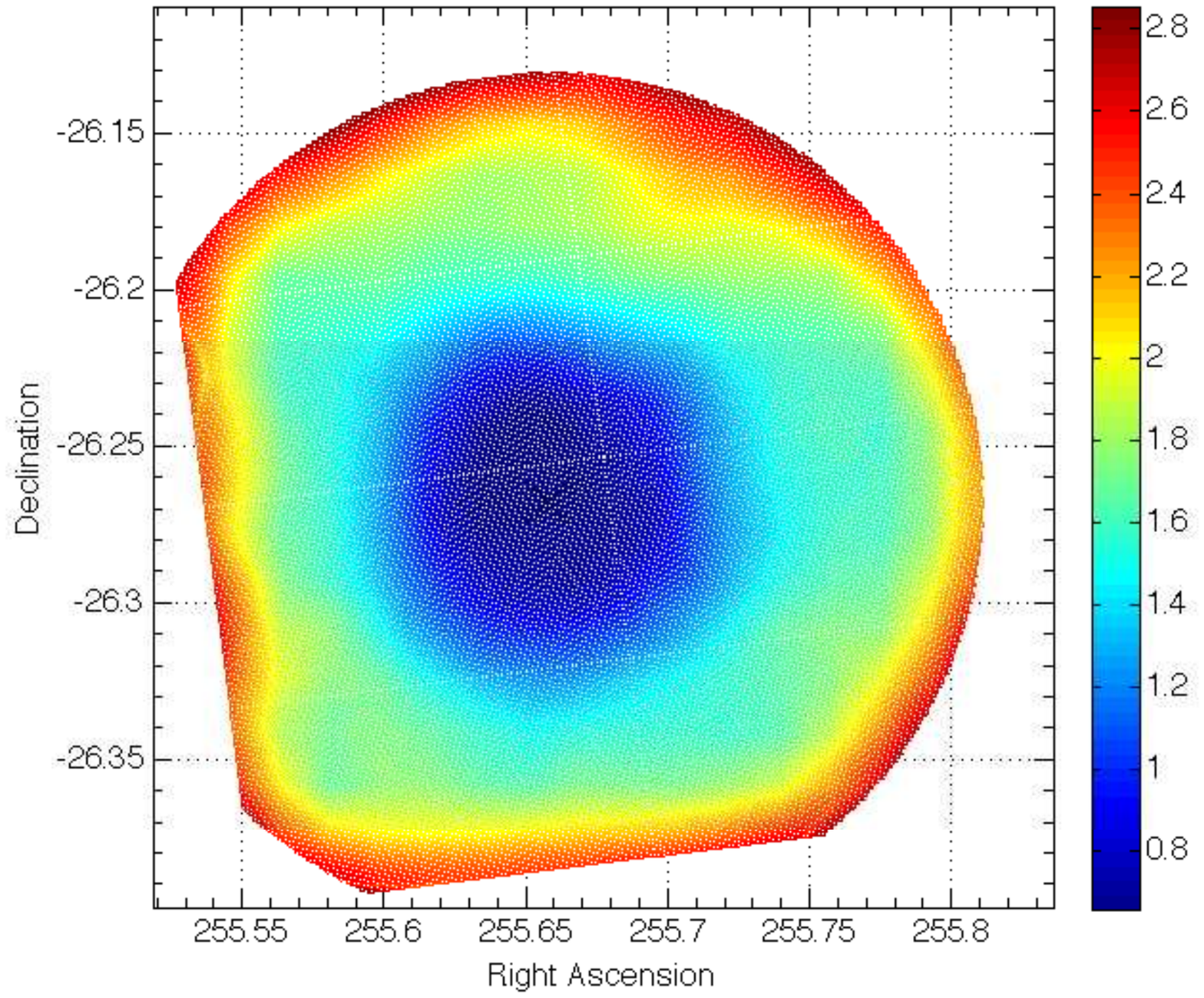}  &
\includegraphics[scale=0.28]{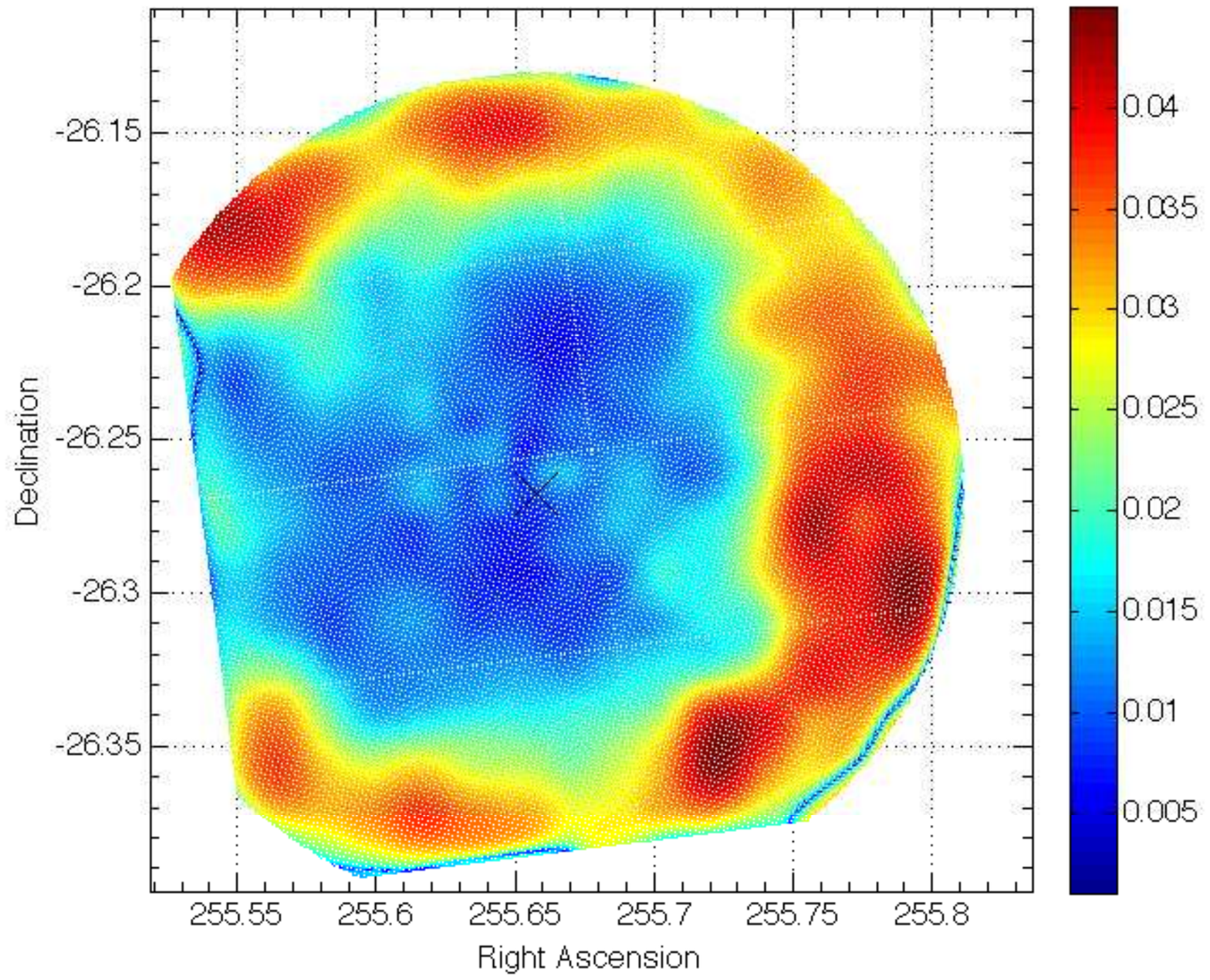} \\
\end{tabular}
\caption{\footnotesize As in Figure \ref{figngc6121}, but for the
  cluster NGC 6273 - M 19. Only our Magellan photometry was used to
  build the CMDs in both colors.}
\label{figngc6273}
\end{figure}

\begin{figure}[htbp]
%\epsscale{0.77}
\plotone{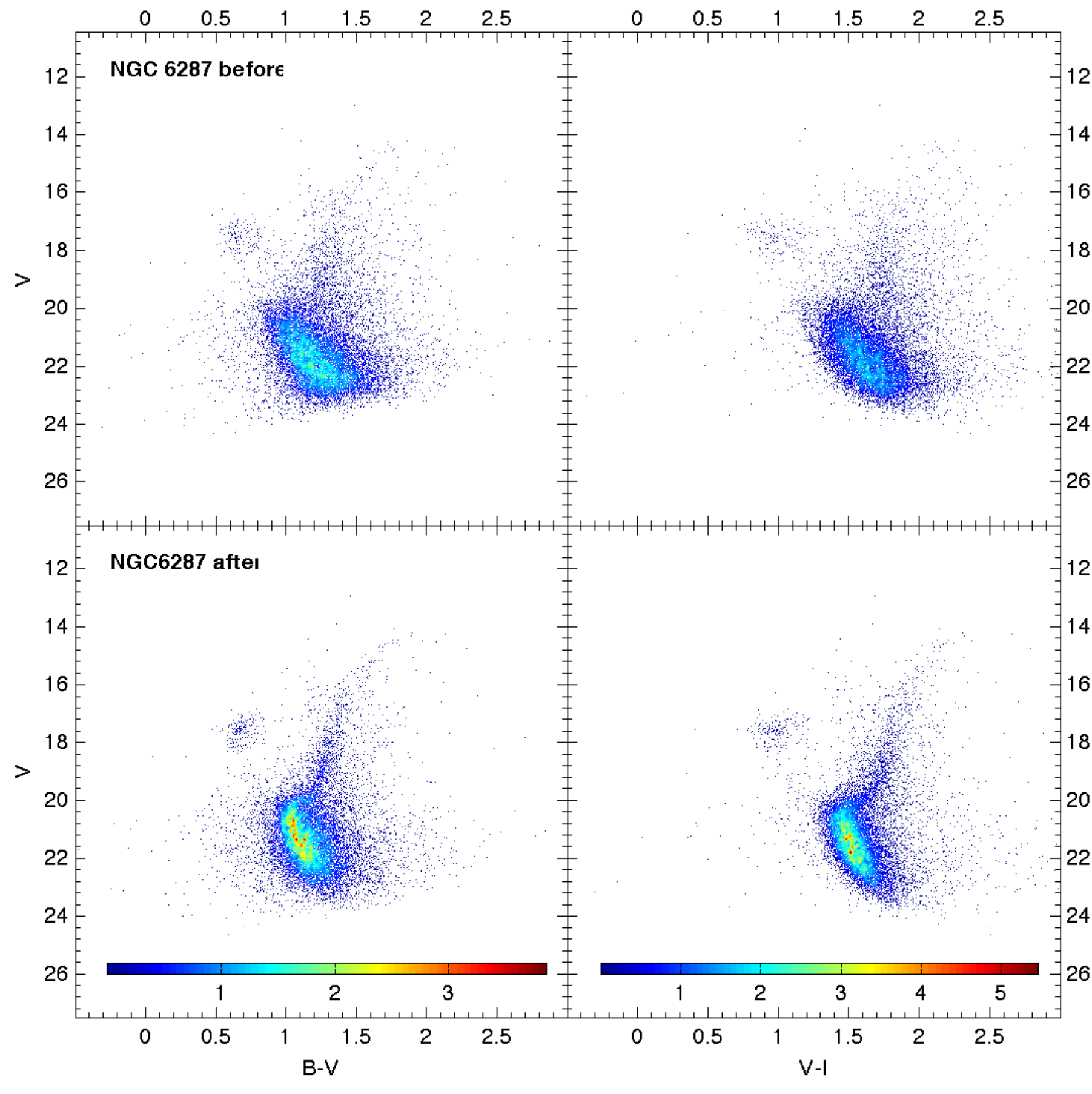}
\begin{tabular}{ccc}
\includegraphics[scale=0.28]{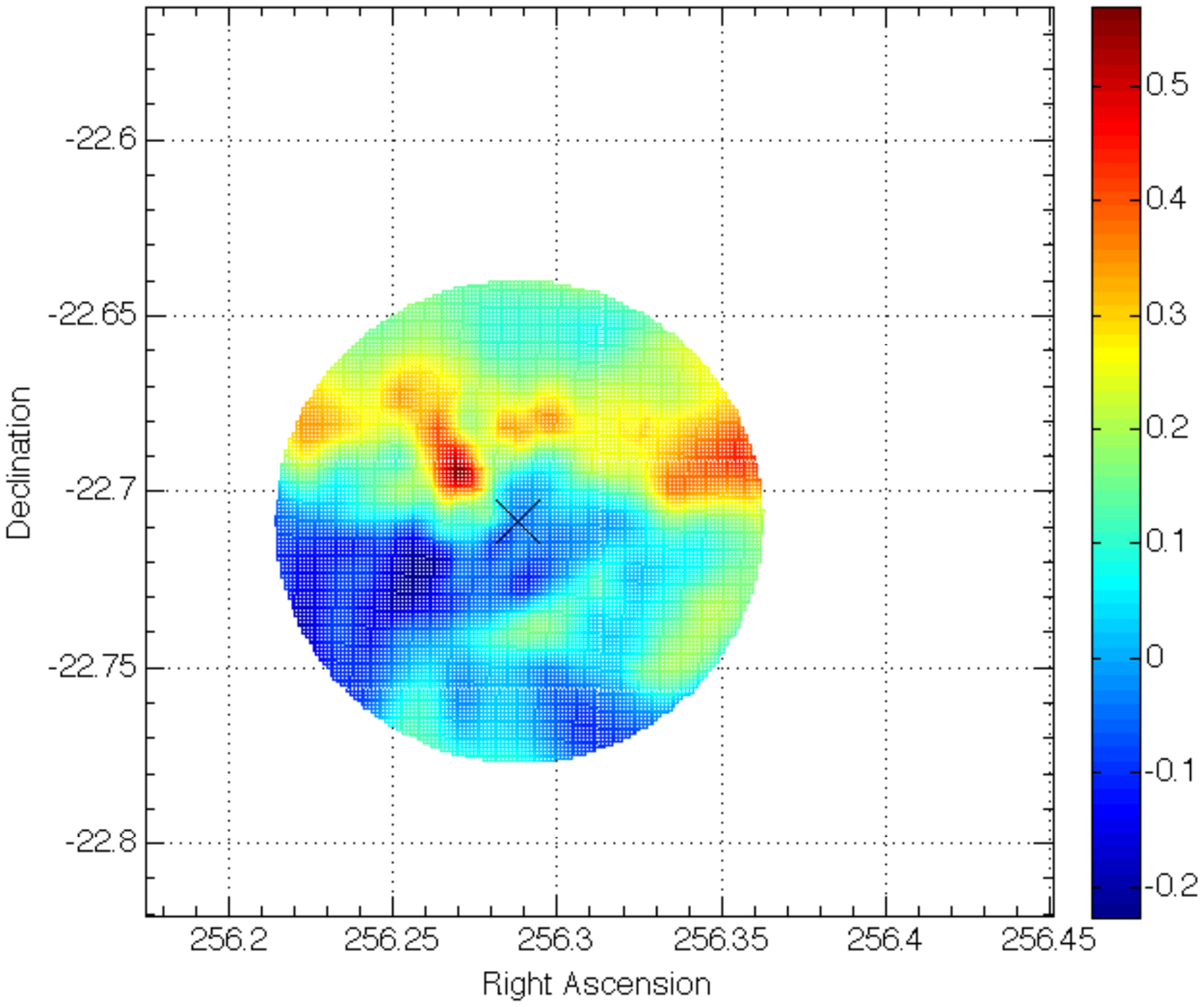}  &
\includegraphics[scale=0.28]{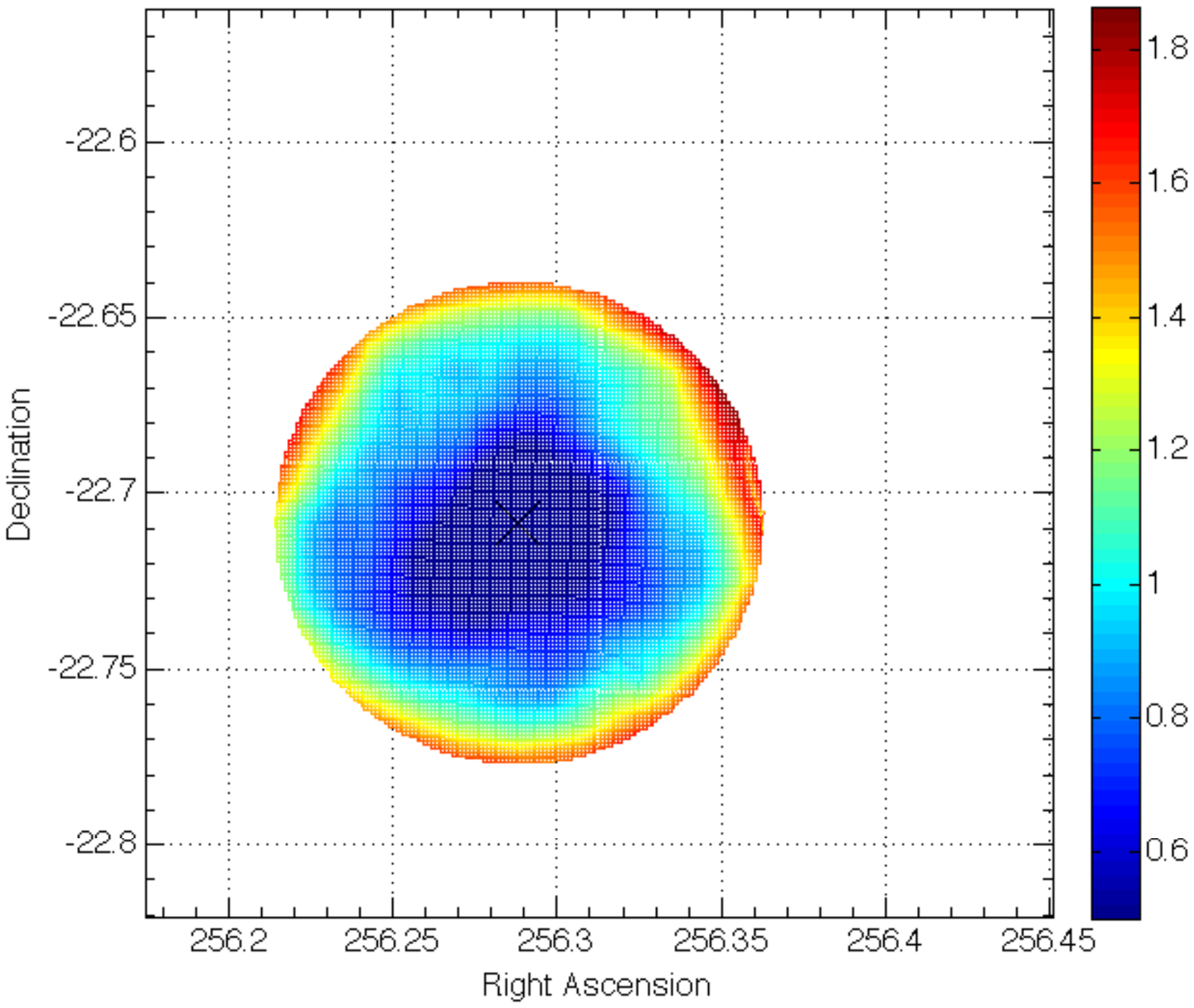}  &
\includegraphics[scale=0.28]{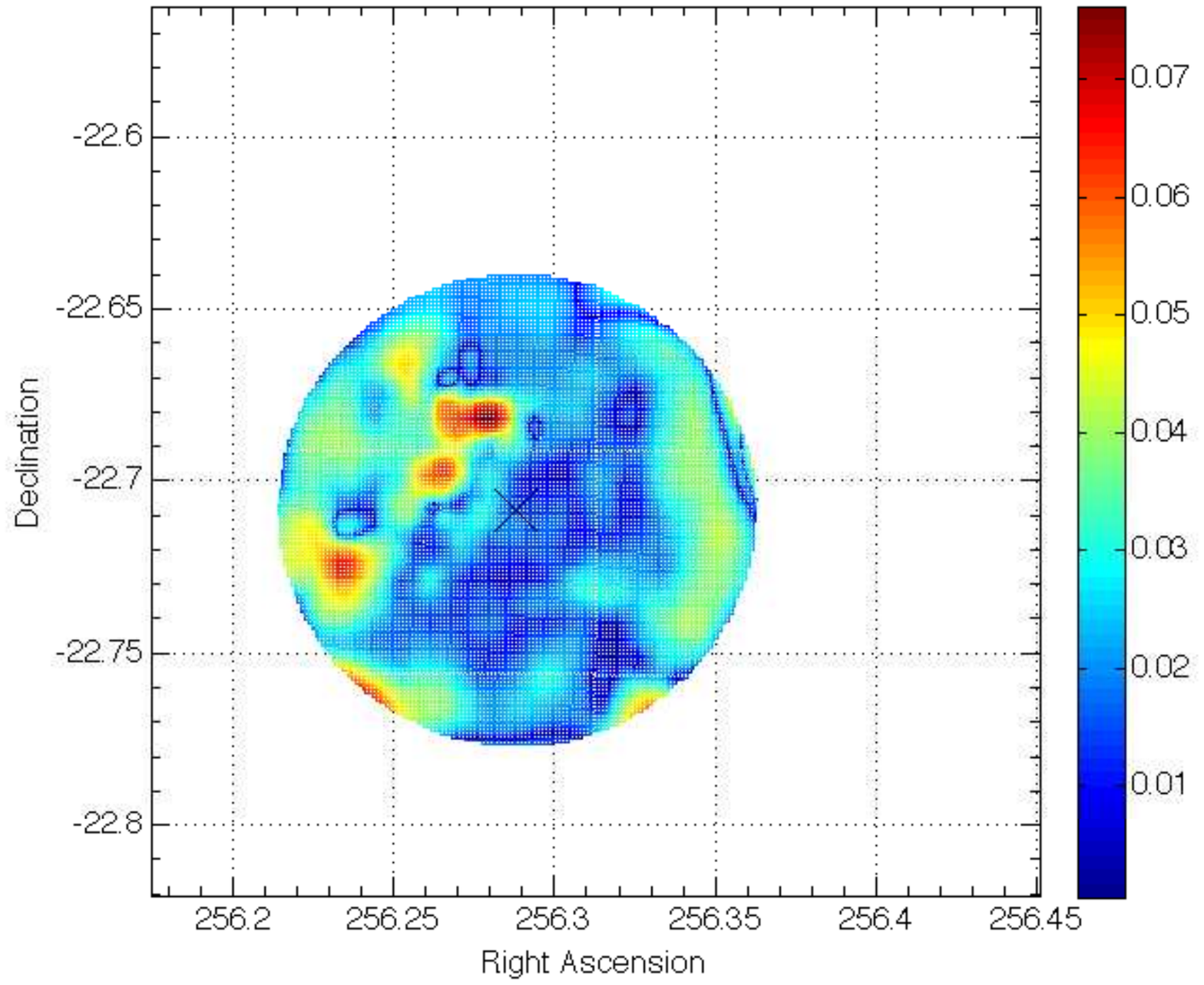} \\
\end{tabular}
\caption{\footnotesize As in Figure \ref{figngc6121}, but for the
  cluster NGC 6287. Only our Magellan photometry was used to build the
  CMDs in both colors. }
\label{figngc6287}
\end{figure}

\begin{figure}[htbp]
%\epsscale{0.77}
\plotone{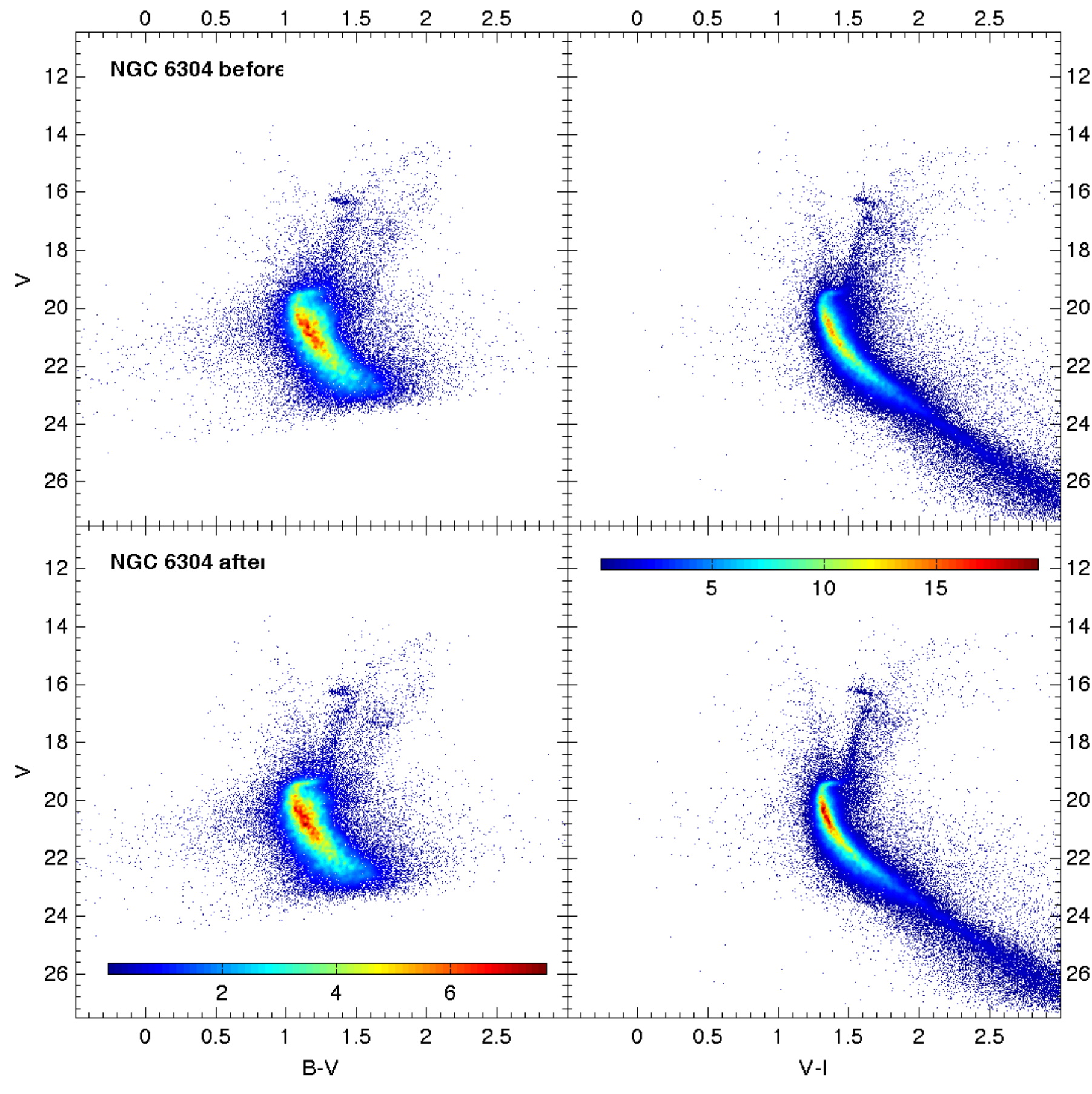}
\begin{tabular}{ccc}
\includegraphics[scale=0.28]{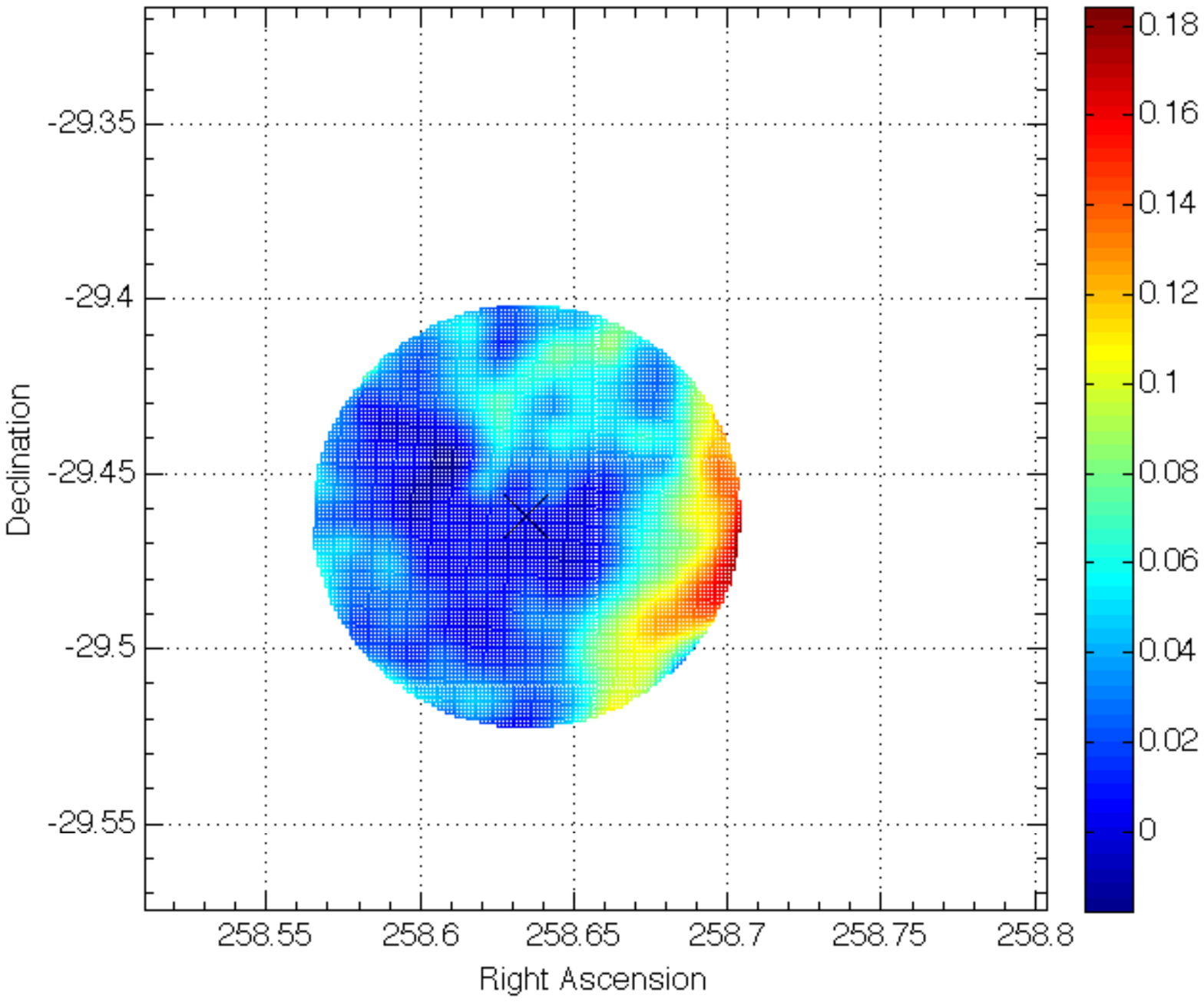}  &
\includegraphics[scale=0.28]{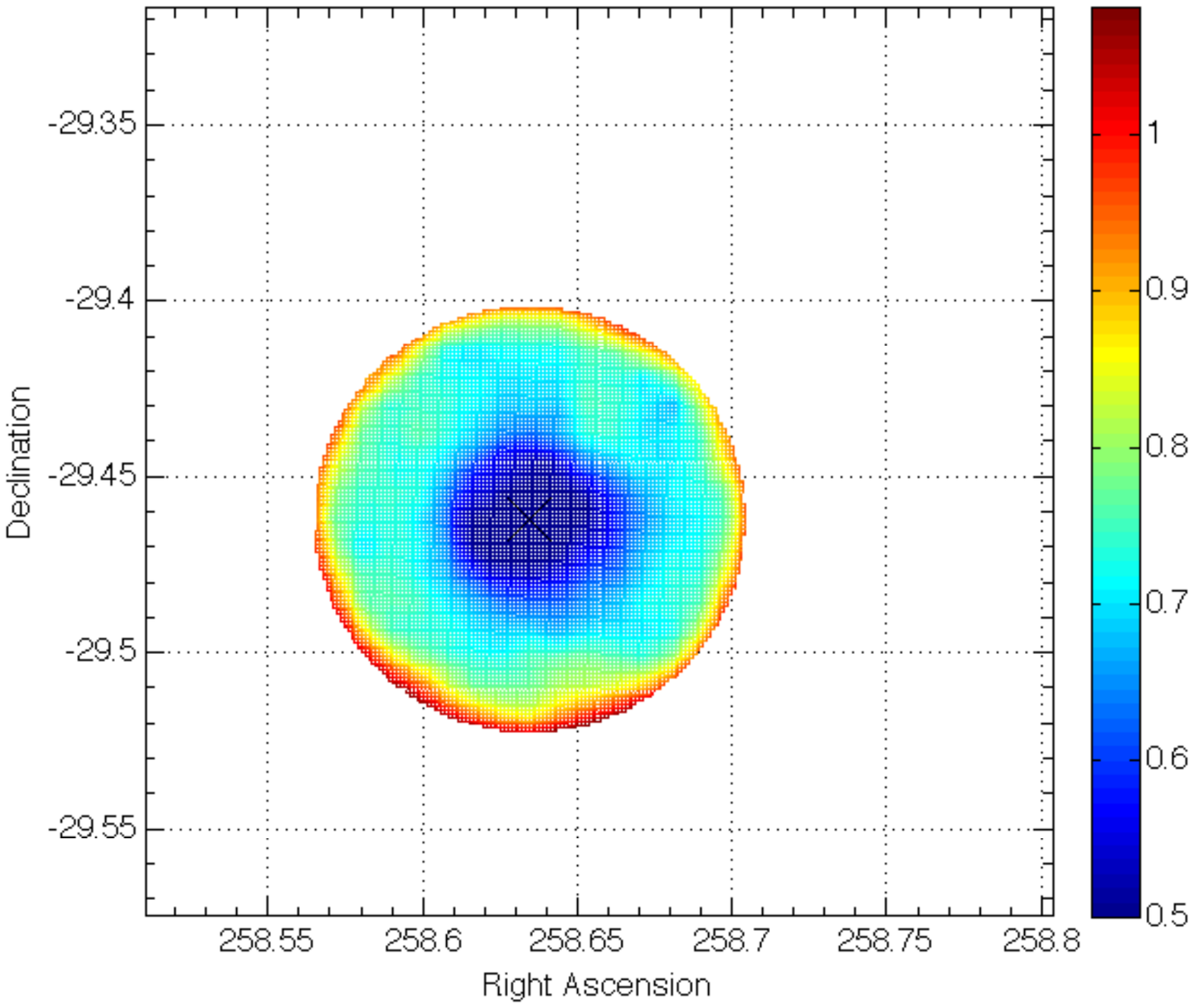}  &
\includegraphics[scale=0.28]{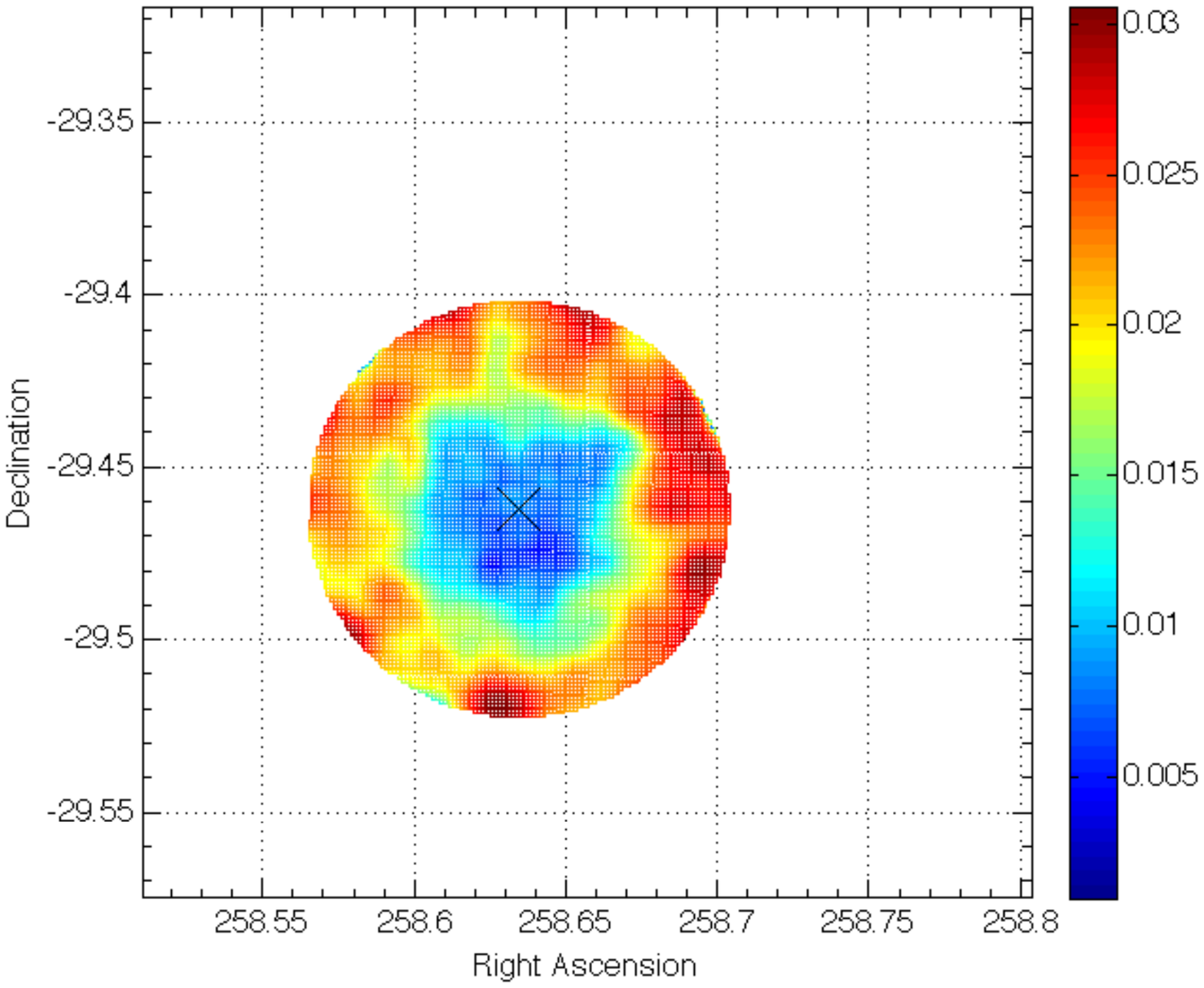} \\
\end{tabular}
\caption{\footnotesize As in Figure \ref{figngc6121}, but for the
  cluster NGC 6304. Only our Magellan photometry was used to build the
  $B-V$ vs. $V$ CMD. ACS photometry (from project 10775) and Magellan
  photometry were used to build the $V-I$ vs. $V$ CMD.}
\label{figngc6304}
\end{figure}

\begin{figure}[htbp]
%\epsscale{0.77}
\plotone{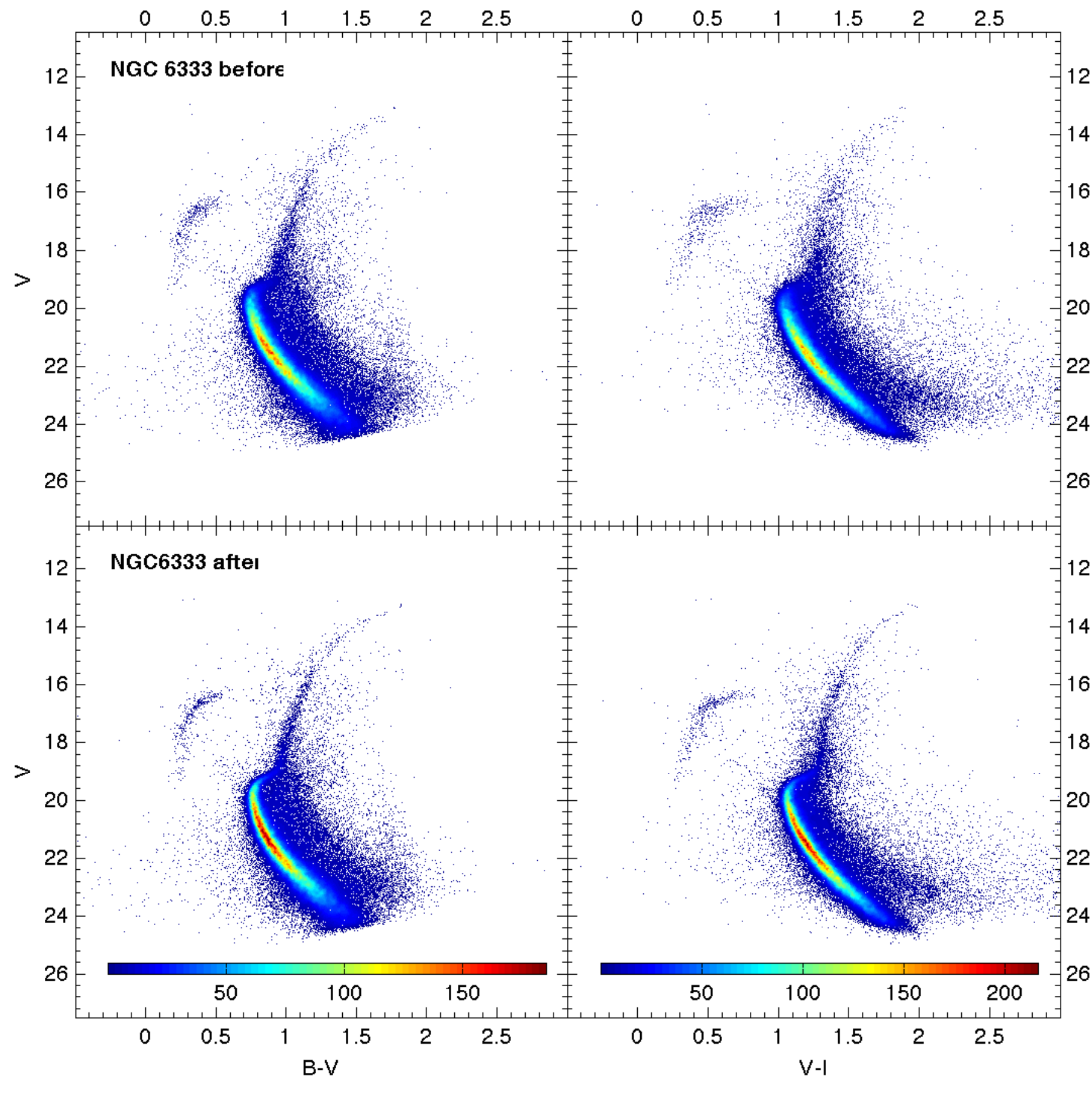}
\begin{tabular}{ccc}
\includegraphics[scale=0.28]{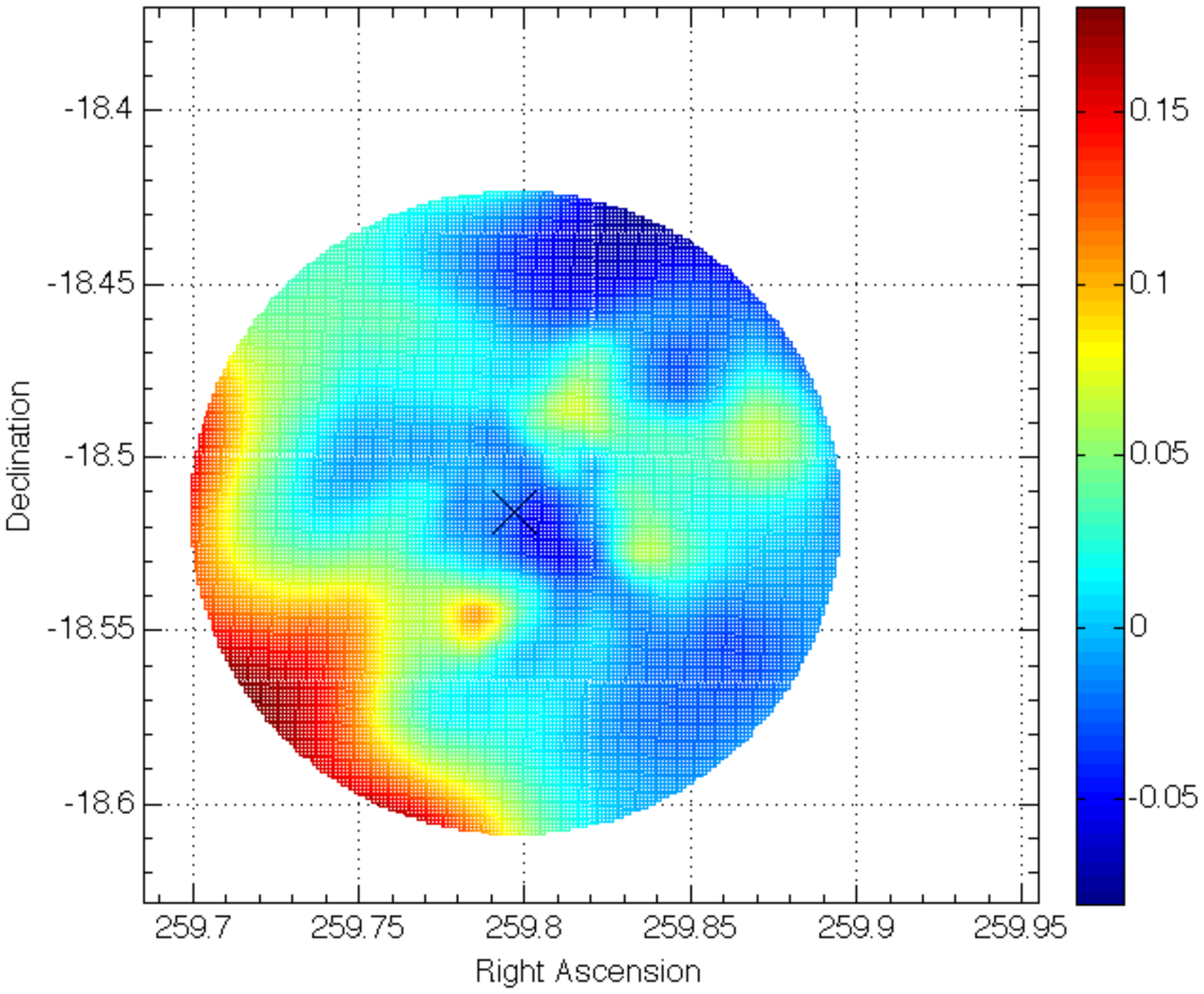}  &
\includegraphics[scale=0.28]{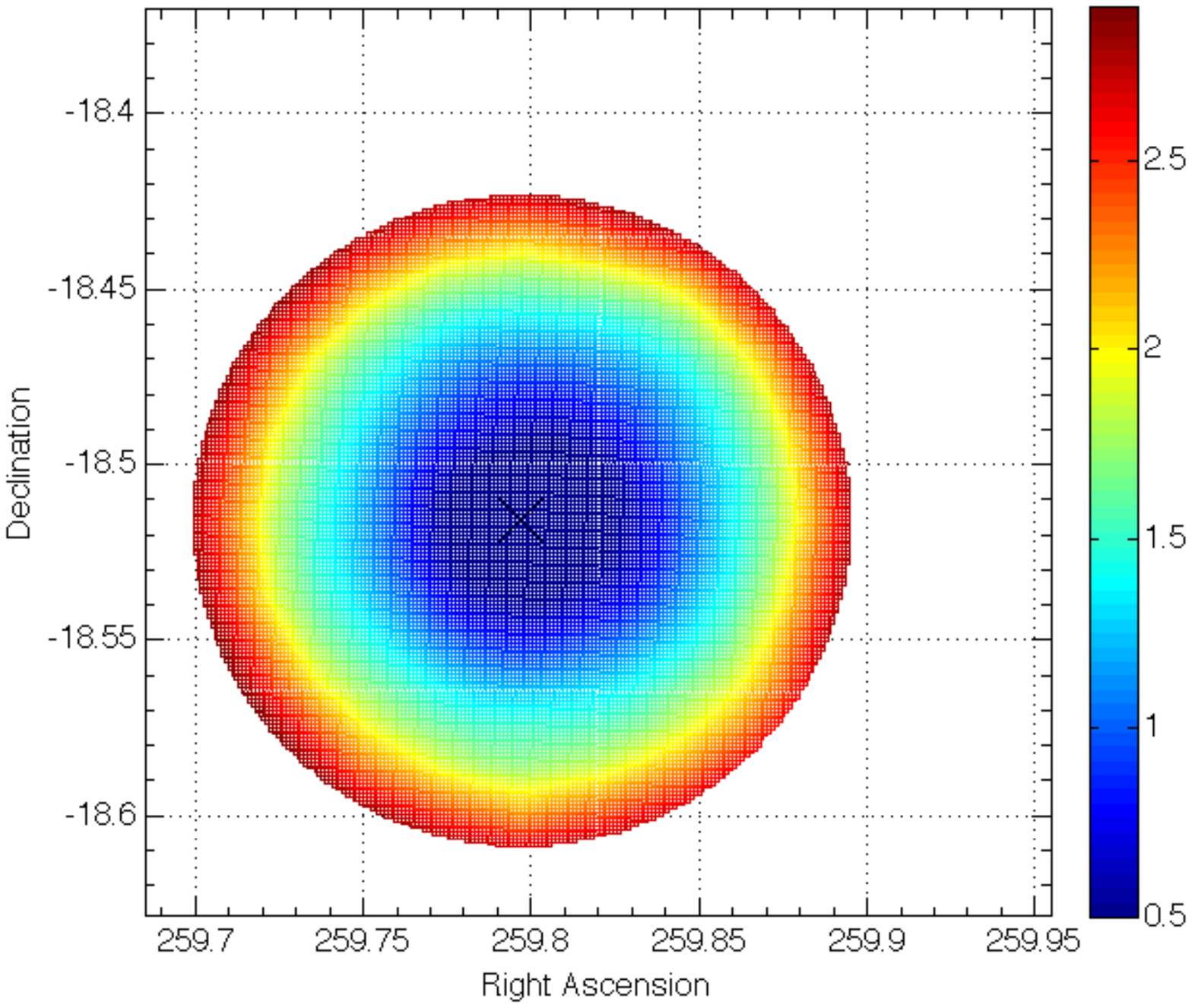}  &
\includegraphics[scale=0.28]{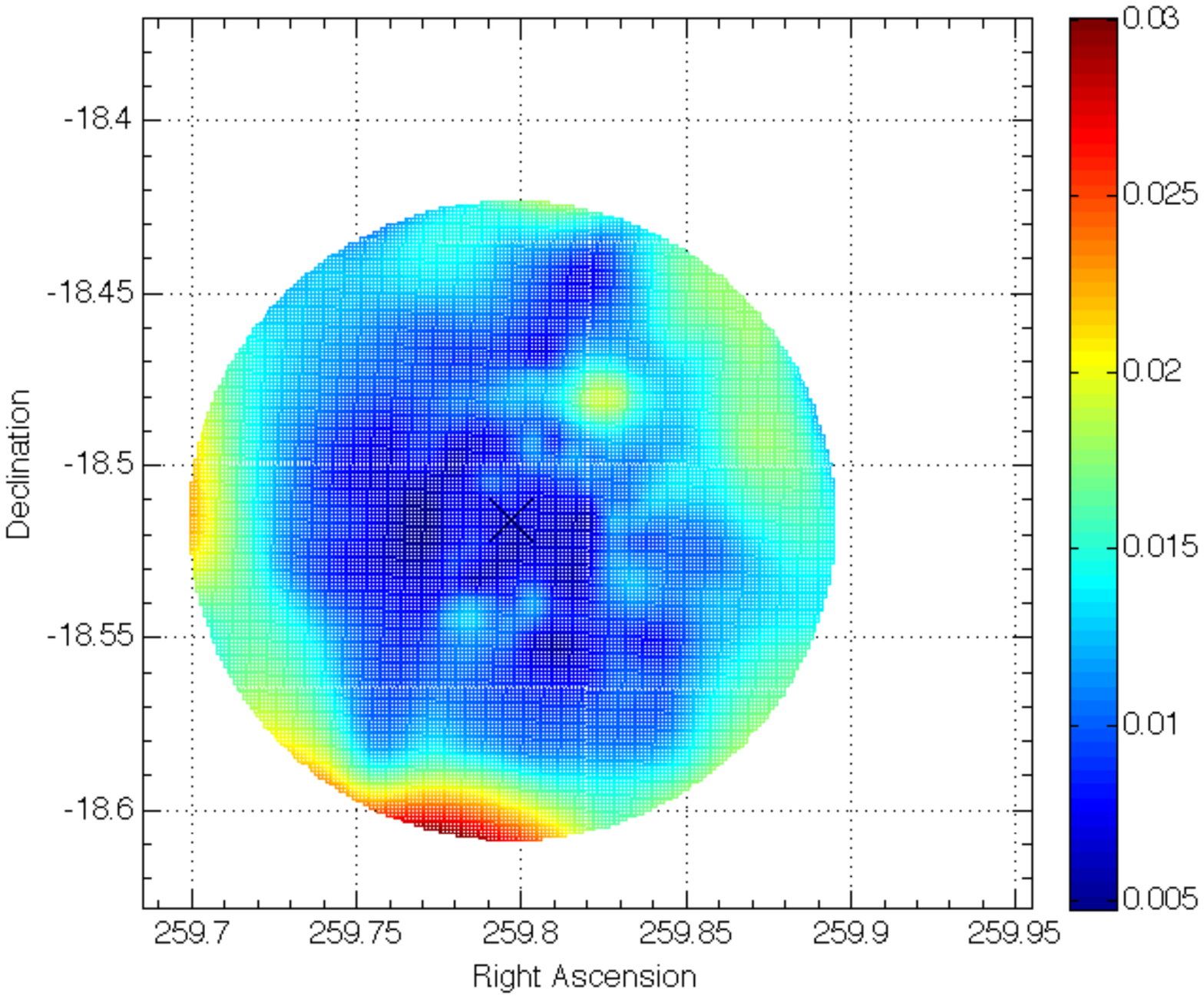} \\
\end{tabular}
\caption{\footnotesize As in Figure \ref{figngc6121}, but for the
  cluster NGC 6333 - M 9. ACS photometry (from our project 10573) and
  our Magellan photometry were used to build the CMDs in both colors.}
\label{figngc6333}
\end{figure}

\begin{figure}[htbp]
%\epsscale{0.77}
\plotone{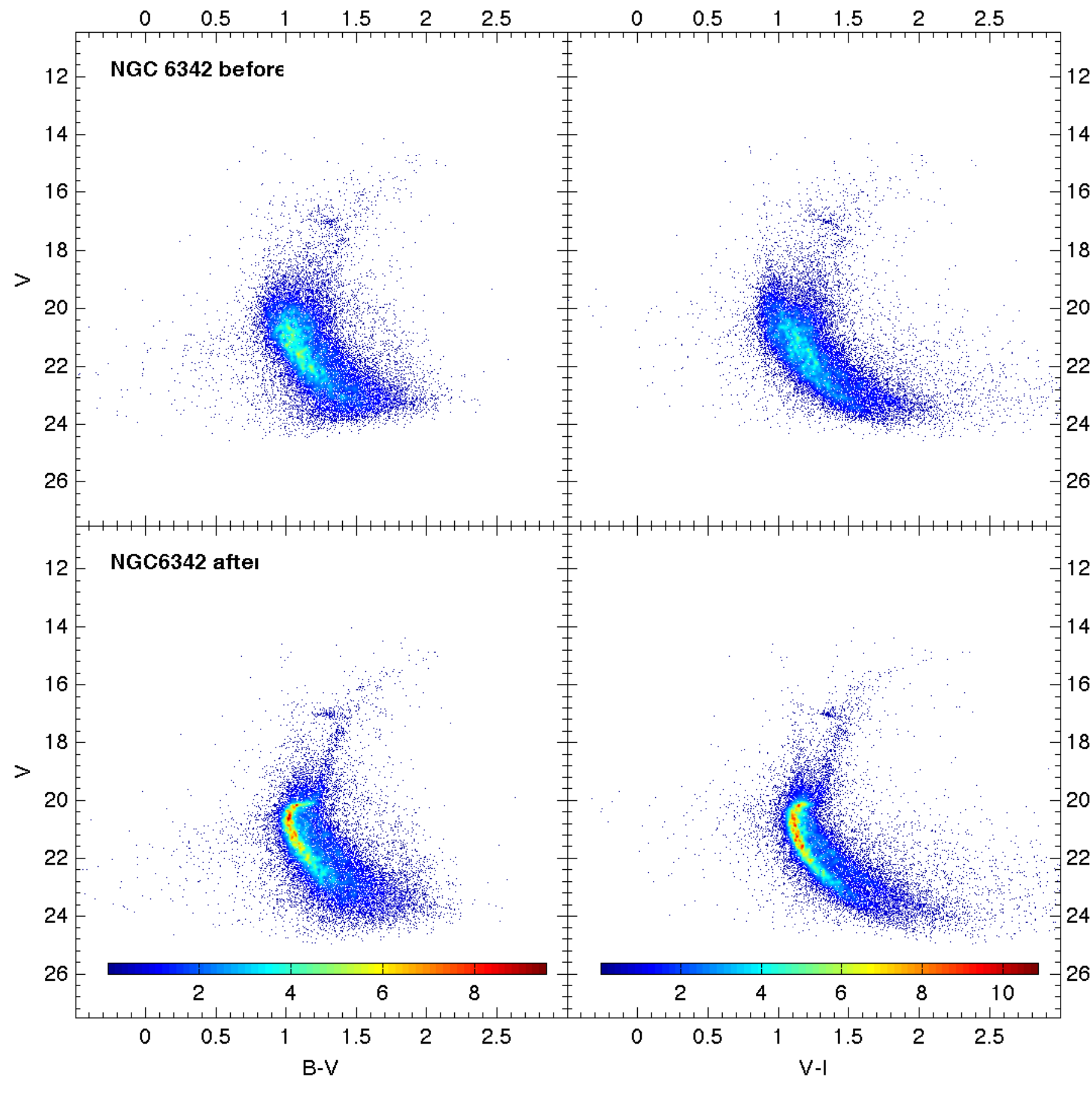}
\begin{tabular}{ccc}
\includegraphics[scale=0.28]{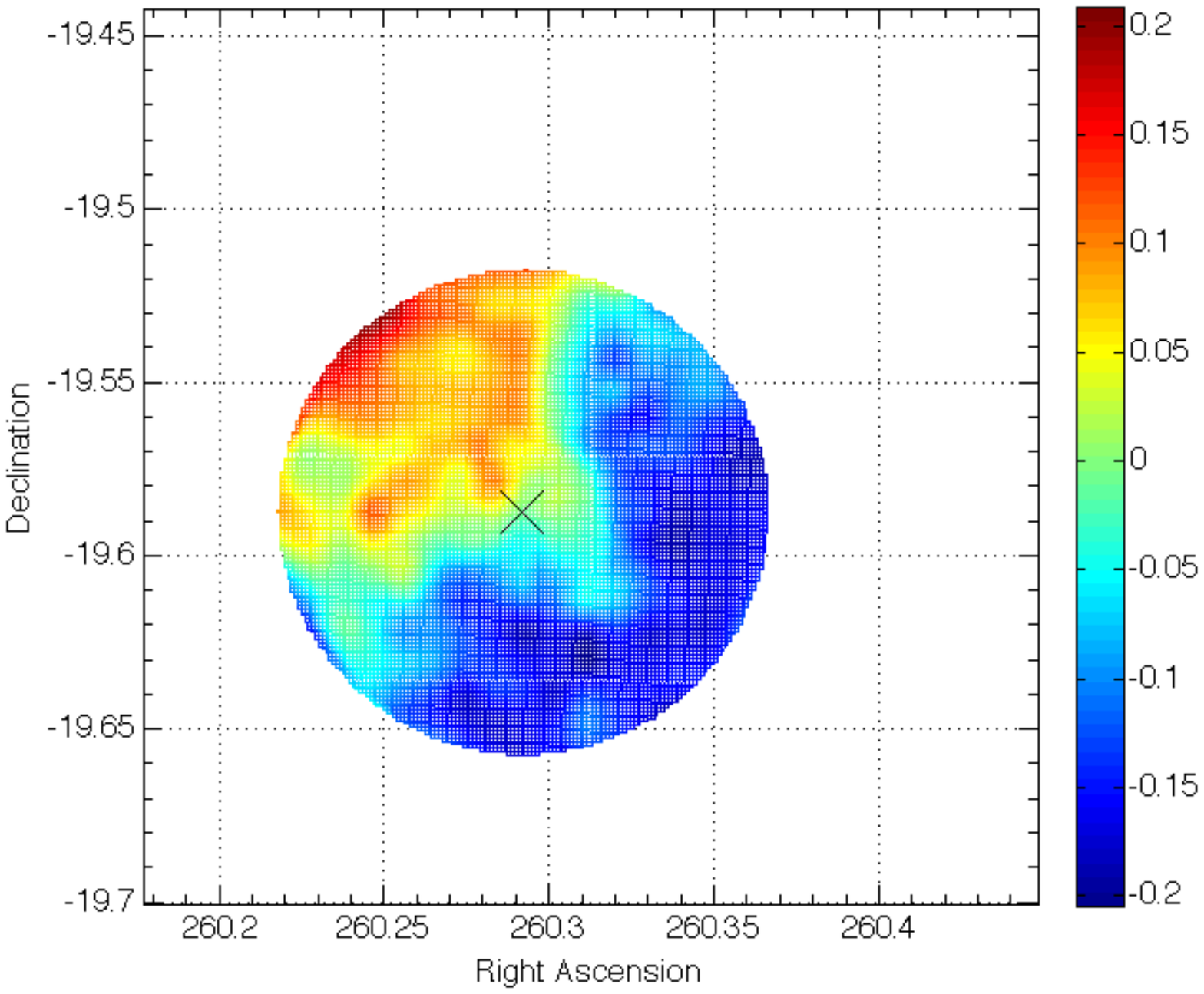}  &
\includegraphics[scale=0.28]{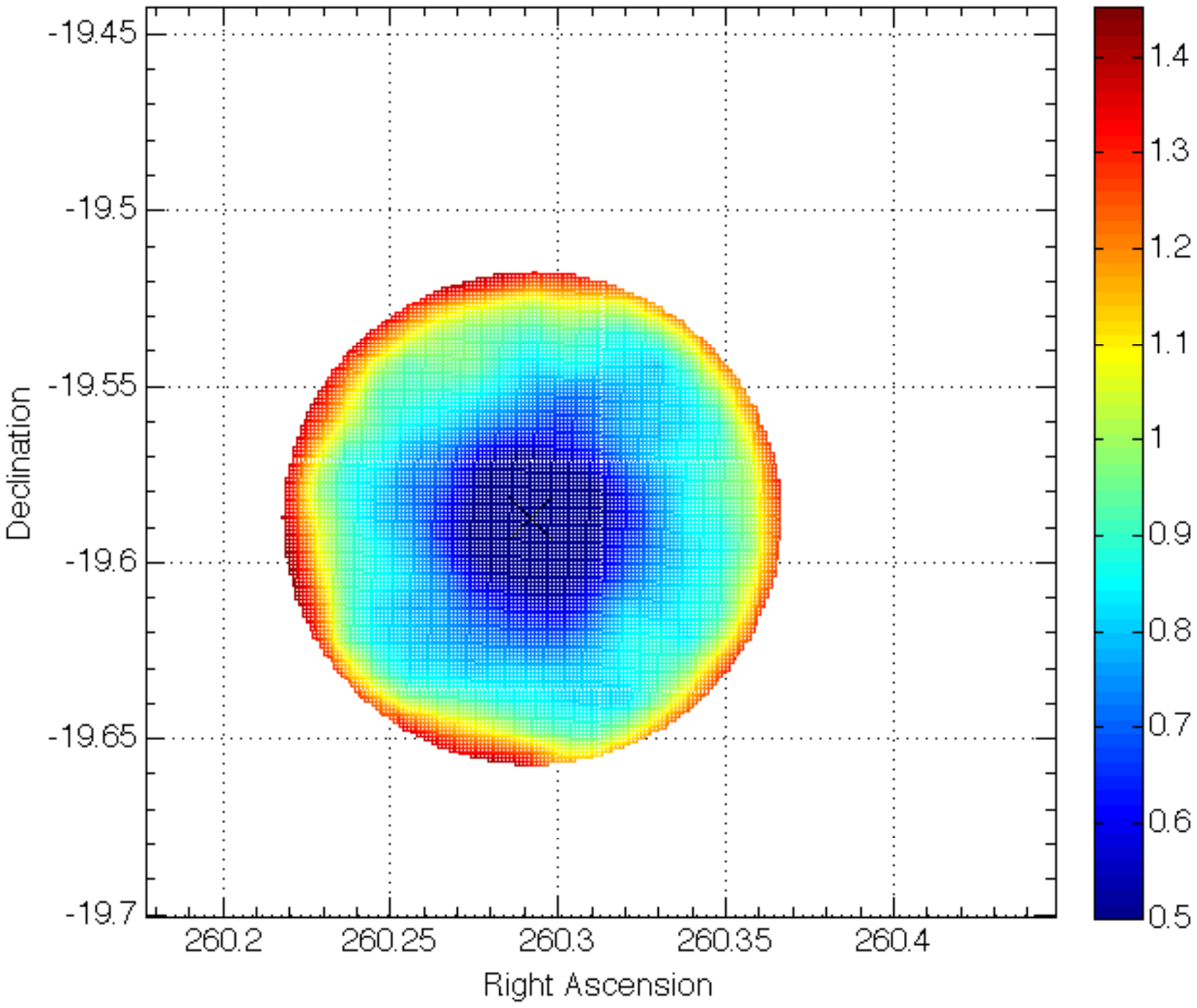}  &
\includegraphics[scale=0.28]{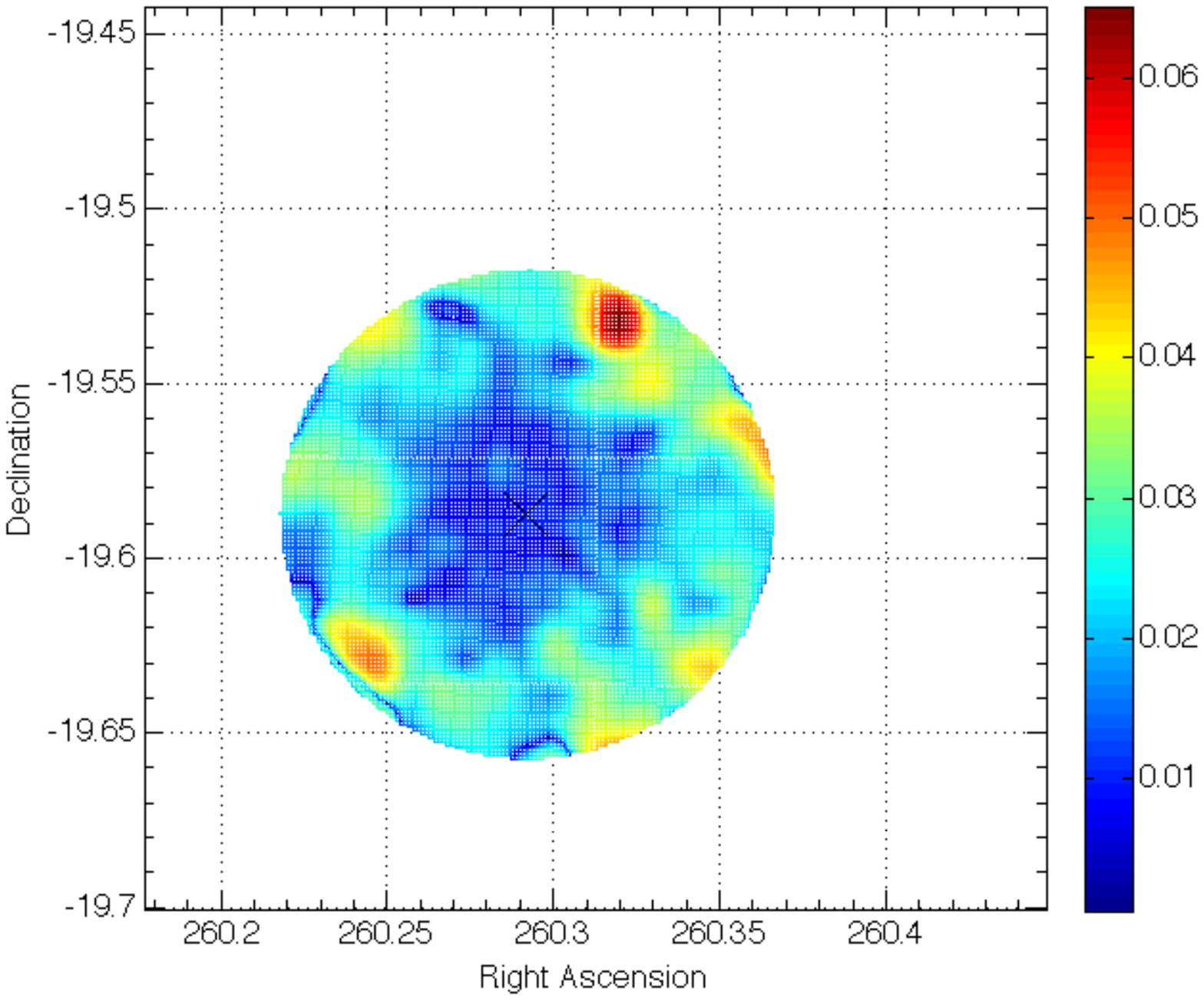} \\
\end{tabular}
\caption{\footnotesize As in Figure \ref{figngc6121}, but for the
  cluster NGC 6342. Only our Magellan photometry was used to build the
  CMDs in both colors. Notice that the $V-I$ vs. $V$ CMD could not be
  correctly calibrated in color using the method described in the text
  because of the lack of calibrating data in the $I$ filter.}
\label{figngc6342}
\end{figure}

\begin{figure}[htbp]
%\epsscale{0.77}
\plotone{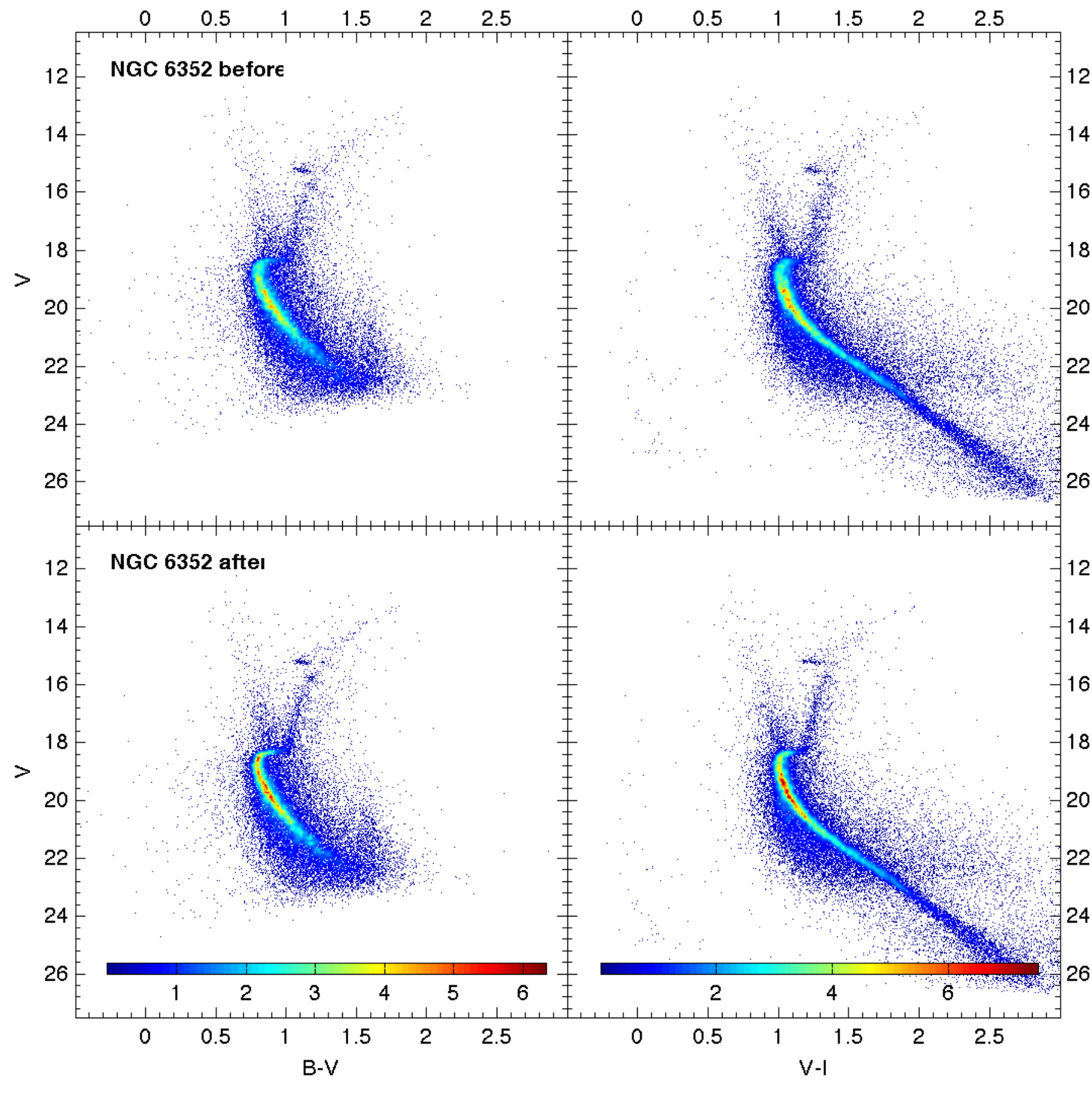}
\begin{tabular}{ccc}
\includegraphics[scale=0.28]{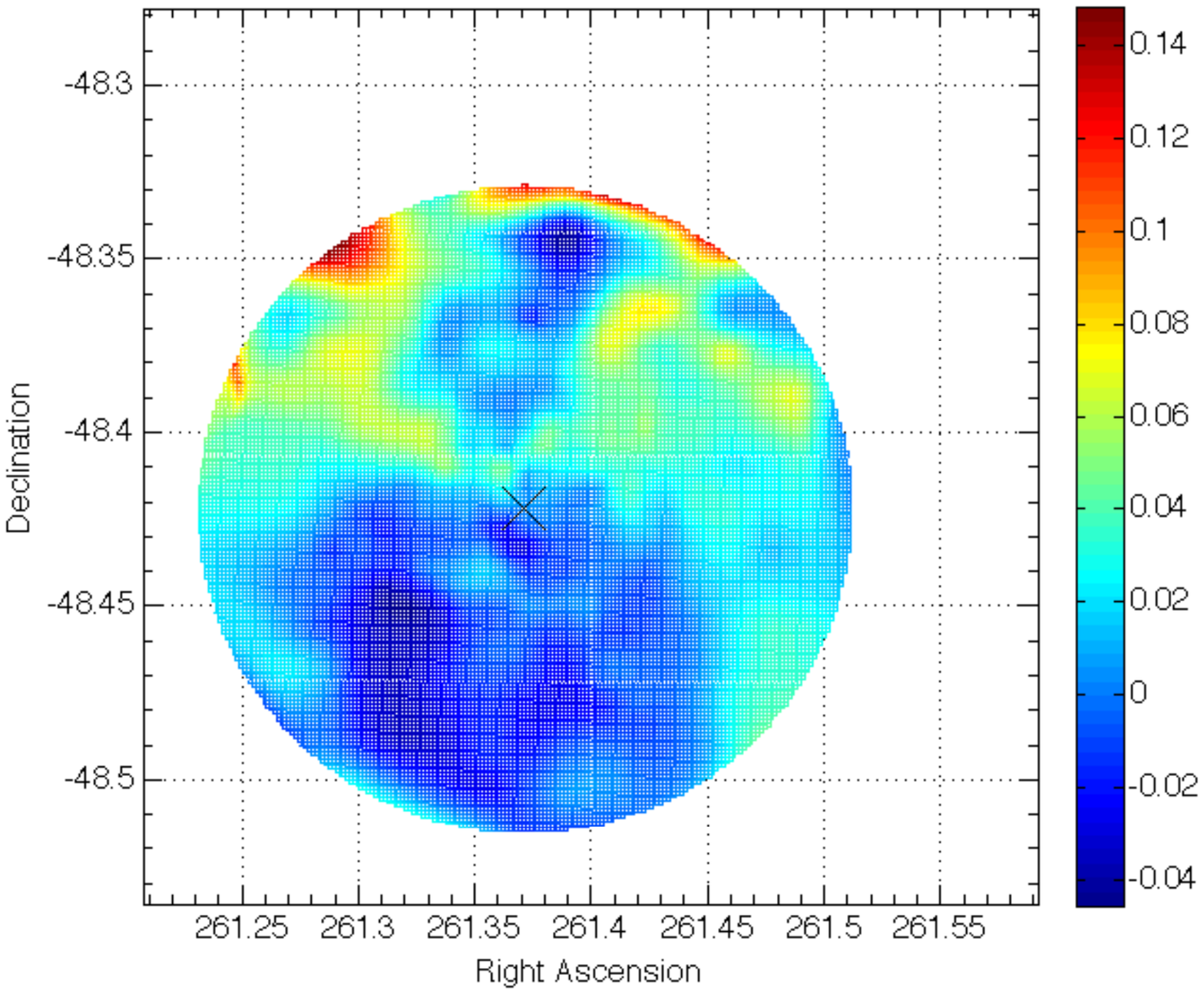}  &
\includegraphics[scale=0.28]{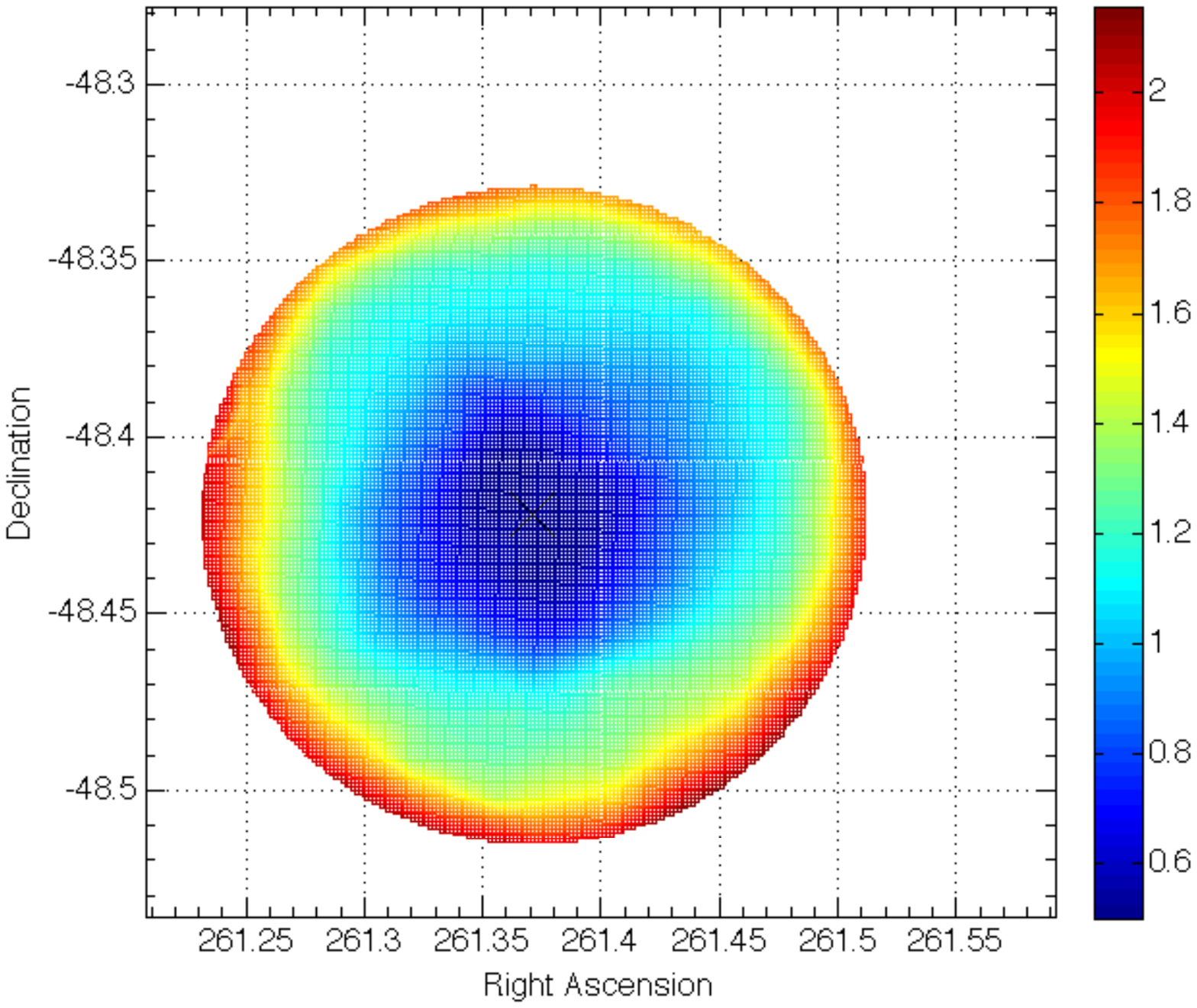}  &
\includegraphics[scale=0.28]{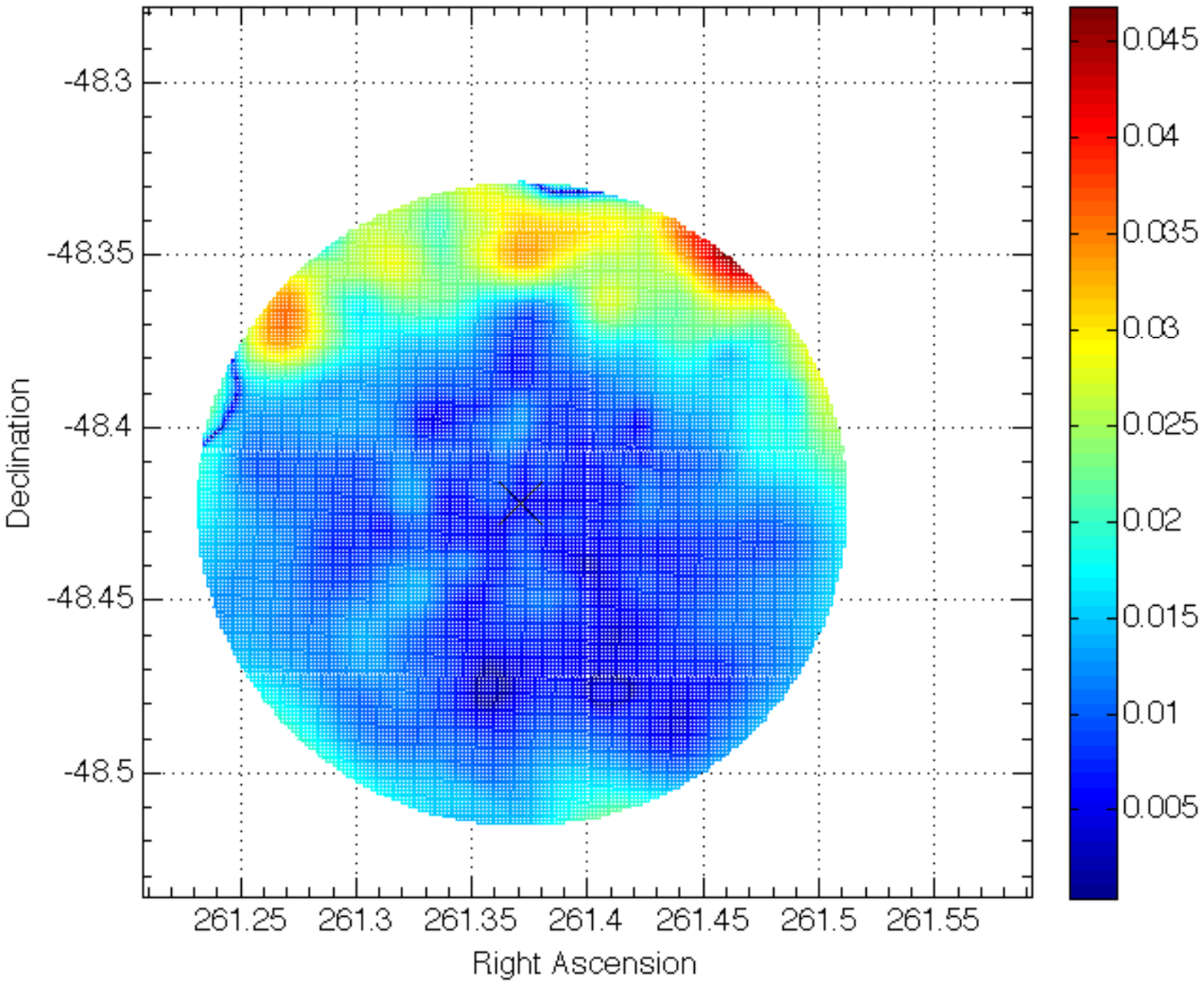} \\
\end{tabular}
\caption{\footnotesize As in Figure \ref{figngc6121}, but for the
  cluster NGC 6352. Only our Magellan photometry was used to build the
  $B-V$ vs. $V$ CMD. ACS photometry (from project 10775) and Magellan
  photometry were used to build the $V-I$ vs. $V$ CMD.}
\label{figngc6352}
\end{figure}

\begin{figure}[htbp]
%\epsscale{0.77}
\plotone{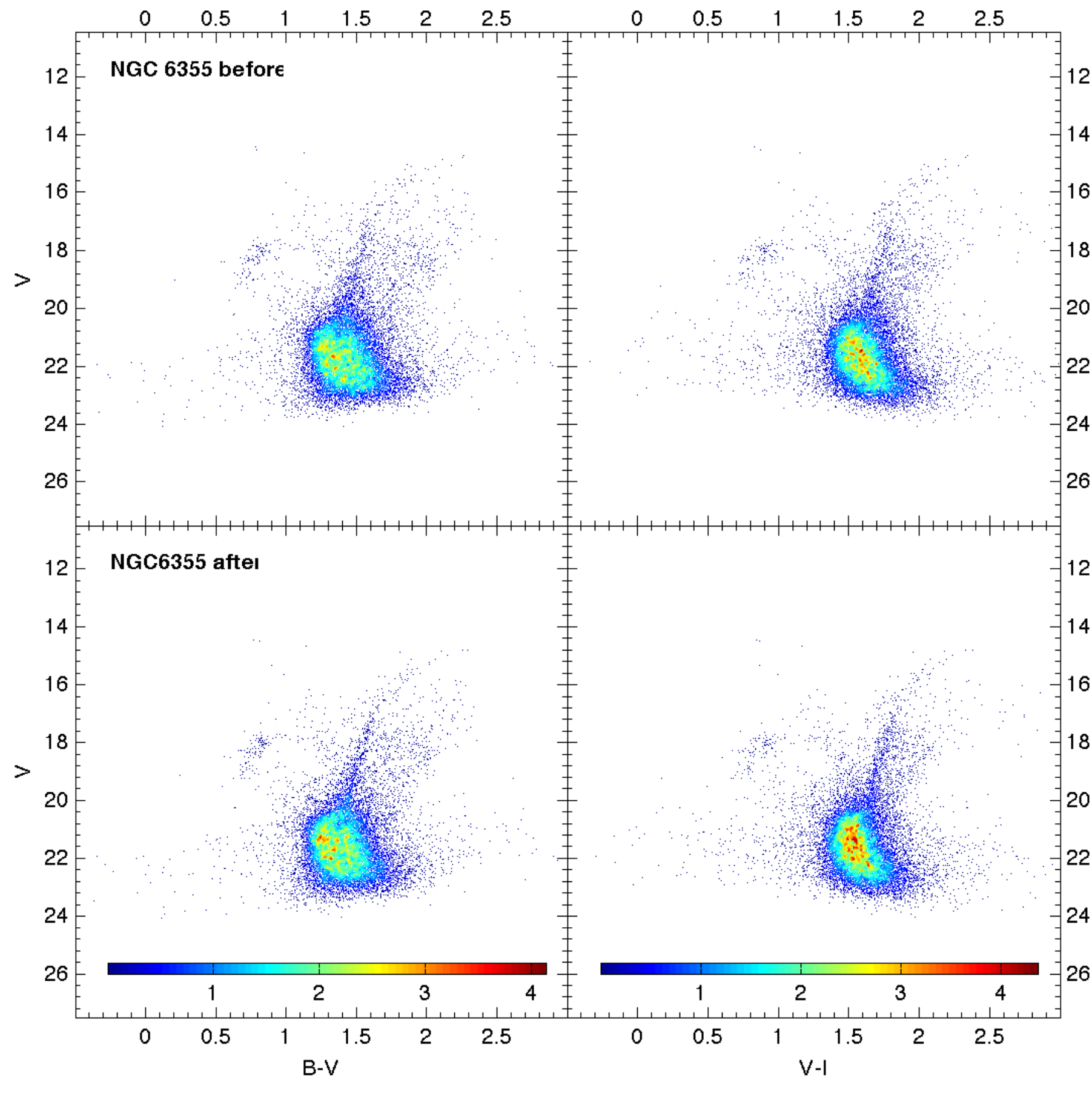}
\begin{tabular}{ccc}
\includegraphics[scale=0.28]{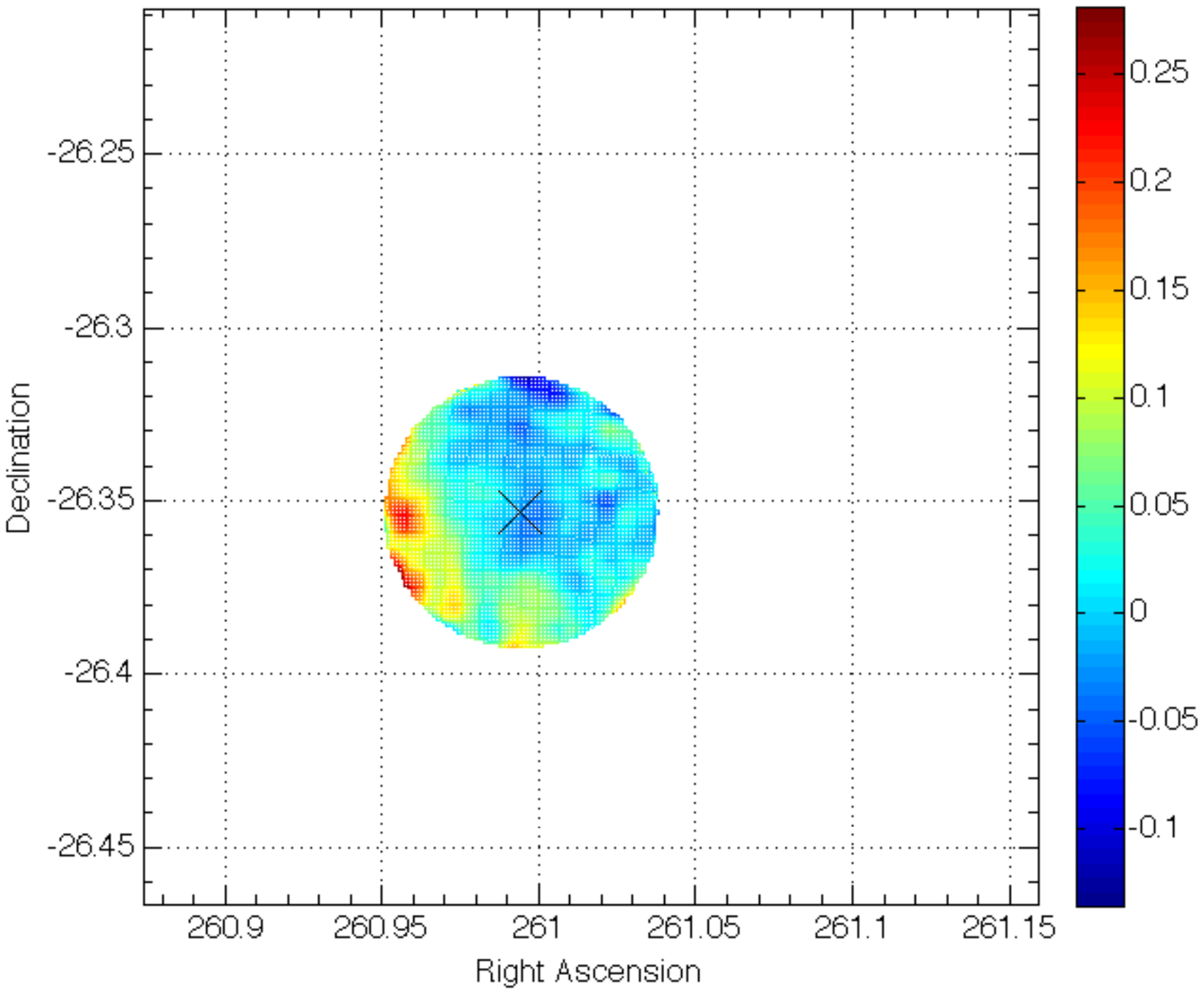}  &
\includegraphics[scale=0.28]{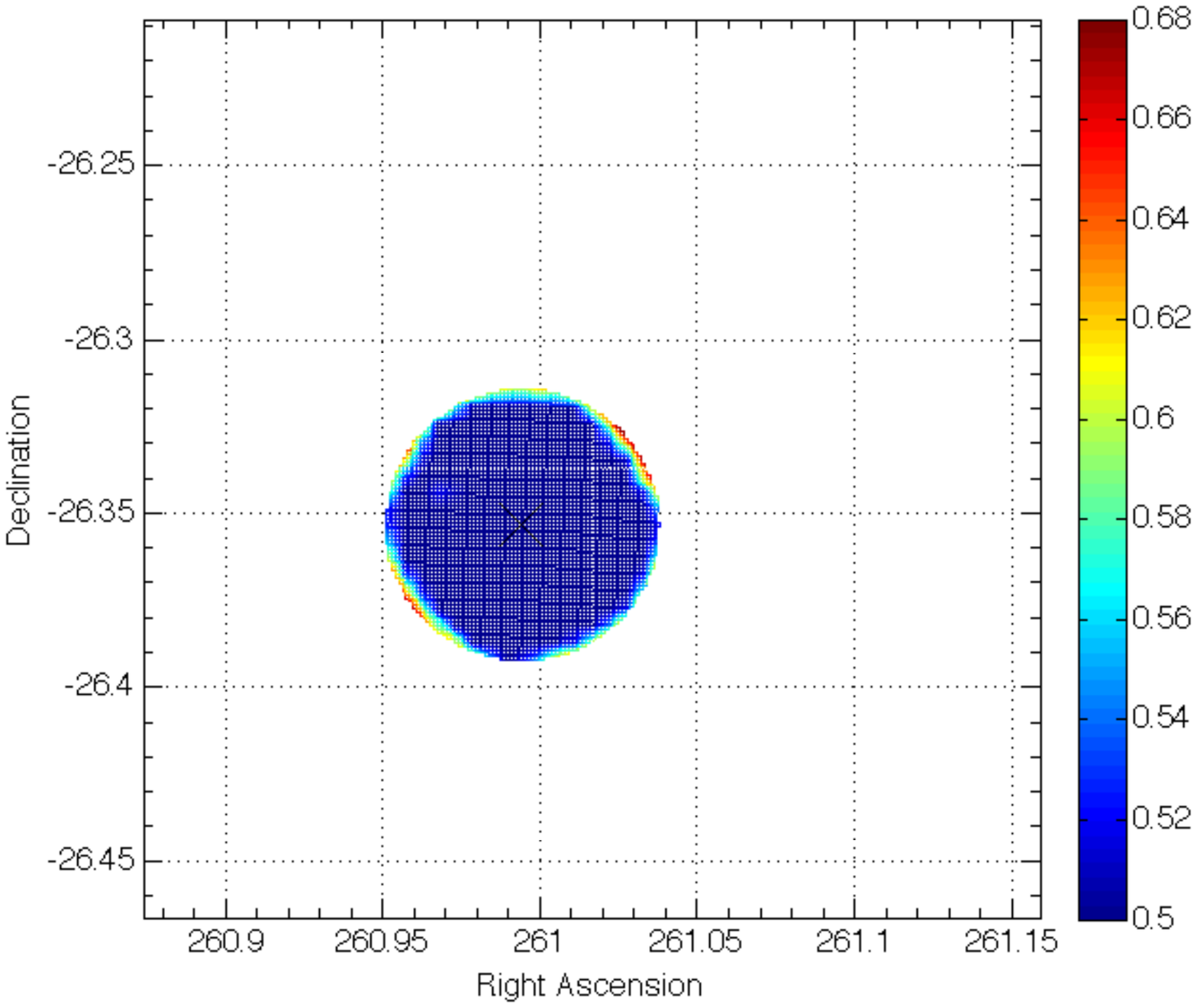}  &
\includegraphics[scale=0.28]{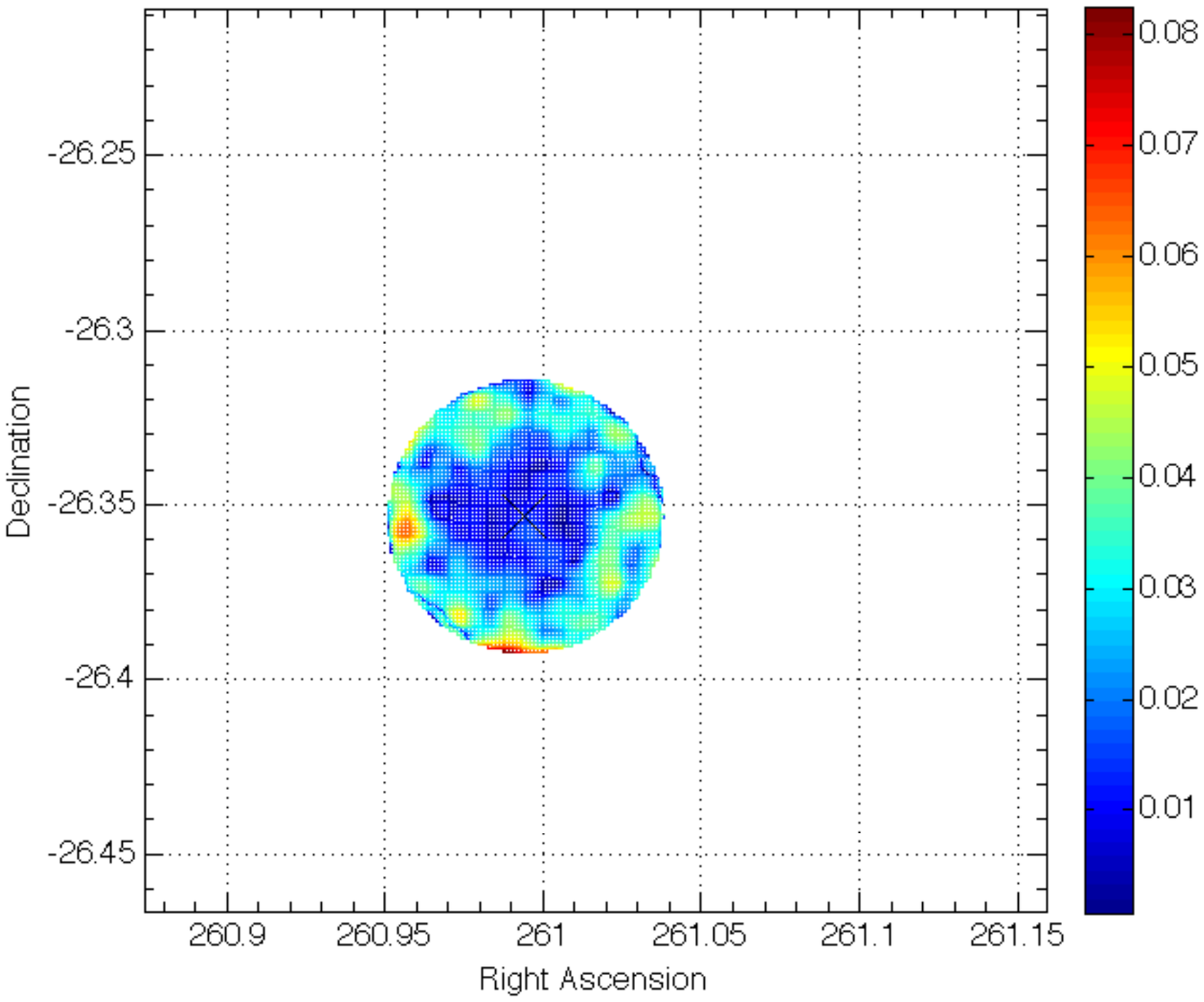} \\
\end{tabular}
\caption{\footnotesize As in Figure \ref{figngc6121}, but for the
  cluster NGC 6355. Only our Magellan photometry was used to build the
  CMDs in both colors. Notice that the $V-I$ vs. $V$ CMD could not be
  correctly calibrated in color using the method described in the text
  because of the lack of calibrating data in the $I$ filter.}
\label{figngc6355}
\end{figure}

\begin{figure}[htbp]
%\epsscale{0.77}
\plotone{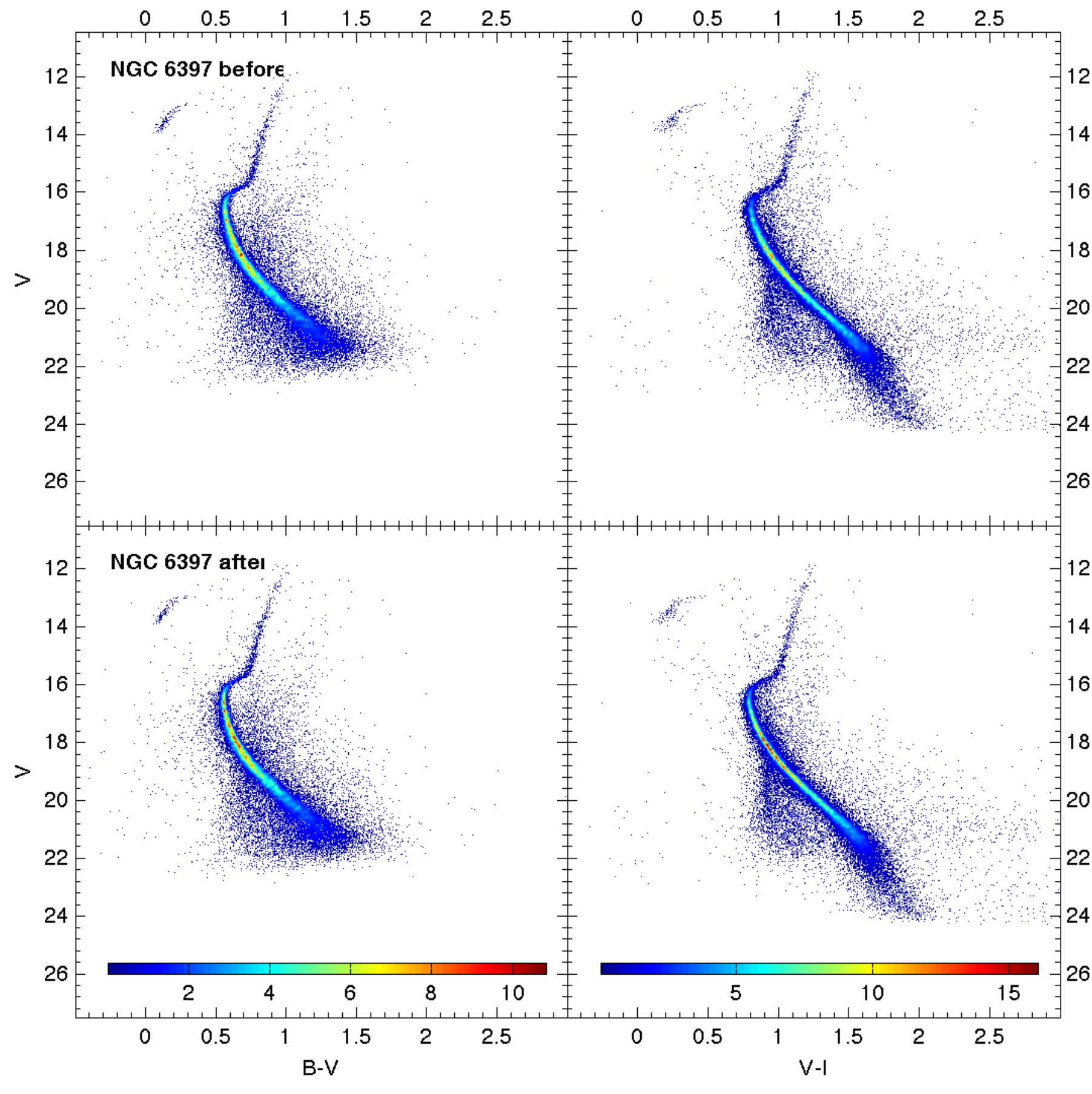}
\begin{tabular}{ccc}
\includegraphics[scale=0.28]{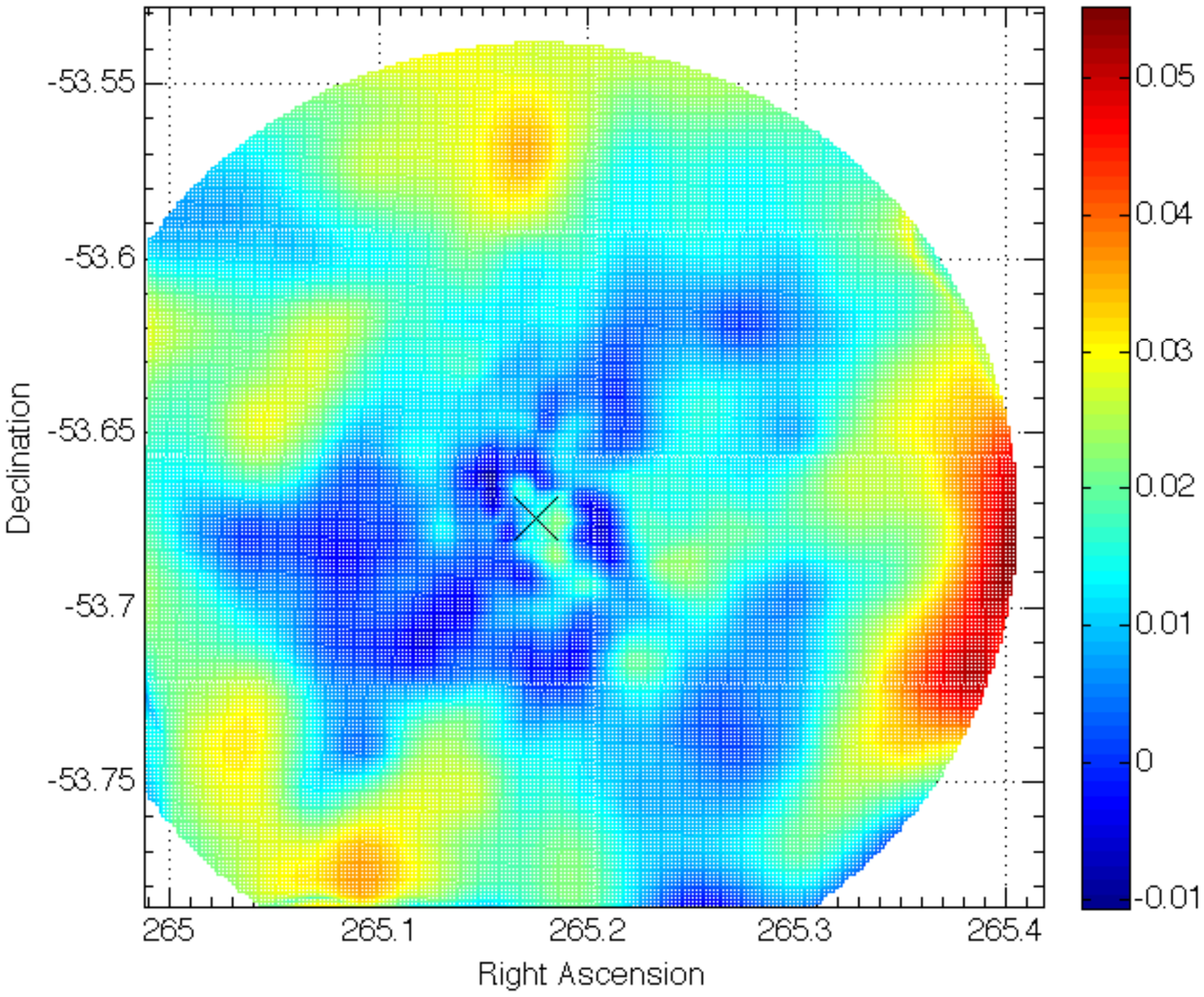}  &
\includegraphics[scale=0.28]{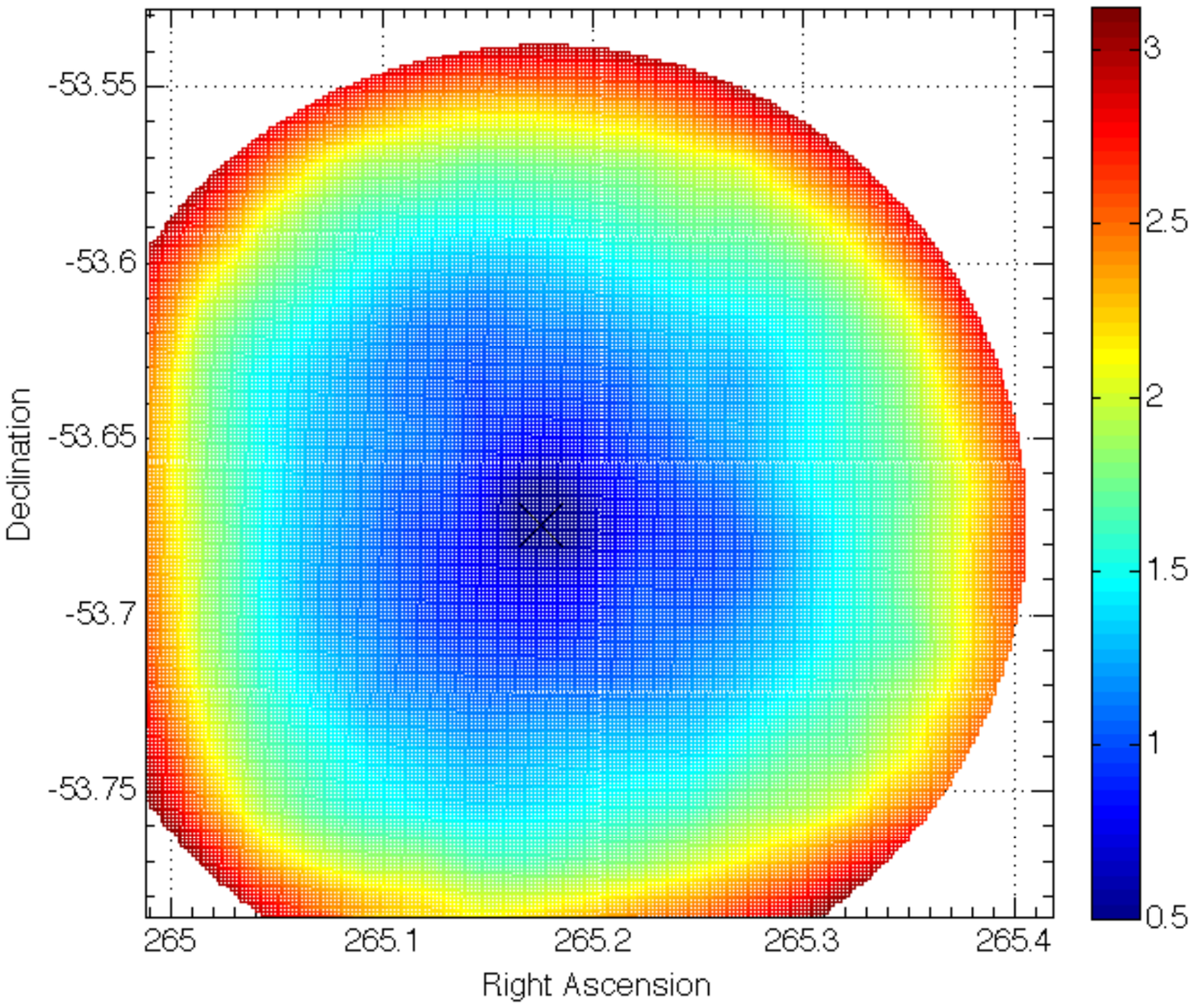}  &
\includegraphics[scale=0.28]{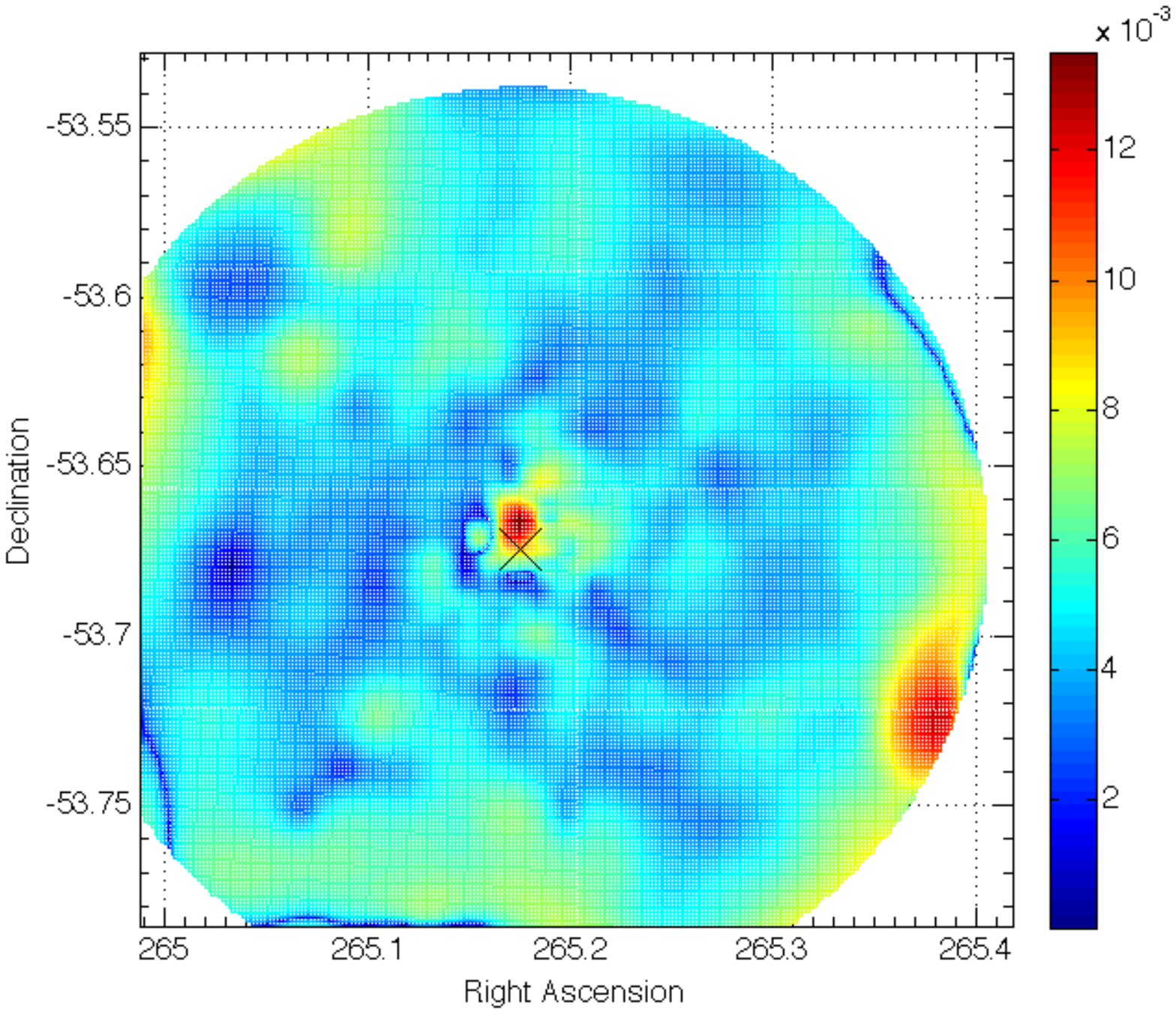} \\
\end{tabular}
\caption{\footnotesize As in Figure \ref{figngc6121}, but for the
  cluster NGC 6397, before and after being differentially
  dereddened. Only our Magellan photometry was used to build the $B-V$
  vs. $V$ CMD. ACS photometry (from project 10775) and Magellan
  photometry were used to build the $V-I$ vs. $V$ CMD.}
\label{figngc6397}
\end{figure}

\begin{figure}[htbp]
%\epsscale{0.77}
\plotone{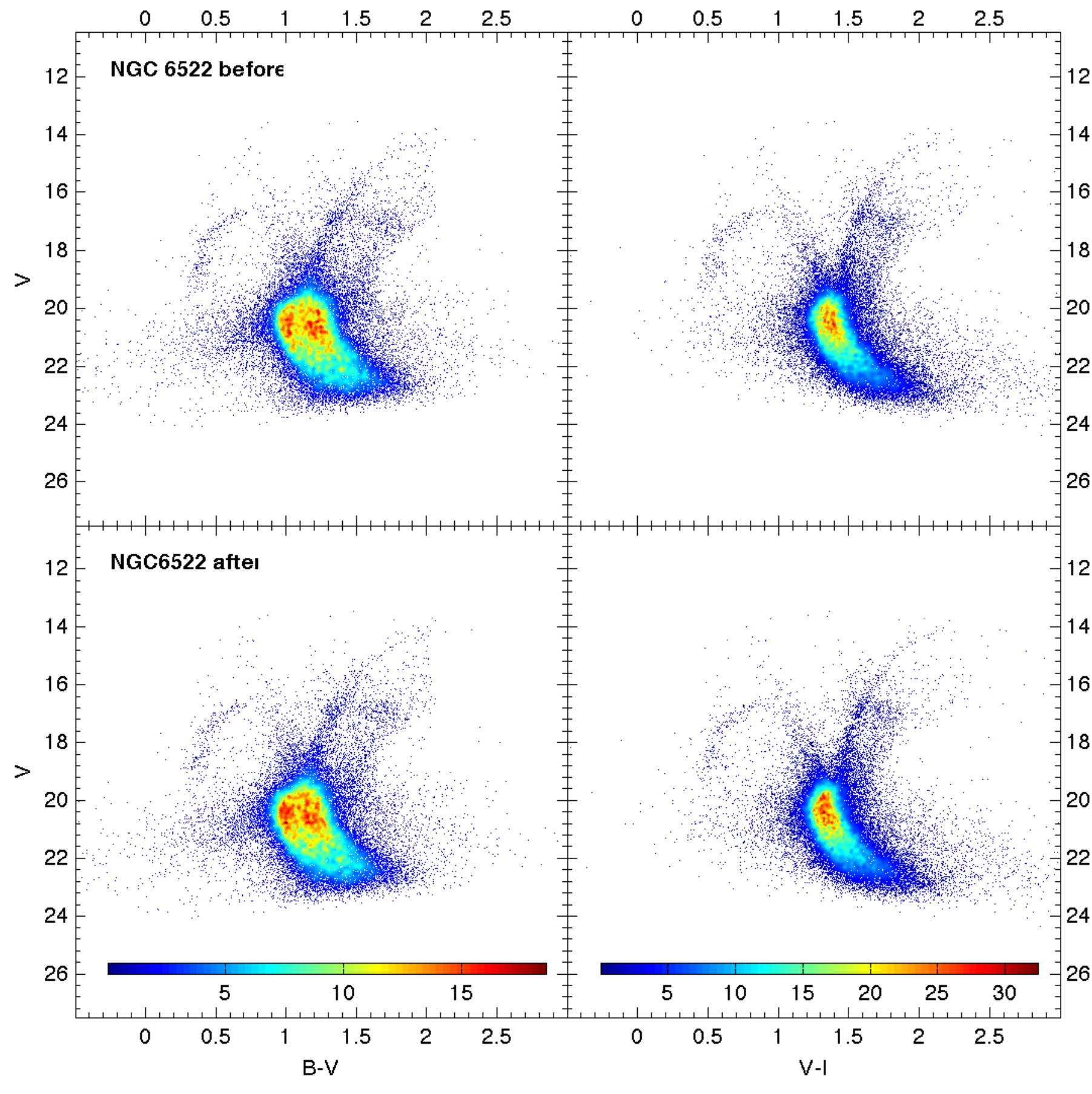}
\begin{tabular}{ccc}
\includegraphics[scale=0.28]{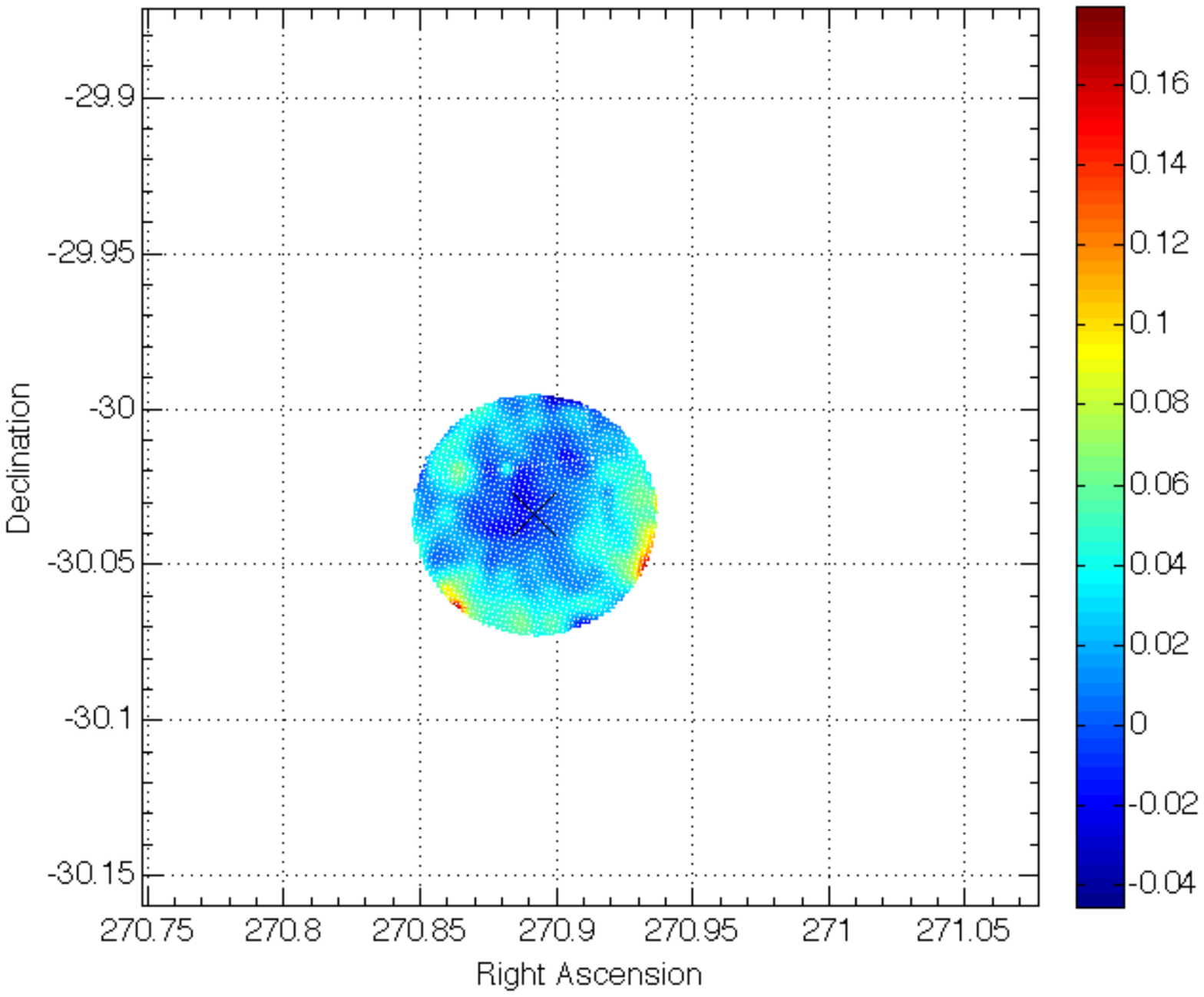}  &
\includegraphics[scale=0.28]{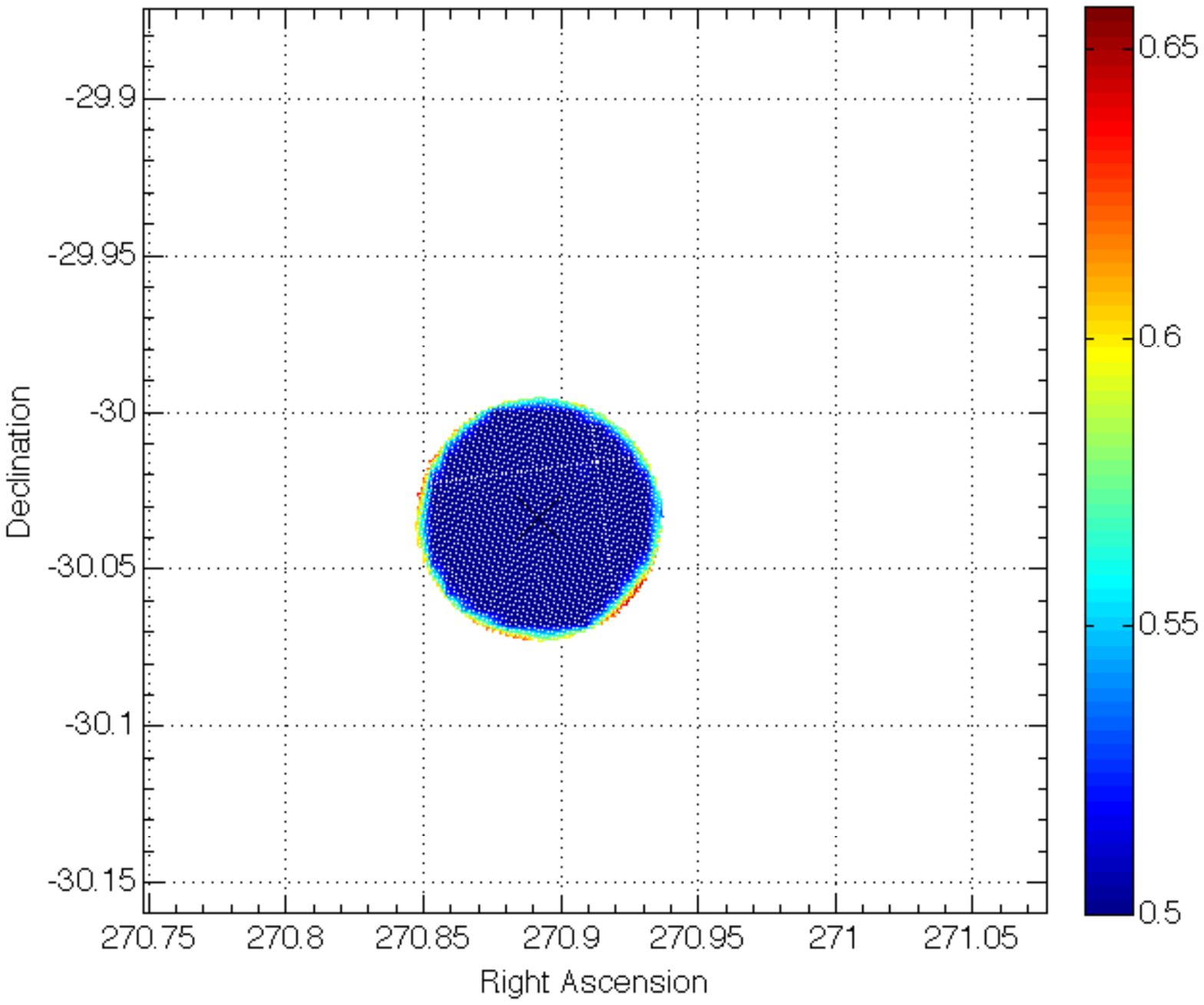}  &
\includegraphics[scale=0.28]{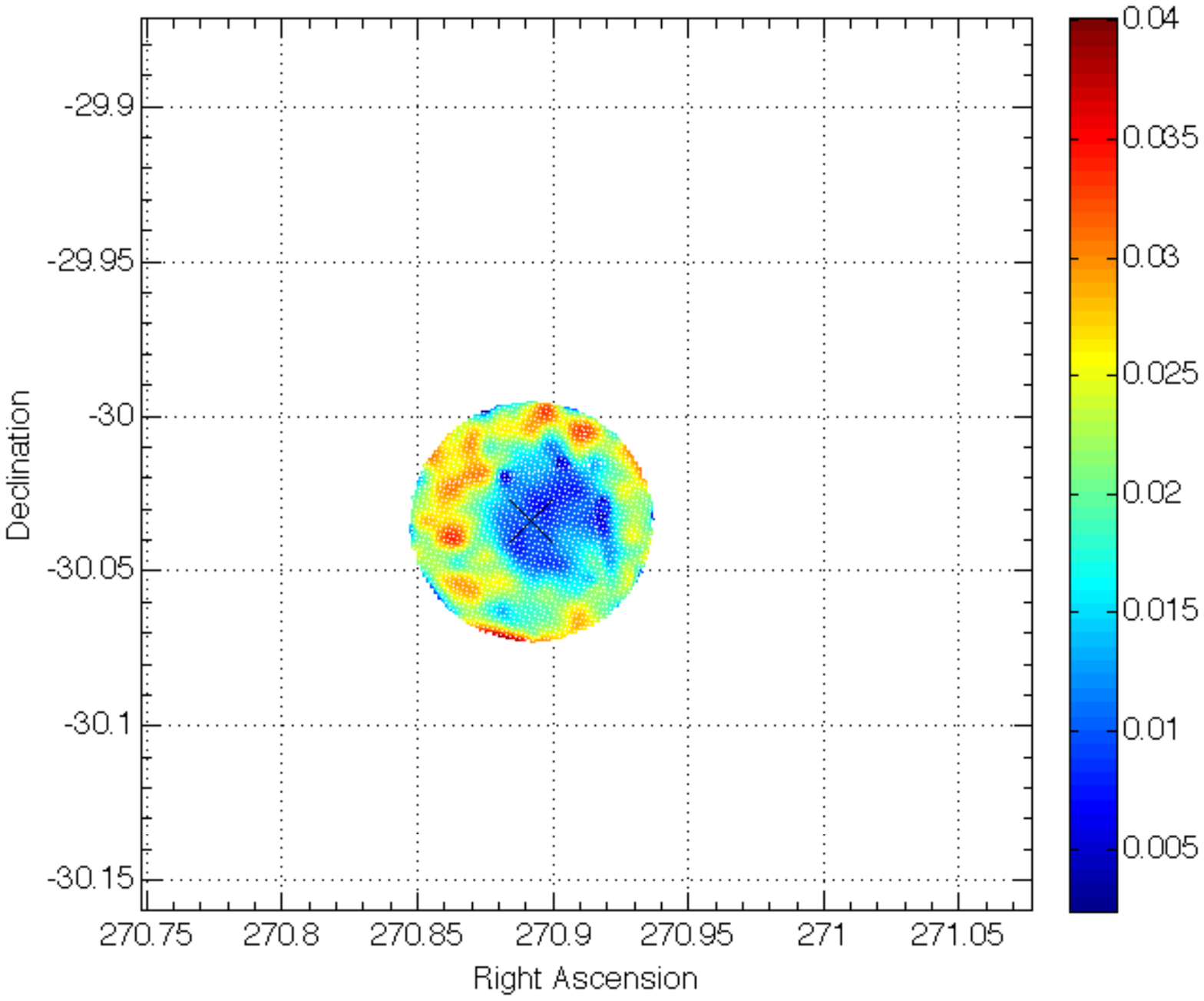} \\
\end{tabular}
\caption{\footnotesize As in Figure \ref{figngc6121}, but for the
  cluster NGC 6522. Only our Magellan photometry was used to build the
  CMDs in both colors.}
\label{figngc6522}
\end{figure}

\begin{figure}[htbp]
%\epsscale{0.77}
\plotone{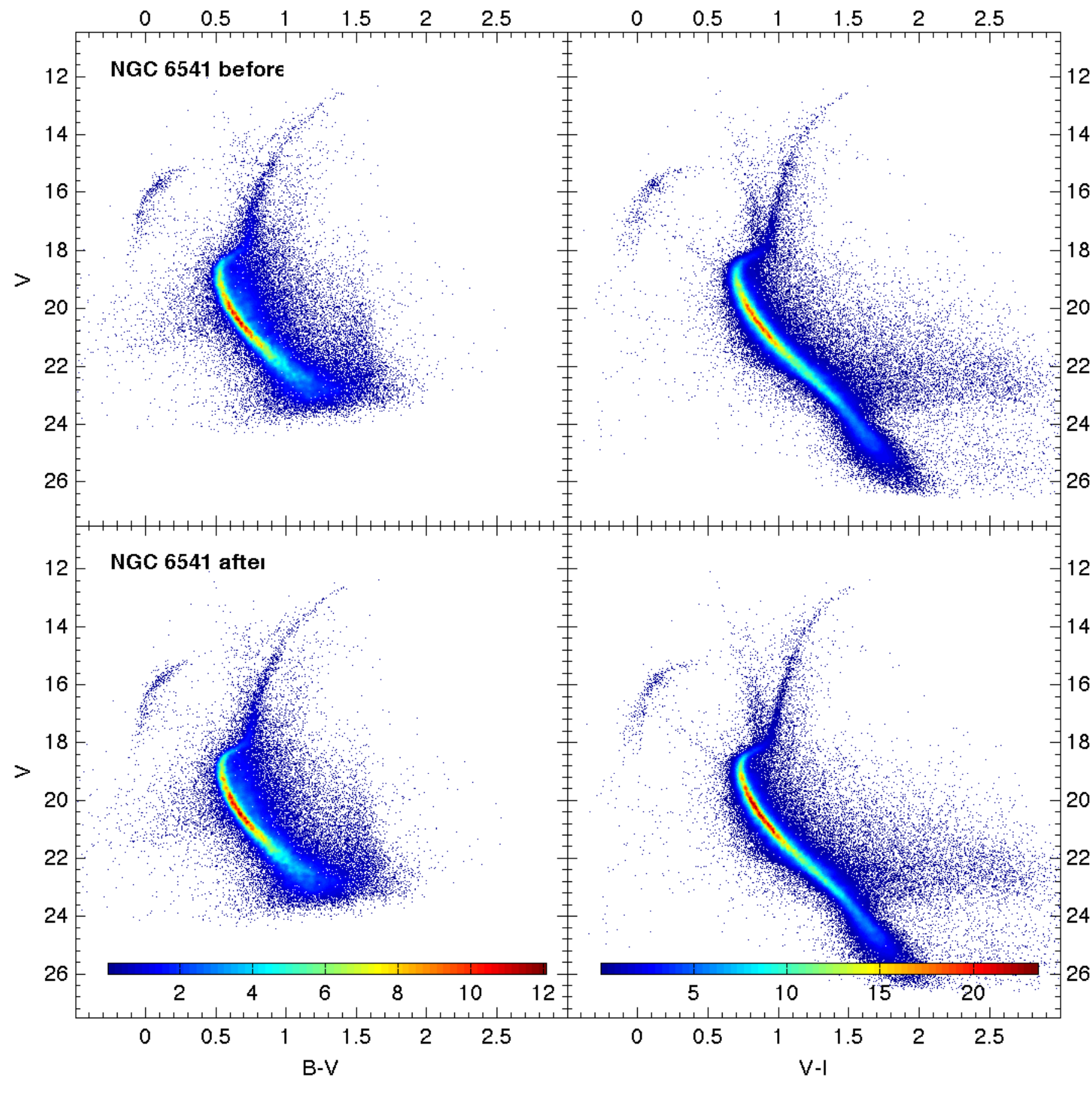}
\begin{tabular}{ccc}
\includegraphics[scale=0.28]{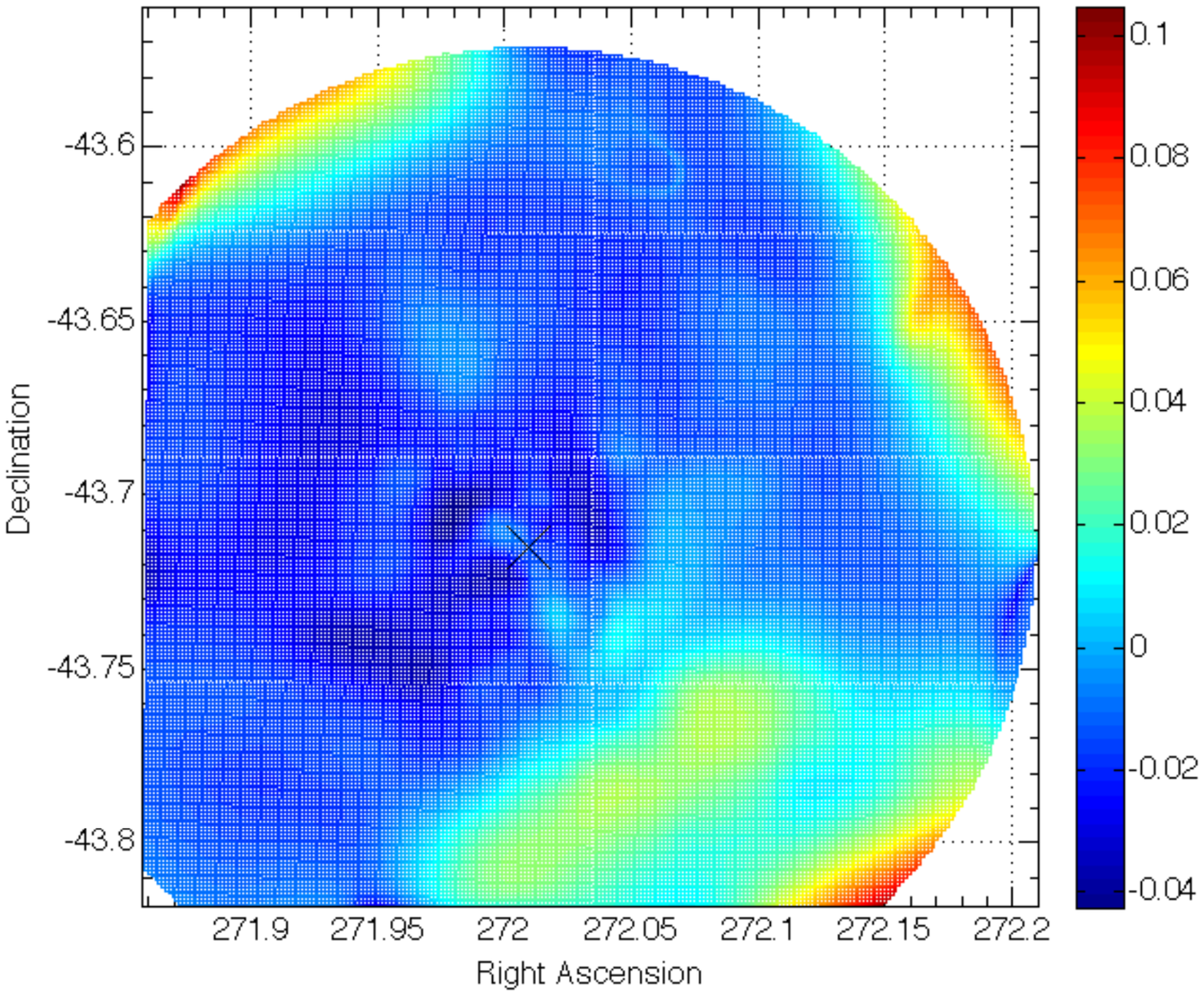}  &
\includegraphics[scale=0.28]{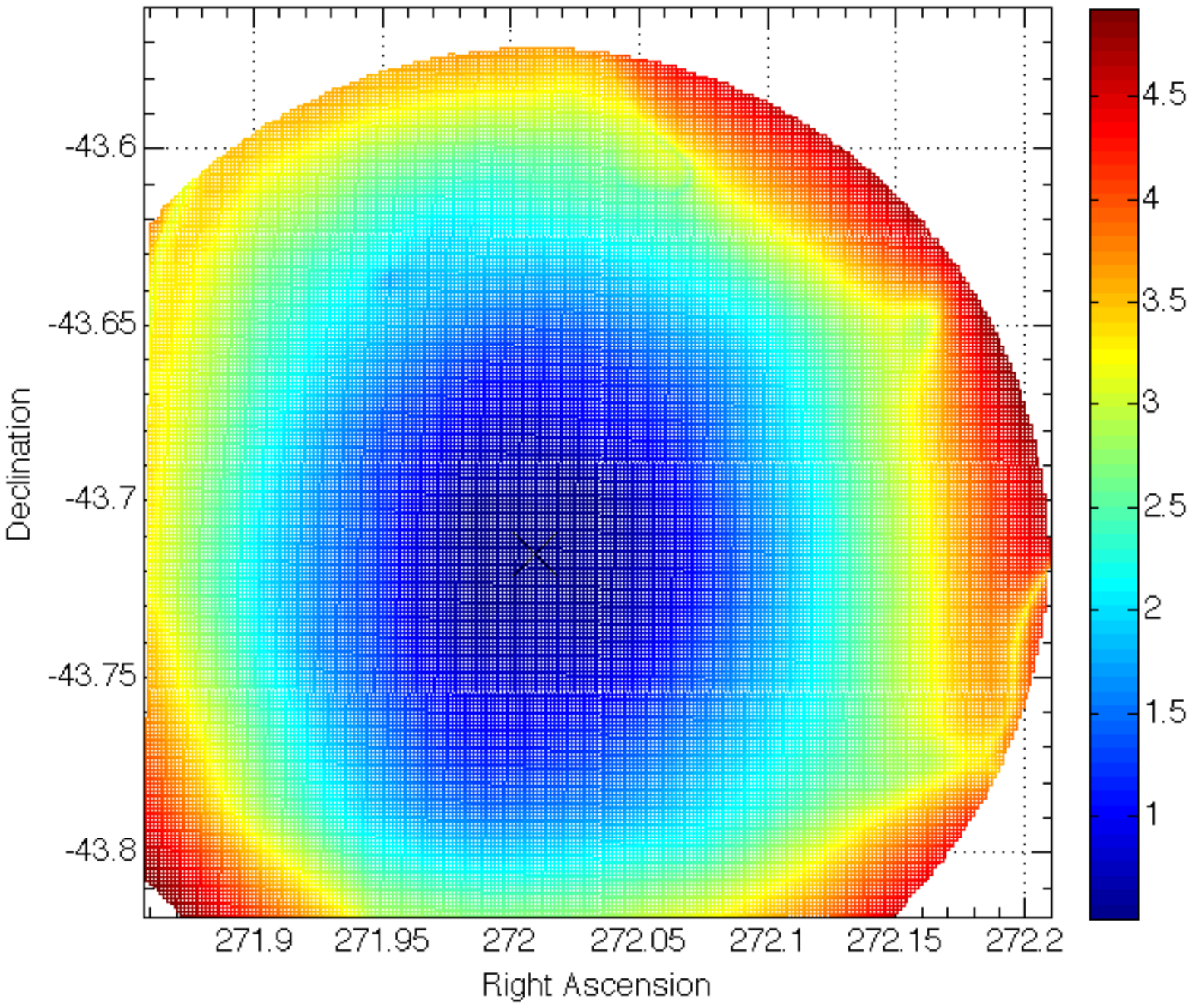}  &
\includegraphics[scale=0.28]{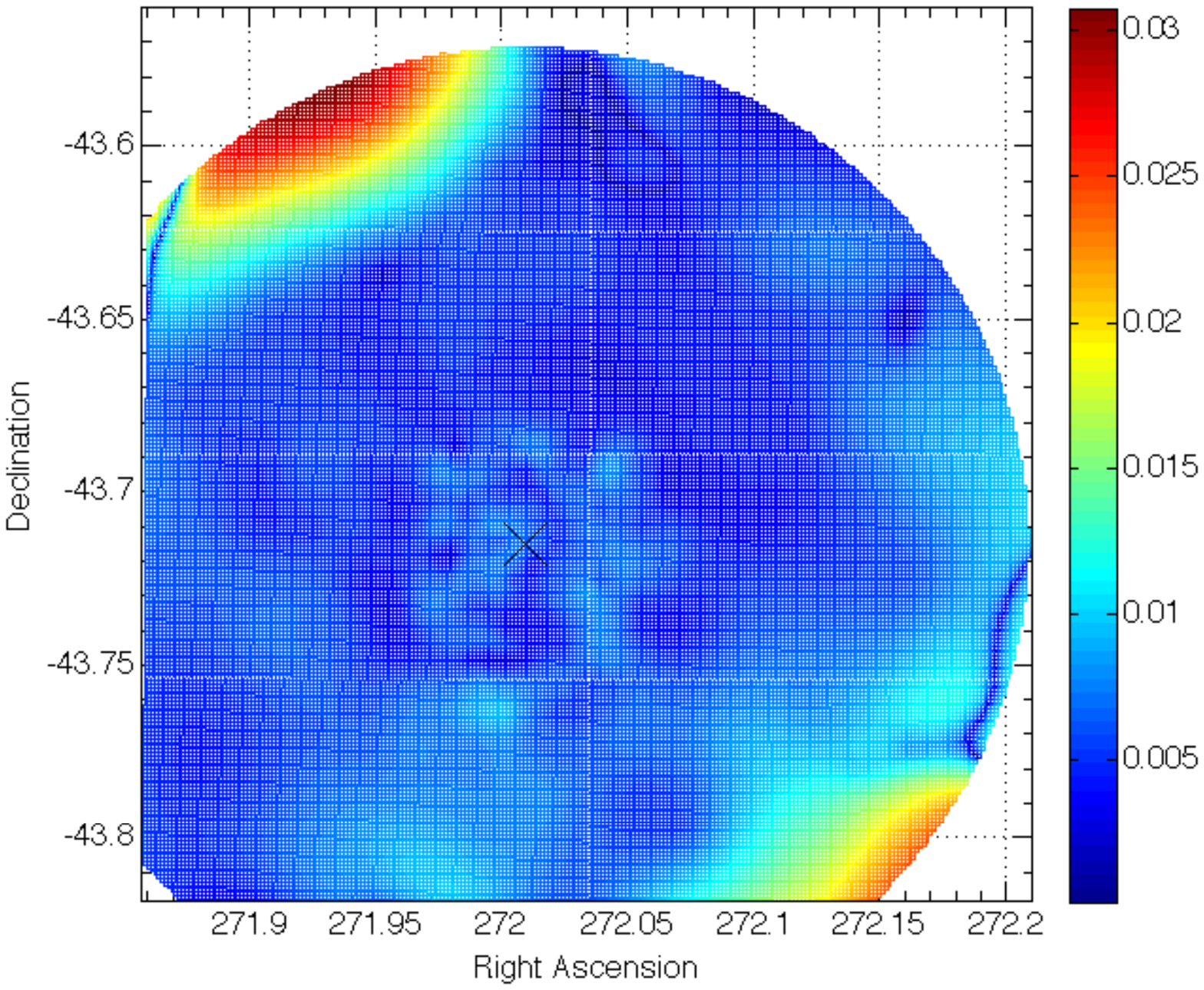} \\
\end{tabular}
\caption{\footnotesize As in Figure \ref{figngc6121}, but for the
  cluster NGC 6541. Only our Magellan photometry was used to build the
  $B-V$ vs. $V$ CMD. ACS photometry (from project 10775) and Magellan
  photometry were used to build the $V-I$ vs. $V$ CMD.}
\label{figngc6541}
\end{figure}

\begin{figure}[htbp]
%\epsscale{0.77}
\plotone{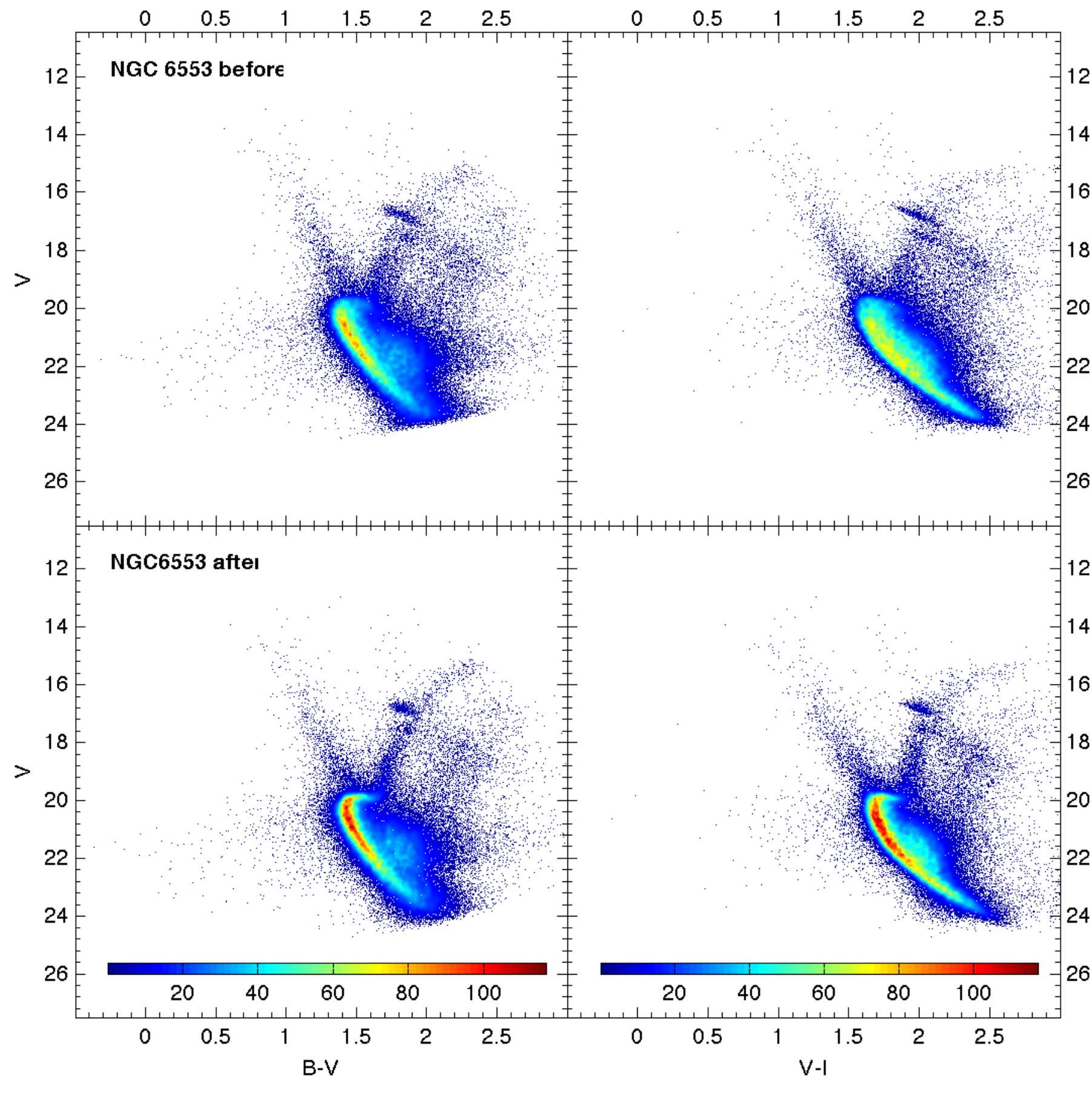}
\begin{tabular}{ccc}
\includegraphics[scale=0.28]{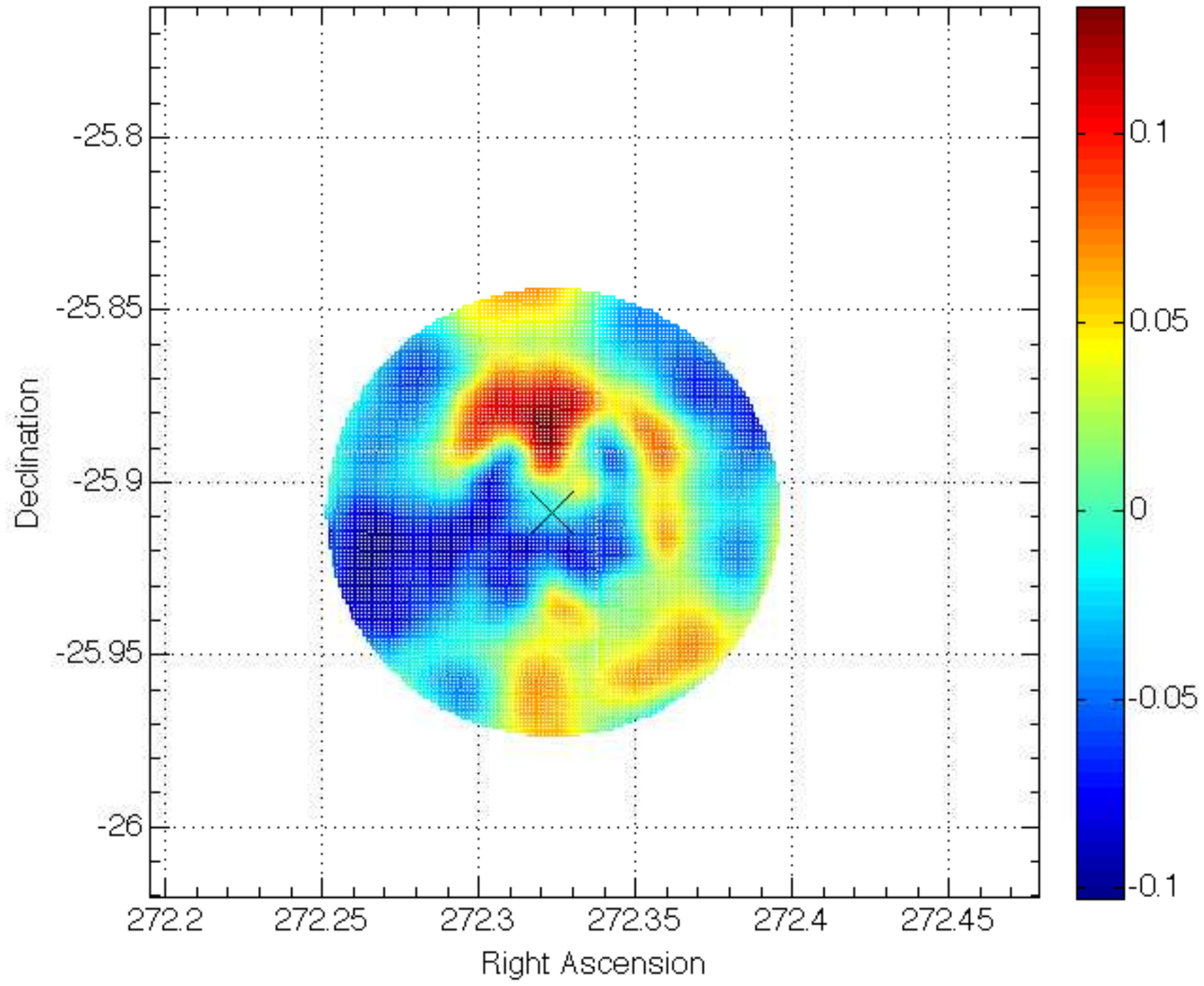}  &
\includegraphics[scale=0.28]{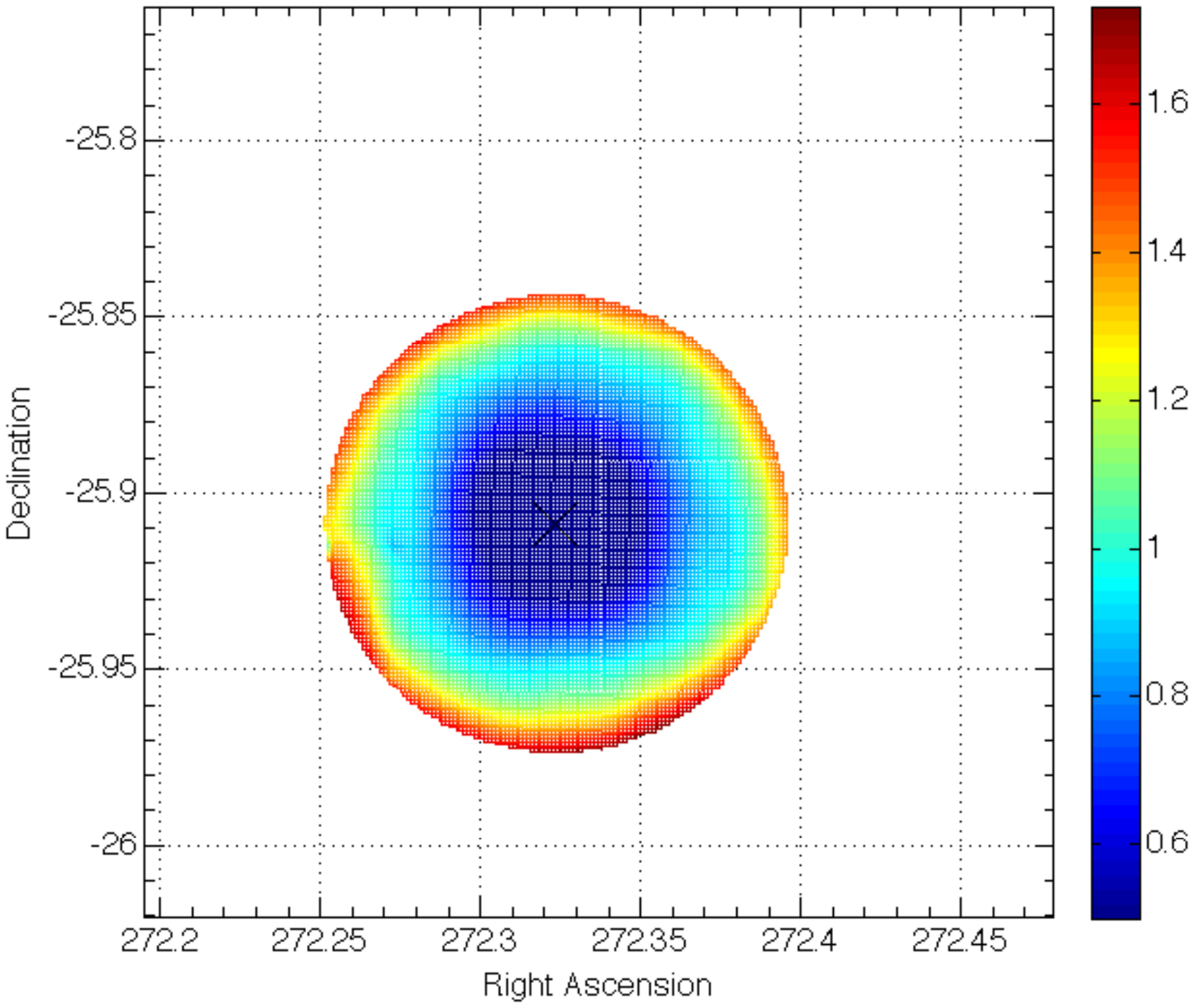}  &
\includegraphics[scale=0.28]{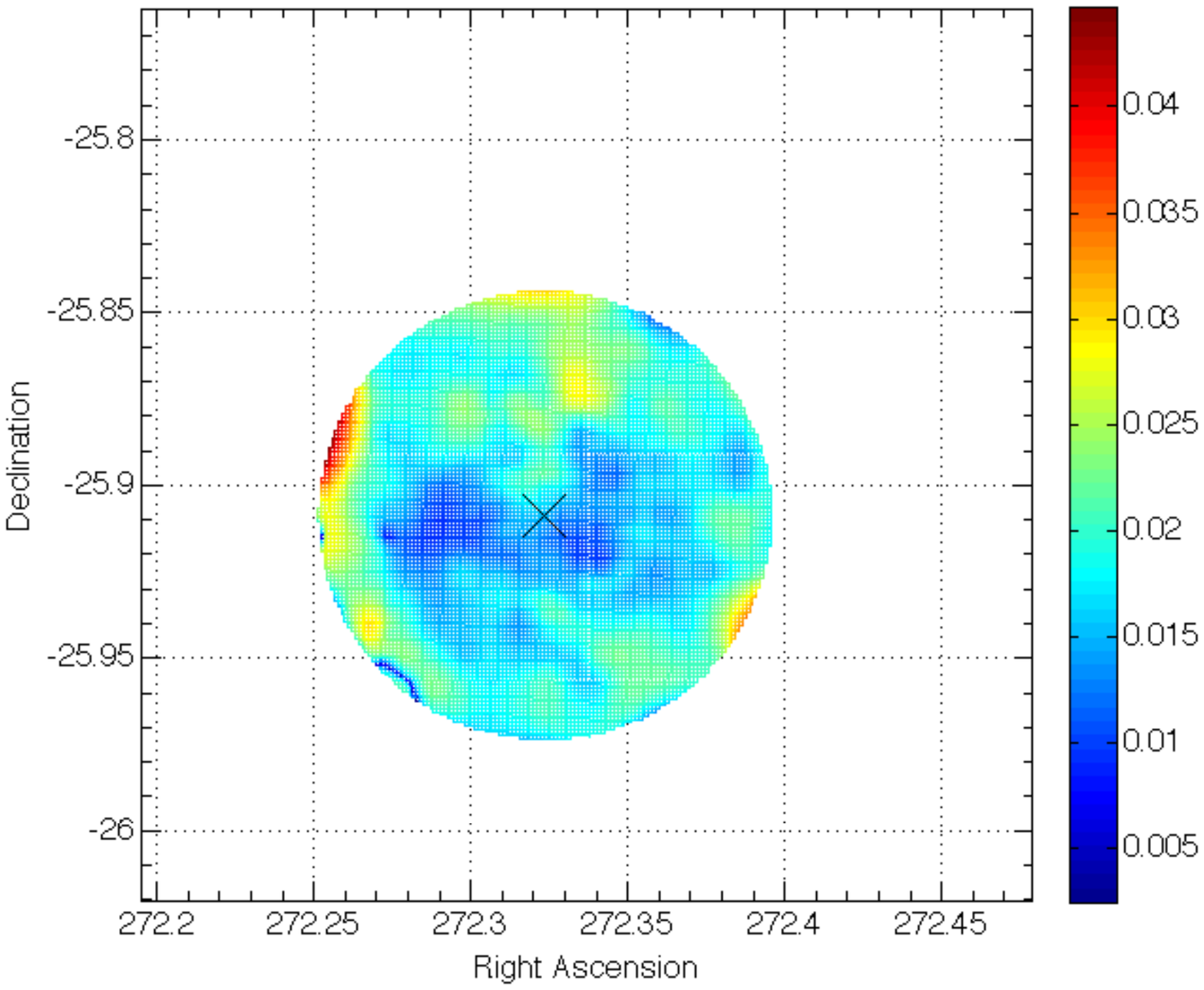} \\
\end{tabular}
\caption{\footnotesize As in Figure \ref{figngc6121}, but for the
  cluster NGC 6553. ACS photometry (from our project 10573) and our
  Magellan photometry were used to build the CMDs in both colors.}
\label{figngc6553}
\end{figure}

\begin{figure}[htbp]
%\epsscale{0.77}
\plotone{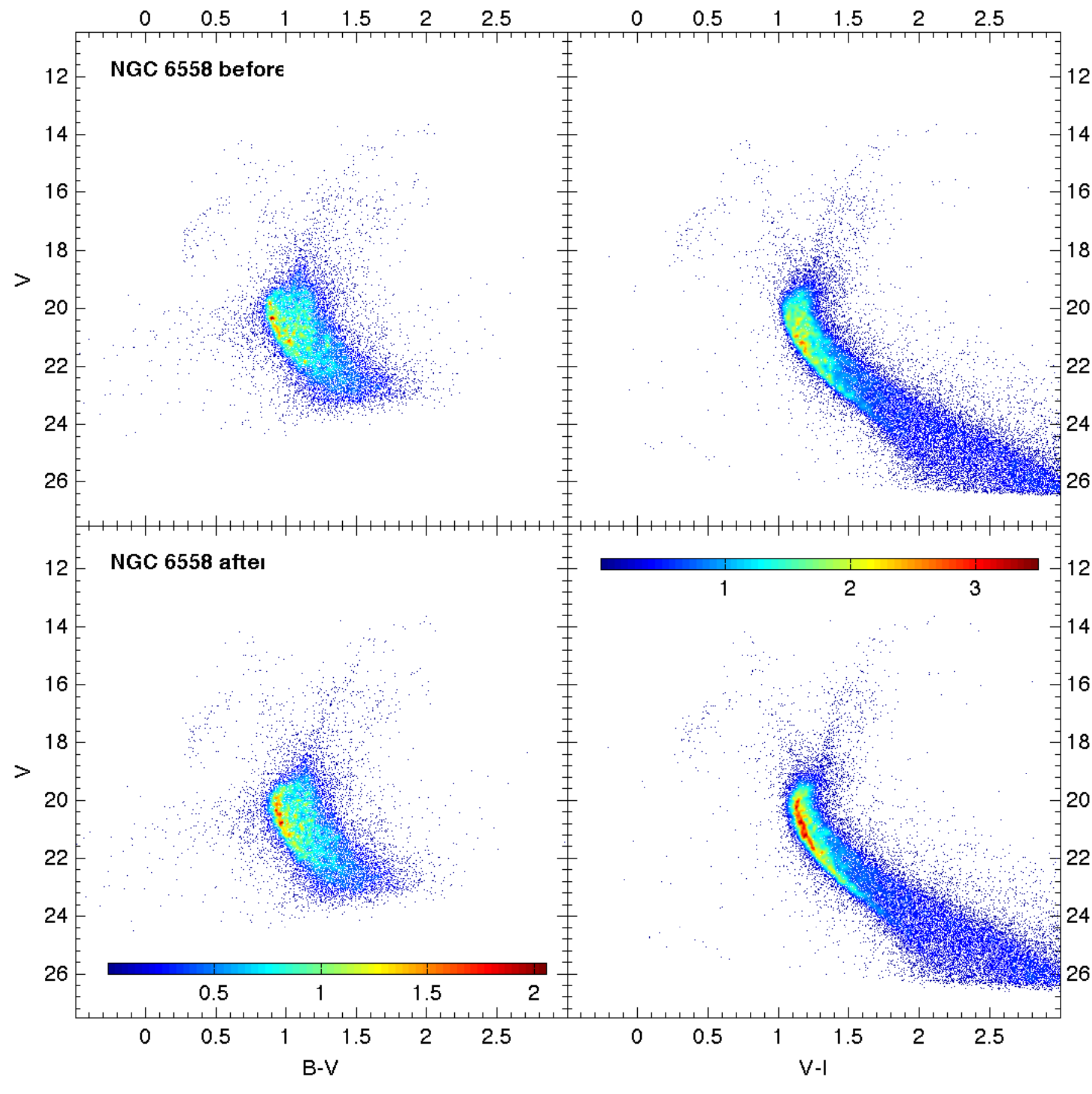}
\begin{tabular}{ccc}
\includegraphics[scale=0.28]{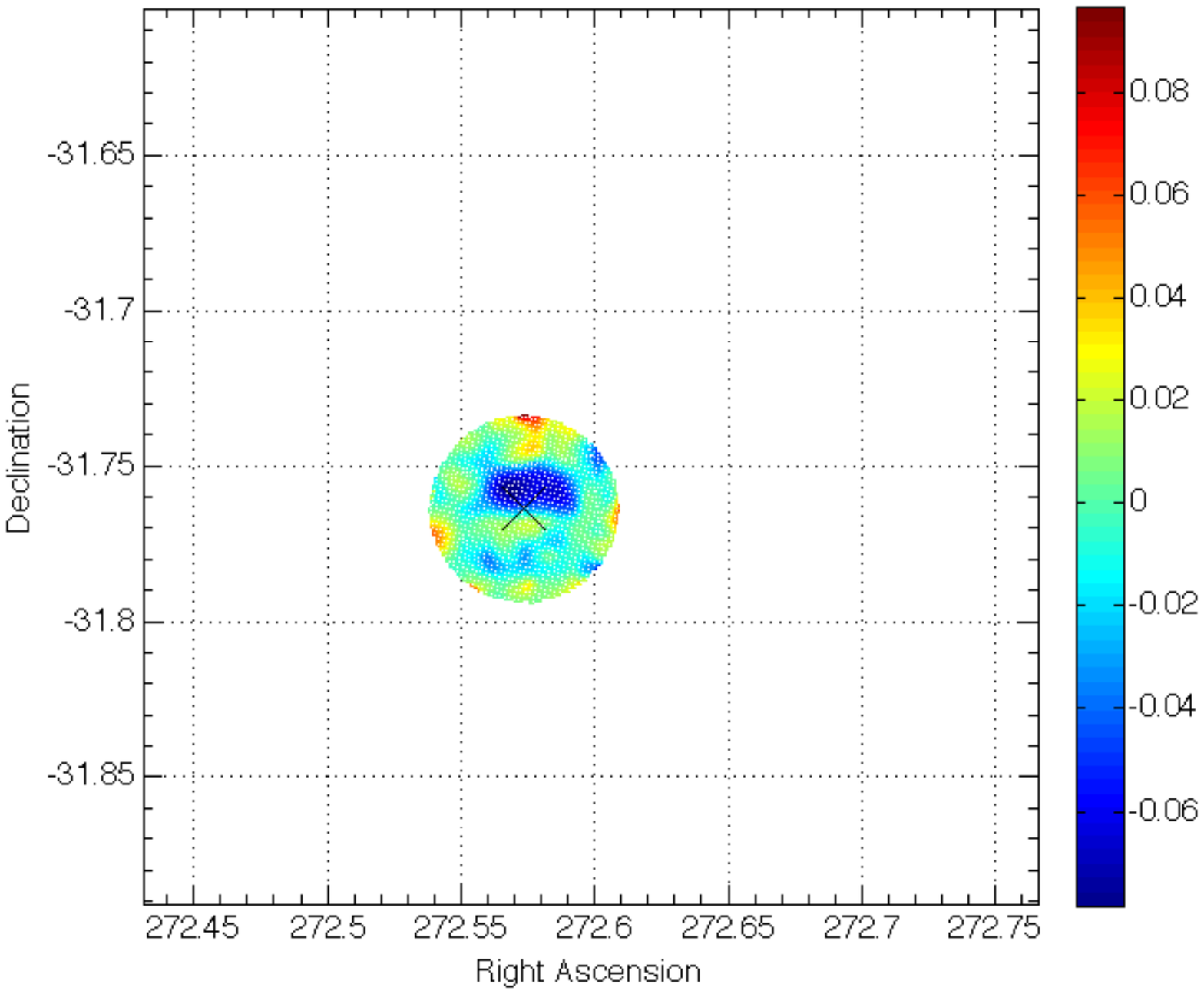}  &
\includegraphics[scale=0.28]{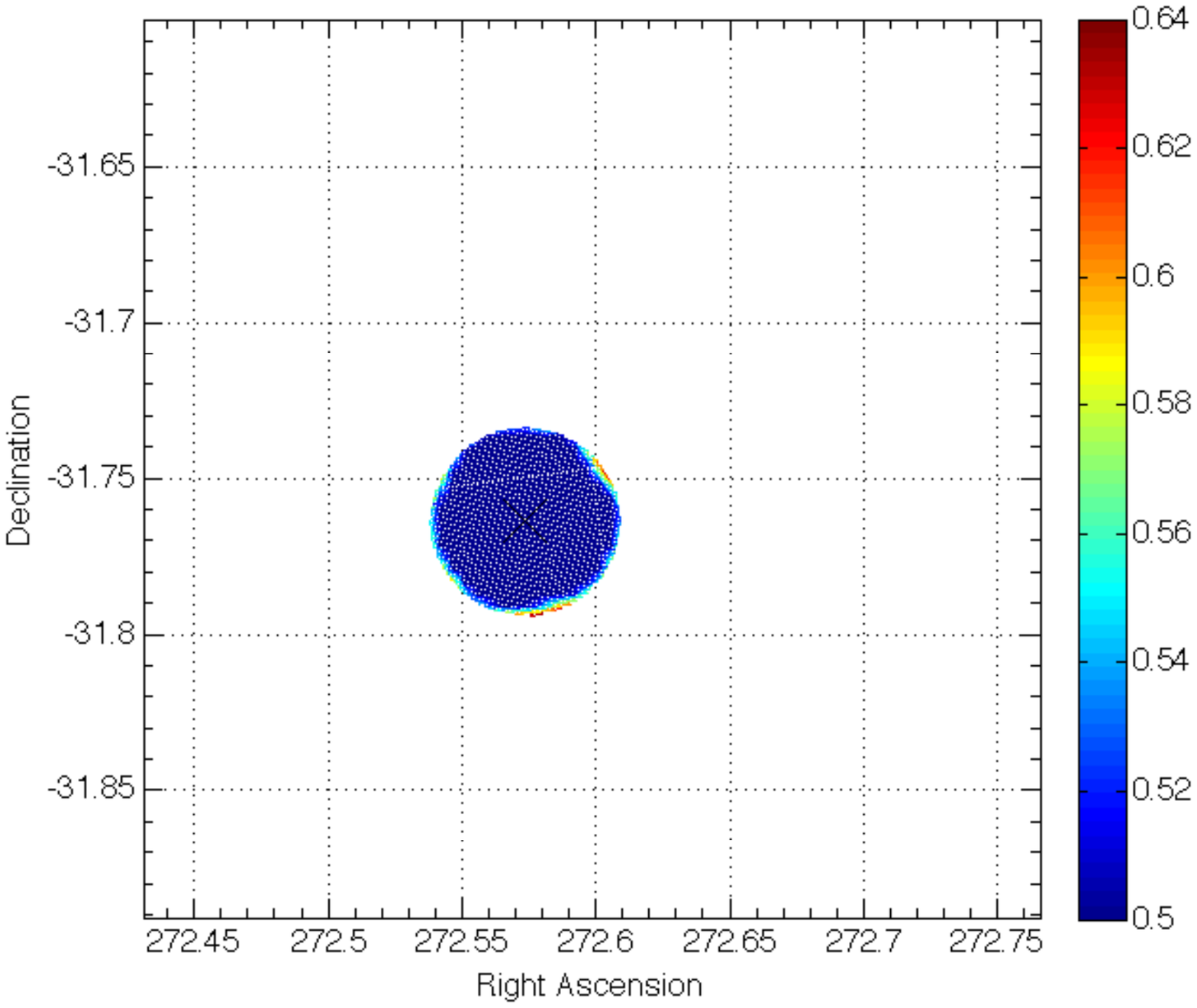}  &
\includegraphics[scale=0.28]{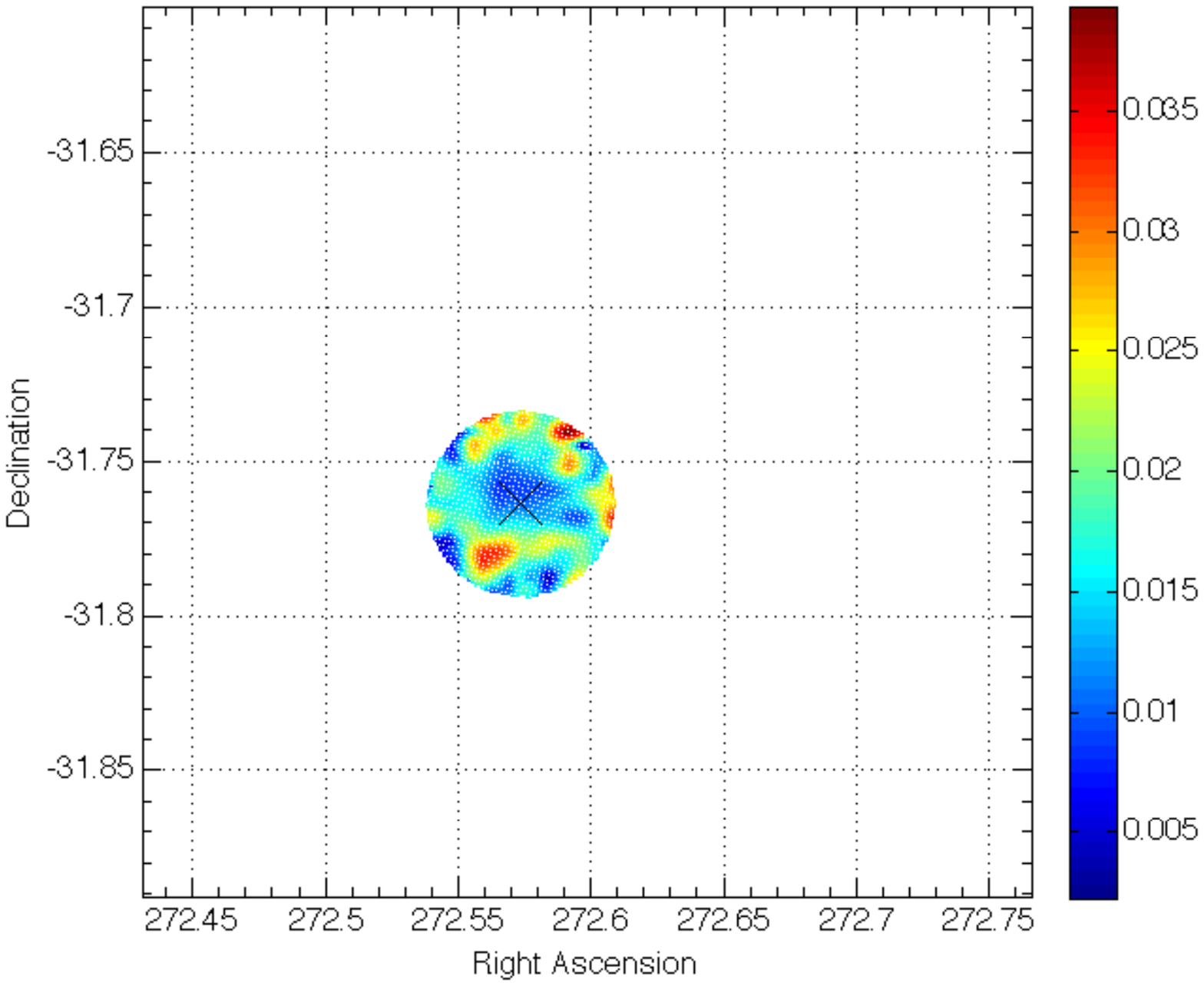} \\
\end{tabular}
\caption{\footnotesize As in Figure \ref{figngc6121}, but for the
  cluster NGC 6558. Only our Magellan photometry was used to build the
  $B-V$ vs. $V$ CMD. ACS photometry (from project 9799) and Magellan
  photometry were used to build the $V-I$ vs. $V$ CMD. Notice that the
  $B-V$ vs. $V$ CMD could not be correctly calibrated in color using
  the method described in the text because of the lack of calibrating
  data in the $B$ filter.}
\label{figngc6558}
\end{figure}

\begin{figure}[htbp]
%\epsscale{0.77}
\plotone{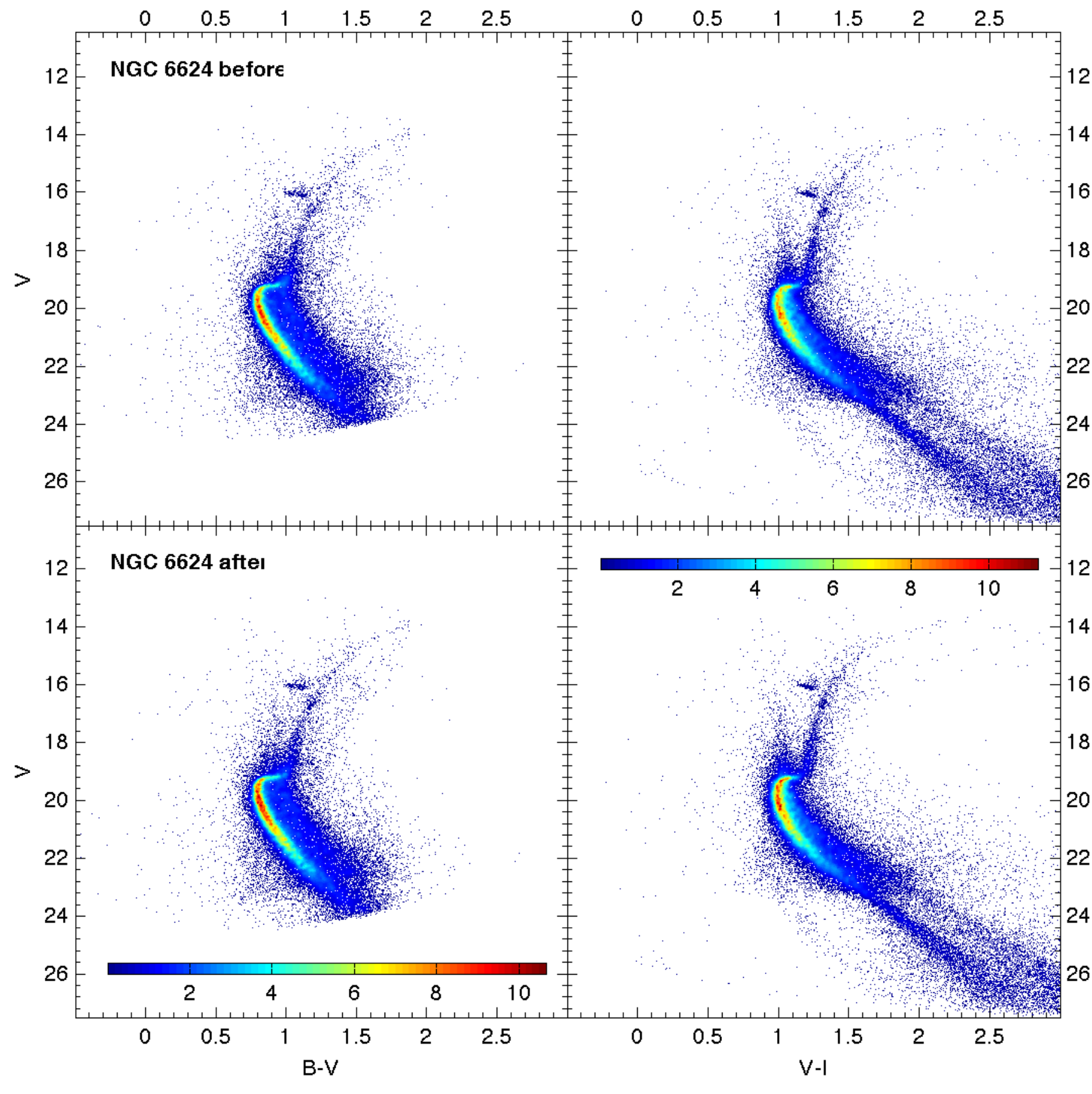}
\begin{tabular}{ccc}
\includegraphics[scale=0.28]{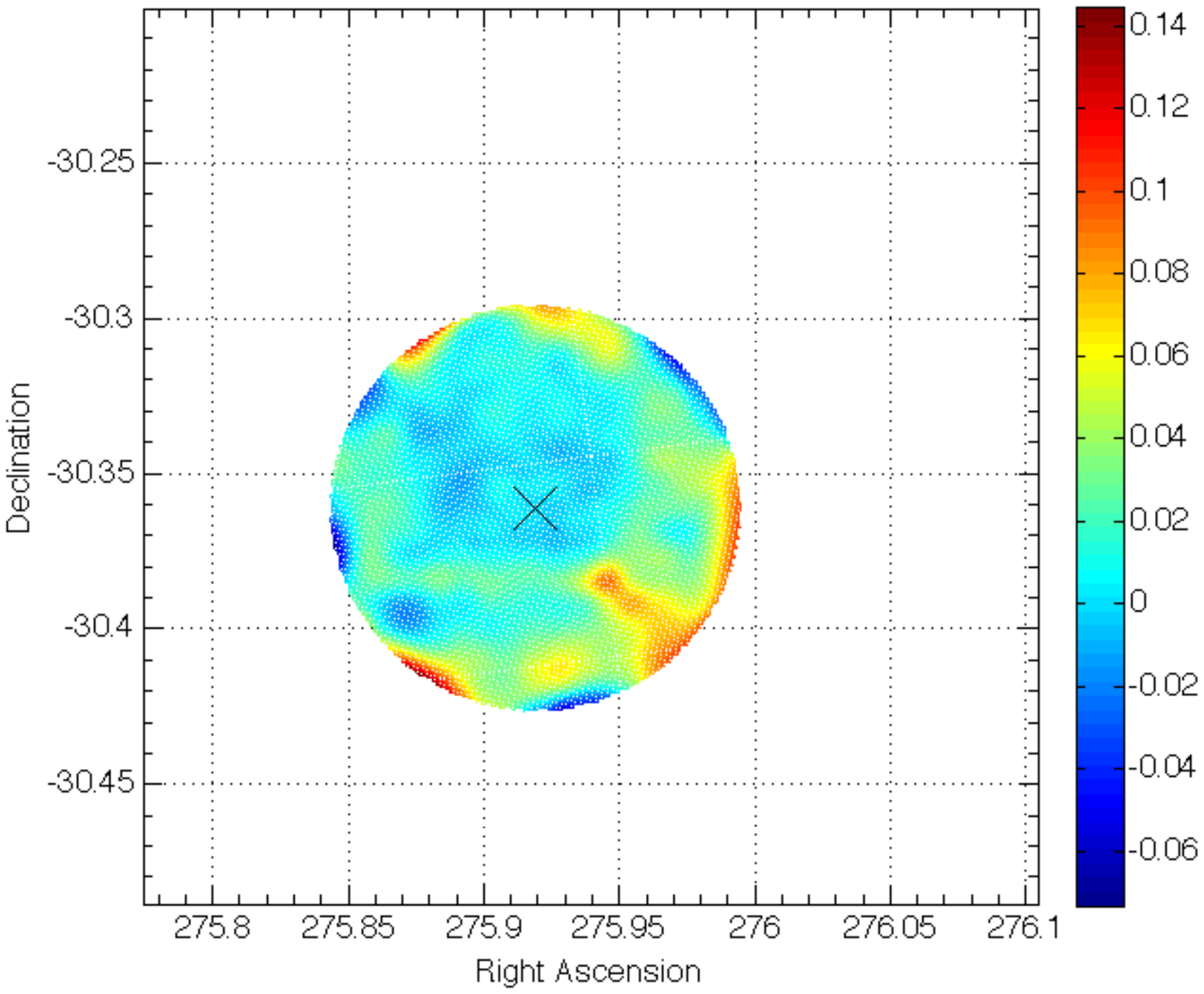}  &
\includegraphics[scale=0.28]{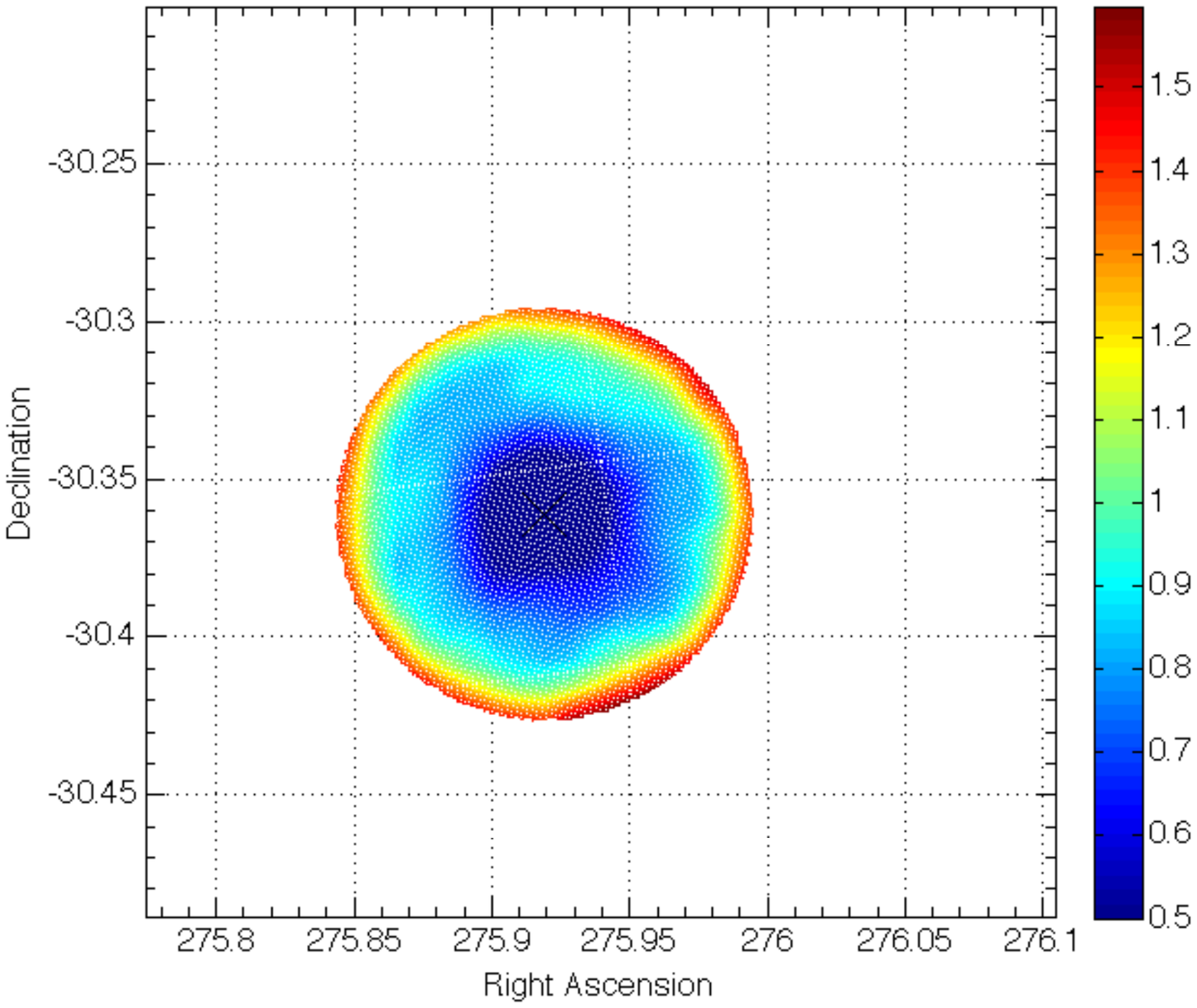}  &
\includegraphics[scale=0.28]{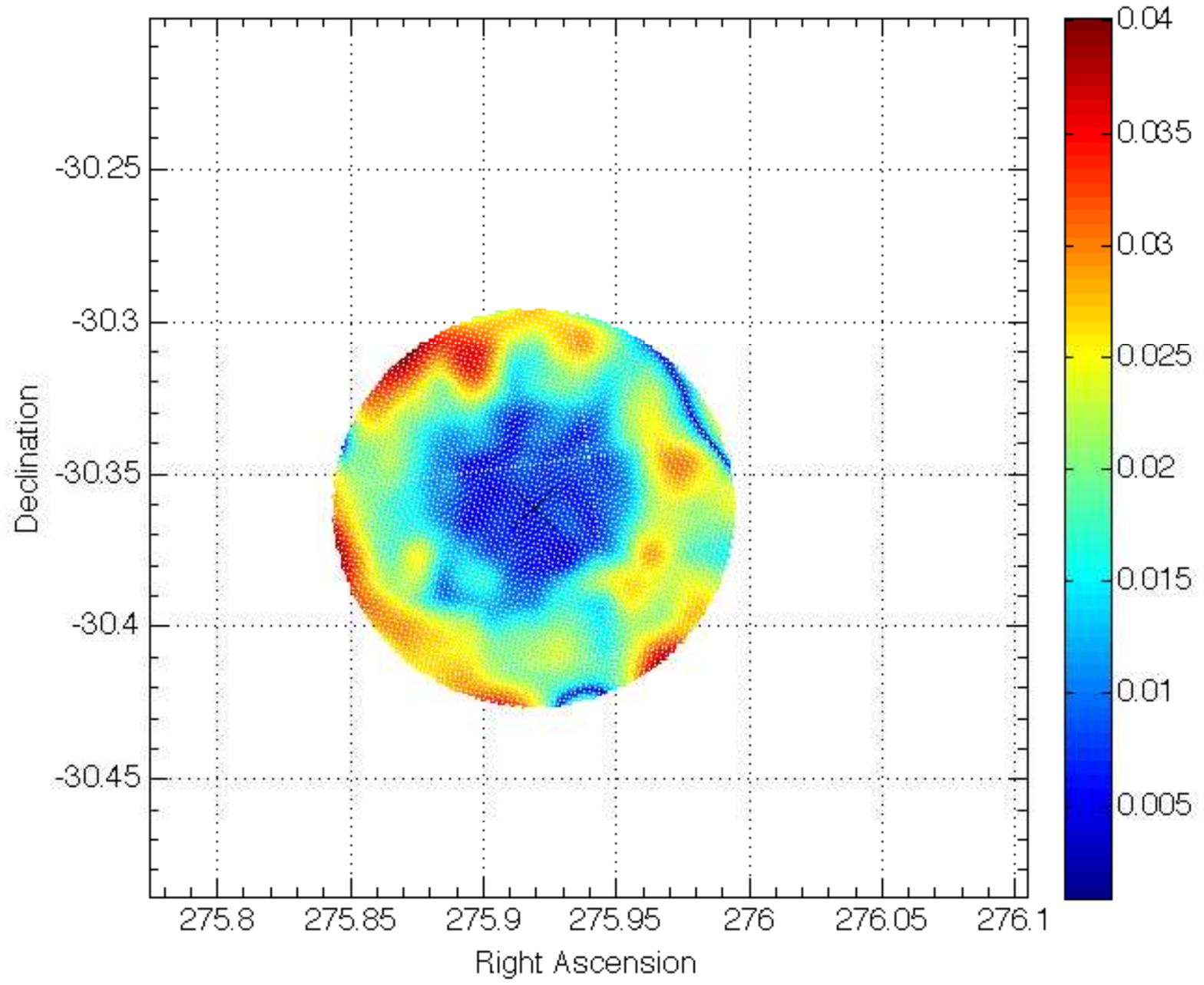} \\
\end{tabular}
\caption{\footnotesize As in Figure \ref{figngc6121}, but for the
  cluster NGC 6624. ACS photometry (from our project 10573) and
  Magellan photometry were used to build the $B-V$ vs. $V$ CMD. ACS
  photometry (from project 10775) and Magellan photometry were used to
  build the $V-I$ vs. $V$ CMD.}
\label{figngc6624}
\end{figure}

\begin{figure}[htbp]
%\epsscale{0.77}
\plotone{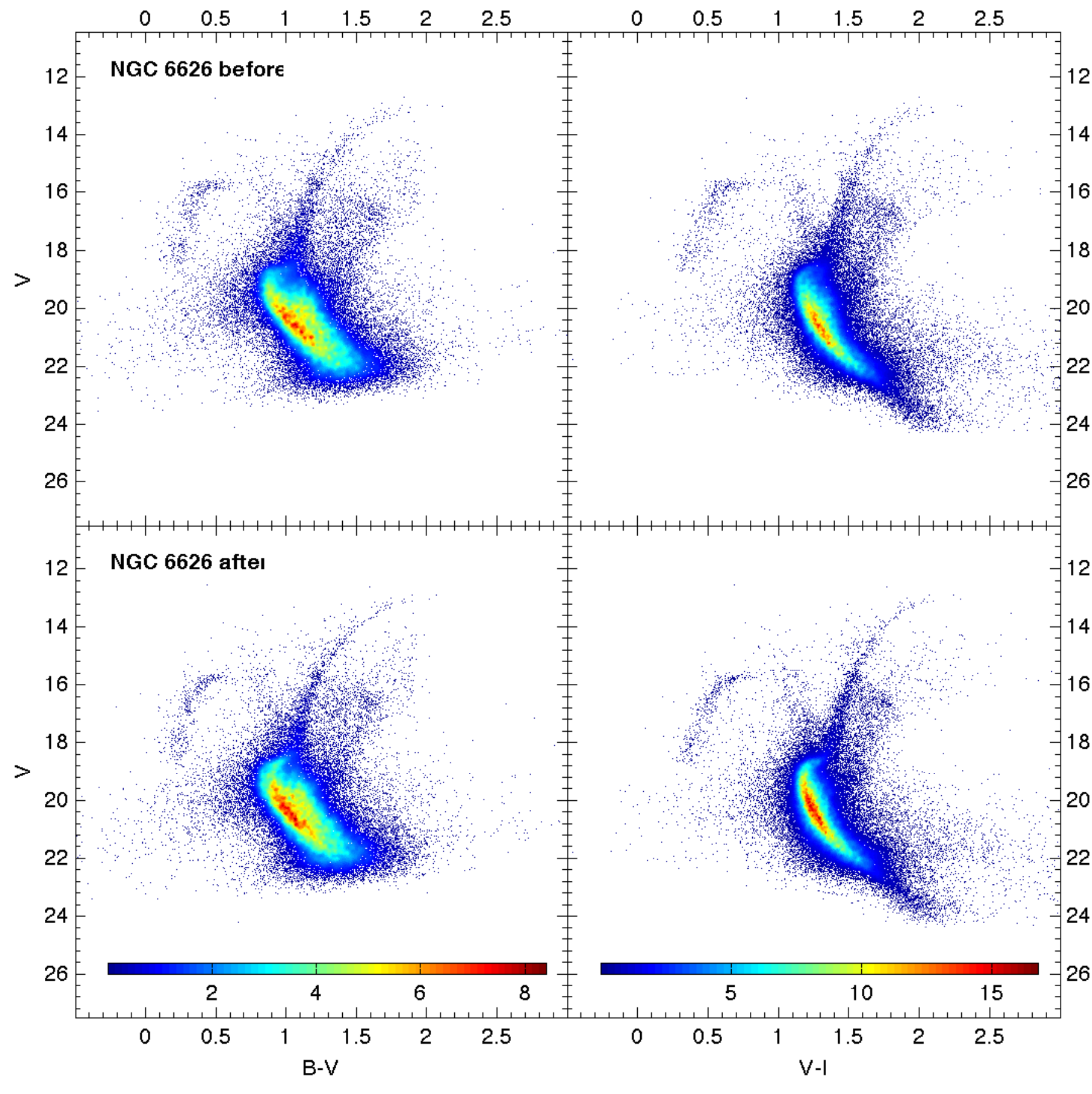}
\begin{tabular}{ccc}
\includegraphics[scale=0.28]{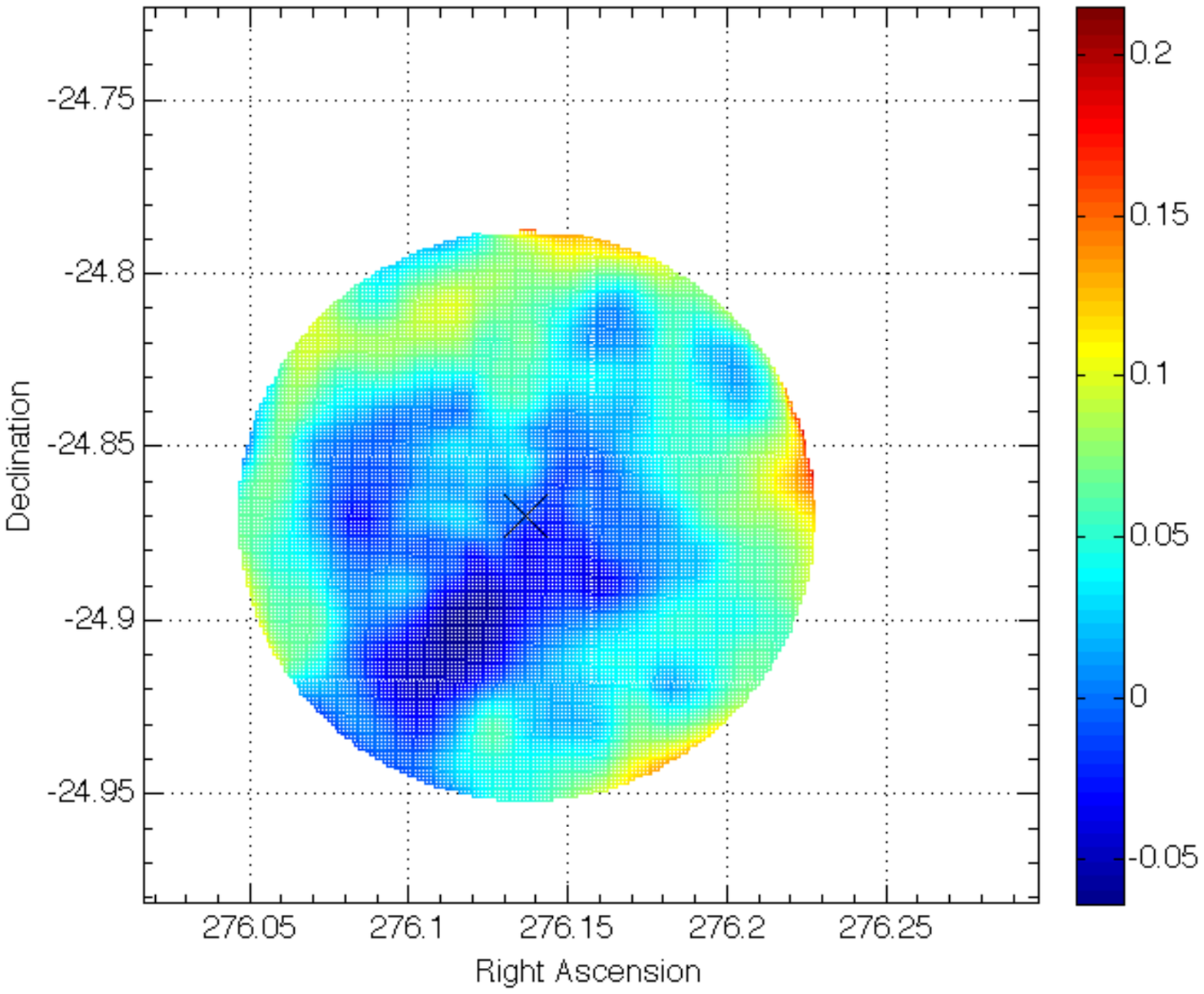}  &
\includegraphics[scale=0.28]{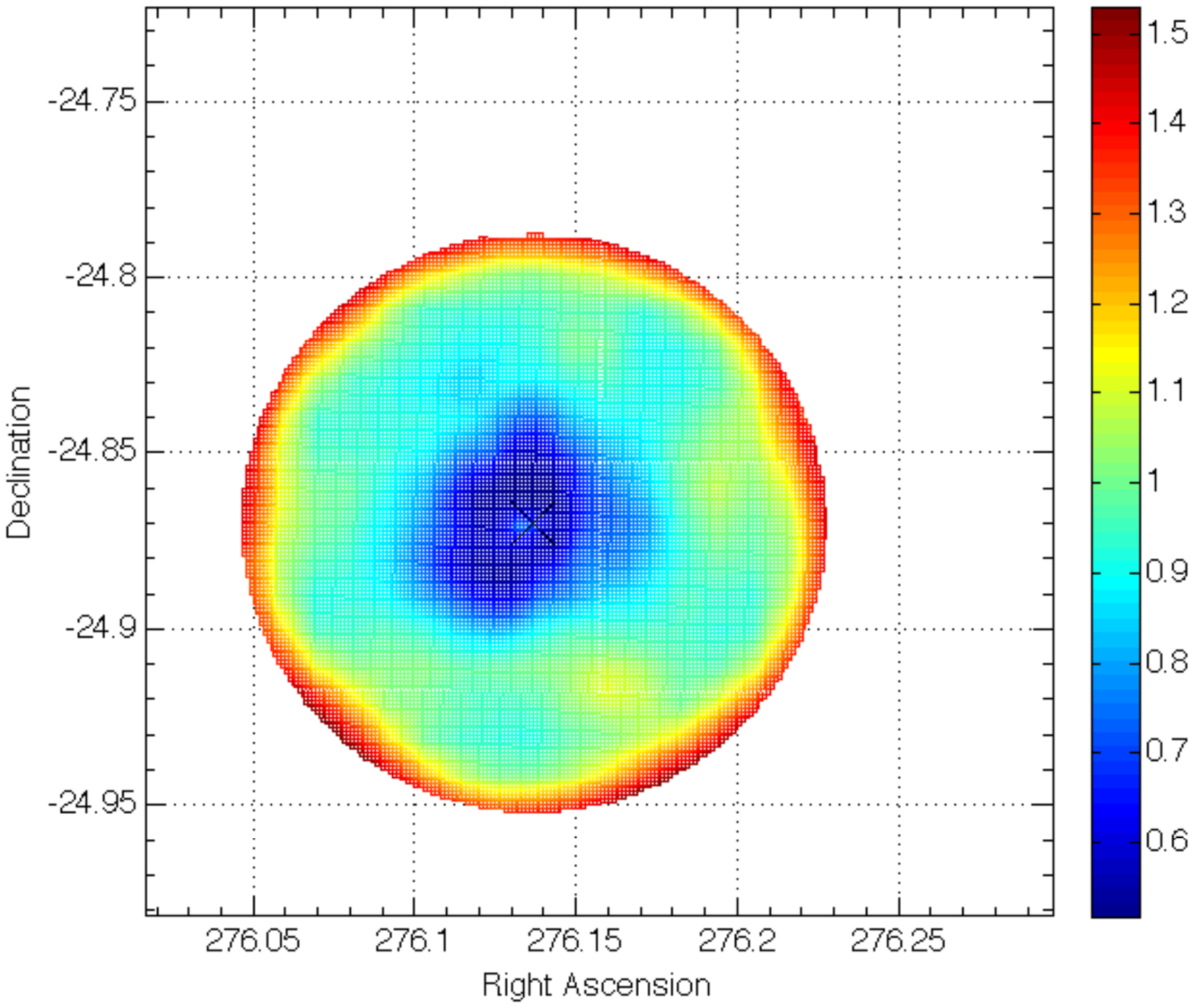}  &
\includegraphics[scale=0.28]{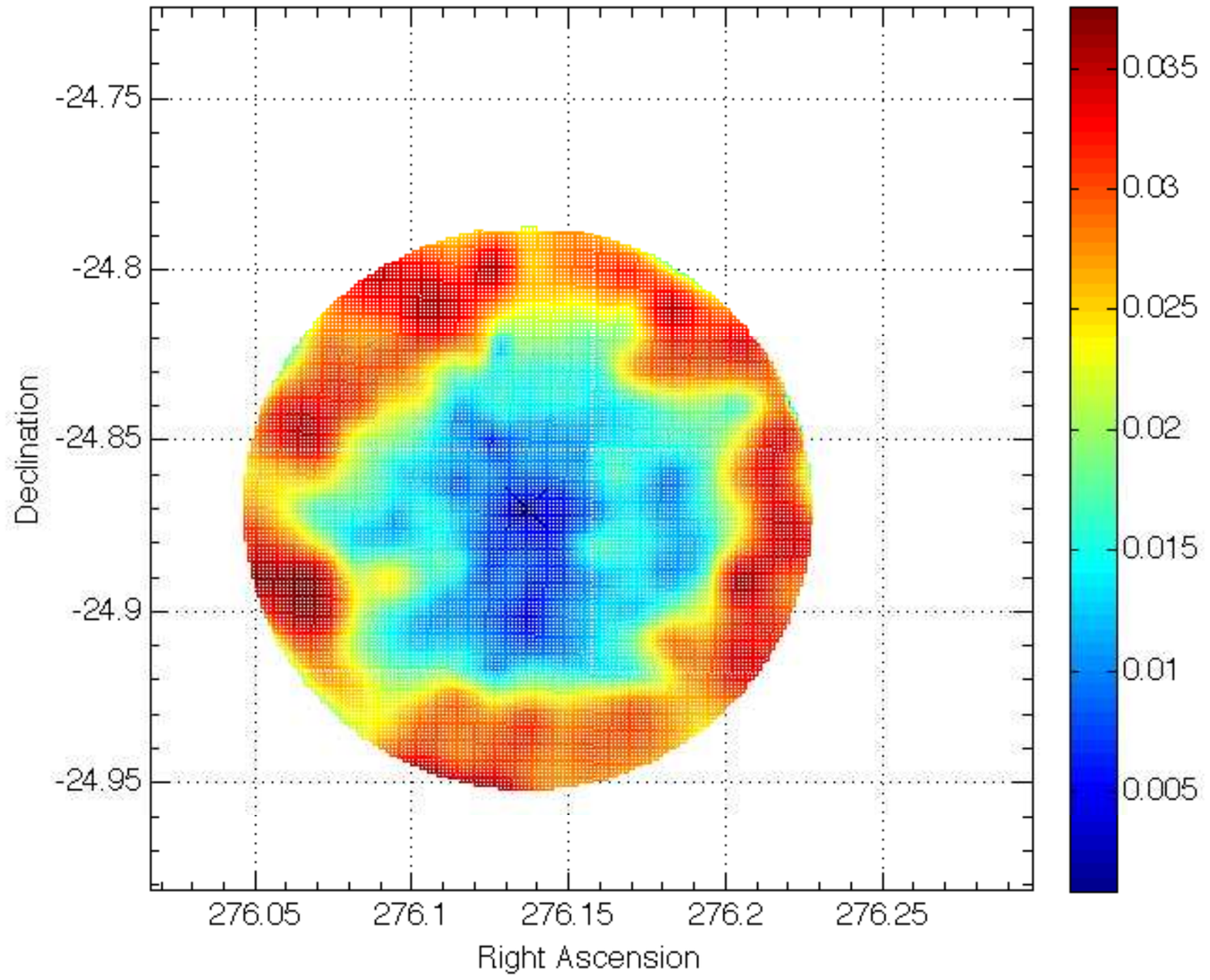} \\
\end{tabular}
\caption{\footnotesize As in Figure \ref{figngc6121}, but for the
  cluster NGC 6626 - M 28. Only our Magellan photometry was used to
  build the $B-V$ vs. $V$ CMD. WFPC2 photometry (from project 6779)
  and Magellan photometry were used to build the $V-I$ vs. $V$
  CMD. Notice that the $B-V$ vs. $V$ CMD could not be correctly
  calibrated in color using the method described in the text because
  of the lack of calibrating data in the $B$ filter.}
\label{figngc6626}
\end{figure}

\begin{figure}[htbp]
%\epsscale{0.77}
\plotone{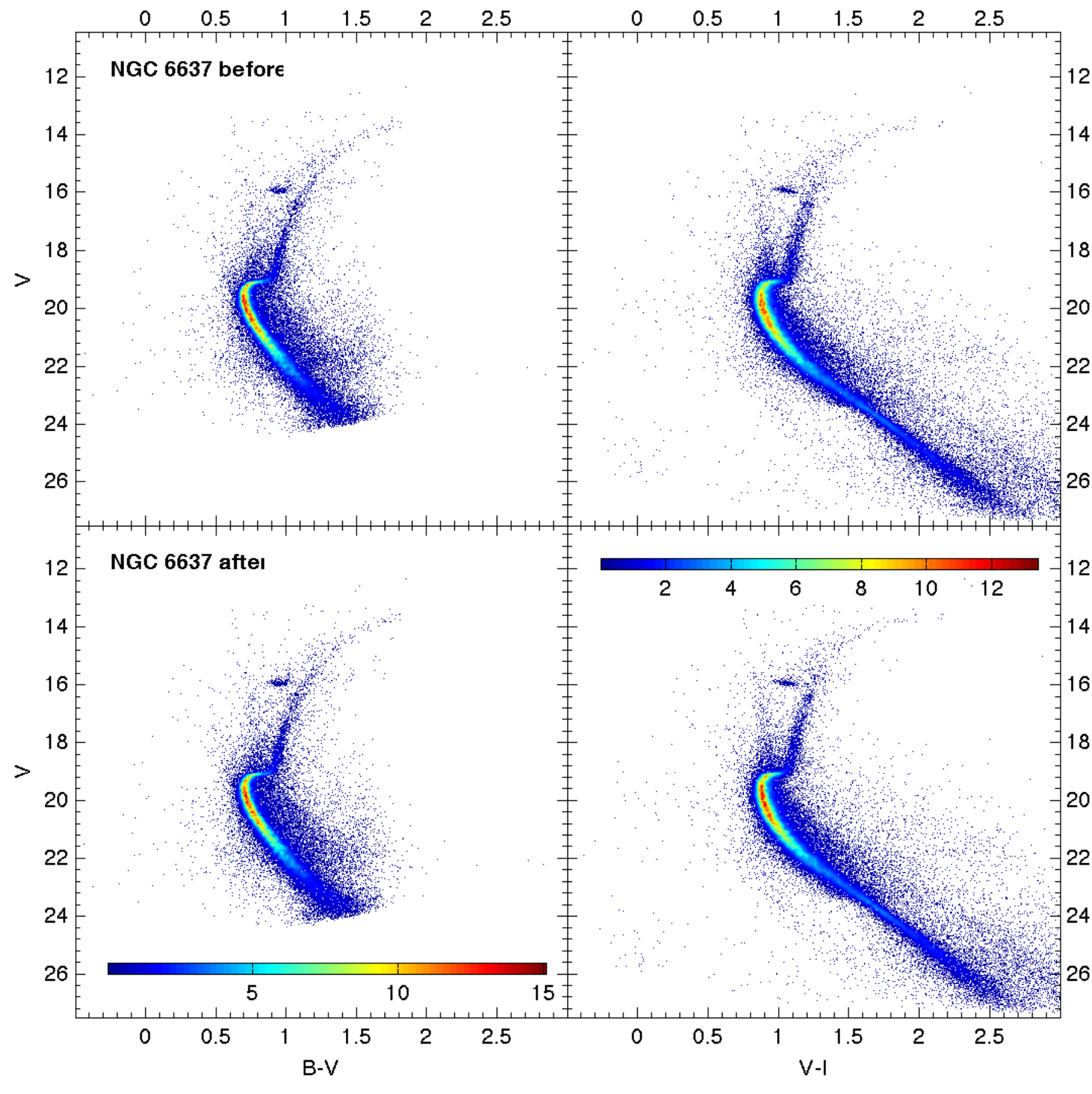}
\begin{tabular}{ccc}
\includegraphics[scale=0.28]{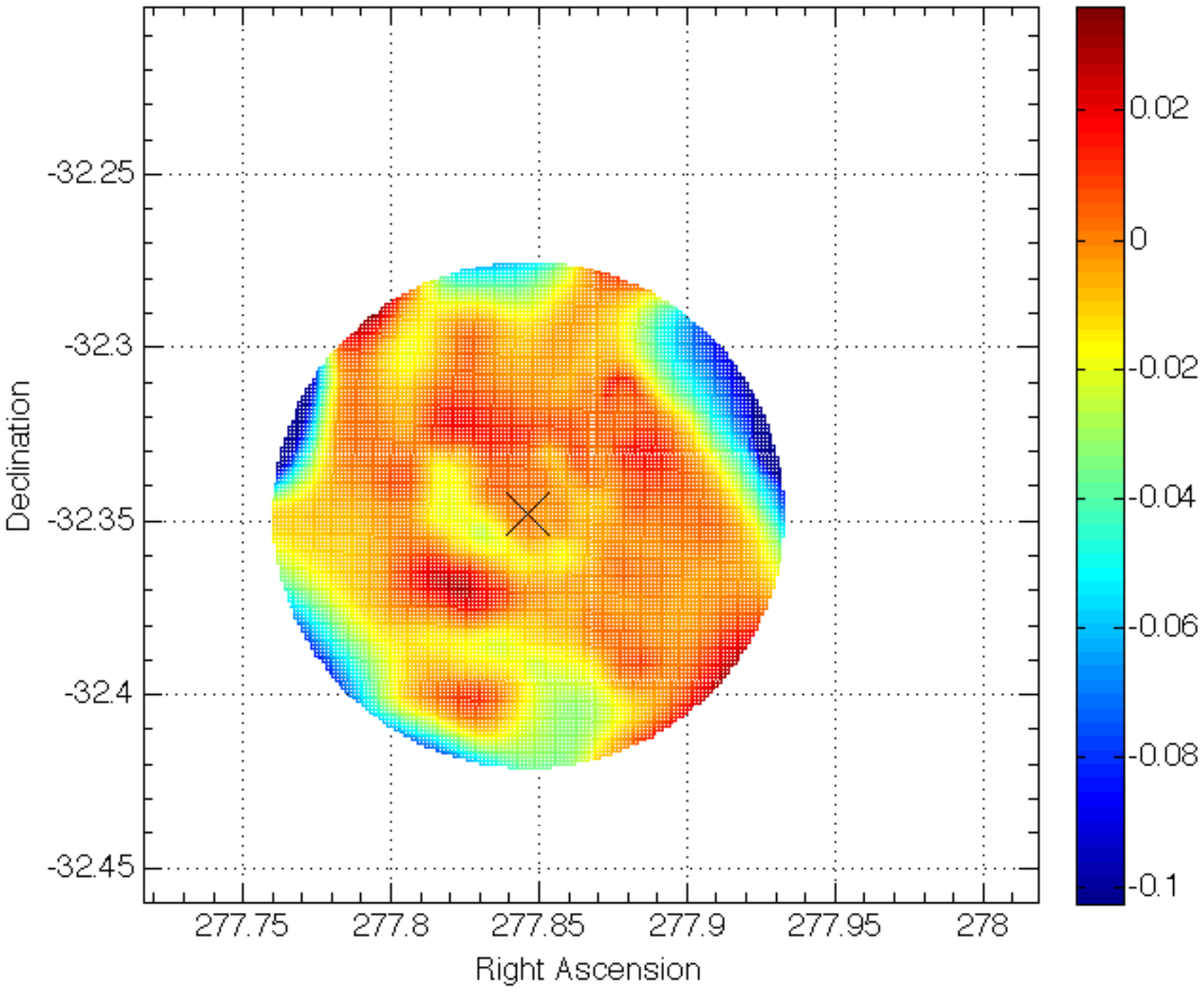}  &
\includegraphics[scale=0.28]{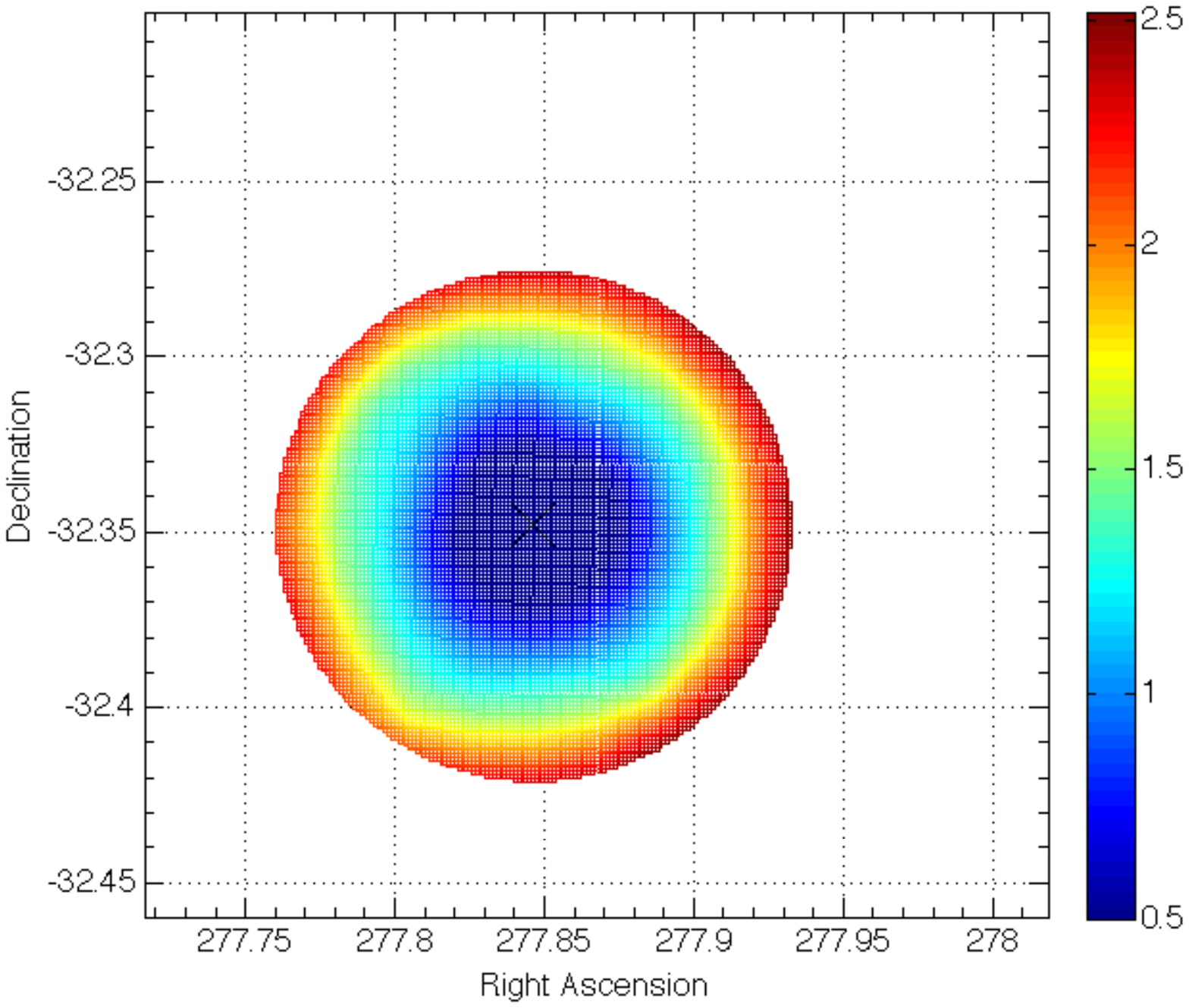}  &
\includegraphics[scale=0.28]{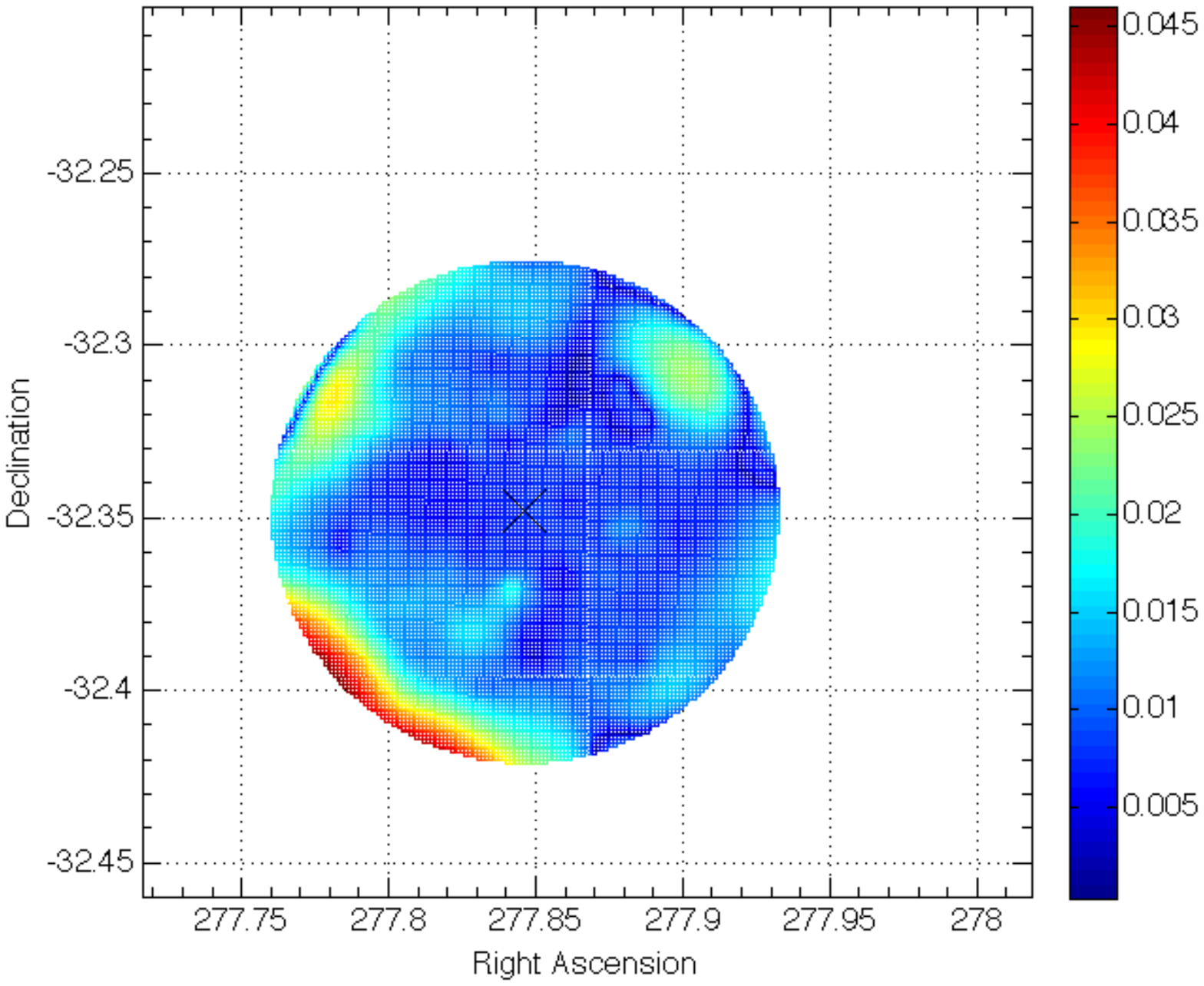} \\
\end{tabular}
\caption{\footnotesize As in Figure \ref{figngc6121}, but for the
  cluster NGC 6637 - M 69. ACS photometry (from our project 10573) and
  Magellan photometry were used to build the $B-V$ vs. $V$ CMD. ACS
  photometry (from project 10775) and Magellan photometry were used to
  build the $V-I$ vs. $V$ CMD.}
\label{figngc6637}
\end{figure}

\begin{figure}[htbp]
%\epsscale{0.77}
\plotone{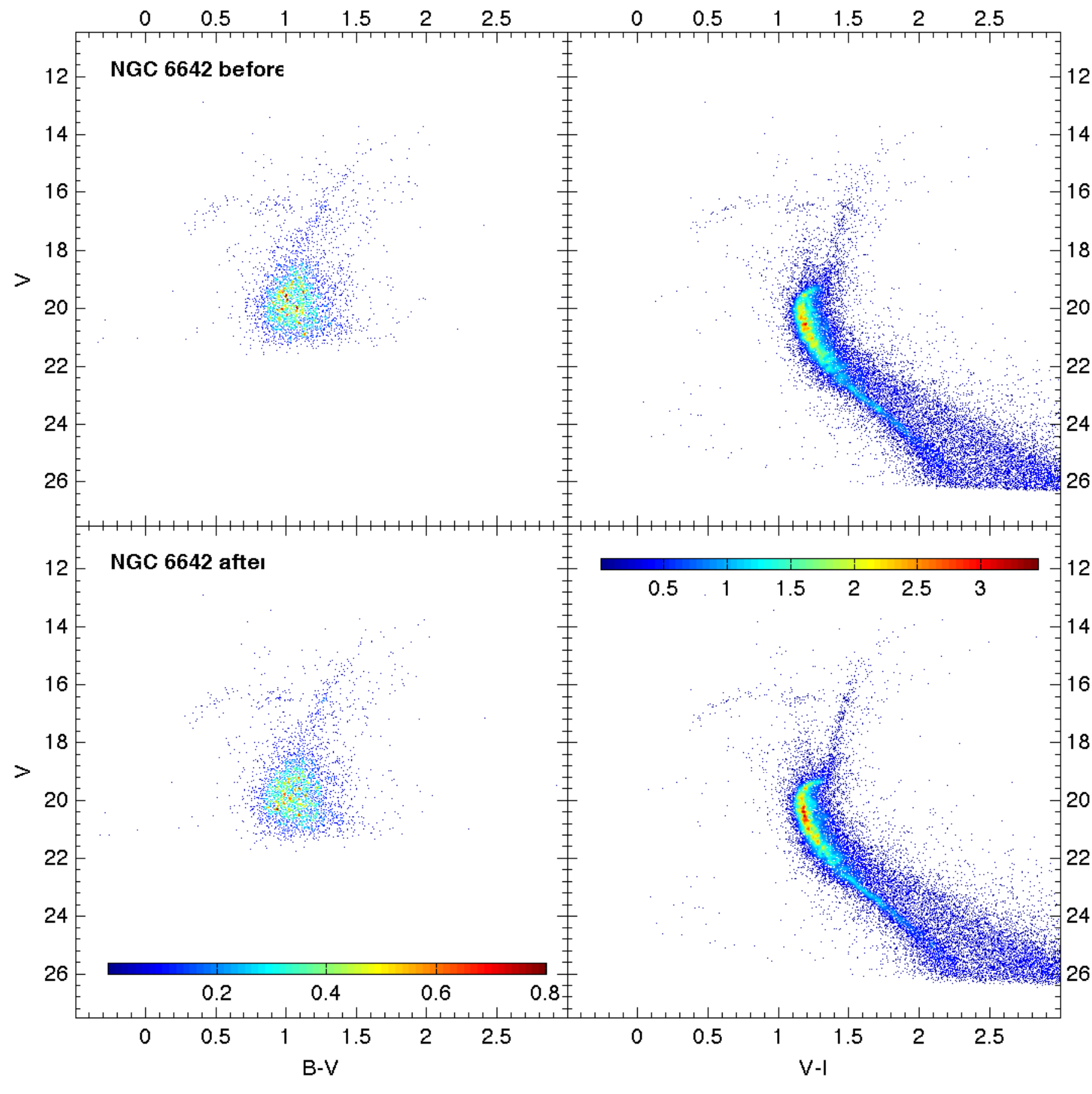}
\begin{tabular}{ccc}
\includegraphics[scale=0.28]{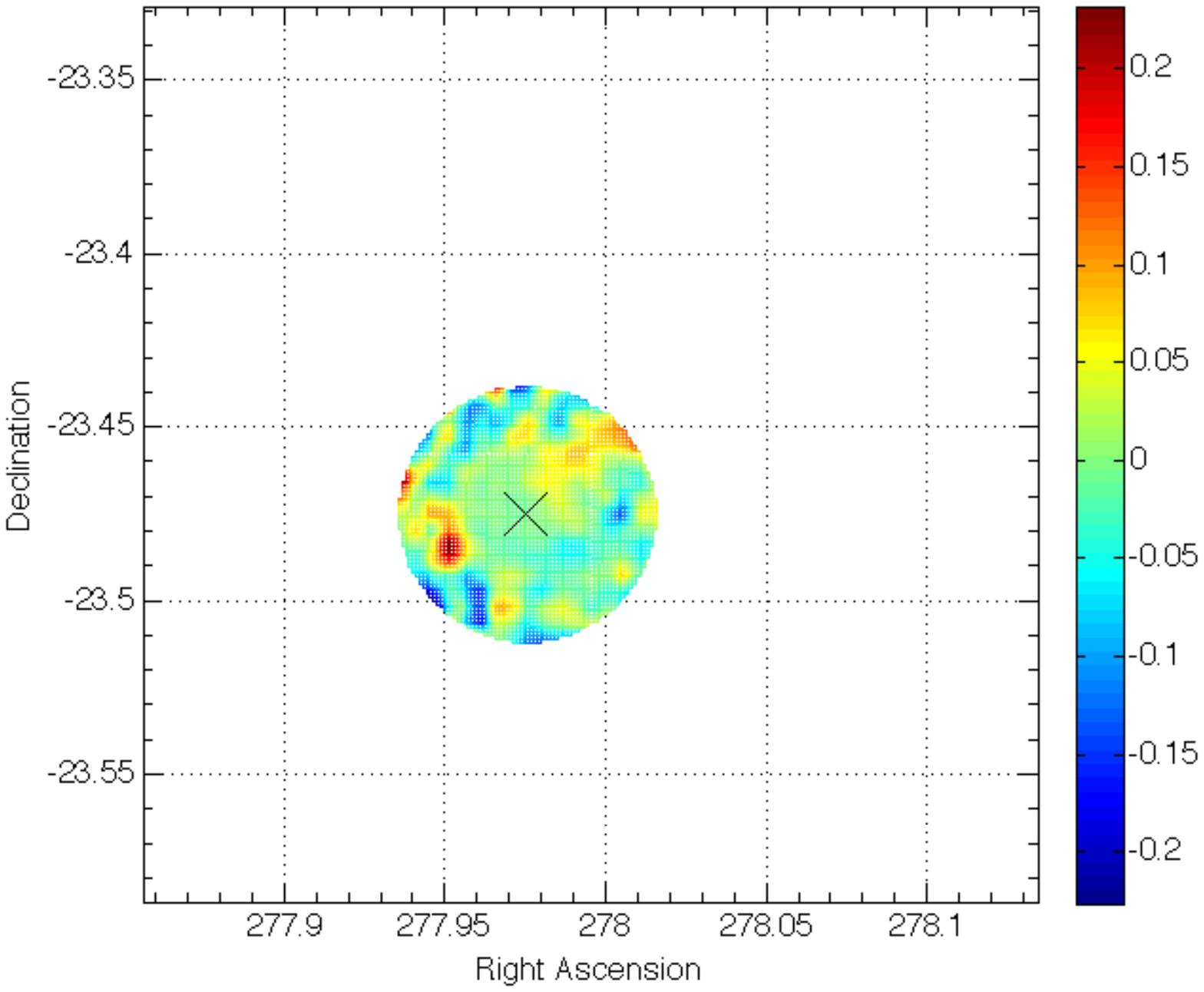}  &
\includegraphics[scale=0.28]{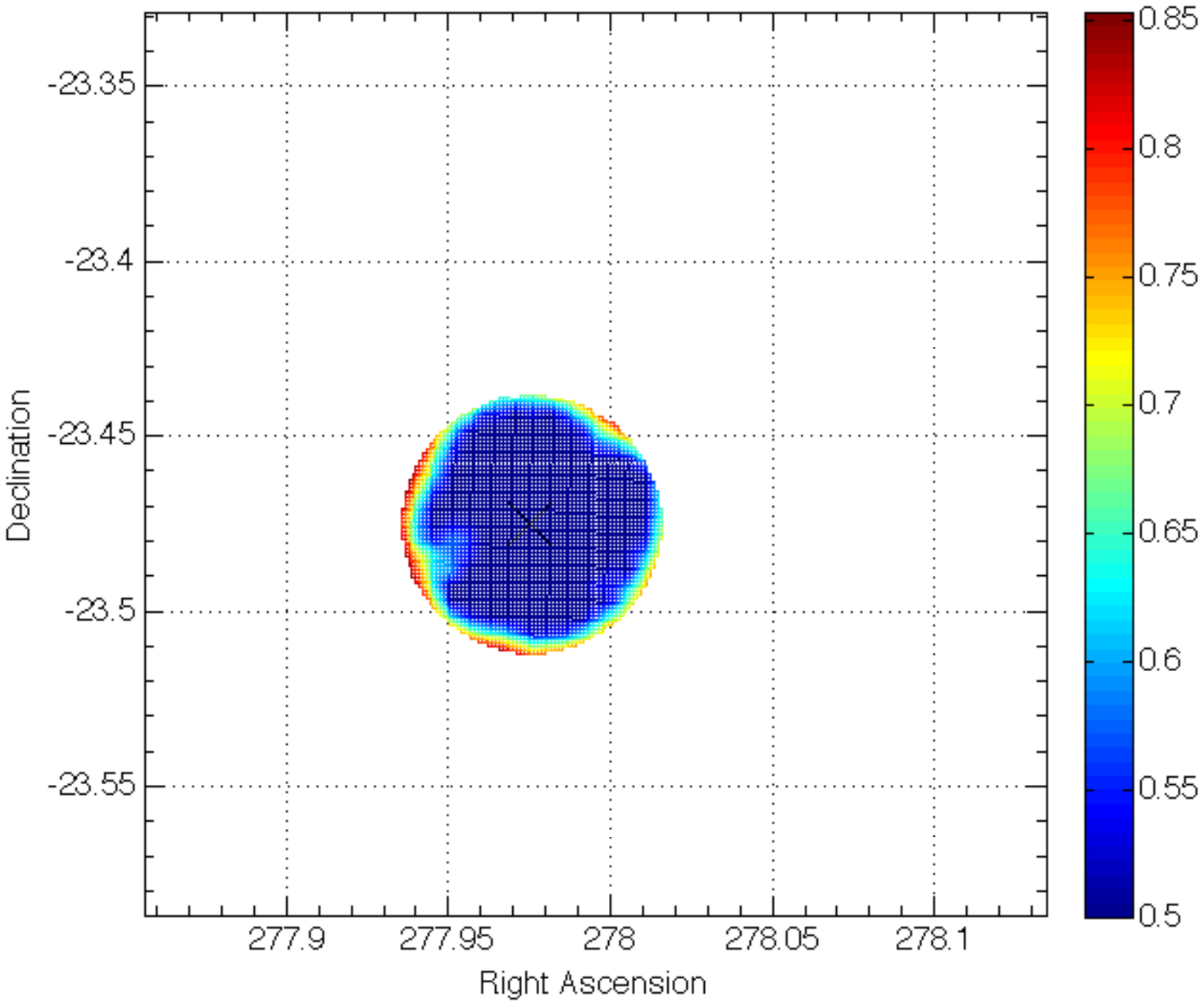}  &
\includegraphics[scale=0.28]{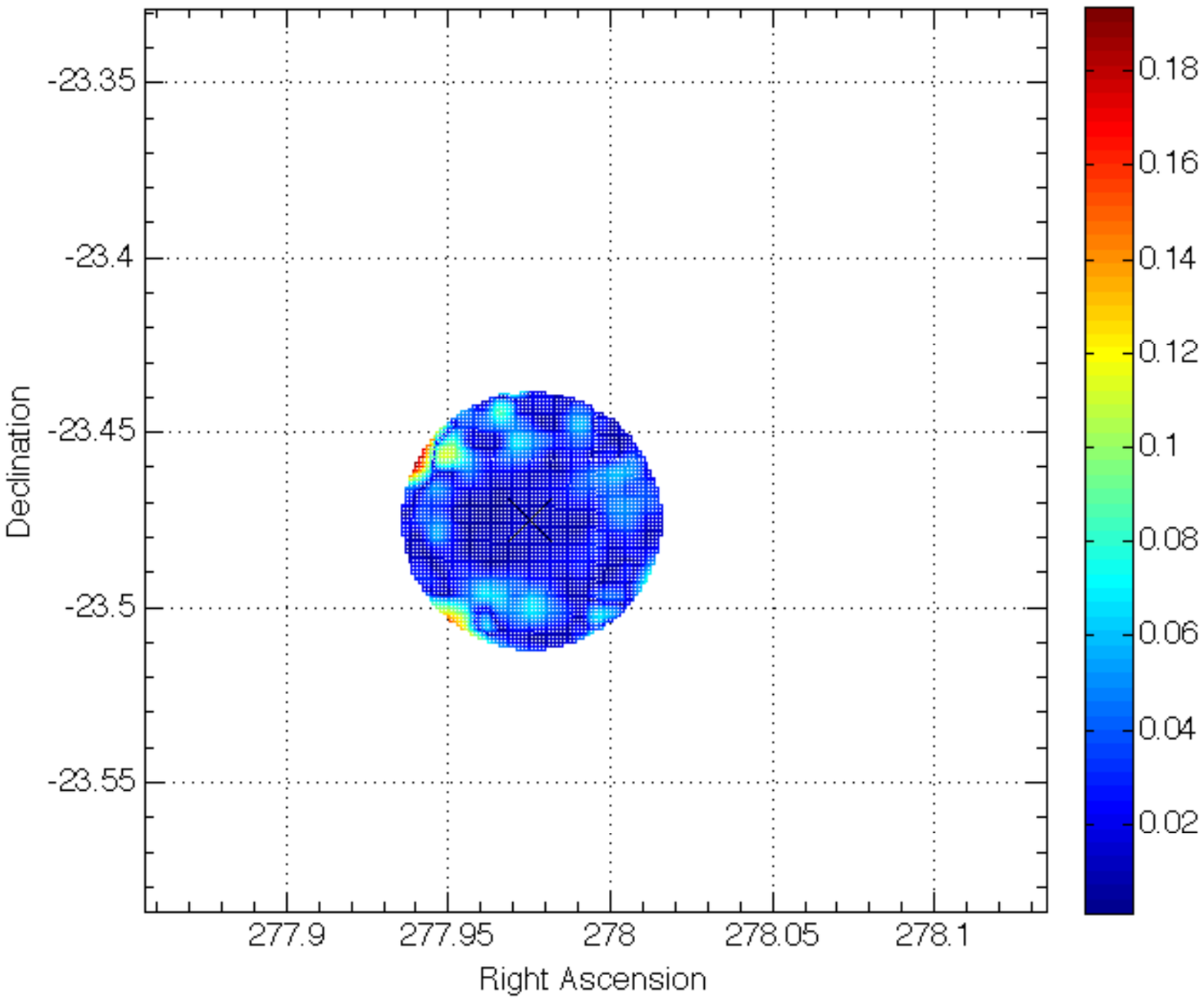} \\
\end{tabular}
\caption{\footnotesize As in Figure \ref{figngc6121}, but for the
  cluster NGC 6642. Only our Magellan photometry was used to build the
  $B-V$ vs. $V$ CMD. ACS photometry (from project 10775) and Magellan
  photometry were used to build the $V-I$ vs. $V$ CMD. }
\label{figngc6642}
\end{figure}

\begin{figure}[htbp]
%\epsscale{0.77}
\plotone{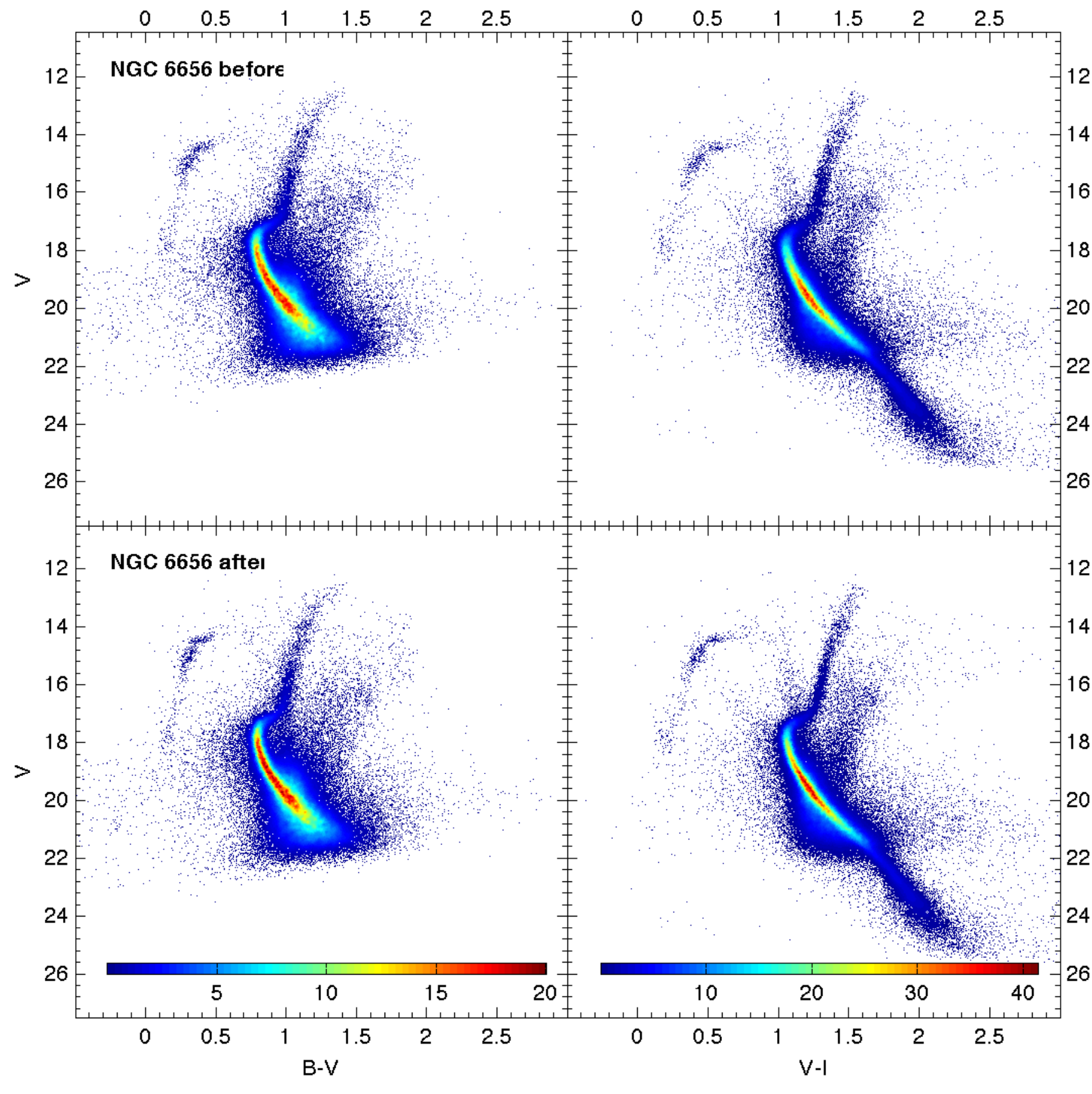}
\begin{tabular}{ccc}
\includegraphics[scale=0.28]{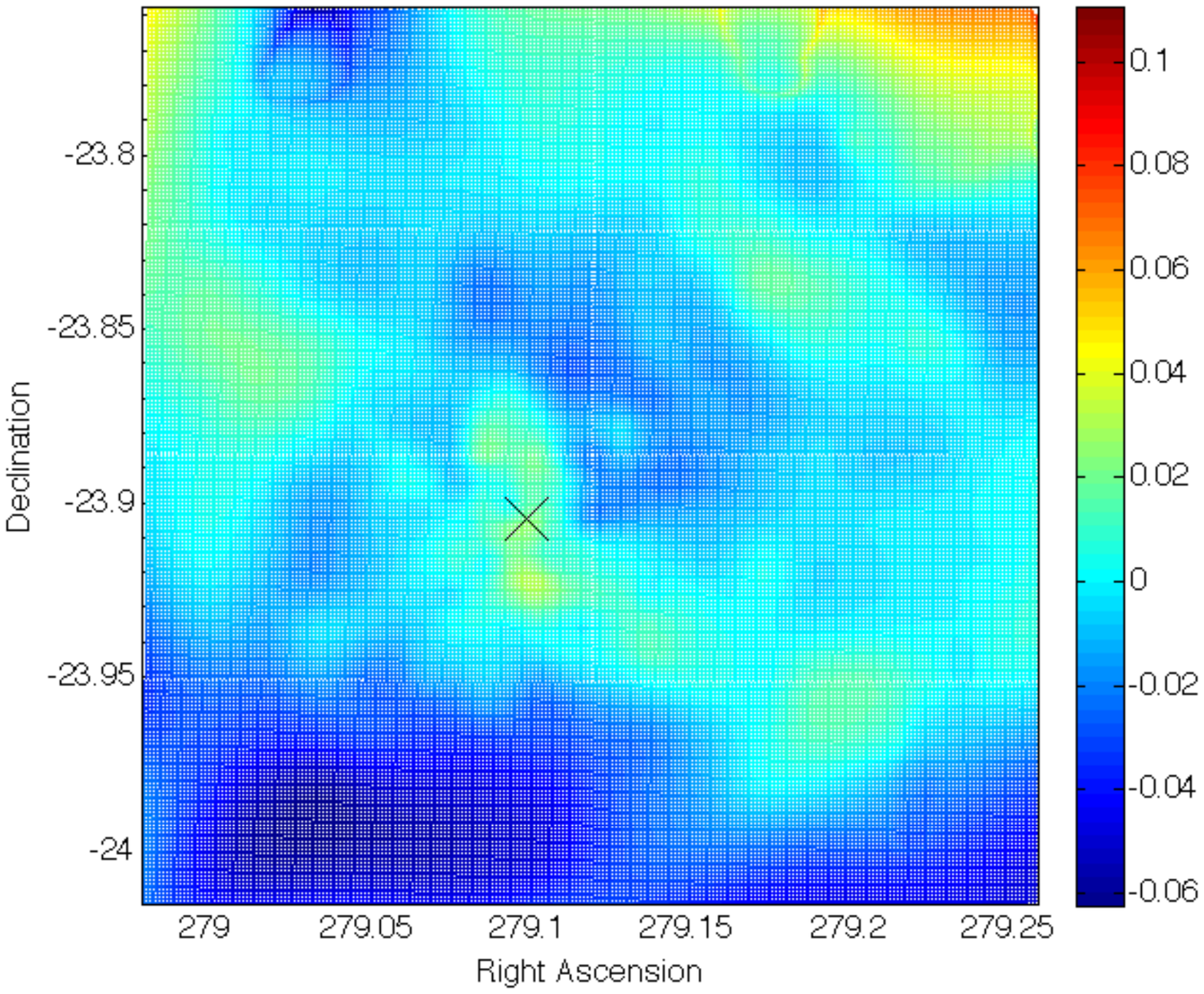}  &
\includegraphics[scale=0.28]{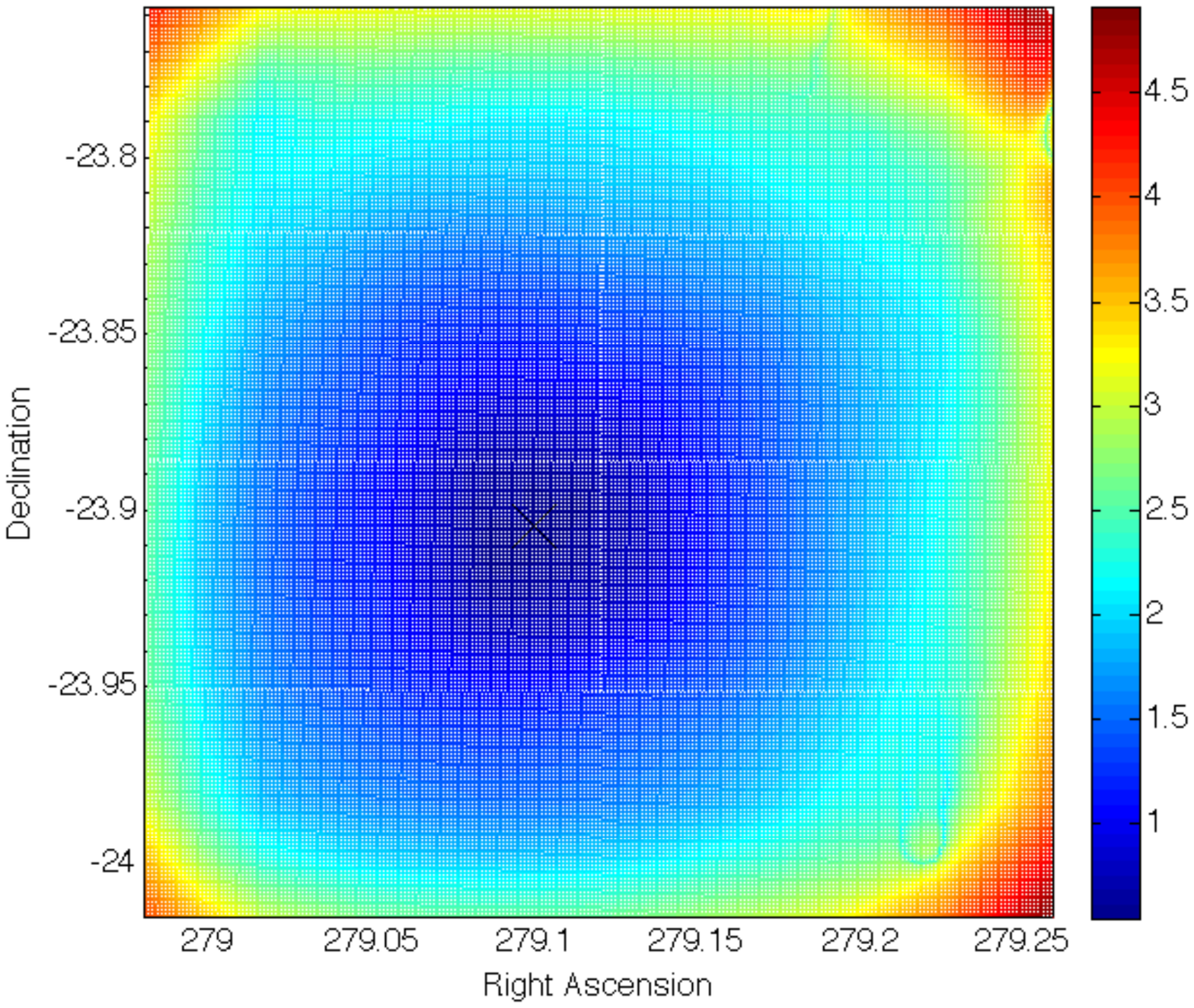}  &
\includegraphics[scale=0.28]{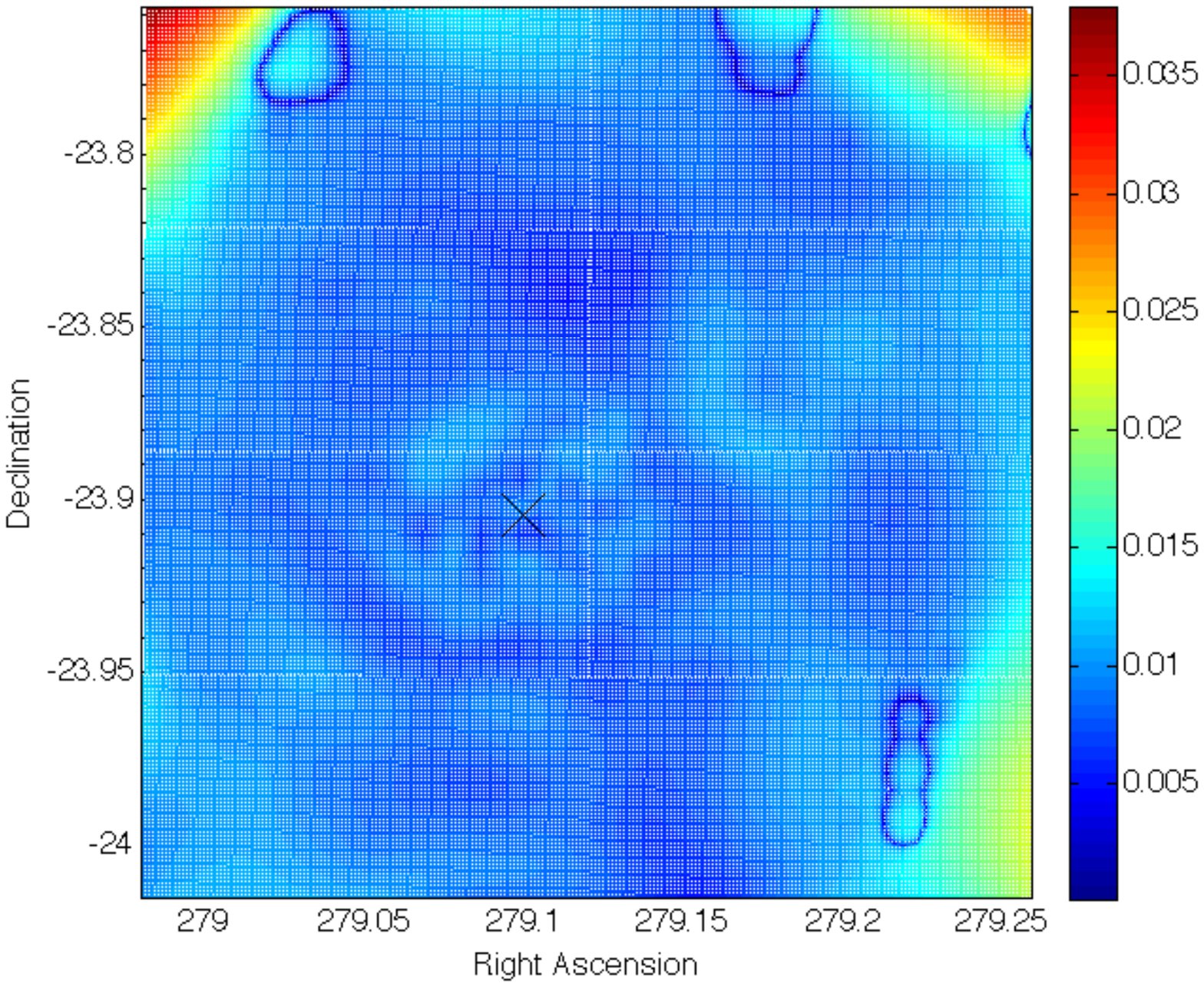} \\
\end{tabular}
\caption{\footnotesize As in Figure \ref{figngc6121}, but for the
  cluster NGC 6656 - M 22. Only our Magellan photometry was used to
  build the $B-V$ vs. $V$ CMD. ACS photometry (from project 10775) and
  Magellan photometry were used to build the $V-I$ vs. $V$ CMD.}
\label{figngc6656}
\end{figure}

\begin{figure}[htbp]
%\epsscale{0.77}
\plotone{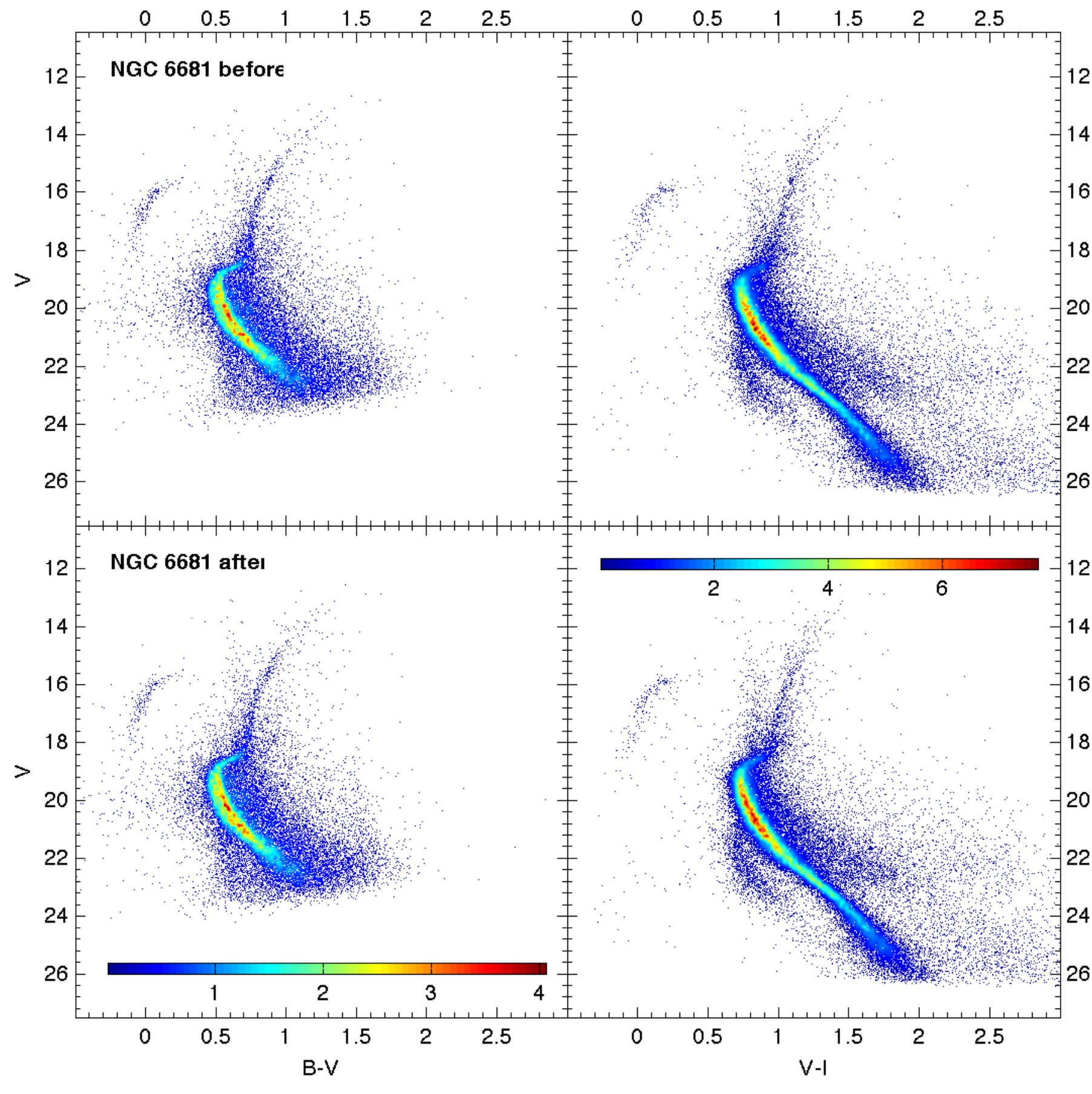}
\begin{tabular}{ccc}
\includegraphics[scale=0.28]{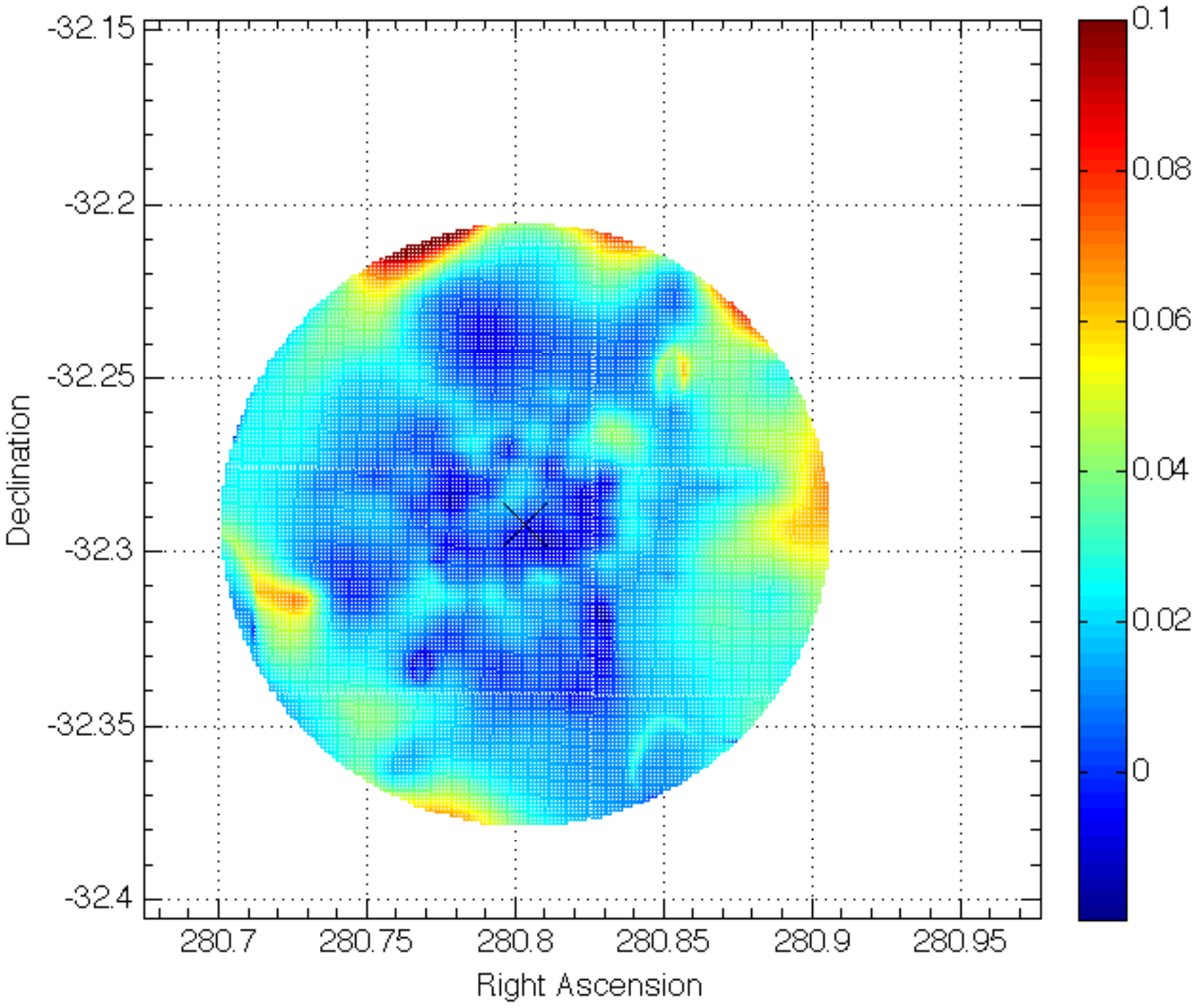}  &
\includegraphics[scale=0.28]{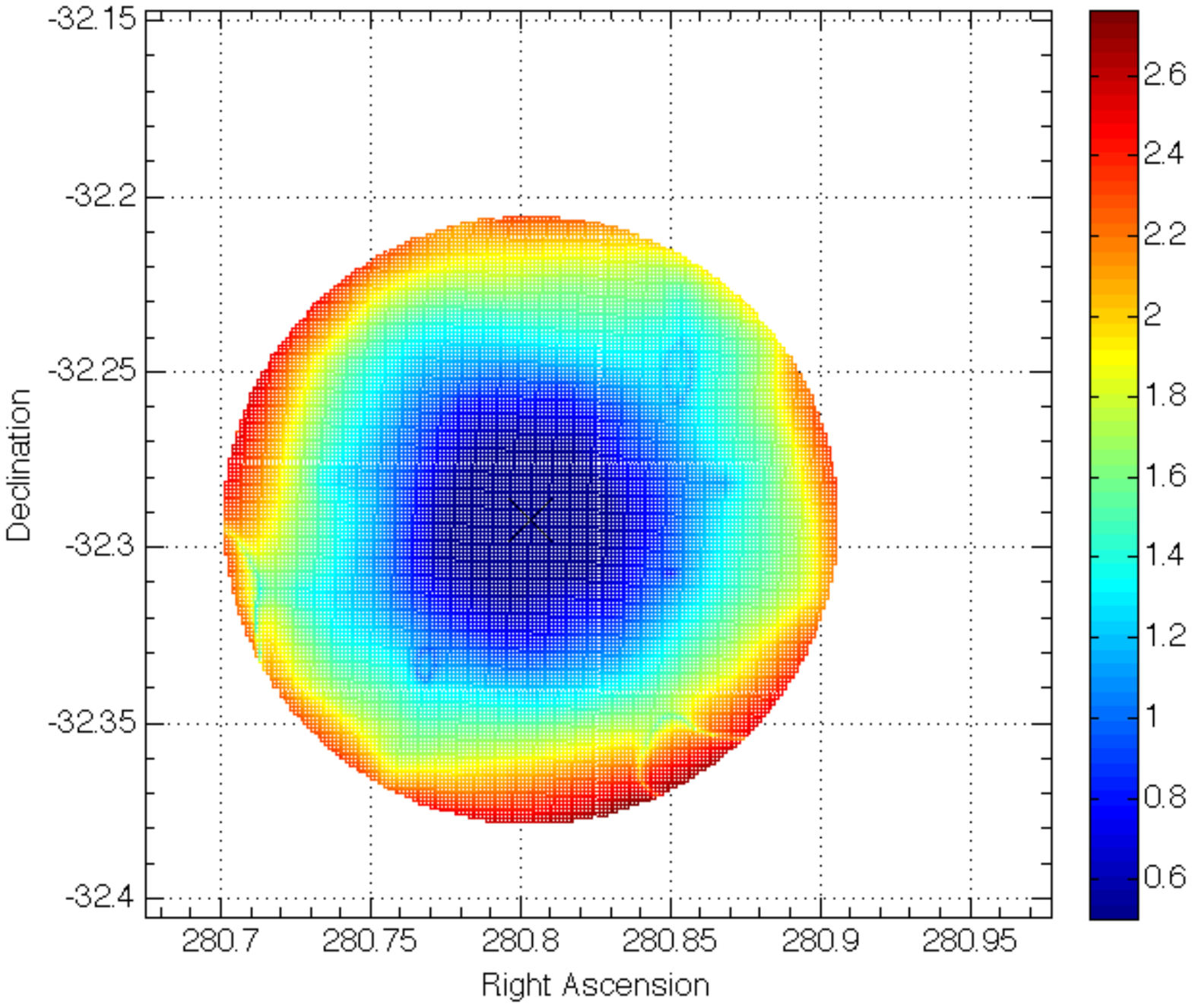}  &
\includegraphics[scale=0.28]{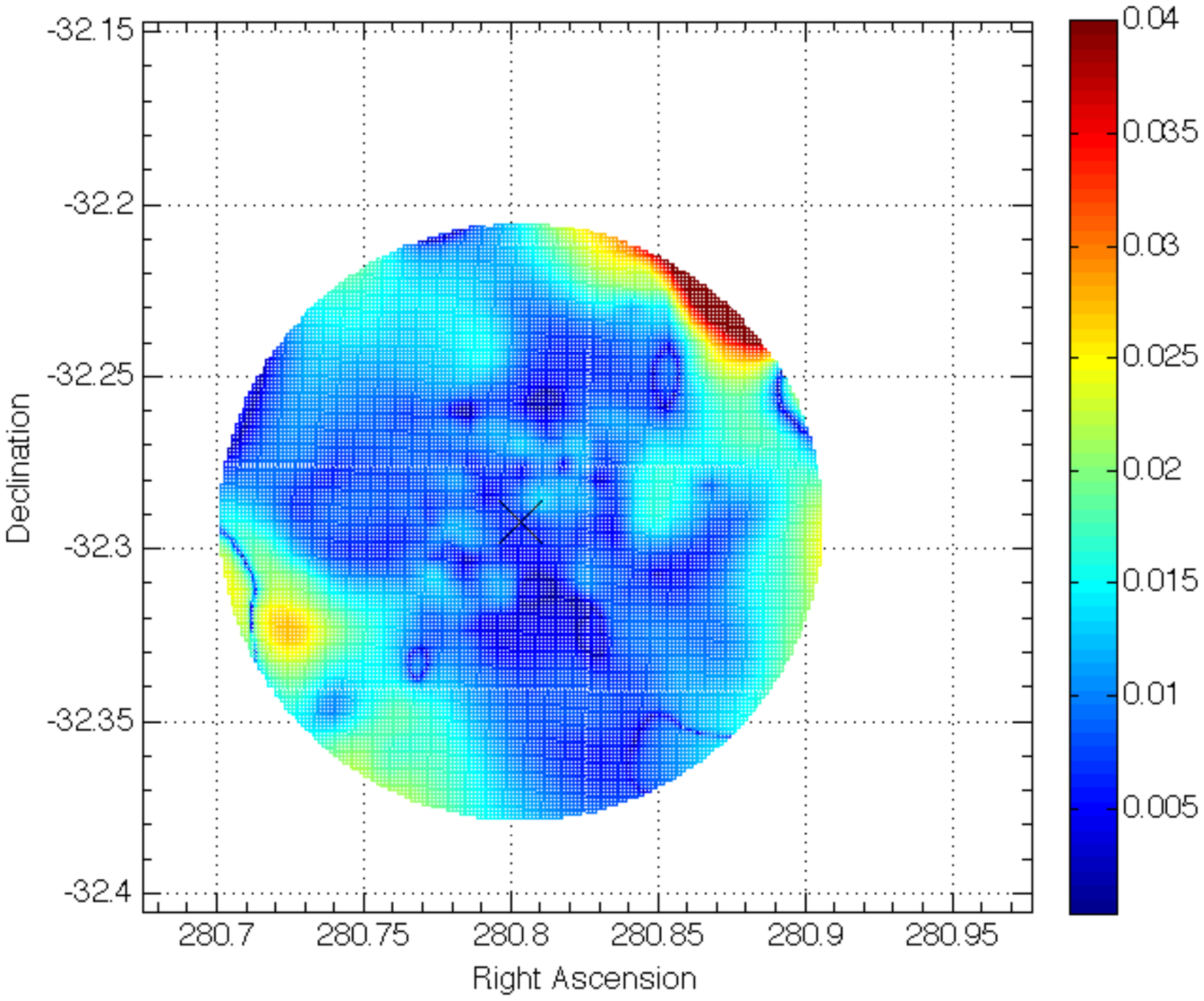} \\
\end{tabular}
\caption{\footnotesize As in Figure \ref{figngc6121}, but for the
  cluster NGC 6681 - M 70. Only our Magellan photometry was used to
  build the $B-V$ vs. $V$ CMD. ACS photometry (from project 10775) and
  Magellan photometry were used to build the $V-I$ vs. $V$ CMD.}
\label{figngc6681}
\end{figure}

\begin{figure}[htbp]
%\epsscale{0.77}
\plotone{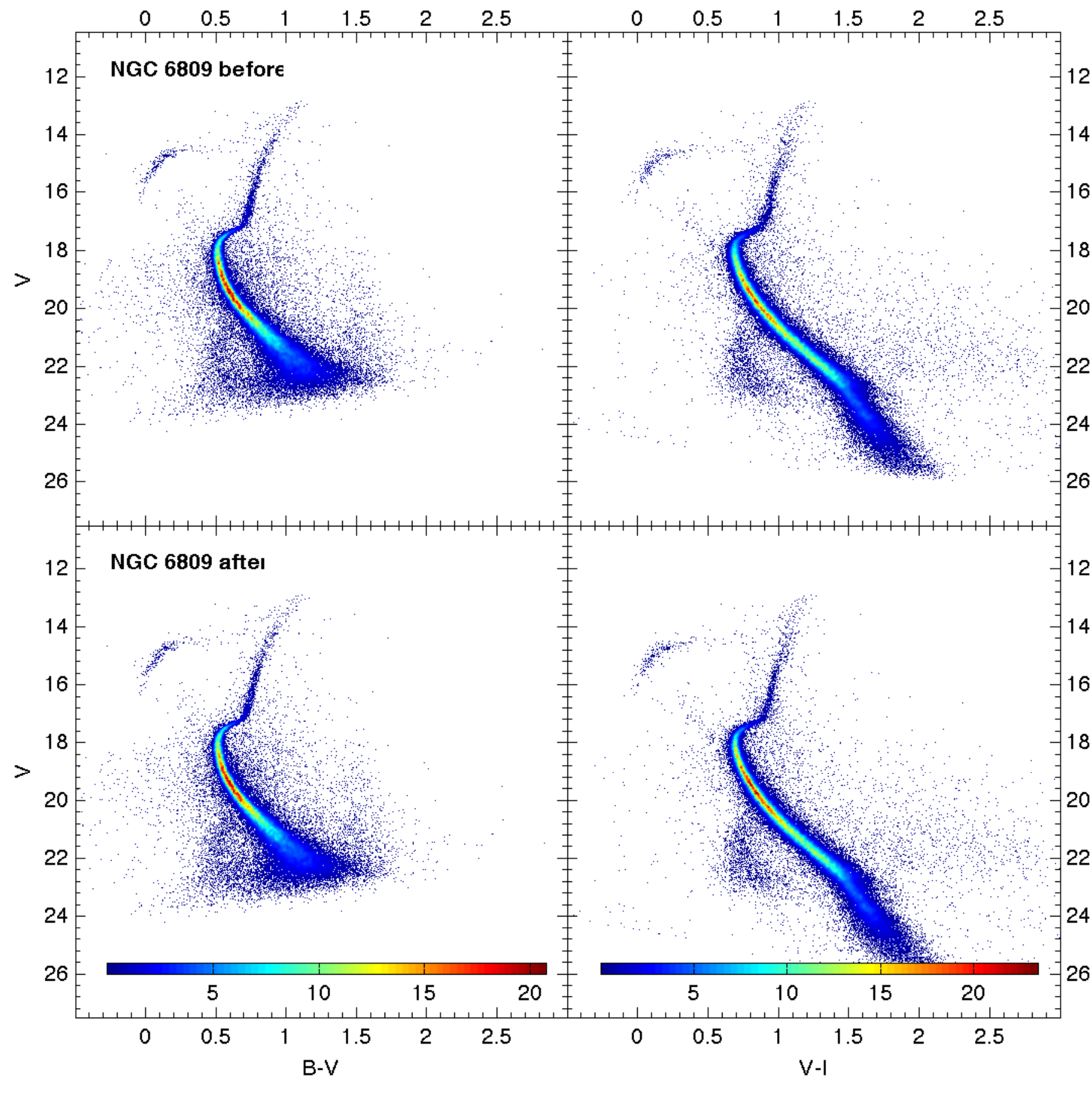}
\begin{tabular}{ccc}
\includegraphics[scale=0.28]{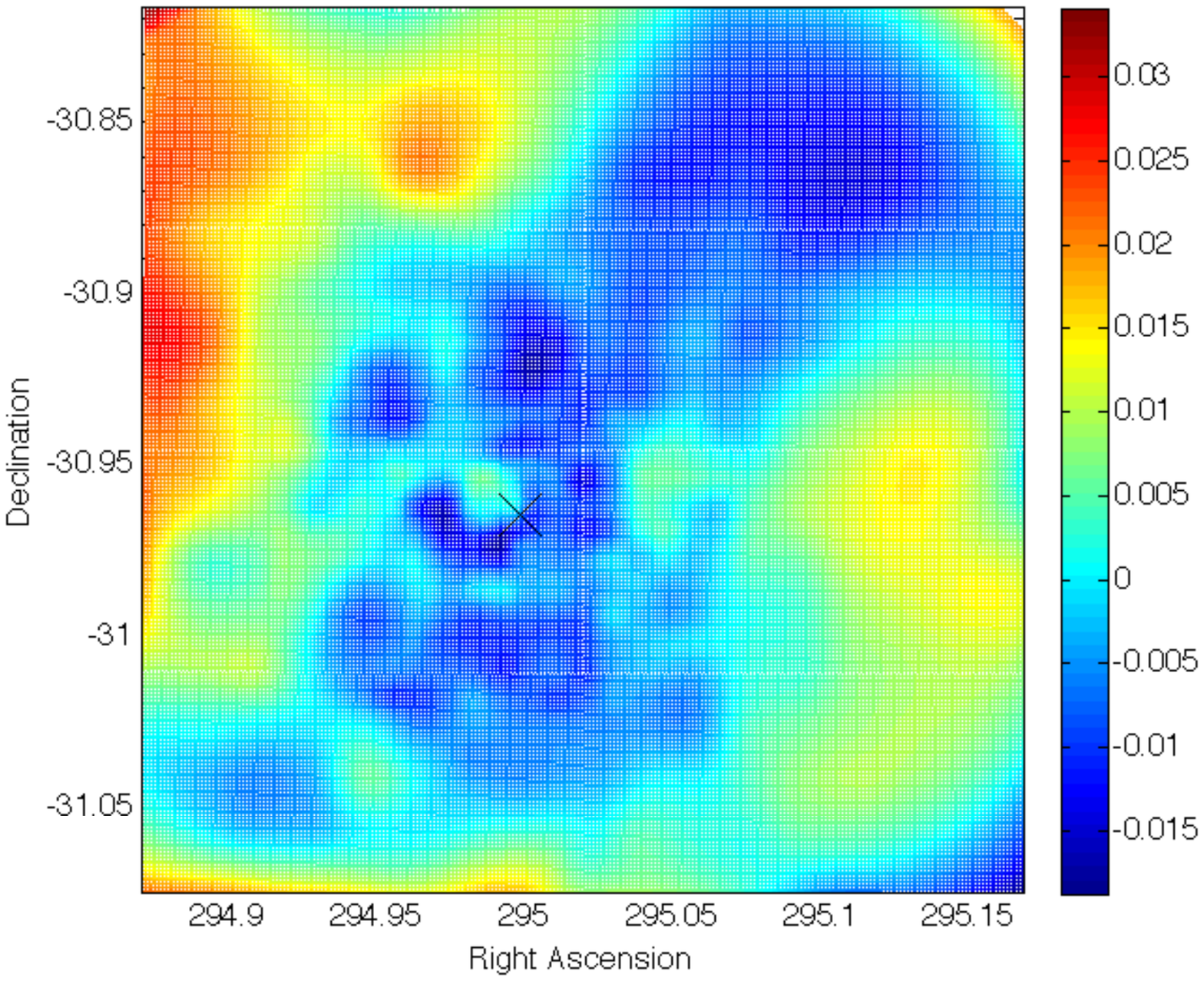}  &
\includegraphics[scale=0.28]{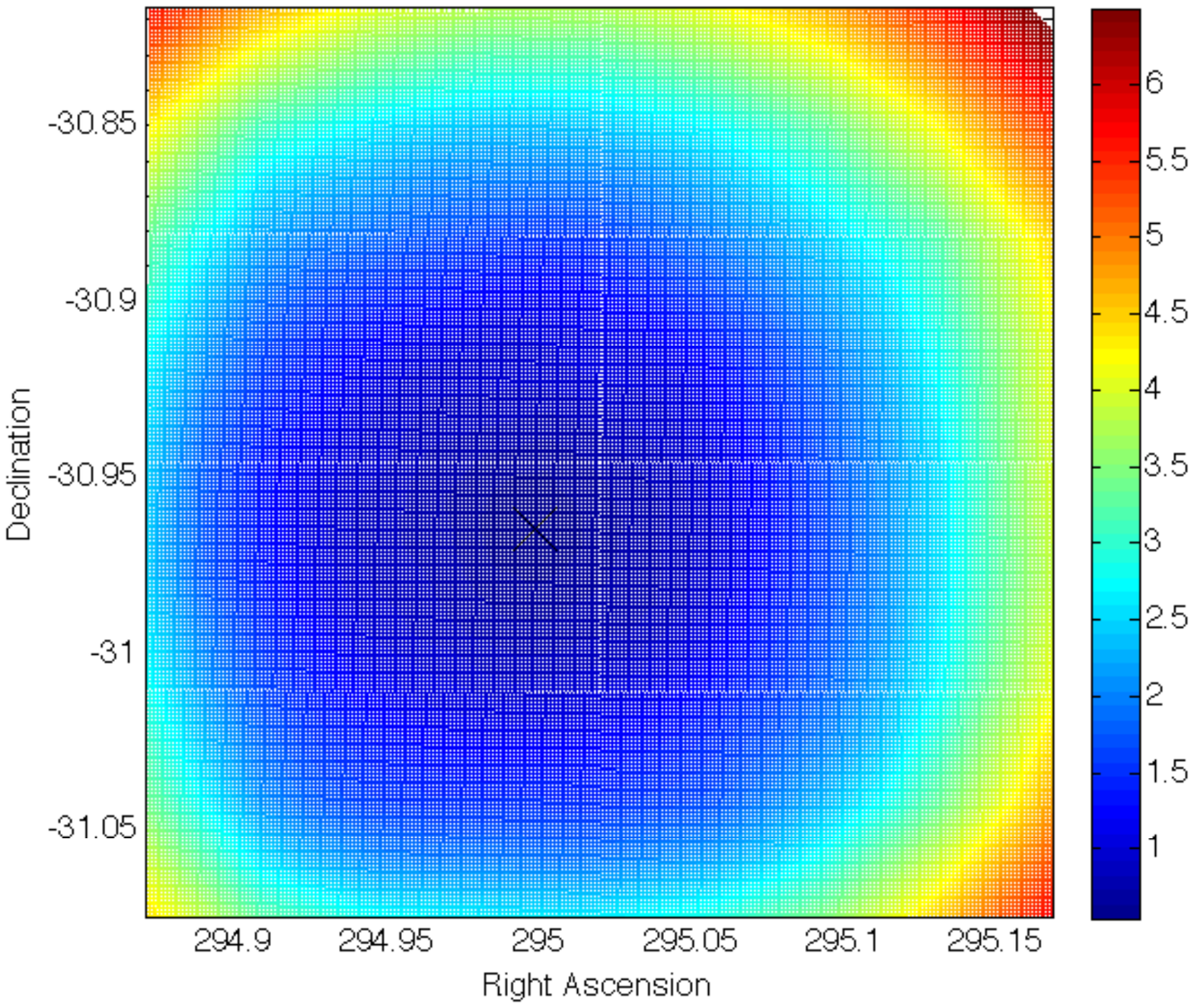}  &
\includegraphics[scale=0.28]{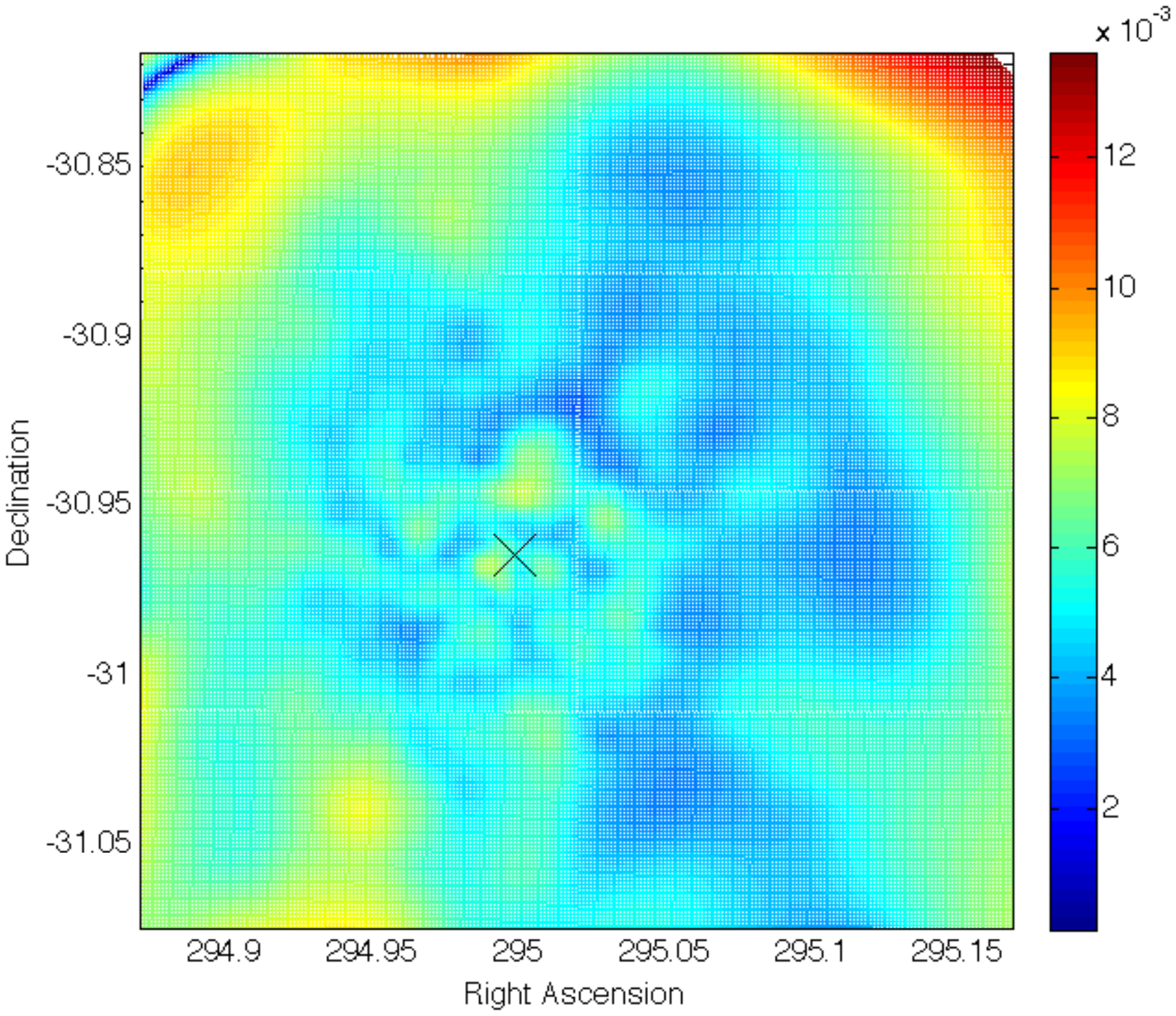} \\
\end{tabular}
\caption{\footnotesize As in Figure \ref{figngc6121}, but for the
  cluster NGC 6809 - M 55. Only our Magellan photometry was used to
  build the $B-V$ vs. $V$ CMD. ACS photometry (from project 10775) and
  Magellan photometry were used to build the $V-I$ vs. $V$ CMD.}
\label{figngc6809}
\end{figure}

Finally, after the process converges for both colors, we average the
extinction maps obtained from both colors, generating a final
extinction map and the dereddened CMDs (see
Figures \ref{figngc6121} to \ref{figngc6809}).

We now describe in more detail the individual cases of the different
clusters in the sample. We display their $B-V$ vs. $V$ and $V-I$
vs. $V$ CMDs, before and after being differentially dereddened (see
Figures \ref{figngc6121} to \ref{figngc6809}), and describe any
improvements in the photometry. The CMDs are plotted using the merged
space and ground data, whenever space-based observations deeper than
ground-based observations exist, or the ground photometry in the rest
of the cases.  To quantify improvements in the photometry, we have
measured the definition of the evolutionary sequences in the CMD by
fitting Gaussian functions to different magnitude cuts in the
different sequences (1.5 mag above and 0.5 mag below the TO, this is,
RGB and upper MS), similar to what we did in Figure 9 of Paper I. To
give an idea of the importance of the field star contamination, we
compare the total number of observed stars in every FOV with the
number of stars provided by the Besan\c{c}on model of the Galaxy for
the same pointing and FOV.

\begin{figure}[tp] \centering
\begin{tabular}{cccc}
NGC 6121 & & NGC 6144 & \\
\includegraphics[scale=0.21]{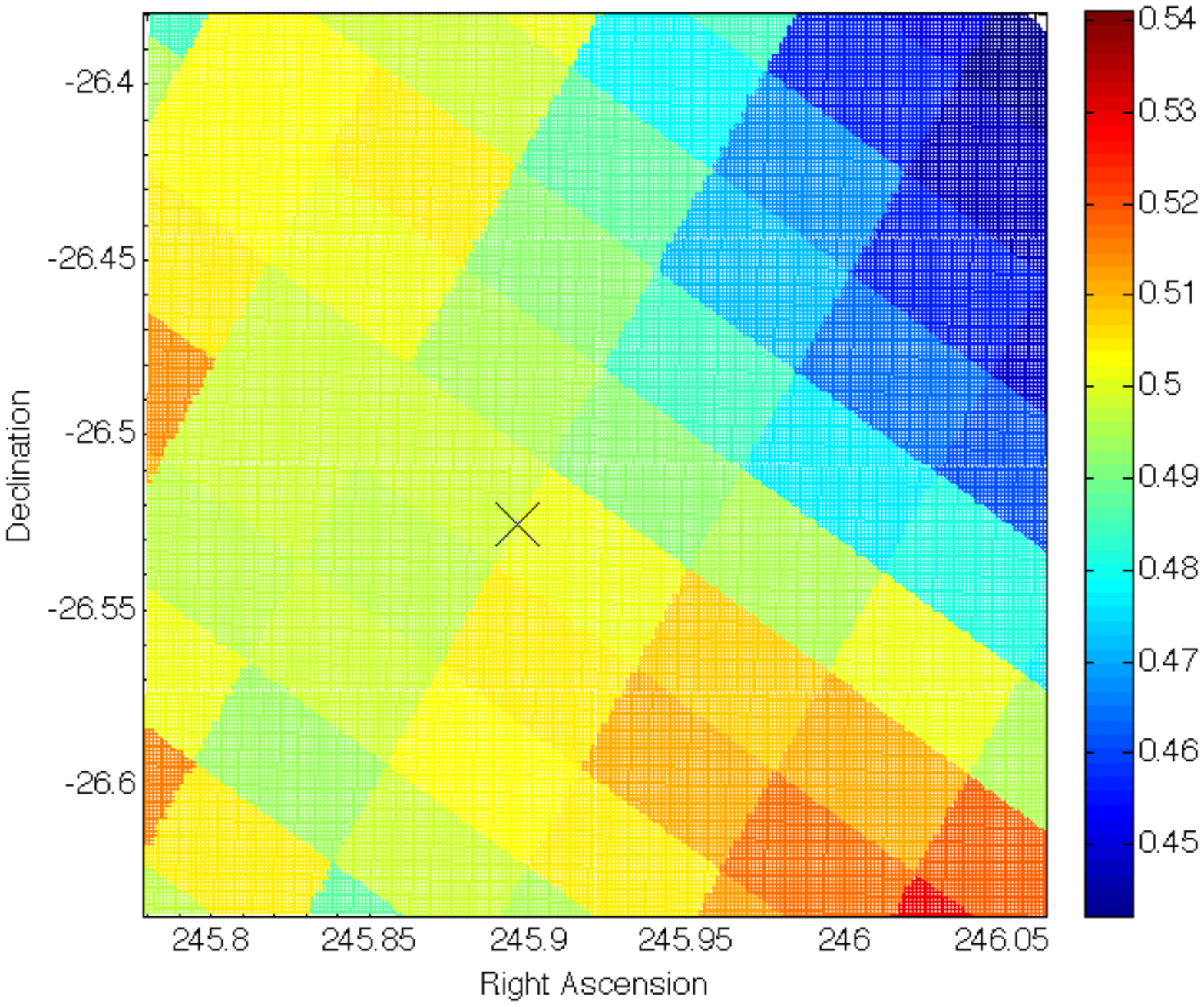}  &
\includegraphics[scale=0.21]{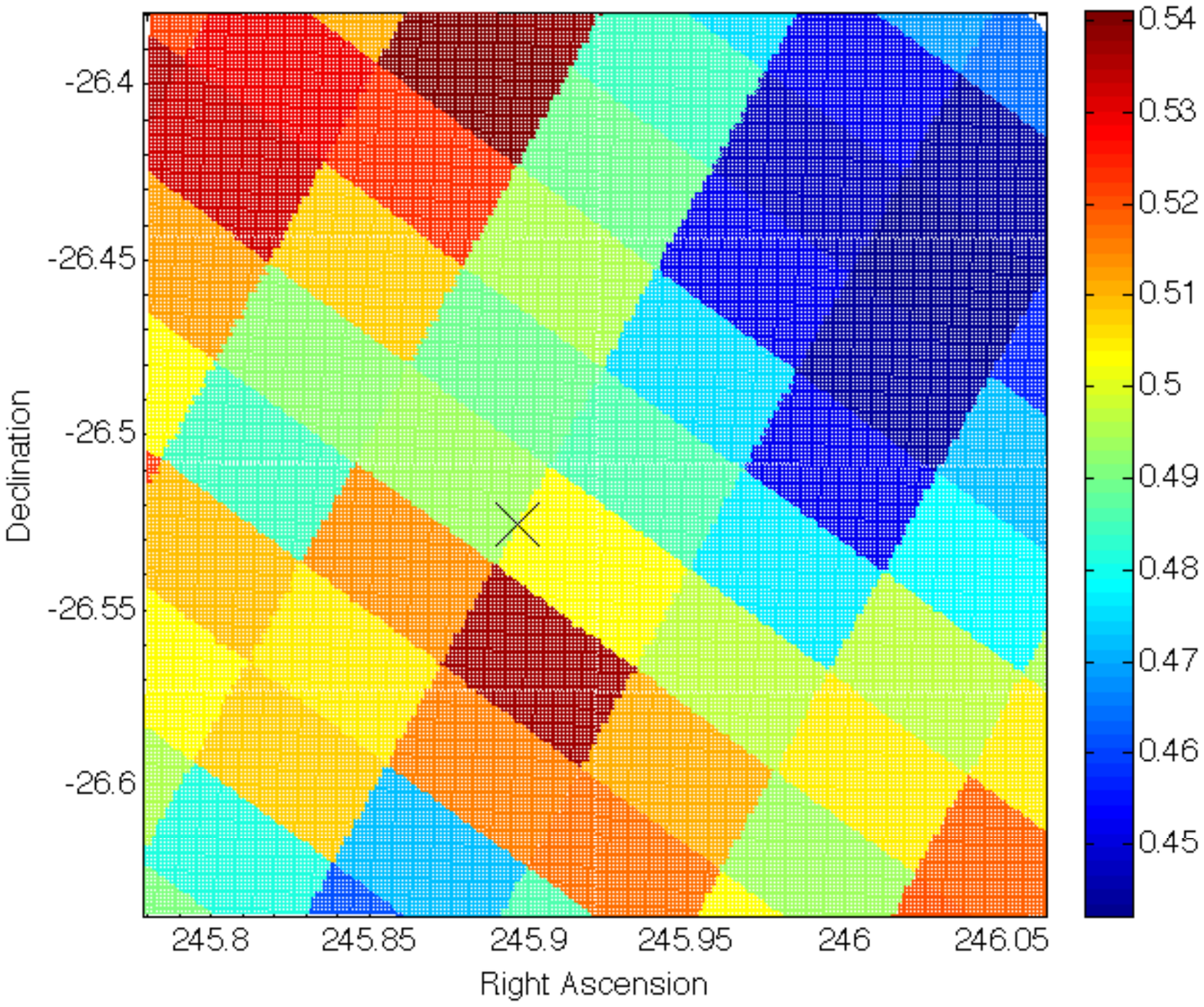}  &
\includegraphics[scale=0.21]{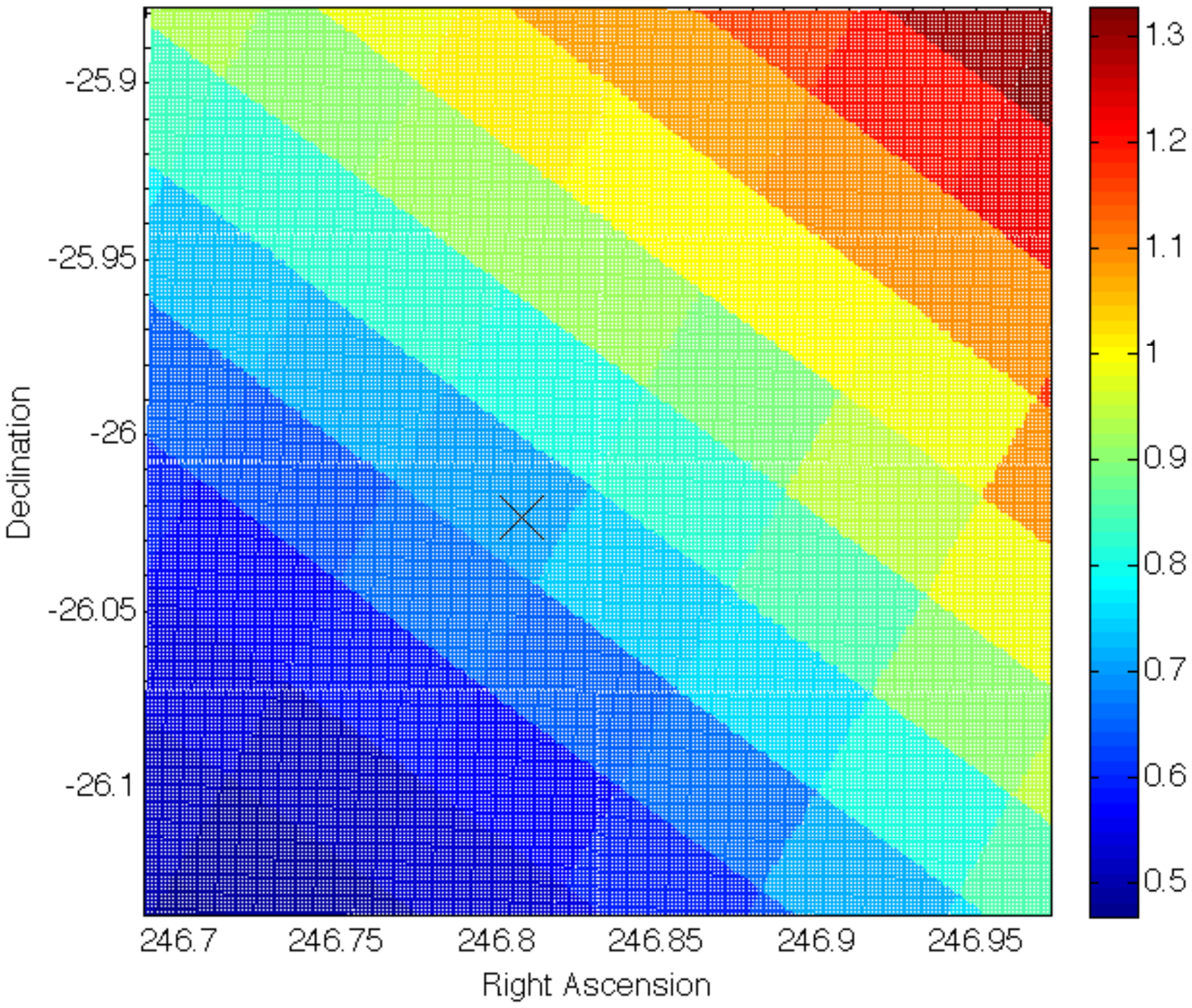}  &
\includegraphics[scale=0.21]{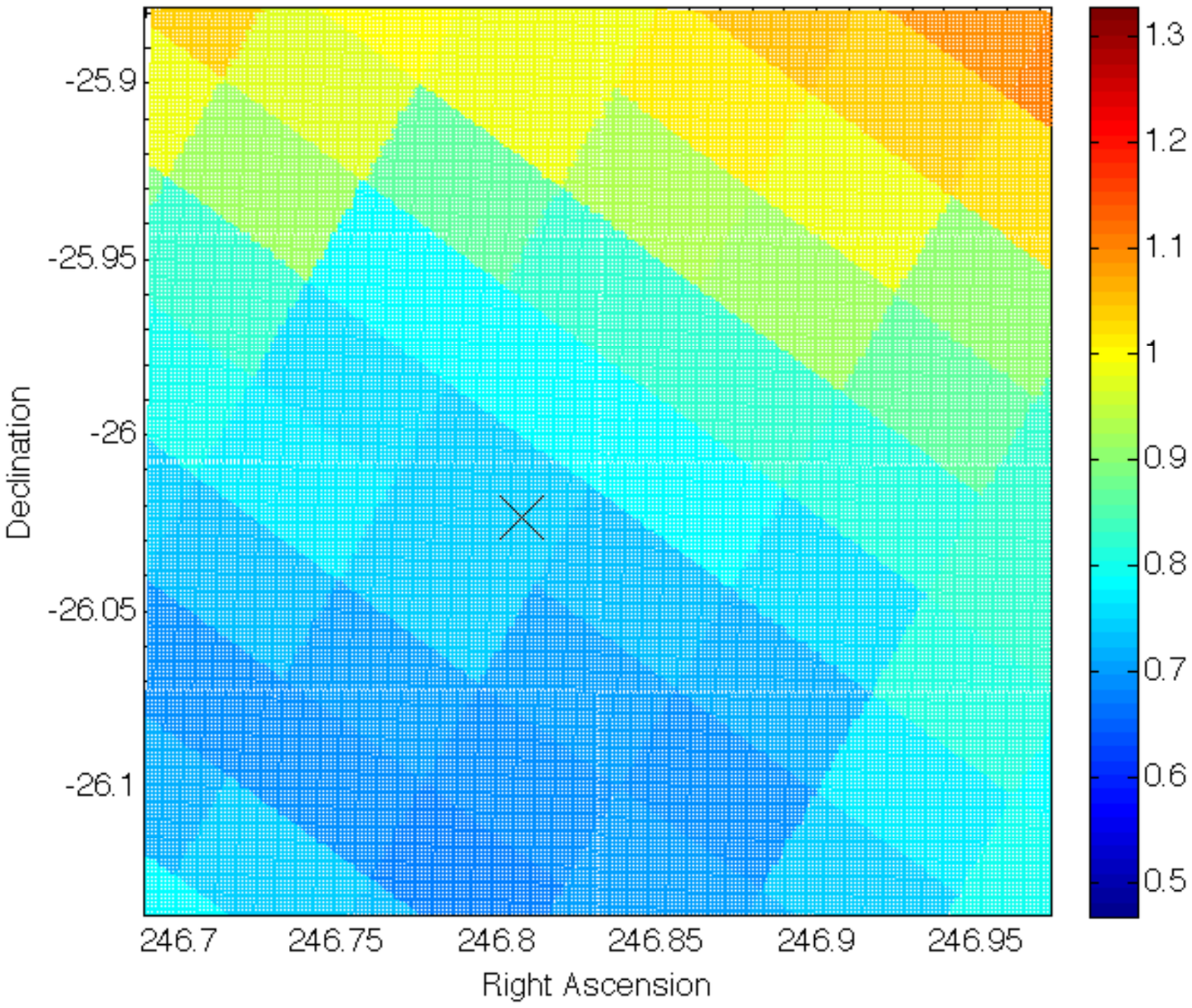} \\
NGC 6218 & & NGC 6235 & \\
\includegraphics[scale=0.21]{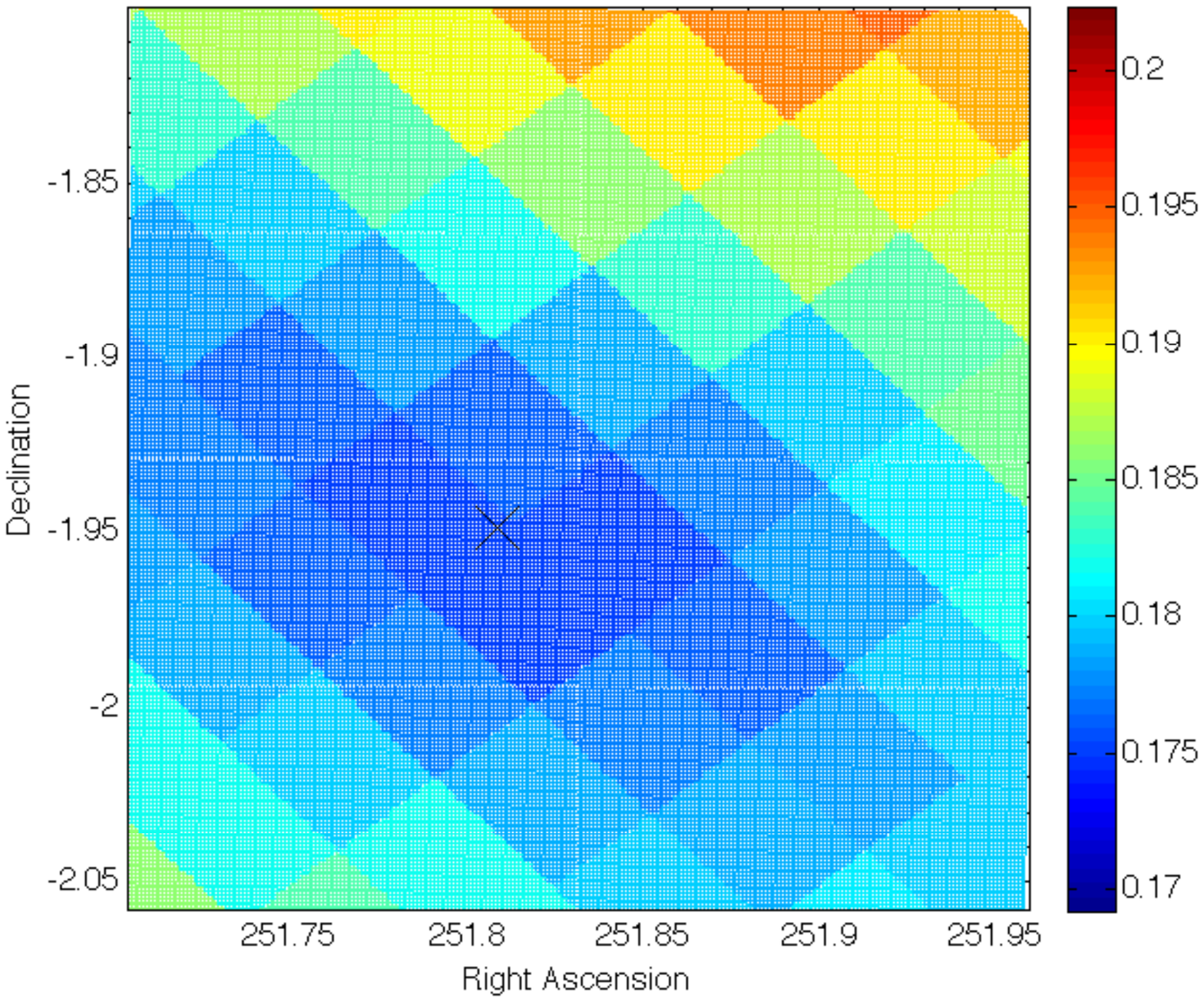}  &
\includegraphics[scale=0.21]{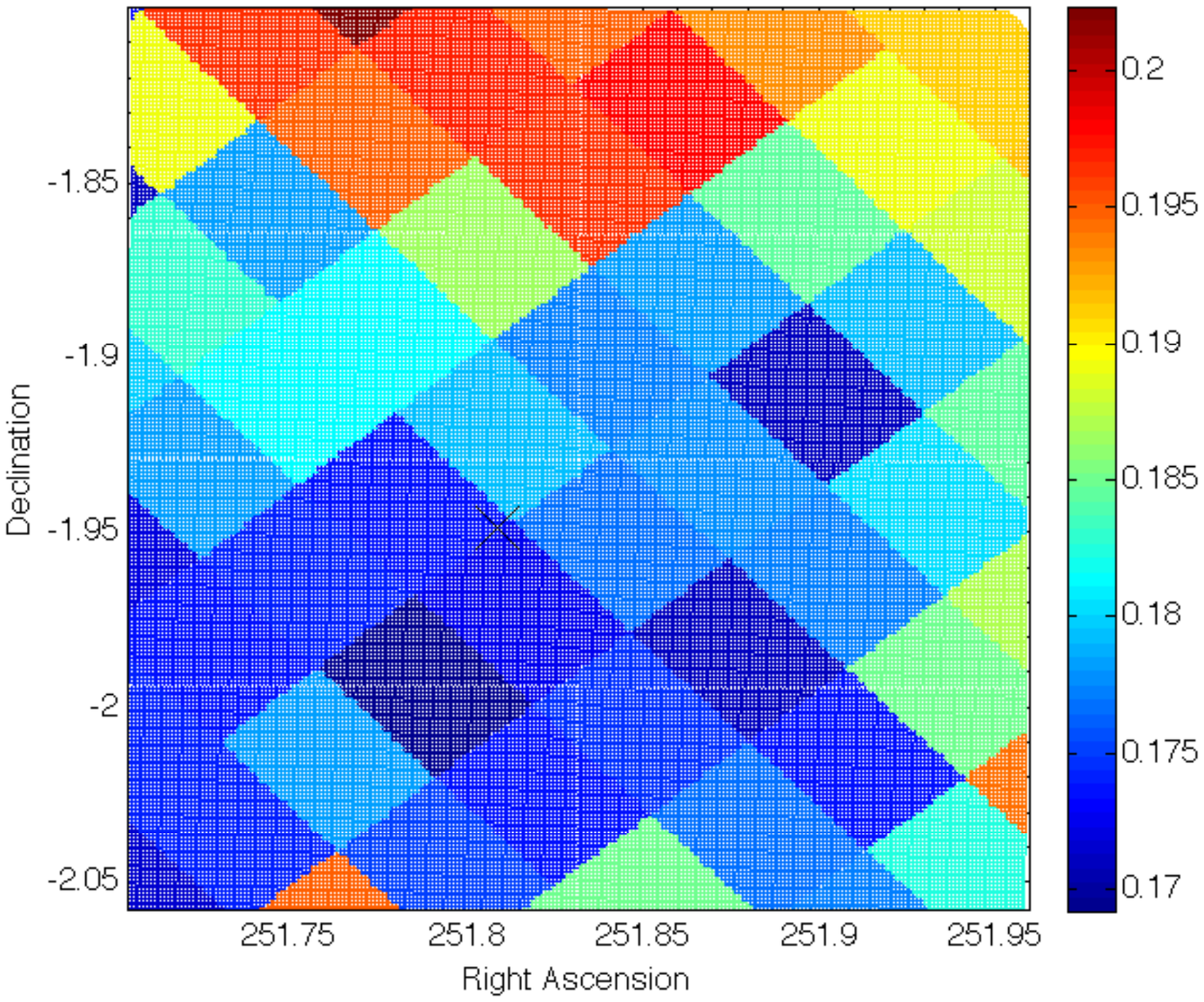}  &
\includegraphics[scale=0.21]{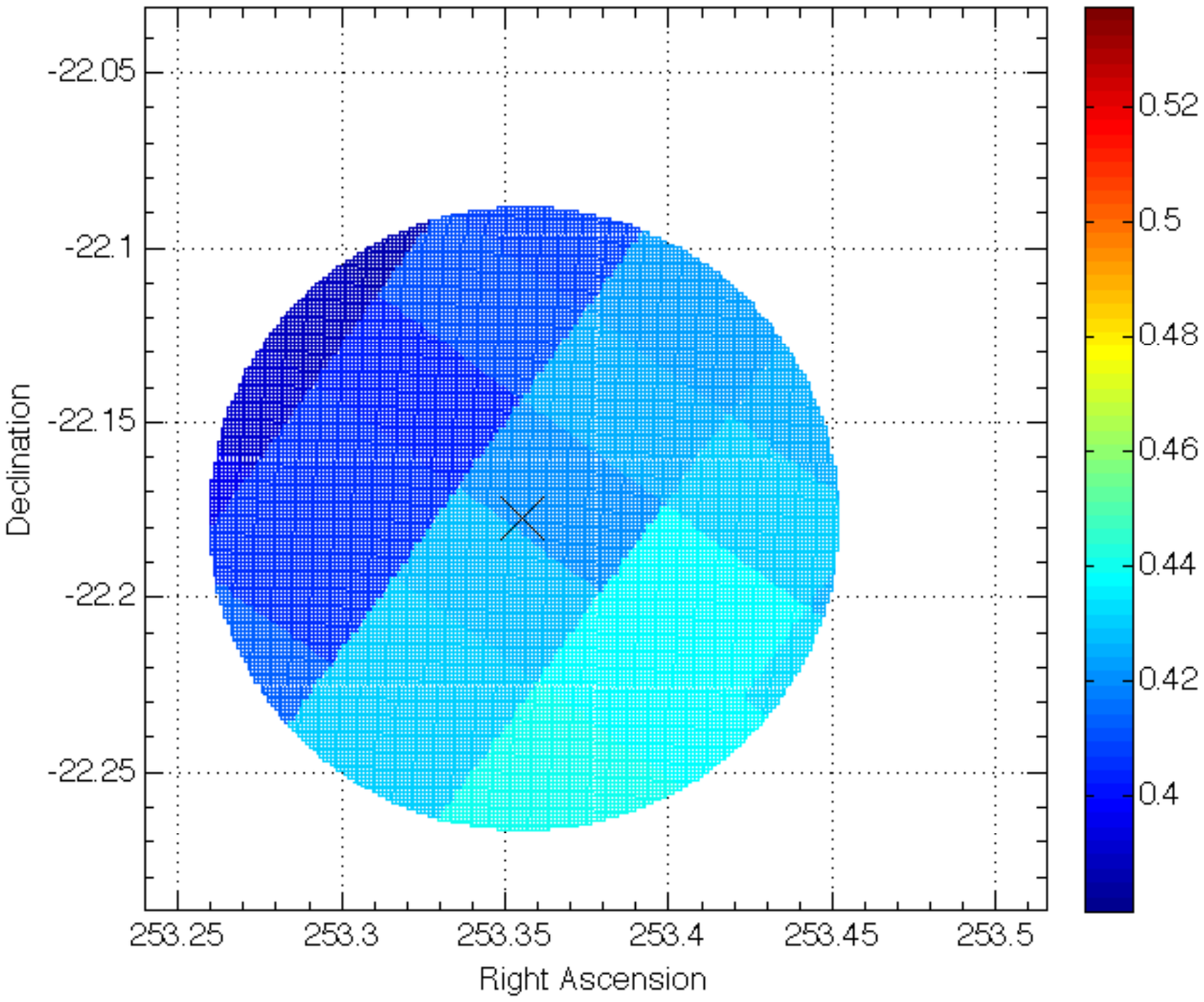}  &
\includegraphics[scale=0.21]{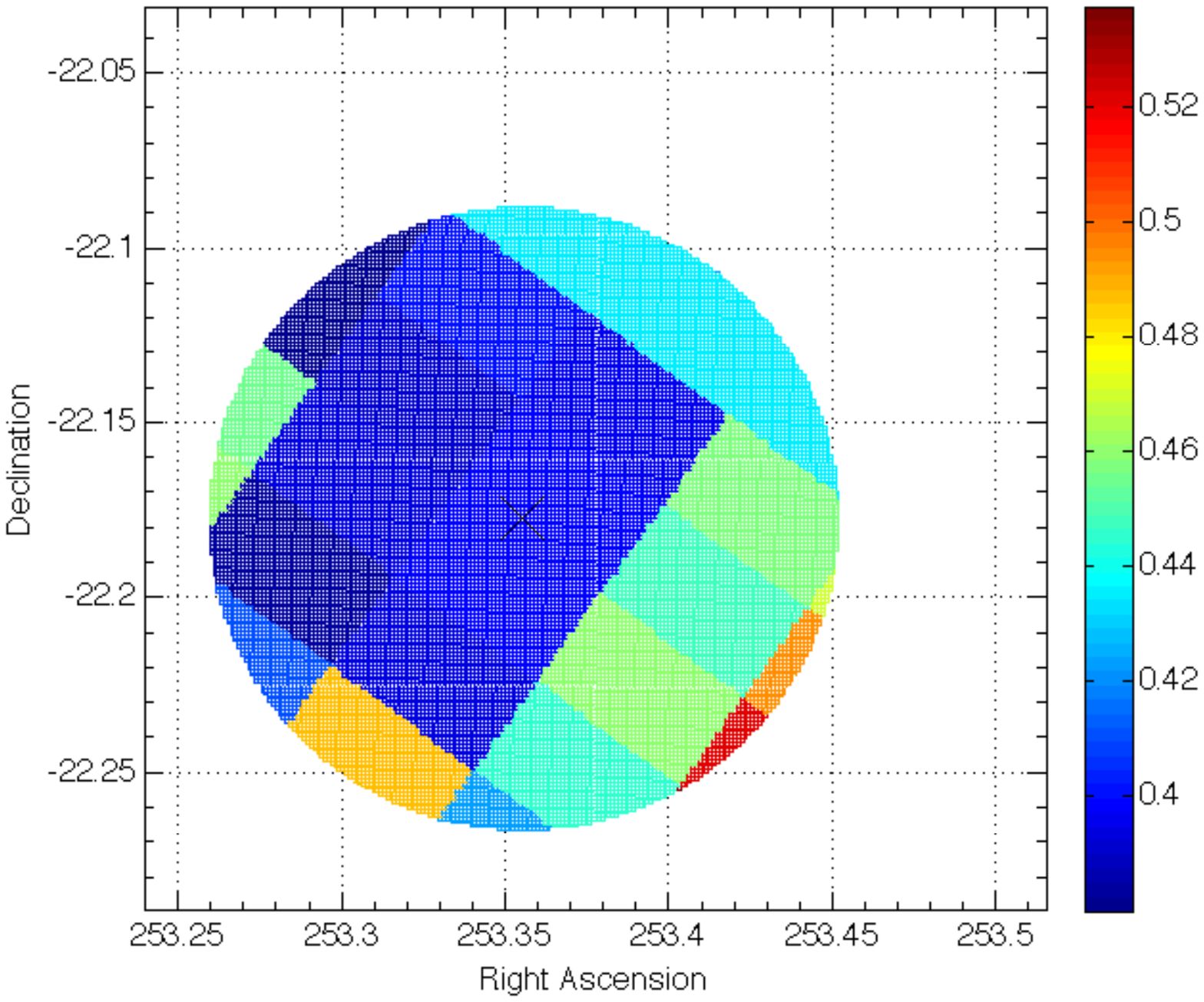} \\
NGC 6254 & & NGC 6266 & \\
\includegraphics[scale=0.21]{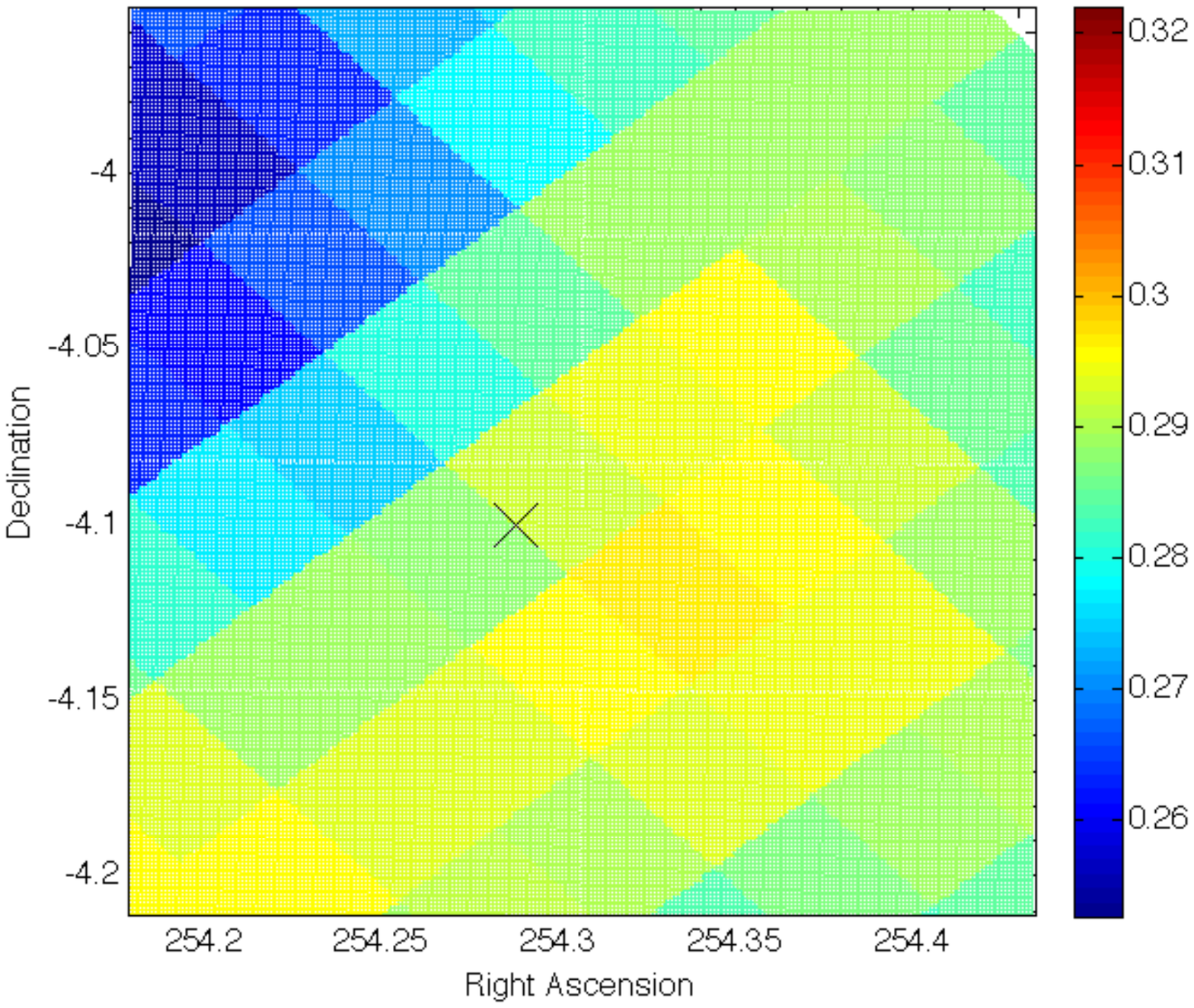}  &
\includegraphics[scale=0.21]{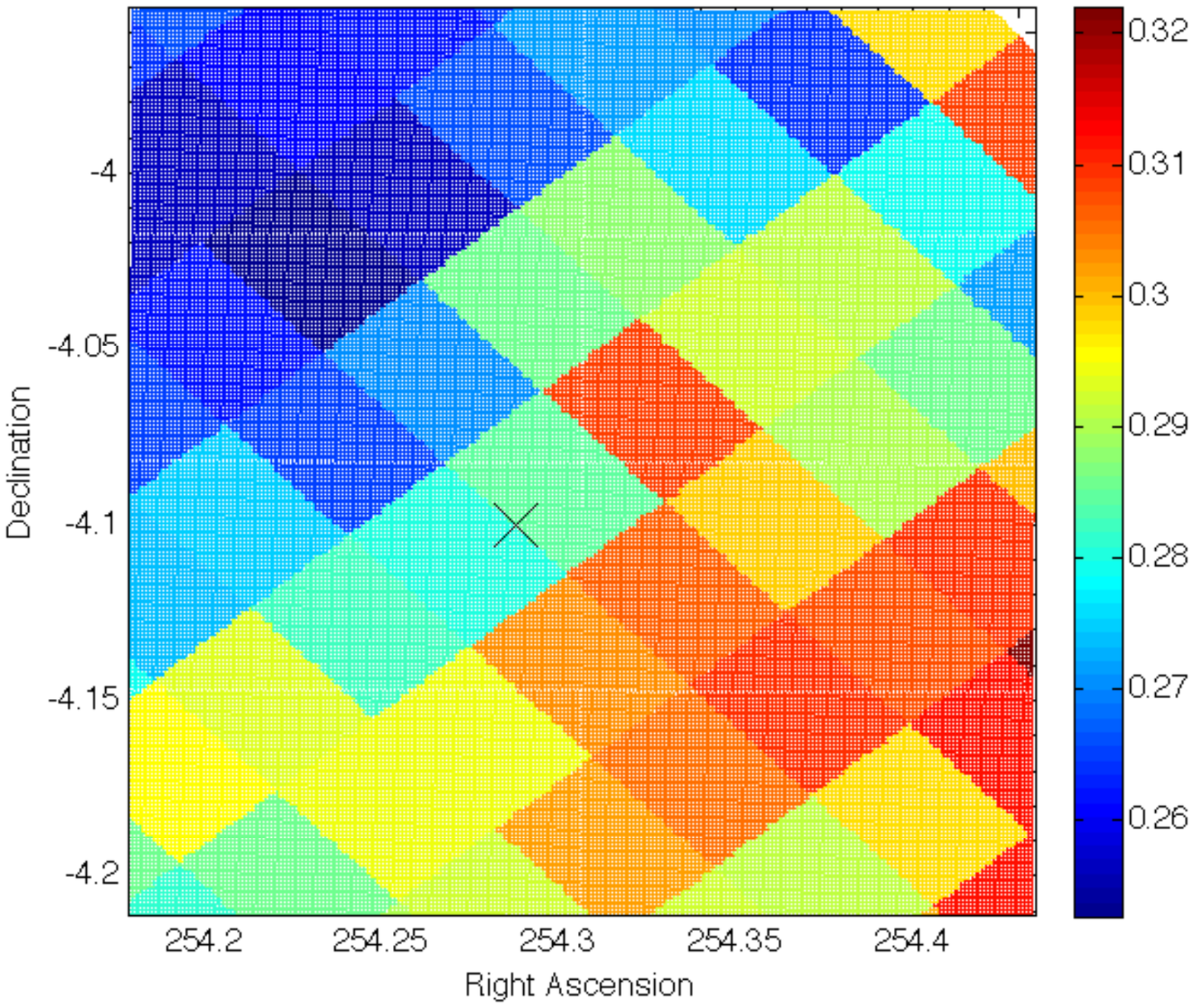}  &
\includegraphics[scale=0.21]{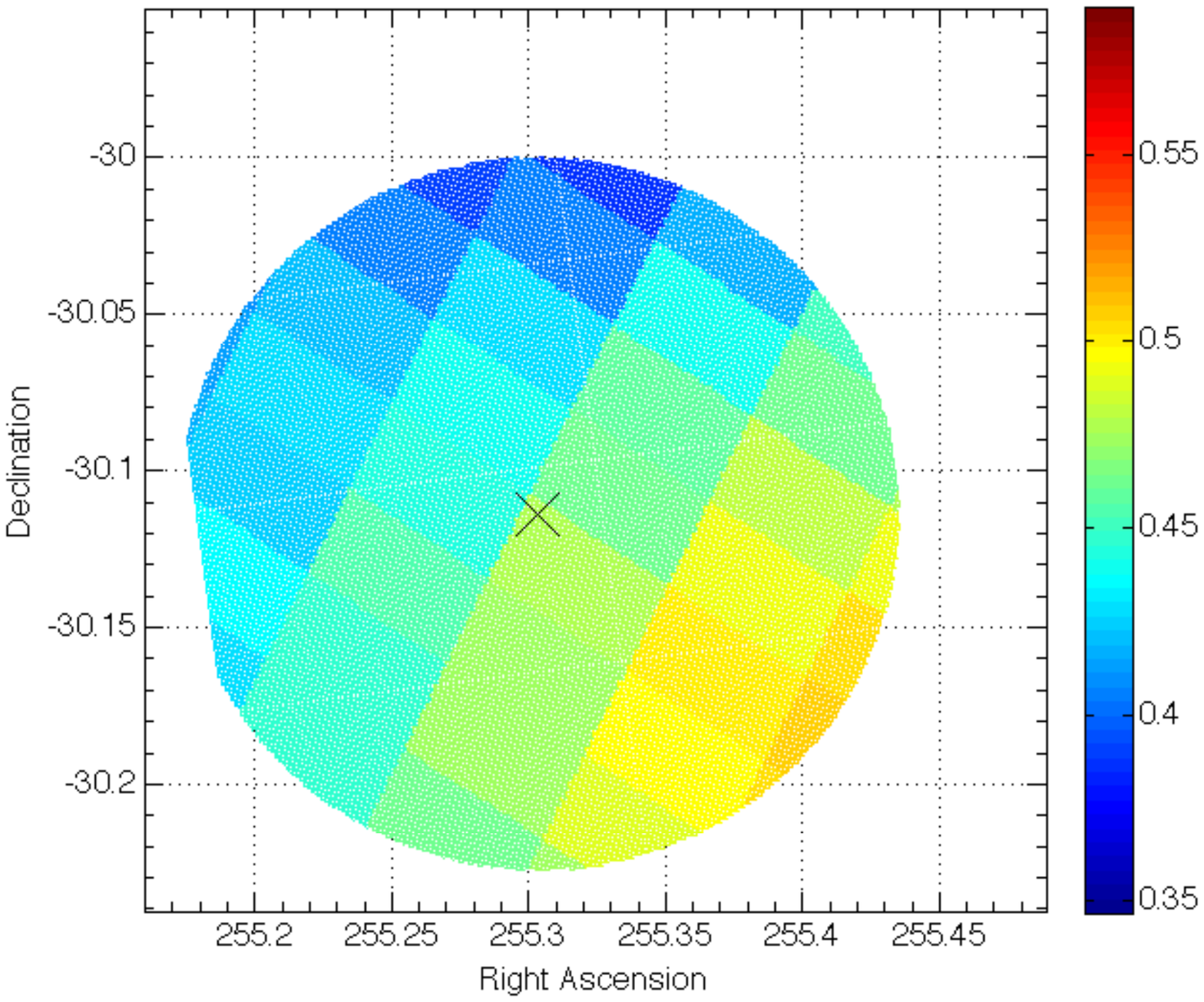}  &
\includegraphics[scale=0.21]{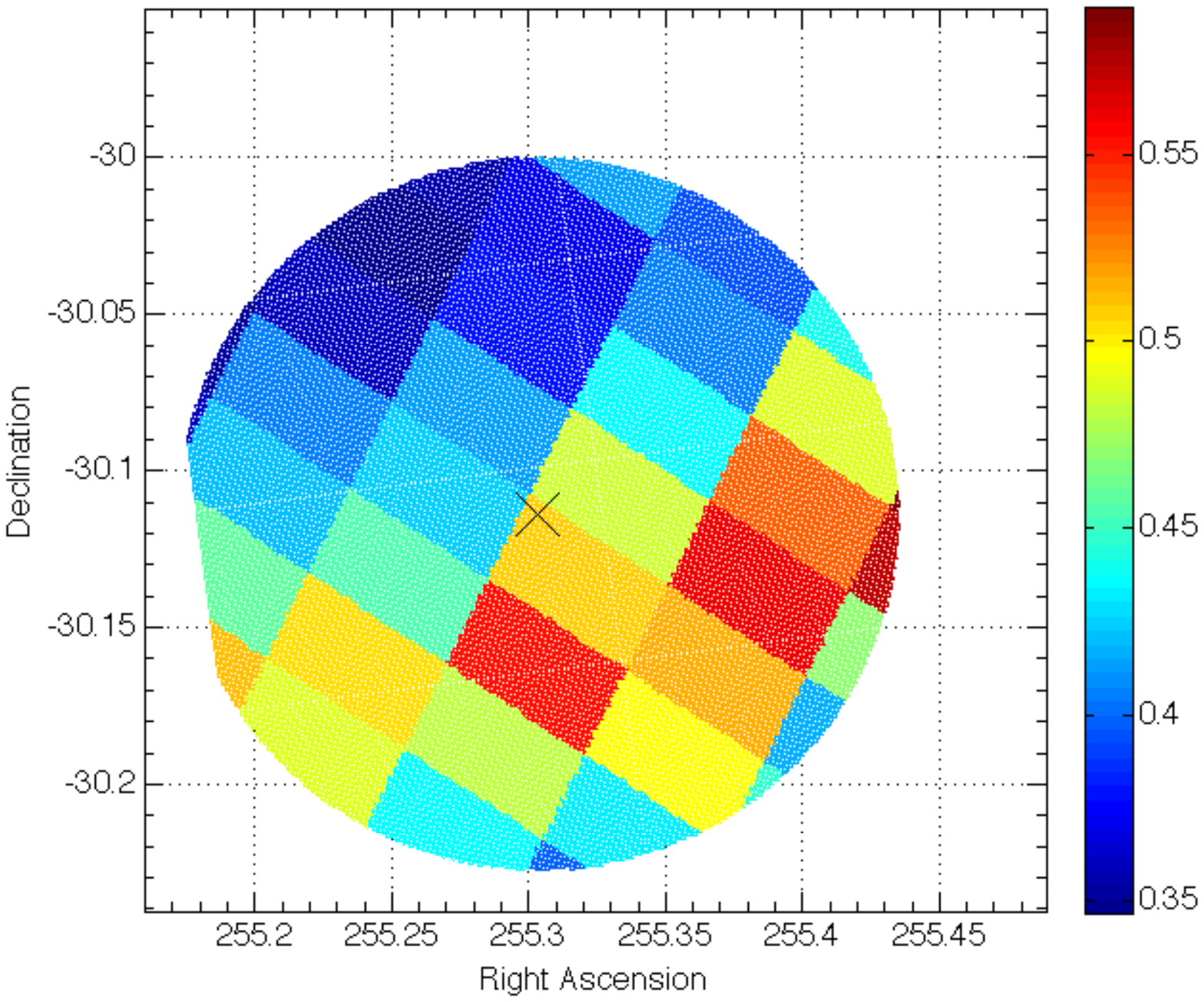} \\
\end{tabular}
\caption{Comparison of the E(B-V) extinction maps provided by our
  technique (left) with the SFD maps (right). We have lowered the
  resolution of our maps (see text) to compare them more easily with
  the SFD maps. From the comparison of both maps for every GC we have
  obtained a reddening zero point for our map, that we have added to
  facilitate the comparison.}
\label{figschelegel}
\end{figure}

We also present the extinction maps across the field of the clusters,
along with their resolution and precision (see Figures
\ref{figngc6121} to \ref{figngc6809}, and Table \ref{tabdifred}). We
compare these maps with the SFD extinction maps (see Figure
\ref{figschelegel}). The extinction maps provided by our dereddening
technique give the differential extinction in the field referred to
the unknown absolute extinction where our ridgelines lie. The
comparison with the SFD maps allows us to establish these absolute
extinction zero points. In order to make the comparison, we need to
bring down our extinction maps to the lower resolution of the SFD
maps. Since our extinction maps are described by non-parametric
continuous functions of the spatial coordinates right ascension and
declination, as mentioned in previous paragraphs, we evaluate these
functions in $\sim1600$ different points in every pixel of the SFD
maps and average the obtained values using our precision maps to
calculate a weighted average that provides us with one extinction
value per SFD pixel. We observe that both our maps and the SFD maps
generally agree in identifying the same regions of higher and lower
extinction, although our maps usually show higher variation (reaching
sometimes a factor of 2) in the extinction values than the SFD
maps. Notice also that the SFD maps give the extinction integrated
along the line of sight (i.e., extinction in the foreground, but also
in the background of the observed clusters), while the extinction maps
provided by our technique show the extinction just in the foreground
of the observed clusters.

\subsection{NGC 6121 - M 4 (Figure \ref{figngc6121})}

The changes in the extinction across the field are significant
(${\Delta}E(B-V)=0.15$). The highest extinction is located in the
northwestern region of the field, while the northeastern region of the
field shows the lowest. There is a small blob of higher
extinction (${\Delta}E(B-V)\sim0.05$) $\sim0.02\deg$ south of the
center of the cluster, which the SFD map fails to show (see Figure
\ref{figschelegel}). The absolute extinction zero point of our 
reddening map is $E(B-V)=0.50$ from comparing our map with the SFD
map. These significant values of extinction, both absolute and
differential, are a consequence of the cluster being located behind the
Ophiucus dust complex, about 120 pc away from us.

M 4 is likely the closest GC to the Sun. Because of this, stars in the
upper RGB are saturated in our images, even in the shortest exposures,
and they do not appear in our CMDs of this cluster.  The region
studied for this cluster encompasses the whole FOV of our CCD, since
the number of field stars present is small ($<20\%$ of the stars in
our CMD). Due to the proximity of M 4 to the Sun, most of the field
stars are located behind the cluster. They are generally dimmer than
the cluster stars, and the evolutionary sequences of the field stars
are easily differentiable from the cluster sequences. The definition
of the different branches in the CMD of the cluster are improved after
the dereddening process, especially in the $V,V-I$ CMD. The
evolutionary sequences get $\sim35\%$ tighter there. A noticeable
feature in the CMDs of this cluster is the presence of a bimodal HB,
with both red and blue components, though the blue HB (BHB) is not
very extended. This bimodality of the HB, noted also in previous
studies (e.g., \citet{ca98}, and references therein), suggests the
presence of several stellar populations in this cluster.

Some recent ground-based optical photometric studies of this cluster
include \citet{ro00a} in $VI$, \citet{mo02} in $UBV$, and \citet{an06}
in $BV$. Recent deep optical {\it HST} CMDs of this
cluster have been presented in $U_{336}V_{555}I_{814}$ by
\citet{ba04}, in $V_{555}I_{814}$ by \citet{ri04}, and in
$V_{606}I_{814}$ by \citet{fe04} and \citet{ha04a}. This cluster is
also a member of the ACS survey of GGCs obtained in the
$V_{606}I_{814}$ bands \citep{an08a}. Recent infrared photometric
studies of this cluster are by \citet{ch02} in $JK$, and by
\citet{fe00} in $VJK$. \citet{pu99} also used the {\it HST} for an
optical and infrared study of the stellar mass function of this
cluster in $V_{555}I_{814}J_{110}H_{160}$. The presence of significant
differential reddening in this cluster has been pointed out in many
studies, and a map of the differential extinction was provided by
\citet{mo02} (see Figure \ref{figmapcomp}). Recently, the presence of
distinct populations has been suggested from spectroscopic studies of
stars in the RGB \citep{ma08,ca10} and in the HB \citep{ma11a}.

\subsection{NGC 6144 (Figure \ref{figngc6144})}

NGC 6144 is also located behind the ${\rho}$ Ophiuchi dust cloud and
very close ($\sim 40'$ northeast) to NGC 6121 (M 4). Unsurprisingly,
NGC 6144 also suffers from high and differential reddening. The
variation of the extinction across the field is one of the highest
shown in our sample (${\Delta}E(B-V)=0.5$). Extinction is higher in
the northern region of the observed field, $\sim0.1\deg$ from the
center of the cluster, and peaks in the northeastern area. The
absolute extinction zero point of our reddening map is $E(B-V)=0.75$
from comparing our map with the SFD map. This is one of the few cases
in which the differential extinction variation across the field is
more important (by a factor of 2) in the SFD maps than in ours.

The region studied for this cluster encompasses the whole FOV of our
CCD, since there is only a moderate field star contamination ($\sim
50\%$ of the stars in our CMD). Also the field stars are redder on
average than the stars in the cluster whose different evolutionary
sequences are clearly distinguishable in the CMD. Even if previous
works (e.g., \citet{ma90,gr92}) give a higher value for the reddening
vector towards ${\rho}$ Ophiuchi dust cloud, with $R_V\sim4.3$, still
with the lower value assumed in our method, $R_V\sim3.1$, the
different evolutionary sequences get better defined ($\sim25\%$
tighter) in the CMD of the cluster after applying the dereddening
process. This cluster shows a short BHB, which suggests the presence
of a single population, or the dominance of one population.

The most recent ground-based optical photometric study of this cluster
we found was done in $BVI$ by \citet{ne00} for the RGB and HB
stars. From {\it HST} data, the first CMD of this cluster was
published by \citet{sa07} as a part of their $V_{606}I_{814}$ ACS
survey of GGCs. \citet{la10} provide also CMDs in
$U_{336}V_{606}H{\alpha}_{656}R_{675}I_{814}$ using HST data. Infrared
upper RGB photometry has been presented for this cluster by
\citet{da00} in $JHK$, and by \citet{mi95} in
$JK$.

\subsection{NGC 6218 - M 12 (Figure \ref{figngc6218})}

The differential extinction across the field of this cluster is very
mild (${\Delta}E(B-V)\sim0.04$). Extinction is higher in the northern
region ($\sim0.1\deg$ of the cluster center).  The absolute extinction
zero point of our reddening map is $E(B-V)=0.18$ from comparing with the
SFD map.

The region studied for this cluster encompasses the whole FOV of our
CCD, since the number of field stars present is small ($<10\%$ of the
stars in our CMD). The mild differential extinction produces only a
small improvement in the CMD of the cluster after being dereddened.
The evolutionary sequences get $\sim10\%$ tighter in $V,V-I$, a little
less in $V,B-V$. Some of the CMD features, like the RGB (Thompson)
bump, are more clearly visible now. A significant feature of this
cluster is its BHB with a blue tail, which suggests the presence of
several populations of stars.

Ground-based $VI$ photometric data on M12 down to a few magnitudes
under the TO have most recently been presented by \citet{vo02} and
\citet{ro00b}, $BV$ data by \citet{br96}, $BVI$ data by \citet{ha04b}
and deeper $VR$ data by \citet{ma06}. {\it HST} $B_{439}V_{555}$ data
on M12 were published by \citet{pi02} using the WFPC2. This cluster is
also a member of the ACS survey of GGCs in $V_{606}I_{814}$
\citep{an08a}. We have found no infrared photometrical study of this
cluster. \citet{vo02} mapped the extinction in this location, finding
little differential reddening in the field (see Figure
\ref{figmapcomp}). Recently, \citet{ca07} have suggested the presence
of two distinct populations with different helium content from a
spectroscopic study of RGB stars. \citet{ca10} also find
evidence for different populations from a spectroscopic study of stars
in the RGB.

\subsection{NGC 6235 (Figure \ref{figngc6235})}

The changes in extinction across the field are moderate
(${\Delta}E(B-V)\sim0.1$), being the southeastern area of the field
the region of highest extinction.  The absolute extinction 
zero point of our reddening map is $E(B-V)=0.39$ from comparing our map with
the SFD map. These moderately high values of extinction, both absolute
and differential, are a consequence of the cluster's projected position,
close to the Ophiucus dust complex.

The region studied for this cluster includes only stars located less than
$5.37\arcmin$ away from the cluster center, due to the high number of
field stars relative to the GC stars ($\sim 85\%$ of the stars when we
look at the whole FOV of IMACS). As with NGC 6144, the field stars are
redder on average than the cluster stars. This allow us to easily
differentiate the cluster and field evolutionary sequences.  These
sequences in the cluster get $\sim15\%$ tighter after being
differentially dereddened. This cluster shows a BHB, not very
extended, which suggests the presence of a single population, or the
dominance of one population.

The most recent ground-based optical photometric study of this cluster
we found was done in $BV$ for RGB and HB stars by \citet{ho03}, who
found a differential extinction relation along the X axis of their
data and used it to correct their photometry.  This cluster has also
been observed with the WFPC2 aboard the {\it HST} and its
$B_{439}V_{555}$ CMD has been published in the GGC survey by
\citet{pi02}. Infrared upper RGB photometry has been presented for
this cluster by \citet{da00} in $JHK$ and by \citet{mi95} in
$JK$. Deeper $JHK$ RGB near-infrared photometry was presented by
\citet{ch10}.

\subsection{NGC 6254 - M 10 (Figure \ref{figngc6254})}

The extinction variation across the field of this cluster is mild
(${\Delta}E(B-V)\sim0.1$). Extinction is higher towards the west,
peaking close ($\sim0.02\deg$) to the cluster center.  The absolute
extinction zero point of our reddening map is $E(B-V)=0.28$ from
comparing our map with the SFD map.

The region studied for this cluster encompasses the whole FOV of our
CCD, since the number of field stars present is small ($<10\%$ of the
stars in our CMD). The CMD of this cluster, with an extended BHB, and
its distance to the center of the Galaxy are very similar to those of
M 12. The evolutionary sequences get $\sim25\%$ tighter after being
differentially dereddened, and features like the the RGB bump are more
clearly visible now. THe BHB with a blue tail suggests the presence of
several stellar populations.

NGC 6254 has previously been studied using ground-based facilities by
\citet{ro00b} and \citet{vo02} in $VI$ and by \citet{po05} in
$BVI$. \citet{pi99b} used both ground-based facilities and {\it HST}
for their study of this cluster in the $VI$ bands. \citet{be10} used
also the {\it HST} for their mass segregation study of this cluster in
the $V_{606}I_{814}$ bands. This cluster is also a member of the ACS
survey of GGCs in $V_{606}I_{814}$ \citep{an08a}. \citet{va04a}
studied this cluster in the infrared $JK$ bands. The presence of
differential reddening has been previously noted, and \citet{vo02}
mapped the extinction in the cluster field, finding a higher
differential extinction than in M 12 (see Figure
\ref{figmapcomp}). \citet{ca10} find evidence for different
populations from a spectroscopic study of stars in the RGB.

\subsection{NGC 6266 - M 62 (Figure \ref{figngc6266})}

A band of material causing a significant increase in extinction
(${\Delta}E(B-V)\sim0.25$) crosses the field in the east-west
direction less than $0.02\deg$ south from the cluster center.
The absolute extinction zero point of our reddening map is
$E(B-V)=0.47$ from comparing our map with the SFD map. In the SFD map
the presence of this band of extinction is not so clear.

The area studied for this cluster is restricted to the inner
$6.84\arcmin$ of the cluster, due to the number of field stars ($\sim
50\%$ of the stars when we look at the whole FOV of IMACS) and the
relatively small tidal radius ($r_{t}=11.28'$). The definition of the
different branches in the CMD of the cluster is highly improved,
approximately by a factor of 2, after the dereddening process. Also
now the presence of an extended blue horizontal branch (EBHB) is
clearly visible and the location of the RGB bump is more easily
identifiable. A striking feature in the CMD of this cluster is the
presence of an extended and well populated HB, with both red and blue
components. The BHB in particular is very extended. The shape of the
HB and the fact that this is one of the most massive clusters in the
Milky Way make this cluster a clear candidate to have several stellar
populations.

Previous work using ground-based data includes that of \citet{ro00a}
in $VI$, and \citet{ge04}, \citet{co05} and \citet{co10} in
$BV$. \citet{be06} studied this cluster using a combination of $BVI$
(ground) and $U_{255}U_{336}V_{555}$ (space) bands. \citet{co08}
studied a pulsar in M 62 and used their {\it HST} data to show the
$B-R$ vs. $R$ CMD of the cluster. NGC 6266 is also included in the
{\it HST} GGCs survey in $B_{439}V_{555}$ published by
\citet{pi02}. \citet{va07} and \citet{ch10} studied this cluster in
the infrared $JHK$ bands. The presence of differential reddening has
been previously noted, and \citet{ge04} mapped the extinction towards
the central region of the cluster (see Figure \ref{figmapcomp}).

\subsection{NGC 6273 - M 19 (Figure \ref{figngc6273})}

The changes in the extinction across the field are significant
(${\Delta}E(B-V)\sim0.3$). Extinction is higher towards the east, and
peaks in a small region located very close, less than $0.02\deg$, of
the center of the cluster. The absolute extinction zero point of our
reddening map is $E(B-V)=0.32$ from comparing our map with the SFD
map.

The area studied for this cluster is restricted to the inner
$8.26\arcmin$ of the cluster, due to the number of field stars ($\sim
40\%$ of the stars when we look at the whole FOV of IMACS).  The
definition of the different branches in the CMD of the cluster is
highly improved after the dereddening process. The different
sequences get tighter by approximately a factor of 2 after being
differentially dereddened, and now the presence of an EBHB is clearly
visible and the location of the RGB bump is more easily
identifiable. The EBHB suggests the presence of several
stellar populations in this cluster.

The most recent optical photometric study of this cluster
we found was done by \citet{pi99a} in
$B_{439}V_{555}$ from {\it HST} images. As part of their study, they
obtained a map of the differential reddening of the central region of
the cluster.  \citet{pi02} included the CMD from this work in their
{\it HST} GGCs survey in $B_{439}V_{555}$. In the infrared
this cluster was studied in $JHK$ by \citet{ch10}, by \citet{da01} and by
\citet{va07}.

\subsection{NGC 6287 (Figure \ref{figngc6287})}

The variation of the extinction across the field is one of the highest
shown in our sample (${\Delta}E(B-V)=0.75$). Extinction is higher in
the northern region. A narrow band of material causing the highest
extinction in the field crosses it in the east-west direction less
than $0.02\deg$ from the cluster center. The absolute extinction
zero point of our reddening map is $E(B-V)=0.67$ from comparing our
map with the SFD map. These significant values of extinction, both
absolute and differential, are a consequence of the cluster's projected
proximity to the Ophiucus dust complex.

The area studied for this cluster is restricted to the inner
$4.13\arcmin$ of the cluster due to the number of field stars ($\sim
75\%$ of the stars when we look at the whole FOV of IMACS).  The
definition of the different branches in the CMD of the cluster is
highly improved, approximately by a factor of 2, after the dereddening
process. Now the RGB and SGB are easily identifiable, which was not
the case before applying the dereddening technique. A short BHB is observed
in the CMD, suggesting the presence of a single population, or the
dominance of one population.

Previous photometric studies in the optical were done in $BV$
by \citet{st94} from ground-based observations and in $VI$ by
\citet{fu99} from space-based observations. NGC 6287 is included in
the {\it HST} GGCs survey in $B_{439}V_{555}$ published by
\citet{pi02}. In the infrared, \citet{ch10} and \citet{da01} studied this cluster in $JHK$, and \citet{le01} in $J_{110}H_{160}$ using the {\it HST}.

\subsection{NGC 6304 (Figure \ref{figngc6304})}

The variation of the extinction across the field is moderate
(${\Delta}E(B-V)=0.15$). The area located $\sim0.05\deg$ east of the
center of the cluster is the region of highest extinction.  The
absolute extinction zero point of our reddening map is $E(B-V)=0.49$
from comparing our map with the SFD map.

The area studied for this cluster only includes stars with distances less
than $3.65\arcmin$ away from the cluster center, due to the high
number of field stars relative to the GC stars ($>90\%$ of the stars
when we look at the whole FOV of IMACS). The improvement of the CMD of
this cluster is only marginal.  The CMD of this cluster shows a very
short, red HB (RHB), typical of high-metallicity GGCs.

This cluster has been previously observed and studied using
ground-based facilities by \citet{or00} in $BV$ and \citet{ro00a} in
$VI$. It is one of the clusters in the GGCs survey by \citet{pi02} in
$B_{439}V_{555}$ and in the ACS GGCs survey in $V_{606}I_{814}$
\citep{an08a} done using {\it HST} data. \citet{va05} studied this
cluster in the infrared $JHK$ bands. \citet{va07} include results
from this study in their infrared compilation of bulge GGCs.

\subsection{NGC 6333 - M 9 (Figure \ref{figngc6333})}

The changes in the extinction across the field are significant
(${\Delta}E(B-V)\sim0.25$). The area located $\sim0.1\deg$ southwest
of the cluster center is the region of highest extinction. 
The absolute extinction zero point of our reddening map is
$E(B-V)=0.43$ from comparing our map with the SFD map.

The area studied for this cluster is restricted to the inner
$5.59\arcmin$ of the cluster due to the number of field stars ($\sim
45\%$ of the stars when we look at the whole FOV of IMACS) and the
relatively small tidal radius ($r_{t}=8.16'$).  The evolutionary
sequences of M 9 get $\sim35\%$ tighter after being differentially
dereddened. This cluster posseses a BHB with a blue tail, which
suggests the presence of several stellar populations in this cluster.

The most recent optical photometric study of this cluster we found
is that of \citet{ja91} in $BV$ for stars brighter than the TO
point. Although this cluster has been previously observed with the
{\it HST} (see Table \ref{tabobsarc}), the first CMD of M 9 using
space data is published in this paper in $BVI$. 
In the infrared \citet{da01} and \citet{ch10} studied this cluster in $JHK$.

\subsection{NGC 6342 (Figure \ref{figngc6342})}

The changes in the extinction across the field are significant
(${\Delta}E(B-V)\sim0.4$).  The northwestern quadrant of the field,
reaching the projected position of the center of the cluster, is the
region of highest extinction.  The absolute extinction 
zero point of our reddening map is $E(B-V)=0.59$ from comparing our map with
the SFD map.

The area studied for this cluster reaches only stars with distances less
than $4.21\arcmin$ away from the cluster center, due to the
significant number of field stars relative to GC stars ($\sim
70\%$ of the stars when we look at the whole FOV of IMACS). The
definition of the different branches in the CMD of the cluster is
highly improved after the dereddening process, getting tighter by a
factor of 2. Now the RGB and SGB are easily identifiable, along with a RHB,
which was not the case before applying the dereddening technique. The RHB is typical of high-metallicity GGCs.

The only recent optical photometric studies found for this cluster are
those of \citet{he99} in $VI$ using ground-based facilities and that
of \citet{pi02} in $B_{439}V_{555}$ using the {\it HST}. In the
infrared \citet{va04b} studied this cluster in $VJHK$. \citet{va07}
include results from this study in their infrared compilation of
bulge GGCs. A map of the differential reddening was produced by
\citet{he99} (see Figure \ref{figmapcomp}).

\subsection{NGC 6352 (Figure \ref{figngc6352})}

The differential extinction variation across the field of this cluster is
only moderate (${\Delta}E(B-V)\sim0.15$). Extinction is higher in the
northern region of the analyzed field. The absolute
extinction zero point of our reddening map is $E(B-V)=0.34$ from
comparing our map with the SFD map.

The area studied for this cluster is restricted to the inner
$5.61\arcmin$ of the cluster, due to the high number of field stars
relative to the GC stars ($>90\%$ of the stars when we look at all the
available FOV).  The evolutionary sequences get $\sim30\%$ tighter
after being differentially dereddened.  The cluster shows a RHB,
typical of high-metallicity GGCs.

This cluster was studied by \citet{sa94} in $BV$ and \citet{ro00a} and \citet{pa10} in
$VI$ from the ground, and from space by \citet{fu95} in $VI_C$, using
pre-COSTAR data, and by \citet{fa02} in $V_{555}I_{814}$ and by
\citet{pu03} in $V_{606}I_{814}$. It is also a member of the ACS
survey of GGCs in $V_{606}I_{814}$ \citep{an08a}. We have found no infrared
photometrical study of this cluster. \citet{pa10} show the presence of two populations of stars with anticorrelated CN and CH band strengths.

\subsection{NGC 6355 (Figure \ref{figngc6355})}

The changes in the extinction across the field are significant
(${\Delta}E(B-V)\sim0.3$). The southwestern area of the cluster is the
region of highest extinction. The absolute extinction zero point of our
reddening map is $E(B-V)=1.20$ from comparing our map with the SFD
map.

The area studied for this cluster is restricted to the inner
$2.36\arcmin$ of the cluster, due to the high number of field stars
relative to the GC stars ($>90\%$ of the stars when we look at all the
available FOV). The evolutionary sequences get $\sim10\%$ tighter
after being differentially dereddened. This cluster posseses a BHB,
although the small number of stars
in the HB, due to the small studied field, makes it difficult to
infer any information about different populations in this cluster.

The only previous recent optical photometric studies found are those by
\citet{or03} in $BVI$ using ground-based facilities, and that by
\citet{pi02} in $B_{439}V_{555}$ using the {\it HST}. In the infrared
\citet{ch10}, \citet{da01}, and \citet{va07} studied this cluster in $JHK$.

\subsection{NGC 6397 (Figure \ref{figngc6397})}

The differential extinction in this cluster is only mild
(${\Delta}E(B-V)\sim0.05$). Extinction is not especially concentrated
in any area of the field. The absolute extinction zero point of our
reddening map is $E(B-V)=0.18$ from comparing our map with the SFD
map.

This is the cluster with the smallest apparent distance modulus,
though it is not the one closest to the Sun. As for M 4, stars in the
upper RGB are saturated in our images, even in the shortest exposures, and
they do not appear in our CMDs of this cluster.  The contamination from
field stars is moderate ($\sim 30\%$ of the stars when we look at all
the available FOV), mainly affecting MS stars in the cluster.  This
moderate contamination allows us to study almost all the available
FOV, up to $8.18\arcmin$ from the cluster center.  The sequences of
NGC 6397 get $\sim10\%$ tighter after being differentially dereddened.
This cluster shows a short BHB, which suggests the presence of a
single population, or the dominance of one population.

Some of the most recent studies of this cluster done with ground-based
facilities are those by \citet{ro00a} in $VI$, \citet{an06} in
$UBVI$, and \citet{ka06} in $BV$.  This cluster has also been
observed with the {\it HST}. A series of papers has been published
dedicated to the study of the characteristics of this cluster using
very deep photometry obtained with the ACS
\citep{an08b,hu08,ri08,da08}. NGC 6397 is also included in the GGC
survey from \citet{pi02} in $B_{439}V_{555}$ and of the ACS GGC survey
in $V_{606}I_{814}$ \citep{an08a}. \citet{ma00} also used the {\it
  HST} for an optical and infrared study of the stellar mass function
of this cluster in $V_{606}I_{814}J_{110}H_{160}$. Recent
spectroscopic studies of this cluster have found different chemistry
among its stars: \citet{ca10} show some O-Na
anti-correlation is present, although the number
statistics is small (only 16 stars studied, the smallest in their
sample of GCs); \citet{li11} claim the presence of two different
stellar populations that have different chemical compositions of N, O,
Na, Mg, and probably Al, and note that the cluster is clearly
dominated ($75\%$) by the second generation.

\subsection{NGC 6522 (Figure \ref{figngc6522})}

The differential extinction variation in this cluster is moderate
(${\Delta}E(B-V)=0.1$), although we should realize that the studied
region is also small (see next paragraph).  Extinction is higher in the
southwestern region of the analyzed field. The absolute extinction 
zero point of our reddening map is $E(B-V)=0.58$ from comparing our map
with SFD map.

This cluster is located in Baade's Window. The high number of
non-cluster member stars ($>90\%$ of the stars when we look at all the
available FOV) restricts the study in this region to stars closer than
$2.33\arcmin$ from the cluster center. The improvement of the CMD of
this cluster after being dereddened is only marginal. It
shows a BHB with a blue tail, which suggests the presence of several
stellar populations in this cluster.

The most recent optical photometric studies found, done with ground-based
data, are those by \citet{ba94} in $BV$, and \citet{te94} in
$BVI$. \citet{te98} show a differential extinction relation with
respect to the center of the cluster and use it to correct their
previous photometry. Using the {\it HST}, the cluster was studied by
\citet{sh98} in the $U_{336}B_{439}$ bands using pre-COSTAR data. It
is also one of the clusters in the \citet{pi02} GCs survey in
$B_{439}V_{555}$. In the infrared, 
\citet{da01} and \citet{va10} studied this cluster in $JHK$.

\subsection{NGC 6541 (Figure \ref{figngc6541})}

The differential extinction in this cluster is moderate
(${\Delta}E(B-V)\sim0.1$). Extinction is not especially concentrated
in any area of the field, although there seems to be an increase towards the
southeast area of the observed field. The absolute extinction
zero point of our reddening map is $E(B-V)=0.16$ from comparing our
map with the SFD map.

The region studied for this cluster encompasses most of the FOV of our
CCD, up to $8.63\arcmin$ from the cluster center, since there is only
a moderate field star contamination ($\sim 40\%$ of the stars in our
CMD). Also the field stars are redder on average than the stars in the
cluster, which lets us distinguish NGC 6541 evolutionary sequences
unambiguosly. The sequences in the cluster get $\sim10\%$ tighter
after being differentially dereddened.  This cluster possesses a BHB
with a blue tail, which suggests the presence of several stellar
populations.

Ground-based studies of this cluster were presented by \citet{al97} in
$BVRI$ and by \citet{ro00a} in $VI$. It has also been studied using {\it
  HST} WFPC2 data by \citet{le06} in $VI$, and it is one of the
clusters in the ACS GGC survey observed in $V_{606}I_{814}$
\citep{an08a}. In the infrared, 
\citet{ki06} studied this cluster in $JHK$.

\subsection{NGC 6553 (Figure \ref{figngc6553})}

There are significant changes in the extinction across the field
(${\Delta}E(B-V)\sim0.25$). In our maps, the highest extinction is
located in the northern region of the field, while the southwestern
region of the field shows the lowest, which is not the trend shown in
the SFD maps, but it does agree well with the trend of
the map provided by \citet{he99} (see Figure \ref{figmapcomp}). The
absolute extinction zero point of our reddening map is $E(B-V)=1.35$
from comparing our map with the SFD map.

The area studied for this cluster is restricted to the inner
$3.91\arcmin$ of the cluster, due to the high number of field stars
relative to GC stars ($>90\%$ of the stars when we look at all the
available FOV).  The upper MS and RGB get $\sim30\%$ tighter after
being differentially dereddened, and the RGB bump now is more clearly
identifiable. The well-populated bump, the very extended upper RGB and
a RHB, suggest that this is a metal-rich GGCs.

Some of the most recent optical photometric studies of this cluster
are that by \citet{sa99} in $VI$ using ground-based data and those by
\citet{be01} and \citet{zo01} in $V_{555}I_{814}$ using {\it HST}
data. \citet{zo01} calculated the extinction map, not shown in their
paper, and applied it to get a clean CMD. \citet{he99} show an
extinction map (see Figure \ref{figmapcomp}) calculated with data from
\citet{sa99}. \citet{gu98} in $VIJK$, and \citet{fe00} in $VJK$
studied this cluster in the optical and in the infrared. \citet{va07}
includes results from this last study in their infrared compilation of
bulge GGCs.

\subsection{NGC 6558 (Figure \ref{figngc6558})}

The extinction variations in this cluster are significant
(${\Delta}E(B-V)\sim0.15$), especially when we consider the small size
of the studied field (see next paragraph). 
The absolute
extinction zero point of our reddening map is $E(B-V)=0.51$ from
comparing our map with SFD map.

The high number of non-cluster member stars ($>90\%$ of the stars when
we look at the whole FOV of IMACS) restrict the study of this region to
stars closer than $1.81\arcmin$ from the cluster center. The
improvement of the CMD is only marginal. This cluster shows a BHB,
although the low definition of the HB due to the small number of stars
in this region caused by the small field FOV studied makes it difficult to
infer any information about different populations in this cluster.

We found only two previous optical photometric studies using
ground-based data, those by \citet{ri98} and by \citet{ba07}, both in
$VI$. Although this cluster has been previously observed with the {\it
  HST}, the first CMD of the deep MS stars using space-based data is
published in this paper in $VI$.  In the infrared
\citet{ch10} studied this cluster in $JHK$.

\subsection{NGC 6624 (Figure \ref{figngc6624})}

The differential extinction across the field is moderate
(${\Delta}E(B-V)\sim0.15$). Extinction is higher in the southeastern
region of the analyzed field. The absolute extinction zero point of our
reddening map is $E(B-V)=0.25$ from comparing our map with the SFD
map.

The area studied for this cluster is restricted to the inner
$3.92\arcmin$ of the cluster, due to the high number of field stars
relative to the GC stars ($>90\%$ of the stars when we look at all the
available FOV). There are also some dim blue field stars in our CMD
that belong to the Sagittarius dwarf galaxy. Stars from this galaxy's
MS are easily observed in the CMD at magnitudes dimmer than $V\sim21$
and colors bluer than the cluster MS. The improvement of the CMD of
the cluster after being dereddened is only marginal.  This cluster
possesses a RHB, characteristic of a metal-rich GGC.

This cluster has been studied using ground-based data by \citet{sa94},
who presented a photometric study of the RGB and HB stars in
$BV$. \citet{ri94} and \citet{ro00a} obtained deeper photometry,
reaching a few magnitudes below the TO in $BV$ and $VI$,
respectively. Using {\it HST}, studies of this cluster were
published by \citet{so95} in $BV$ using pre-COSTAR data, and by
\citet{gu96} in $U_{336}B_{439}V_{555}$ and \citet{he00} in $VI$. This
cluster is also a member of the \citet{pi02} $B_{439}V_{555}$ GGCs survey,
and of the $V_{606}I_{814}$ ACS GGCs survey \citep{an08a}. In the
infrared \citet{va04b} studied this cluster in $VJHK$. \citet{va07}
include results from this study in their infrared compilation of
bulge GGCs.

\subsection{NGC 6626 - M 28 (Figure \ref{figngc6626})}

The extinction variations across the field of this cluster are
significant (${\Delta}E(B-V)\sim0.2$). Extinction is lower at the
projected center and southwestern region of the observed field, which
is not the trend observed in the SFD maps. The absolute extinction 
zero point of our reddening map is $E(B-V)=0.46$ from comparing our map
with the SFD map.

The area studied for this cluster is restricted to the inner
$4.95\arcmin$ of the cluster, due to the high number of field stars
relative to the GC stars ($>90\%$ of the stars when we look at all the
available FOV).  Due to its proximity, the cluster's brightest stars
are brighter on average than the field's, and we can identify without
much difficulty stars from the RGB. The evolutionary sequences of M28
get $\sim35\%$ tighter after being differentially dereddened. This
cluster shows a BHB with a blue tail, which suggests the presence of several
stellar populations.

This cluster has been studied in the optical using ground-based data
by \citet{ro00a} in $VI$, and using {\it HST} data by \citet{te01} and
by \citet{go01} in $VI$. \citet{da96b} used a combination of optical
and infrared bands ($BVJK$) to study this cluster. In the infrared
\citet{ch10} studied this cluster in $JHK$.

\subsection{NGC 6637 - M 69 (Figure \ref{figngc6637})}

The differential extinction across most of the observed field is
moderate (${\Delta}E(B-V)\sim0.1$). Extinction is not especially
concentrated in any area of the field.
The absolute extinction zero point of our reddening map is
$E(B-V)=0.17$ from comparing our map with the SFD map.

The area studied for this cluster is restricted to the inner
$4.39\arcmin$ of the cluster, due to the high number of field stars
relative to the GC stars ($>90\%$ of the stars when we look at all the
available FOV).  There are also some dim blue field stars in our CMD
that belong to the Sagittarius dwarf galaxy. Stars from this galaxy's
MS are easily observed in the CMD at magnitudes dimmer than $V\sim21$
and colors bluer than the cluster MS. The improvement of the CMD of
this cluster is only marginal. This cluster posseses a RHB, typical of high-metallicity GGCs.

This cluster has been the subject of several studies, usually along
with the also metal-rich NGC 6624. Using ground-based data
\citet{fe94}, in $BVJK$, and \citet{sa94}, in $BV$, present a
photometric study of the RGB and HB stars of this cluster, and
\citet{ri94} and \citet{ro00a} got deeper photometry, reaching a
few magnitudes below the TO, in $BV$ and $VI$ respectively. Using {\it
  HST} data, \citet{he00} studied this cluster in $VI$. This cluster
is also a member of the \citet{pi02} $B_{439}V_{555}$ GGCs survey, and of
the $V_{606}I_{814}$ ACS GGCs survey \citep{an08a}. In the infrared
this cluster was studied by \citet{fe00} in $VJK$ and by \citet{va05}
in $JHK$. \citet{va07} include results from these studies in their
infrared compilation of bulge GGCs.

\subsection{NGC 6642 (Figure \ref{figngc6642})}

The differential extinction in this cluster is high
(${\Delta}E(B-V)\sim0.2$). The absolute extinction zero point of our
reddening map is $E(B-V)=0.36$ from comparing our map with the SFD
map.

During the observation of this cluster with the Magellan telescope, we
had the least photometric conditions of our run (see Table
\ref{taboffsets}). This resulted in the shallowest CMD of our study,
reaching only stars down to the very upper parts of the MS. The high
number of non-cluster member stars ($>90\%$ of the stars when we look
to all the available FOV) restricts the study in this area to stars
closer than $2.22\arcmin$ from the cluster center. The improvement of
the CMD of this cluster after being dereddened is only marginal. A
prominent feature in the CMDs of this cluster is the
presence of a bimodal HB, with both red and blue components, though
the blue BHB is not very extended. This bimodality of the HB can suggest the presence of several stellar populations.

The most recent optical photometric study of NGC 6642 using
ground-based data was done by \citet{ba06} in $BVI$. Using {\it HST},
\citet{ba09} studied this cluster in $V_{606}I_{814}$. NGC 6642
is also a member of the \citet{pi02} $B_{439}V_{555}$ GGCs survey. In the
infrared 
\citet{da01}, \citet{ki06}, and \citet{va07} studied this cluster in $JHK$.

\subsection{NGC 6656 - M 22 (Figure \ref{figngc6656})}

The extinction variation across the field of this cluster is moderate
(${\Delta}E(B-V)\sim0.1$). Extinction is lower in the southern area
of the field.
The absolute extinction zero point of our reddening map is
$E(B-V)=0.33$ from comparing our map with the SFD map.

The relative proximity of M22 explains why the upper RGB stars are
saturated in our observations and are excluded from the CMDs of this
cluster. The region studied for this cluster encompasses the whole FOV
of our CCD, since there is only a moderate field star contamination
($\sim 40\%$ of the stars in our CMD). Due to the proximity of M 22,
most of the field stars are located behind the cluster. They are
generally dimmer than the cluster stars, and the evolutionary
sequences of the field stars are easily differentiable from the
brighter sequences of the cluster.  After being dereddened, the
sequences of the cluster get $\sim25\%$ tighter in $V,V-I$, a little
less in $V,B-V$.  An EBHB suggests the presence of several
stellar populations in this cluster.

There are several optical photometric studies for this cluster. Using
ground-based data, it was studied by \citet{mo04} in $BVI$, by
\citet{we04} in $UV$, by \citet{ka01} in $BV$, by \citet{ro00a} in
$VI$, and by \citet{an95}, by \citet{ri99} and by \citet{le09} in Str\"{o}mgrem $uvbyCa$, $vby$ and $VbyCa$ bands, respectively.  \citet{pi99b} used both ground
facilies and the {\it HST} for their study of this cluster in $VI$,
and \citet{an03} show CMDs for the cluster in
$B_{435}V_{606}R_{675}I_{814}$ bands. This cluster is also a member of
the ACS survey of GGCs in $V_{606}I_{814}$ \citep{an08a}. \citet{pi09} shows a double SGB in his study of M 22 using $V_{606}I_{814}$.  In the
infrared \citet{mi95}, \citet{da96a} and \citet{ch02} studied this
cluster in $JK$. The presence of a metallicity spread in this cluster has been the subject of a long debate, complicated by the presence of insterstellar reddening (see \citet{pa10} for a brief historic review). 
Some recent studies seem to confirm the presence of at least two
stellar populations with different chemical compositions, both
photometrically \citep{pi09} and spectroscopically
\citep{ma09b,da09,ca10,ma11b}, although the studies by \citet{ka08}
and by \citet{pa10} do not find clear CH and CN anti-correlations or
bimodalities for the stars in their sample.

\subsection{NGC 6681 - M 70 (Figure \ref{figngc6681})}

The differential extinction across most of the observed field is mild
(${\Delta}E(B-V)\sim0.05$). Extinction is not especially concentrated
in any area of the field. The absolute extinction zero point of our
reddening map is $E(B-V)=0.09$ from comparing our map with the SFD
map.

The area studied for this cluster is restricted to the inner
$5.21\arcmin$ of the cluster, due to the high number of Galactic field
stars relative to the GC stars ($\sim 70\%$ of the stars when we look
to all the available FOV).  The field stars are redder on average than
the cluster stars, which allows us to easily distinguish the different
evolutionary sequences. There are also some dim blue field stars in
our CMD that belong to the Sagittarius dwarf galaxy. Stars from this
galaxy's MS are easily observed in the CMD at magnitudes dimmer than
$V\sim21$ and colors bluer than the cluster MS.  The improvement of
the CMD of this cluster is only marginal. M 70 posseses a BHB with a
blue tail, suggesting the presence of several stellar populations in
this cluster.

The most recent optical photometric studies we found, using ground-based data,
are those by \citet{br96} in $BV$, and by \citet{ro00a} in $VI$. Using
{\it HST} data, \citet{pi02} included it in their GGCs survey in
$B_{435}V_{555}$. This cluster is also a member of the
$V_{606}I_{814}$ ACS survey of GGCs \citep{an08a}. In the infrared
\citet{ki06} studied this cluster in $JHK$.

\subsection{NGC 6809 - M55 (Figure \ref{figngc6809})}

The extinction variation across the field of this cluster is only mild
(${\Delta}E(B-V)\sim0.05$). Extinction is higher in the northwestern
area of the observed field. The improvement of the CMD of this cluster
is only marginal. The absolute extinction zero point of our reddening
map is $E(B-V)=0.14$ from comparing our map with the SFD map.

The region studied for this cluster encompasses the whole FOV of
our CCD, since the number of field stars present is small
($<10\%$ of the stars in our CMD), though there is
clear contamination from stars in the background Sagittarius dwarf
galaxy. Stars from its MS are easily observed in the CMD at magnitudes
dimmer than $V\sim21$ and colors bluer than the cluster MS.  The tip
of the RGB is saturated in our CMDs of this cluster, due to the small
apparent distance modulus of M 55. The improvement of the CMD of this
cluster is only marginal. The cluster shows a BHB, not very extended,
suggesting the presence of a single population, or the dominance of
one population.

The latest optical photometric studies of this cluster done using
ground-based data are by \citet{ka05} in $UBVI$, by \citet{py01} and
by \citet{zl11} in $BV$, and by \citet{ro00a} in $VI$. Using a
combination of ground-based and {\it HST} data, this cluster was
studied by \citet{la07} in $BVI$, and by \citet{ba08} in
$BVRIH_{\alpha}$.  The ACS $VI$ photometry used by \citet{ba08} comes
from \citet{an08a}, since this cluster is one of the members of the
ACS GGC survey. In the infrared this cluster was studied by
\citet{fe00} in $VJK$. The presence of more than one stellar
population has been suggested by \citet{ca10} based on the Na-O
anti-correlation found in their study, but the studies by \citet{ka08}
and by \citet{pa10} do not find a clear CH and CN anti-correlations or
bimodalities on the stars in their sample.

\section{Summary}
This article is the second of a series devoted to the study of the
inner GGCs. High and patchy differential extintion in the line of
sight makes difficult to extract accurate and meaningful information
about the physical parameters of the GCs located in the Galactic bulge
and inner disk and halo. In this article we present a highly
homogeneous and complete optical sample of the brightest GCs located
towards the center of the Milky Way.
We produce some of the deepest CMDs of the 25 inner GGCs that 
comprise our sample, using Magellan 6.5m telescope and {\it HST}
observations. We apply a new dereddening technique to map the
differential extinction in front of these GCs and eliminate their
effects in the CMDs\footnote{We are working on making this photometric
  database easily accesible to the community through a webpage}. The
GGCs in our sample show tighter evolutionary sequences, up to a factor
of 2 in some cases, in the CMDs built after applying the dereddening
method. These cleaned CMDs will be used in subsequent papers to
analyze the astrophysical parameters of the GCs and their stellar
populations. The high-definition extinction maps will be also used in
future papers to study the properties of the obscuring material along
the GGCs' lines of sight.

\acknowledgments This study was supported by grants 0206081 from NSF
and GO10573.01-A from STScI. STcI is operated by NASA under contract
to AURA. Support was also provided by the Ministry for
the Economy, Development, and Tourism's Programa Iniciativa
Cient\'{i}fica Milenio through grant P07-021-F, awarded to The Milky
Way Millennium Nucleus, by Proyecto Fondecyt Regular 1110326, by Basal
PFB-06, FONDAP 15010003, and Anillos ACT-86. This publication makes
use of data products from the Two Micron All Sky Survey, which is a
joint project of the University of Massachusetts and the Infrared
Processing and Analysis Center/California Institute of Technology,
funded by the National Aeronautics and Space Administration and the
National Science Foundation. This research used the facilities of the
Canadian Astronomy Data Centre operated by the National Research
Council of Canada with the support of the Canadian Space Agency.

\appendix

\section{DoPHOT}
\label{ap1}
In this work, we have obtained the photometry using a new version of
DoPHOT. DoPHOT is a software created to extract PSF photometry from
astronomical images. A complete and deep explanation of its main
characteristics, and a description of how it works, is given by \citet{sc93}

The main new characteristics of the distribution of DoPHOT we
developed for this work are:
\begin{itemize}
\item All the program subroutines and functions have been rewritten to
  make them compatible with the standards of Fortran90.
\item We have tried to make the subroutines and functions more easily
  readable.  To achieve this, we have eliminated all the common
  statements present in previous versions and named all the variables
  used by the subroutine and function in the call statement. Also with
  the same spirit of making the program more understandable we have
  created, combined or eliminated some of the subroutines, and
  included, at the beginning of each file, a brief explanation of what
  every function and subroutine does.
\item The program works now with real values for the pixel values, so
  there is no need to worry about conversions to integers as in
  previous distributions.
\item The stars are masked, obliterated, added, and subtracted across
  the different subroutines of the program in circular sections, and
  not in rectangular sections as before. This way fewer usable pixels
  are obliterated, and fewer pixels outside an interesting radius are
  dealt with.
\end{itemize}

In addition to this, this distribution shares some of the characteristics of 
the program that were present in previous distributions, but that were not 
mentioned in the original paper:
\begin{itemize}
\item DoPHOT allows for a variable PSF model in the image.
\item DoPHOT can perform an aperture correction to the magnitudes
  obtained, according to their positions and magnitudes. The program
  checks for variations in the aperture correction in magnitude and in
  position in the image, and corrects the fitted magnitude
  accordingly.
\end{itemize}

This new version of DoPHOT is publicly available through:\\
http://www.astro.puc.cl/$\sim$jalonso/dophot.tar

\section{Instrumental calibration}
\label{ap2}
The camera in IMACS consists of eight chips, each with its own slightly
different mean quantum efficiency and color sensitivity.  To calibrate
our data, we must first transform the photometry from each of the
individual chips to a common zero point, then transform this system to
a standard system--in our case, the Johnson-Cousins system
\citep{be79}.  In this appendix, we shall focus on the first step of
transforming the individual chips to a common instrumental system.

For this purpose, we arbitrarily chose one of the chips, chip 2 (see
Figure \ref{figchipcal}), as the one defining the instrumental system.
The goal was to tie all the other chips to this chip's photometric zero
point.  We attempted to do this in two ways.  First, we observed one
\citet{la92} field (PG 1047+003) in every one of the chips and
in all three filters (BVI) in rapid succession on every night of our
run. There were very few stars in this field, and the scatter we found
in the mean photometric offsets -- about 0.05 mag between chips -- was
too large given our need to bring all our photometry to a common
system to ~1-2\% precision.  We surmise that the relatively poor
results from this approach reflects the quality of the nights during
the 2005 run, and also possible flat-field variations at the 2-3\%
level for this run.

We tried a second approach during a sequence of observations during a
2006 Magellan run (see Table \ref{tabobsmag}) in which we tried to
auto-calibrate the IMACS chip array using stars from the cluster NGC
6397.  The basic idea was to obtain a minimal number of exposures in
each filter so that at least part of all chips overlapped with chip 2,
the chip defining our instrumental photometric system.  We found that
a sequence of three exposures could achieve the desired overlap.
Figure \ref{figchipcal} illustrates the approach.  In the left panel,
the darker outline of the IMACS chip shows the location of a first
exposure, in this case at the same location as the actual field we
observed in 2005 for the cluster NGC 6397.  The second exposure, shown
as a lighter (red) outline is offset in both RA and Dec so that four
of its chips (1, 2, 5 and 6) overlap with the position of chip 2 in the
first exposure. The right panel again shows the initial pointing in
the darker outline, with a third pointing as a lighter (green)
outline.  Now, chips 3, 4, 7 and 8 of the last exposure overlap with
chip 2 from the first exposure.  This sequence, repeated for each
filter, generated a large list of stars that have been observed on
chip2 {\it and} on every other chip in the array.  Because only three
exposures were required and because we took these in rapid succession,
we can safely ignore airmass variations as we calculate the offsets
from a given chip to the system defined by chip 2.  In addition, we
have exposures from the 2005 and 2006 runs of the {\it same} fields
(the dark outlines in Figure \ref{figchipcal}) that allow us to
determine any possible variations in chip sensitivities between runs.

The model we used to transform each chip to the chip 2 system has the
form
\begin{equation} 
m_2 = m_i + Z_i + \gamma_i c_i ,
\end{equation}
where $m_i$ is the instrumental magnitude of chip $i$ (excluding
$i=2$), $Z_i$ is the zero-point offset from chip $i$ to chip 2, and
$\gamma_i$ is the color coefficient for instrumental color $c_i$.  In
practice, we carried out this analysis separately for each filter, B,
V, and I, with the colors defined appropriately for each filter
($(B-V)$ for the B-band transformation and $(V-I)$ for the V and
I-band transformations).  We found that the color terms are all
negligibly small in the chip-to-chip transformations for each filter,
so the final model adopts $\gamma_i = 0$ for all chips and all
colors. Thus we only had to detemine the mean values of the $Z_i$
coefficients as the weighted mean of the differences of the
instrumental magnitudes, $m_{i} - m_{2}$, for all stars in common
between chip $i$ and chip 2.  Weights for each measurement were
assigned according to the photometric errors and the dispersion in the
offset ditribution.  A first application of this approach, to the
calibration sequence of NGC 6397 obtained in 2006 (Figure
\ref{figchipcal}), allowed us to place all the data from that run on
the 2006 chip 2 instrumental system.

Since the relative sensitivities of the chips might have changed
between the 2005 and 2006 runs, we then took an additional step to
define the chip-to-chip offsets for the 2005 run from which all our
ground-based cluster photometry is defined. This process involves the
observations of NGC 6397 from our 2006 run (described above) and the
observations of the same field for the same cluster obtained in 2005.

For this purpose we adopted the following transformation: 
\begin{equation} 
m_{i,06} = m_{i,05} + Y_i + \delta_i c_i + k_{05}X_{05} -k_{06}X_{06}.
\end{equation} 
The zero-point offset from the 2005 observations required to place
them on the 2006 system for chip $i$ is given by $Y_i$ (this now
includes the case $i=2$), for a color term $\delta_i$, and an
instrumental color on the 2005 system of $c_i$.  The final two terms
account for the different airmasses and possibly different extinction
coefficients for each run and each observation.  We do not know this
product (extinction coefficient $k$ times airmass $X$) for the 2006
run, because we did not observe standard stars on that run. But we
know that the sum of these products is constant when comparing the
2005 and 2006 NGC 6397 observations. Thus, we can rewrite the equation
above as
\begin{equation} 
m_{i,06} = m_{i,05} + Y'_i +\delta_i c_i ,
\end{equation} 
in close analogy to Equation B1.  Note that the new zero point offset
now implicitly includes the airmass terms:
\begin{equation} 
Y'_i = Y_i + k_{05}X_{05} - k_{06}X_{06}.
\end{equation} 
We can simplify this due to the fact that we found that the color
terms in Equation B3 are all negligibly small. Thus we only had to
detemine the mean values of the $Y'_i$ coefficients as the weighted
mean of the differences of the instrumental magnitudes, $m_{i,06} -
m_{i,05}$, for all stars from chip $i$. As before, weights for each
measurement were assigned according to the photometric errors and the
dispersion in the offsets distribution.

At this stage, we can transform the 2005 NGC 6397 data to a common
instrumental system defined by chip 2 from the 2006 observations
of this same cluster.  But what we really require are the offsets in
the 2005 system that bring all data from the IMACS chips obtained in
2005 to a common system.  That is, the final chip-to-chip offsets for
the 2005 data are simply
\begin{equation} 
\Delta Y'_i = Y'_i - Y'_2 .
\end{equation} 

The values of $Z$ and $\Delta Y'$ required to tie the various chips
systems to a common system defined by chip 2 are listed in Table
\ref{tabzp}. The errors in the values of $Z$ and $\Delta Y'$ are
${\sigma}\sim0.001$, the error of the mean of the distributions used
to obtain them (see Figure \ref{figzp}). The dispersions of the
distributions from which these $Z$ and $\Delta Y'$ are derived, a
measure of the internal precision of our photometry, are
${\sigma}\sim0.01$.

\section{Absolute calibration}
\label{ap3}
We placed our photometry in the Johnson-Cousins photometric system by
observing \citet{la92} fields over a range of airmasses during
the nights of our 2005 observing run. Since the first two nights were
the most photometric of our run, we only used the standard fields
observed those two nights to obtain the following transformations:
\begin{equation}
B = 3.417 + b - 0.283 X + 0.130 (b-v) ,
\end{equation}
\begin{equation}
V = 3.706 + v - 0.166 X - 0.017 (b-v) ,
\end{equation}
\begin{equation}
I = 3.065 + i - 0.049 X - 0.020 (v-i) ,
\end{equation}
where $X$ is the airmass, $b,v,$ and $i$ are the instrumental
magnitudes in the instrumental system and $B,V,$ and $I$ the
corresponding magnitudes in the Johnson-Cousins photometric system.

To fine-tune our calibration we decided to compare our photometry with
\citet{st00} photometric standard stars. We found in some
cases significant absolute offsets between Stetson's and our
photometry.  These zero points could be easily calculated (see Table
\ref{tabcalstars} and Figure \ref{figstetmag}). However, only thirteen
of our clusters had stars in common with Stetson's catalog, and only
twelve had stars calibrated in all three filters.

To extend the comparison to more clusters, we looked for members of
our sample already observed with at least one of the two {\it HST}'s
optical wide-field imaging instruments, WFPC2 and ACS. We found that
all of the clusters have already been observed at least in two of our
filters, and most of them in all three (see Table
\ref{tabcalstars}). We retrieved the available data from the {\it HST}
archive and derived photometry from the stars in these data using the
programs HSTPHOT for the WFPC2 data \citep{do00} and DolPHOT for the
ACS data \citep{do00}. These programs have been specifically tailored
to analyze data from these cameras.

The comparison between the {\it HST} and ground-based data is
complicated by the large resolution difference in the datasets and the
highly crowded fields. When the offsets in the photometries of
individual stars of a cluster are plotted, there is a clear,
asymmetric spread in the data (see Figure \ref{fighstmag}). Due to the
higher resolution of the {\it HST}, individual stars in the {\it HST}
data are blends in the ground-based data.  To obtain the absolute
offset between the two systems eliminating the blended stars effect we
calculated a weighted mean of the offsets of the stars in common (see
Figure \ref{fighstmag}), where the weights were assigned according to
the photometric errors and the dispersion in the offset
distribution. The weighted mean was iteratively calculated, ignoring
stars 2${\sigma}$ away from the mean value at each iteration. Whenever
we have zero points calculated separately from independent WFPC2 and
ACS data, a mean value was adopted.

The results were then compared with the ones we found from the Stetson
comparison, since our aim is to put all the photometry in the Stetson
system (see Figure \ref{fighststet}). A small offset for each filter
was found and added to the individual cluster zero points
(${\Delta}B=-0.045\pm0.012$ mag, ${\Delta}V=-0.056\pm0.007$ mag, and
${\Delta}I=-0.025\pm0.008$ mag). These offsets may be due to the fact that the
{\it HST} observations used in our absolute calibration are located
closer to the center of the cluster than the Stetson stars used, or
they may reflect small system zero point differences. The final offsets
applied to the individual clusters can be seen in Table
\ref{taboffsets}.

\clearpage

\begin{deluxetable}{cccccccccc}
  \tablewidth{0pc} 
  \tablecolumns{10}
  \tablecaption{Characteristics of the GCs in our sample according to the 2010 version of the \citet{ha96} GGC catalog. \label{tabgc}}
  
  \tablehead{\colhead{Cluster name} &
    \colhead{l\tablenotemark{1}} &
    \colhead{b\tablenotemark{2}} & 
    \colhead{$R_{GC}$\tablenotemark{3}} &
    \colhead{$R_{\odot}$\tablenotemark{4}} &
    \colhead{$[Fe/H]$} &
    \colhead{$E(B-V)$} &
    \colhead{$(m-M)_{V}$} &
    \colhead{$V$} &
    \colhead{$r_{t}$\tablenotemark{5}}}
  
  \startdata
NGC 6121 (M 4)&350.97&15.97&5.9&2.2&-1.16&0.35&12.82&-7.19&51.82\\
NGC 6144&351.93&15.70&2.7&8.9&-1.76&0.36&15.86&-6.85&33.35\\
NGC 6218 (M 12)&15.72&26.31&4.5&4.8&-1.37&0.19&14.01&-7.31&17.28\\
NGC 6235&358.92&13.52&4.2&11.5&-1.28&0.31&16.26&-6.29&11.18\\
NGC 6254 (M 10)&15.14&23.08&4.6&4.4&-1.56&0.28&14.08&-7.48&18.47\\
NGC 6266 (M 62)&353.57&7.32&1.7&6.8&-1.18&0.47&15.63&-9.18&11.28\\
NGC 6273 (M 19)&356.87&9.38&1.7&8.8&-1.74&0.38&15.90&-9.13&14.57\\
NGC 6287&0.13&11.02&2.1&9.4&-2.10&0.60&16.72&-7.36&6.96\\
NGC 6304&355.83&5.38&2.3&5.9&-0.45&0.54&15.52&-7.30&13.25\\
NGC 6333 (M 9)&5.54&10.71&1.7&7.9&-1.77&0.38&15.67&-7.95&8.00\\
NGC 6342&4.90&9.72&1.7&8.5&-0.55&0.46&16.08&-6.42&15.81\\
NGC 6352&341.42&-7.17&3.3&5.6&-0.64&0.22&14.43&-6.47&10.44\\
NGC 6355&359.59&5.43&1.4&9.2&-1.37&0.77&17.21&-8.07&15.81\\
NGC 6397&338.17&-11.96&6.0&2.3&-2.02&0.18&12.37&-6.64&15.81\\
NGC 6522&1.02&-3.93&0.6&7.7&-1.34&0.48&15.92&-7.65&15.81\\
NGC 6541&349.29&-11.19&2.1&7.5&-1.81&0.14&14.82&-8.52&13.04\\
NGC 6553&5.26&-3.03&2.2&6.0&-0.18&0.63&15.83&-7.77&7.66\\
NGC 6558&0.20&-6.02&1.0&7.4&-1.32&0.44&15.70&-6.44&9.49\\
NGC 6624&2.79&-7.91&1.2&7.9&-0.44&0.28&15.36&-7.49&18.97\\
NGC 6626 (M 28)&7.80&-5.58&2.7&5.5&-1.32&0.40&14.95&-8.16&11.23\\
NGC 6637 (M 69)&1.72&-10.27&1.7&8.8&-0.64&0.18&15.28&-7.64&7.92\\
NGC 6642&9.81&-6.44&1.7&8.1&-1.26&0.40&15.79&-6.66&9.77\\
NGC 6656 (M 22)&9.89&-7.55&4.9&3.2&-1.70&0.34&13.60&-8.50&31.90\\
NGC 6681 (M 70)&2.85&-12.51&2.2&9.0&-1.62&0.07&14.99&-7.12&9.49\\
NGC 6809 (M 55)&8.79&-23.27&3.9&5.4&-1.94&0.08&13.89&-7.57&15.32\\
  \enddata
  \tablenotetext{1}{Galactic longitude, in degrees.}
  \tablenotetext{2}{Galactic latitude, in degrees.}  
  \tablenotetext{3}{Galactocentric distance, in kiloparsecs.}  
  \tablenotetext{4}{Distance from the Sun, in kiloparsecs.}
  \tablenotetext{5}{Tidal radius, in arcminutes.}
\end{deluxetable}

\begin{deluxetable}{ccccccc}
  \tablewidth{0pc} \tablecolumns{7} \tablecaption{Summary of the
    observations with the Magellan telescope. \label{tabobsmag}}
  
  \tablehead{\colhead{Cluster} &
    \colhead{$RA_{2000}$\tablenotemark{1}} &
    \colhead{$Dec_{2000}$\tablenotemark{2}} & \colhead{UT Date} &
    \colhead{Filter} & \colhead{Exp. Time\tablenotemark{3}} &
    \colhead{Airmass}}
  
  \startdata
  NGC6121&16 23 40.0&-26 30 29.8&2005 Jun 2&$B$&$3\times1,3\times25$&1.37-1.42\\
  \nodata&\nodata&\nodata&\nodata&$V$&$3\times1,3\times20$&1.43-1.48\\
  \nodata&\nodata&\nodata&\nodata&$I$&$3\times1,3\times15$&1.50-1.57\\
  NGC6144&16 27 18.7&-26 00 28.4&2005 May 31&$B$&$3\times5,3\times120$&1.38-1.52\\
  \nodata&\nodata&\nodata&\nodata&$V$&$3\times5,3\times120$&1.55-1.65\\
  \nodata&\nodata&\nodata&\nodata&$I$&$1\times1,3\times3,3\times75$&1.77-1.92\\
  NGC6218&16 47 18.6&-01 55 51.6&2005 Jun 1&$B$&$3\times5,3\times120$&1.34-1.44\\
  \nodata&\nodata&\nodata&\nodata&$V$&$3\times5,3\times120$&1.46-1.55\\
  \nodata&\nodata&\nodata&\nodata&$I$&$1\times1,3\times3,3\times75$&1.56-1.66\\
  NGC6235&16 53 29.8&-22 09 37.8&2005 May 31&$B$&$3\times5,3\times120$&1.19-1.27\\
  \nodata&\nodata&\nodata&\nodata&$V$&$3\times5,3\times120$&1.28-1.34\\
  \nodata&\nodata&\nodata&\nodata&$I$&$1\times1,3\times3,3\times75$&1.36-1.44\\
  NGC6254&16 57 13.0&-04 04 58.3&2005 Jun 1&$B$&$3\times5,3\times120$&1.14-1.16\\
  \nodata&\nodata&\nodata&\nodata&$V$&$3\times5,3\times120$&1.17-1.20\\
  \nodata&\nodata&\nodata&\nodata&$I$&$1\times1,3\times3,3\times75$&1.21-1.25\\
  NGC6266&17 01 17.2&-30 05 42.8&2005 May 30&$B$&$3\times5,3\times120$&1.01-1.02\\
  \nodata&\nodata&\nodata&\nodata&$V$&$3\times5,3\times120$&1.02-1.03\\
  \nodata&\nodata&\nodata&\nodata&$I$&$1\times1,3\times3,3\times75$&1.03-1.04\\
  NGC6273&17 02 41.5&-26 15 17.8&2005 May 30&$B$&$3\times5,3\times120$&1.00\\
  \nodata&\nodata&\nodata&\nodata&$V$&$3\times5,3\times120$&1.00-1.01\\
  \nodata&\nodata&\nodata&\nodata&$I$&$1\times1,3\times3,3\times75$&1.01\\
  NGC6287&17 05 13.8&-22 41 28.2&2005 Jun 2&$B$&$3\times5,3\times120$&1.24-1.32\\
  \nodata&\nodata&\nodata&\nodata&$V$&$3\times5,3\times120$&1.33-1.42\\
  \nodata&\nodata&\nodata&\nodata&$I$&$1\times1,3\times3,3\times75$&1.44-1.53\\
  NGC6304&17 14 37.1&-29 26 43.2&2005 May 31&$B$&$3\times5,3\times120$&1.06-1.08\\
  \nodata&\nodata&\nodata&\nodata&$V$&$3\times5,3\times120$&1.09-1.11\\
  \nodata&\nodata&\nodata&\nodata&$I$&$1\times1,3\times3,3\times75$&1.12-1.15\\
  NGC6333&17 19 16.1&-18 30 00.0&2005 Jun 2&$B$&$3\times5,3\times120$&1.02-1.03\\
  \nodata&\nodata&\nodata&\nodata&$V$&$3\times5,3\times120$&1.02\\
  \nodata&\nodata&\nodata&\nodata&$I$&$1\times1,3\times3,3\times75$&1.02\\
  NGC6342&17 21 14.5&-19 34 13.1&2005 May 31&$B$&$3\times5,3\times120$&1.04-1.06\\
  \nodata&\nodata&\nodata&\nodata&$V$&$3\times5,3\times120$&1.02-1.04\\
  \nodata&\nodata&\nodata&\nodata&$I$&$1\times1,3\times3,3\times75$&1.02\\
  NGC6355&17 24 03.1&-26 20 12.2&2005 Jun 2&$B$&$3\times5,3\times120$&1.03-1.06\\
  \nodata&\nodata&\nodata&\nodata&$V$&$3\times5,3\times120$&1.06-1.09\\
  \nodata&\nodata&\nodata&\nodata&$I$&$1\times1,3\times3,3\times75$&1.09-1.12\\
  NGC6352&17 25 35.1&-48 24 20.40&2005 Jun 1&$B$&$3\times5,3\times120$&1.06\\
  \nodata&\nodata&\nodata&\nodata&$V$&$3\times5,3\times120$&1.06-1.07\\
  \nodata&\nodata&\nodata&\nodata&$I$&$1\times1,3\times3,3\times75$&1.08\\
  NGC6397&17 40 48.2&-53 39 23.4&2005 May 31&$B$&$3\times1,3\times25$&1.11\\
  \nodata&\nodata&\nodata&\nodata&$V$&$3\times1,3\times20$&1.10\\
  \nodata&\nodata&\nodata&\nodata&$I$&$3\times1,3\times15$&1.10\\
%  \nodata&\nodata&\nodata&2006 Apr 20&$B$&$3\times15$&1.10\\
%  \nodata&\nodata&\nodata&\nodata&$V$&$3\times15$&1.10\\
%  \nodata&\nodata&\nodata&\nodata&$I$&$3\times15$&1.10\\
  NGC6522&18 03 38.5&-30 00 59.6&2005 May 30&$B$&$3\times5,3\times120$&1.01-1.02\\
  \nodata&\nodata&\nodata&\nodata&$V$&$3\times5,3\times120$&1.02-1.03\\
  \nodata&\nodata&\nodata&\nodata&$I$&$1\times1,3\times3,3\times75$&1.04-1.05\\
  NGC6541&18 08 07.8&-43 41 20.7&2005 May 31&$B$&$3\times5,3\times120$&1.15-1.18\\
  \nodata&\nodata&\nodata&\nodata&$V$&$3\times5,3\times120$&1.18-1.22\\
  \nodata&\nodata&\nodata&\nodata&$I$&$1\times1,3\times3,3\times75$&1.23-1.26\\
  NGC6553&18 09 20.1&-25 53 26.6&2005 May 31&$B$&$3\times5,3\times120$&1.02-1.03\\
  \nodata&\nodata&\nodata&\nodata&$V$&$3\times5,3\times120$&1.03-1.06\\
  \nodata&\nodata&\nodata&\nodata&$I$&$1\times1,3\times3,3\times75$&1.07-1.09\\
  NGC6558&18 10 23.1&-31 44 48.0&2005 May 30&$B$&$3\times5,3\times120$&1.07-1.10\\
  \nodata&\nodata&\nodata&\nodata&$V$&$3\times5,3\times120$&1.11-1.13\\
  \nodata&\nodata&\nodata&\nodata&$I$&$1\times1,3\times3,3\times75$&1.14-1.18\\
  NGC6624&18 23 45.2&-30 20 39.7&2005 May 30&$B$&$3\times5,3\times100$&1.21-1.24\\
  \nodata&\nodata&\nodata&\nodata&$V$&$3\times5,3\times100$&1.25-1.30\\
  \nodata&\nodata&\nodata&\nodata&$I$&$1\times1,3\times3,3\times60$&1.31-1.36\\
  NGC6626&18 24 37.3&-24 51 10.8&2005 Jun 2&$B$&$3\times5,3\times120$&1.02-1.04\\
  \nodata&\nodata&\nodata&\nodata&$V$&$3\times5,3\times120$&1.04-1.05\\
  \nodata&\nodata&\nodata&\nodata&$I$&$1\times1,3\times3,3\times75$&1.06-1.07\\
  NGC6637&18 31 28.0&-32 19 55.6&2005 Jun 2&$B$&$3\times5,3\times120$&1.20-1.24\\
  \nodata&\nodata&\nodata&\nodata&$V$&$3\times5,3\times120$&1.25-1.30\\
  \nodata&\nodata&\nodata&\nodata&$I$&$1\times1,3\times3,3\times75$&1.17-1.19\\
  NGC6642&18 31 58.7&-23 27 36.0&2005 Jun 2&$B$&$3\times5,3\times120$&1.36-1.43\\
  \nodata&\nodata&\nodata&\nodata&$V$&$3\times5,3\times120$&1.45-1.53\\
  \nodata&\nodata&\nodata&\nodata&$I$&$1\times1,3\times3,3\times75$&1.56-1.67\\
  NGC6656&18 36 28.6&-23 53 13.5&2005 May 31&$B$&$3\times1,3\times35$&1.52-1.58\\
  \nodata&\nodata&\nodata&\nodata&$V$&$3\times1,3\times30$&1.45-1.51\\
  \nodata&\nodata&\nodata&\nodata&$I$&$3\times1,3\times20$&1.38-1.43\\
  NGC6681&18 43 17.4&-32 16 29.7&2005 Jun 2&$B$&$3\times5,3\times120$&1.00-1.01\\
  \nodata&\nodata&\nodata&\nodata&$V$&$3\times5,3\times120$&1.01-1.02\\
  \nodata&\nodata&\nodata&\nodata&$I$&$1\times1,3\times3,3\times75$&1.02-1.03\\
  NGC6809&19 40 04.1&-30 56 43.8&2005 May 31&$B$&$3\times1,3\times35$&1.07-1.08\\
  \nodata&\nodata&\nodata&\nodata&$V$&$3\times1,3\times30$&1.09-1.10\\
  \nodata&\nodata&\nodata&\nodata&$I$&$3\times1,3\times20$&1.11-1.12\\
  \enddata

  \tablenotetext{1}{Units of right ascension are hours, 
    minutes, and seconds.}
  \tablenotetext{2}{Units of declination are
    degrees, arcminutes, and arcseconds.}  
  \tablenotetext{3}{Number of exposures $\times$ exposure time, in seconds.}  
 
\end{deluxetable}

\begin{deluxetable}{cccccc}
  \tablewidth{0pc} \tablecolumns{6} \tablecaption{Summary of ACS
    observations from Snapshot program 10573. \label{tabobsacs}}
  
  \tablehead{\colhead{Cluster} &
    \colhead{$\alpha_{2000}$\tablenotemark{1}} &
    \colhead{$\delta_{2000}$\tablenotemark{2}} &
    \colhead{UT Date} &
    \colhead{Filter} &
    \colhead{Exp. Time\tablenotemark{3}}}
  
  \startdata
  NGC6218&16 47 14.50&-01 56 52.0&2006 Feb 1&$B_{435}$&$120$\\
  \nodata&\nodata&\nodata&\nodata&$V_{555}$&$50$\\
  \nodata&\nodata&\nodata&\nodata&$I_{814}$&$20$\\
  NGC6333&17 19 11.80&-18 30 59.0&2006 May 31&$B_{435}$&$2\times340$\\
  \nodata&\nodata&\nodata&\nodata&$V_{555}$&$240$\\
  \nodata&\nodata&\nodata&\nodata&$I_{814}$&$90$\\
  NGC6553&18 09 17.50&-25 54 28.0&2006 Apr 4&$B_{435}$&$3\times340$\\
  \nodata&\nodata&\nodata&\nodata&$V_{555}$&$300$\\
  \nodata&\nodata&\nodata&\nodata&$I_{814}$&$60$\\
  NGC6624&18 23 40.70&-30 21 39.0&2006 Jun 5&$B_{435}$&$360$\\
  \nodata&\nodata&\nodata&\nodata&$V_{555}$&$160$\\
  \nodata&\nodata&\nodata&\nodata&$I_{814}$&$65$\\
  NGC6637&18 31 23.20&-32 20 53.0&2006 Jun 6&$B_{435}$&$300$\\
  \nodata&\nodata&\nodata&\nodata&$V_{555}$&$120$\\
  \nodata&\nodata&\nodata&\nodata&$I_{814}$&$50$\\
  \enddata

  \tablenotetext{1}{Units of right ascension are hours, 
    minutes, and seconds.}
  \tablenotetext{2}{Units of declination are
    degrees, arcminutes, and arcseconds.}  
  \tablenotetext{3}{Number of exposures $\times$ exposure time, in seconds.}  
 
\end{deluxetable}

\begin{deluxetable}{cccc}
  \tablewidth{0pc} \tablecolumns{4} \tablecaption{Summary of {\it HST}
    programs whose archival data have been used in our analysis.
    \label{tabobsarc}}
  
  \tablehead{\colhead{Cluster} &
    \colhead{Camera} &
    \colhead{Program} &
    \colhead{Filters}}
  
  \startdata
  NGC6121&WFPC2&6116&$V_{555},I_{814}$\\
%  \nodata&ACS&10120&$B_{435},R_{625}$\\
  \nodata&ACS&10775&$V_{606},I_{814}$\\
  NGC6144&WFPC2&11014&$B_{439},R_{675}$\\
  \nodata&ACS&10775&$V_{606},I_{814}$\\
  NGC6218&WFPC2&8118&$B_{439},V_{555}$\\
%  \nodata&ACS&10005&$B_{435},R_{625}$\\
  \nodata&ACS&10775&$V_{606},I_{814}$\\
  NGC6235&WFPC2&7470&$B_{439},V_{555}$\\
  \nodata&ACS&\nodata&\nodata\\
  NGC6254&WFPC2&6113&$V_{606},I_{814}$\\
  \nodata&ACS&10775&$V_{606},I_{814}$\\
  NGC6266&WFPC2&8118&$B_{439},V_{555}$\\
  \nodata&WFPC2&8709&$V_{555},I_{814}$\\
%  \nodata&ACS&10120&$B_{435},R_{625}$\\
  \nodata&ACS&\nodata&\nodata\\
  NGC6273&WFPC2&7470&$B_{439},V_{555}$\\
  \nodata&ACS&\nodata&\nodata\\
  NGC6287&WFPC2&6561&$B_{439},V_{555},I_{814}$\\
  \nodata&ACS&\nodata&\nodata\\
  NGC6304&WFPC2&7470&$B_{439},V_{555}$\\
  \nodata&ACS&10775&$V_{606},I_{814}$\\
  NGC6333&WFPC2&5366&$V_{555},I_{814}$\\
  \nodata&ACS&\nodata&\nodata\\
  NGC6342&WFPC2&7470&$B_{439},V_{555}$\\
  \nodata&ACS&\nodata&\nodata\\
  NGC6352&WFPC2&5366&$V_{555},I_{814}$\\
  \nodata&ACS&10775&$V_{606},I_{814}$\\
  NGC6355&WFPC2&7470&$B_{439},V_{555}$\\
  \nodata&ACS&\nodata&\nodata\\
  NGC6397&WFPC2&5929&$B_{439},V_{555}$\\
  \nodata&WFPC2&7335&$V_{555},I_{814}$\\
  \nodata&ACS&10775&$V_{606},I_{814}$\\
  NGC6522&WFPC2&6095&$B_{439},V_{555}$\\
%  \nodata&ACS&9690&$B_{435},R_{625}$\\
  \nodata&ACS&\nodata&\nodata\\
  NGC6541&WFPC2&5366&$V_{555},I_{814}$\\
%  \nodata&ACS&10120&$B_{435},R_{625}$\\
  \nodata&ACS&10775&$V_{606},I_{814}$\\
  NGC6553&WFPC2&7307&$V_{555},I_{814}$\\
  \nodata&ACS&\nodata&\nodata\\
  NGC6558&WFPC2&\nodata&\nodata\\
  \nodata&ACS&9799&$V_{606},I_{814}$\\
  NGC6624&WFPC2&5324&$B_{439},V_{555}$\\
  \nodata&WFPC2&5366&$V_{555},I_{814}$\\
  \nodata&ACS&10775&$V_{606},I_{814}$\\
  NGC6626&WFPC2&6779&$V_{555},I_{814}$\\
  \nodata&ACS&\nodata&\nodata\\
  NGC6637&WFPC2&8118&$B_{439},V_{555}$\\
  \nodata&WFPC2&5366&$V_{555},I_{814}$\\
  \nodata&ACS&10775&$V_{606},I_{814}$\\
  NGC6642&WFPC2&8118&$B_{439},V_{555}$\\
  \nodata&ACS&9799&$V_{606},I_{814}$\\
  NGC6656&WFPC2&5344&$B_{439},R_{675}$\\
  \nodata&WFPC2&7615&$V_{606},I_{814}$\\
  \nodata&ACS&10775&$V_{606},I_{814}$\\
  NGC6681&WFPC2&8723&$B_{439},V_{555}$\\
  \nodata&ACS&10775&$V_{606},I_{814}$\\
  NGC6809&WFPC2&\nodata&\nodata\\
  \nodata&ACS&10775&$V_{606},I_{814}$\\
  \enddata

\end{deluxetable}

\begin{deluxetable}{cccc}
  \tablewidth{0pc} \tablecolumns{4} \tablecaption{Limits for the stars used in our analysis. \label{tabclimit}}

  \tablehead{\colhead{Cluster} &
    \colhead{$V_{CL}$\tablenotemark{a}} &
    \colhead{$\Delta V_{CL}$\tablenotemark{b}} &
    \colhead{$R_{0.1}$\tablenotemark{c}}}
  
  \startdata
NGC 6121&20.70&1.9&\nodata\\
NGC 6144&22.33&1.7&\nodata\\
NGC 6218&21.72&1.8&\nodata\\
NGC 6235&22.65&2.1&5.37\\
NGC 6254&22.07&1.9&\nodata\\
NGC 6266&21.51&2.2&6.84\\
NGC 6273&21.80&2.3&8.26\\
NGC 6287&21.90&2.2&4.13\\
NGC 6304&21.70&2.9&3.65\\
NGC 6333&22.29&2.2&5.59\\
NGC 6342&22.42&2.3&4.21\\
NGC 6352&21.86&2.3&5.61\\
NGC 6355&21.99&2.2&2.36\\
NGC 6397&20.80&2.1&8.18\\
NGC 6522&21.04&3.4&2.33\\
NGC 6541&22.30&2.0&8.63\\
NGC 6553&21.35&2.4&3.91\\
NGC 6558&21.48&3.1&1.81\\
NGC 6624&21.54&2.7&3.92\\
NGC 6626&21.04&2.7&4.95\\
NGC 6637&20.98&2.2&4.39\\
NGC 6642&19.99&1.9&2.22\\
NGC 6656&20.76&2.4&\nodata\\
NGC 6681&21.98&2.2&5.21\\
NGC 6809&21.89&2.0&\nodata\\
  \enddata

  \tablenotetext{a}{$V$ magnitude where the completeness limit is reached.}
  \tablenotetext{b}{$\Delta V$ between the completeness limit and our dimmest observation for that cluster.}
  \tablenotetext{c}{Distance, in arcmin, from the center of the cluster where $P(X=1|r)=0.1$. No distance given if the limit of the camera FOV is reached.}
 
\end{deluxetable}

\begin{deluxetable}{cccccc}
  \tablewidth{0pc} \tablecolumns{6} \tablecaption{Paremeters used to obtain and fit the empirical \citet{ki62} models. \label{tabking}}

  \tablehead{\colhead{Cluster} &
    \colhead{$r_c$\tablenotemark{a}} &
    \colhead{$r_t$\tablenotemark{b}} &
    \colhead{$k$\tablenotemark{c}} &
    \colhead{$c_f$\tablenotemark{d}} &
    \colhead{$c_B$\tablenotemark{e}}}
  
  \startdata
NGC 6121&1.16&51.82&1069.8&22.4&14.2\\
NGC 6144&0.94&33.35&1067.8&11.9&36.5\\
NGC 6218&0.79&17.28&5336.6&4.4&8.1\\
NGC 6235&0.33&11.18&6140.2&56.2&82.3\\
NGC 6254&0.77&18.47&8905.6&13.0&13.3\\
NGC 6266&0.22&11.28&153639.3&221.3&222.8\\
NGC 6273&0.43&14.57&35673.5&162.9&154.8\\
NGC 6287&0.29&6.96&8549.7&62.8&66.1\\
NGC 6304&0.21&13.25&37443.9&586.8&548.4\\
NGC 6333&0.45&8.00&27214.7&143.0&125.8\\
NGC 6342&0.05&15.81&241666.3&165.2&145.6\\
NGC 6352&0.83&10.44&3042.5&123.7&177.5\\
NGC 6355&0.05&15.81&127282.0&374.7&650.0\\
NGC 6397&0.05&15.81&771500.0&60.4&40.5\\
NGC 6522&0.05&15.81&321899.2&972.0&1277.3\\
NGC 6541&0.18&13.04&190598.4&85.3&119.5\\
NGC 6553&0.53&7.66&11152.9&428.5&681.1\\
NGC 6558&0.03&9.49&376029.8&609.4&903.8\\
NGC 6624&0.06&18.97&232671.2&310.2&415.1\\
NGC 6626&0.24&11.23&57055.1&377.1&414.6\\
NGC 6637&0.33&7.92&11542.9&115.4&146.5\\
NGC 6642&0.10&9.77&4603.9&111.9&152.4\\
NGC 6656&1.33&31.90&4607.3&159.8&134.6\\
NGC 6681&0.03&9.49&1660360.2&101.5&118.6\\
NGC 6809&1.80&15.32&3253.0&17.9&17.0\\
  \enddata

  \tablenotetext{a}{Core radius, in arcminutes, according to 2010 version of \citet{ha96} catalog.}
  \tablenotetext{b}{Tidal radius, in arcminutes, according to 2010 version of \citet{ha96} catalog.}
  \tablenotetext{c}{$k$ constant from the fitted \citet{ki62} profile.}
  \tablenotetext{c}{Density of non-member field stars from the fit.}
  \tablenotetext{c}{Density of non-member field stars according to the Besan\c{c}on model.}
 
\end{deluxetable}

\begin{deluxetable}{cc}
  \tablewidth{0pc} 
  \tablecolumns{2}
  \tablecaption{Parameters used to obtain the Besan\c{c}on model for the non-cluster stars in the observed field of NGC 6121, as an example of the technique.\label{tabbesancon}}
\tablehead{\colhead{} &
    \colhead{}}
\startdata
\sidehead{{\bf NGC 6121}}
\hline
\sidehead{{\bf Field of view}}
Field&Small field\\
& l=$350.97^{\circ}$; b=$15.97^{\circ}$\\
& Solid angle=2.100 square degree\\
\sidehead{{\bf Extinction law}}
Diffuse extinction  & 0.0 mag/kpc\\
Discrete clouds & $A_{v}$=1.30; Distance=0pc\\
\sidehead{{\bf Selection on}}
Intervals of magnitude & $15.04 \leq B \leq 24.48$ \\
& $14.56 \leq V \leq 22.58$\\
& $13.18 \leq I \leq 22.53$ \\
Photometric errors & Error function: Exponential \\
& Band=$B$;   A=0.006, B=21.75, C=0.861\\
& Band=$V$;   A=0.006, B=22.24, C=0.901\\
& Band=$I$;   A=0.007, B=23.66, C=0.997\\
\hline
\enddata
\end{deluxetable}

\begin{deluxetable}{ccccc}
  \tablecolumns{5}
  \tablewidth{0pc} 
  \tablecaption{Differential reddening, with respect to our ridgeline
    extinction zeropoints, of a selected set of coordinates across the
    field of our sampled clusters. Table \ref{tabdifred} is published
    in its entirety in the electronic edition of \aj. A portion is
    shown here for guidance regarding its form and
    content.\label{tabdifred}}

  \tablehead{\colhead{$RA_{2000}$\tablenotemark{1}} &
    \colhead{$Dec_{2000}$\tablenotemark{2}} & 
    \colhead{${\Delta}E(B-V)$} &
    \colhead{${\sigma}_{{\Delta}E(B-V)}$} &
    \colhead{Bandwidth\tablenotemark{3}}}

  \startdata
  \hline
  \sidehead{\sc NGC 6121 - M 4}
  \hline
   245.77881 &   -26.63669 & -0.003 &  0.009 &  4.06 \\
   245.77881 &   -26.63571 & -0.004 &  0.008 &  4.03 \\
   245.77882 &   -26.63472 & -0.005 &  0.008 &  4.00 \\
   245.77882 &   -26.63373 & -0.007 &  0.007 &  3.97 \\
   245.77883 &   -26.63274 & -0.008 &  0.006 &  3.93 \\
   245.77883 &   -26.63175 & -0.010 &  0.005 &  3.90 \\
   245.77883 &   -26.63077 & -0.012 &  0.002 &  3.87 \\
   245.77884 &   -26.62978 & -0.012 &  0.003 &  3.84 \\
  \enddata

  \tablenotetext{1}{Units of right ascension are degrees.}
  \tablenotetext{2}{Units of declination are degrees.}  
  \tablenotetext{3}{Units of the bandwidths used in the resolution map are arcminutes.}  
 
\end{deluxetable}

\begin{deluxetable}{ccccccc}
  \tablewidth{0pc} \tablecolumns{7} \tablecaption{Zero points to bring
    the photometry to the chip 2 system. \label{tabzp}}

  \tablehead{\colhead{Chip} &
    \colhead{$Z_B$} &
    \colhead{$\Delta Y'_B$} &
    \colhead{$Z_V$} &
    \colhead{$\Delta Y'_V$} &
    \colhead{$Z_I$} &
    \colhead{$\Delta Y'_I$}}
  
  \startdata
  1 & -0.164 & 0.063 & -0.160 & 0.075 & -0.036 & 0.077\\
  2 & 0.0 & 0.0 & 0.0 & 0.0 & 0.0 & 0.0\\
  3 & -0.107 & 0.005 & -0.119 & 0.008 & -0.017 & 0.016\\
  4 & -0.767 & 0.048 & -0.706 & 0.043 & -0.445 & 0.078\\
  5 & -0.010 & 0.019 & 0.013 & 0.014 & 0.083 & 0.008\\
  6 & -0.137 & 0.010 & -0.173 & 0.006 & -0.118 & 0.021\\
  7 & -0.268 & 0.023 & -0.260 & 0.013 & -0.185 & 0.052\\
  8 & -0.212 & 0.055 & -0.169 & 0.042 & -0.028 & 0.044\\
  \enddata

\end{deluxetable}

\begin{deluxetable}{ccccccc}
  \tablecolumns{7}
  \tablewidth{0pc}
  \tablecaption{Calibrating-star photometry available for the sampled clusters. \label{tabcalstars}}
  
  \tablehead{\colhead{Cluster} &
    \colhead{Stetson\tablenotemark{1}} &
    \colhead{WFPC2\tablenotemark{2}} &
    \colhead{ACS\tablenotemark{3}} &
    \colhead{ACS-10573\tablenotemark{4}} &
    \colhead{{\it HST} missing\tablenotemark{5}} &
    \colhead{Missing\tablenotemark{6}}}
  
  \startdata
  NGC6121 & BVI & VI & VI & \nodata & B & \nodata \\
  NGC6144 & \nodata & B & VI & \nodata & \nodata & \nodata \\
  NGC6218 & BVI & BV & VI & BVI & \nodata & \nodata \\
  NGC6235 & \nodata & BV & \nodata & \nodata & I & I \\
  NGC6254 & BVI & VI & VI & \nodata & B & \nodata \\
  NGC6266 & BVI & BVI & \nodata & \nodata & \nodata & \nodata \\
  NGC6273 & BVI & BV & \nodata & \nodata & I & \nodata \\
  NGC6287 & BV & BVI & \nodata & \nodata & \nodata & \nodata \\
  NGC6304 & \nodata & BV & VI & \nodata & \nodata & \nodata \\
  NGC6333 & \nodata & VI & \nodata & BVI & \nodata & \nodata \\
  NGC6342 & \nodata & BV & \nodata & \nodata & I & I \\
  NGC6355 & \nodata & BV & \nodata & \nodata & I & I \\
  NGC6352 & BVI & VI & VI & \nodata & B & \nodata \\
  NGC6397 & BVI & BVI & VI & \nodata & \nodata & \nodata \\
  NGC6522 & BVI & BV & \nodata & \nodata & I & \nodata \\
  NGC6541 & BVI & VI & VI & \nodata & B & \nodata \\
  NGC6553 & \nodata & VI & \nodata & BVI & \nodata & \nodata \\
  NGC6558 & \nodata & \nodata & VI & \nodata & B & B \\
  NGC6624 & \nodata & BVI & VI & BVI & \nodata & \nodata \\
  NGC6626 & \nodata & BVI & \nodata & \nodata & \nodata & \nodata \\
  NGC6637 & \nodata & BVI & VI & BVI & \nodata & \nodata \\
  NGC6642 & \nodata & BV & VI & \nodata & \nodata & \nodata \\
  NGC6656 & BVI & BVI & VI & \nodata & \nodata & \nodata \\
  NGC6681 & BVI & BV & VI & \nodata & \nodata & \nodata \\
  NGC6809 & BVI & \nodata & VI & \nodata & B & \nodata \\
  \enddata

  \tablenotetext{1}{\citet{st00} calibrating stars observed in 
    these specific Johnson-Cousin filters.}
  \tablenotetext{2}{{\it HST} WFPC2 photometry from archival data available 
    in the specified filters. The filters are $f439w$, $f555w$, $f606w$, 
    and $f814w$, and they are transformed by HSTPHOT to the Johnson-Cousins 
    $B$, $V$, and $I$.}  
  \tablenotetext{3}{{\it HST} ACS photometry from archival data available in 
    the specified filters. The filters are $f606w$ and $f814w$, and 
    they are transformed by DolPHOT to the Johnson-Cousins $B$, $V$, and $I$.}  
  \tablenotetext{4}{{\it HST} ACS photometry from our project (SNAP 10573) available 
    in the specified filters. The filters are $f435w$, $f555w$, and $f814w$, and 
    they are transformed by DolPHOT to the Johnson-Cousins $B$, $V$, and $I$.}
  \tablenotetext{5}{{\it HST} photometry for the cluster in the 
    filter specified is missing.}
  \tablenotetext{6}{{\it HST} photometry and Stetson calibrating 
    stars for the cluster in the filter specified are missing.}
  
\end{deluxetable}

\begin{deluxetable}{cccc}
  \tablecolumns{4}
  \tablewidth{0pc} 
  \tablecaption{Offsets applied to transform to Johnson-Cousins system photometry.\label{taboffsets}}

  \tablehead{\colhead{ }&
    \colhead{Offset in $B$} &
    \colhead{Offset in $V$} &
    \colhead{Offset in $I$}}

  \startdata
  \sidehead{\sc Night 1}
  NGC 6266 & $\it{-0.44\pm0.02}$ & $\it{-0.40\pm0.02}$ & $\it{-0.28\pm0.02}$ \\
  NGC 6273 & $\it{0.06\pm0.02}$ & $\it{-0.10\pm0.03}$ & $\it{0.00\pm0.03}$ \\
  NGC 6522 & $\it{-0.07\pm0.03}$ & $\it{-0.06\pm0.01}$ & $\it{0.04\pm0.02}$ \\
  NGC 6558 & \nodata & $-0.06\pm0.01$ & $-0.05\pm0.01$ \\
  NGC 6624 & $0.01\pm0.02$ & $-0.03\pm0.01$ & $-0.04\pm0.01$ \\
  \sidehead{\sc Night 2}
  NGC 6144 & $-0.01\pm0.02$ & $-0.13\pm0.02$ & $-0.10\pm0.02$ \\
  NGC 6235 & $-0.10\pm0.02$ & $-0.05\pm0.02$ & \nodata \\
  NGC 6304 & $0.02\pm0.02$ & $-0.02\pm0.01$ & $0.01\pm0.02$ \\
  NGC 6342 & $0.02\pm0.02$ & $-0.13\pm0.03$ & \nodata \\
  NGC 6397 & $\it{0.01\pm0.02}$ & $\it{0.01\pm0.01}$ & $\it{-0.03\pm0.01}$ \\
  NGC 6541 & $\it{0.05\pm0.04}$ & $\it{0.01\pm0.02}$ & $\it{0.00\pm0.02}$ \\
  NGC 6553 & $0.16\pm0.07$ & $0.05\pm0.03$ & $0.06\pm0.02$ \\
  NGC 6656 & $\it{0.07\pm0.02}$ & $\it{0.02\pm0.02}$ & $\it{-0.02\pm0.02}$ \\
  NGC 6809 & $\it{0.04\pm0.02}$ & $\it{0.04\pm0.01}$ & $\it{0.05\pm0.02}$ \\
  \sidehead{\sc Night 3}
  NGC 6218 & $\it{0.08\pm0.01}$ & $\it{-0.06\pm0.01}$ & $\it{-0.01\pm0.01}$ \\
  NGC 6254 & $\it{-0.02\pm0.02}$ & $\it{-0.02\pm0.01}$ & $\it{-0.05\pm0.01}$ \\
  NGC 6352 & $\it{-0.05\pm0.02}$ & $\it{-0.15\pm0.02}$ & $\it{-0.15\pm0.01}$ \\
  \sidehead{\sc Night 4}
  NGC 6121 & $\it{-0.02\pm0.02}$ & $\it{-0.38\pm0.02}$ & $\it{-0.21\pm0.02}$ \\
  NGC 6287 & $\it{-0.14\pm0.03}$ & $\it{-0.38\pm0.03}$ & $-0.05\pm0.02$ \\
  NGC 6333 & $-0.03\pm0.03$ & $-0.05\pm0.02$ & $-0.17\pm0.02$ \\
  NGC 6355 & $-0.20\pm0.02$ & $-0.09\pm0.02$ & \nodata \\
  NGC 6626 & $-0.13\pm0.02$ & $-0.21\pm0.03$ & $-0.46\pm0.02$ \\
  NGC 6637 & $-0.27\pm0.03$ & $-0.46\pm0.02$ & $-0.20\pm0.02$ \\
  NGC 6642 & $-0.64\pm0.03$ & $-1.25\pm0.02$ & $-1.42\pm0.02$ \\
  NGC 6681 & $\it{-0.13\pm0.01}$ & $\it{-0.25\pm0.02}$ & $\it{-0.10\pm0.02}$ \\
\enddata

\tablecomments{The numbers in italic correspond to the straight
  comparison between the data from \citet{st00} and our data. The numbers
  in standard type correspond to the comparison between our data and
  the {\it HST} data, plus the offset between {\it HST} data and
  Stetson's data. The errors reported are the dispersions of the
  distributions from which the offsets were calculated, which inform
  about the absolute precision of the photometry of individual stars.}

\end{deluxetable}

\clearpage

\clearpage

\clearpage

\clearpage

\begin{figure}[tp] \centering
\begin{tabular}{cccc}
NGC 6273 & & NGC 6287 & \\
\includegraphics[scale=0.21]{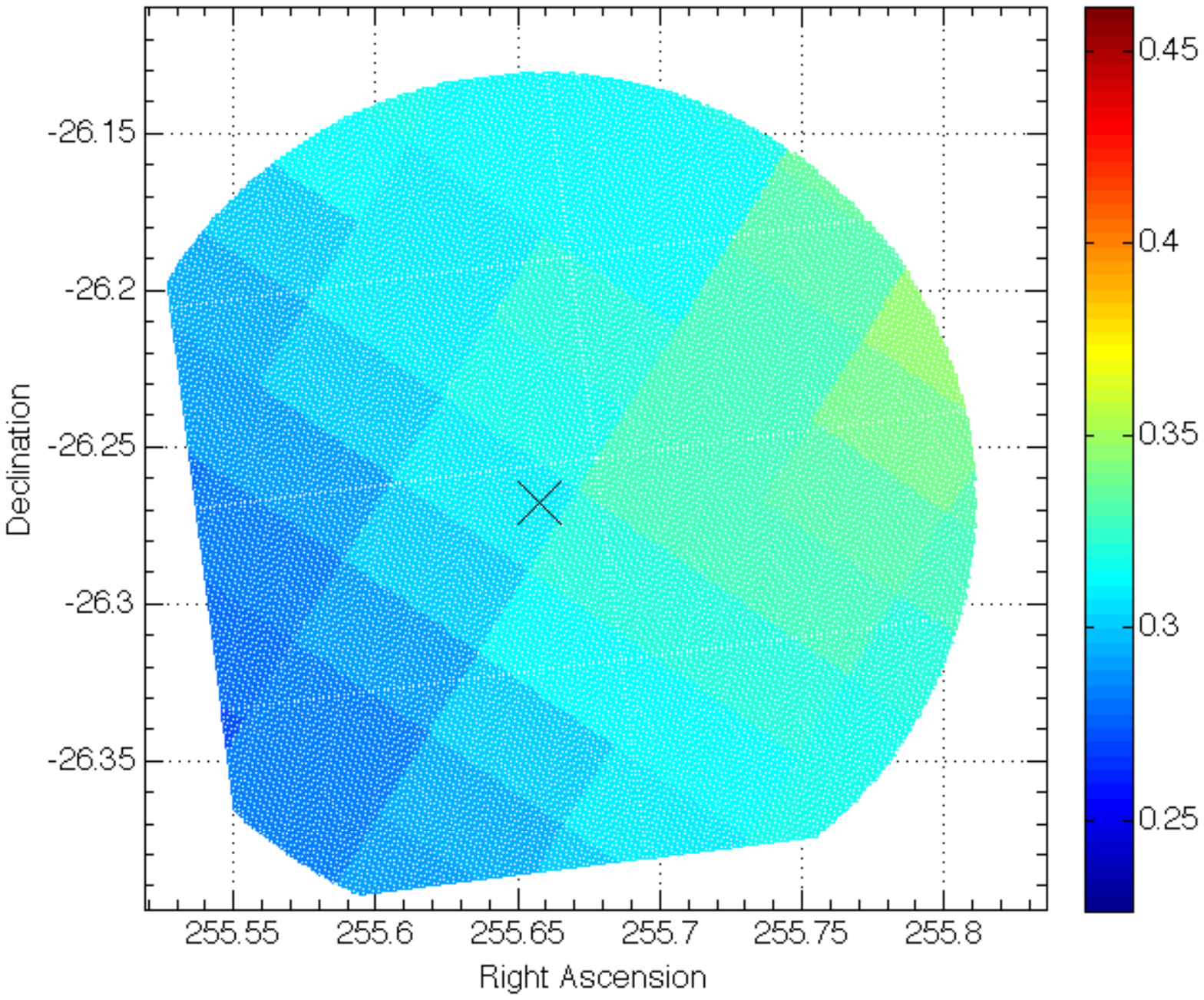}  &
\includegraphics[scale=0.21]{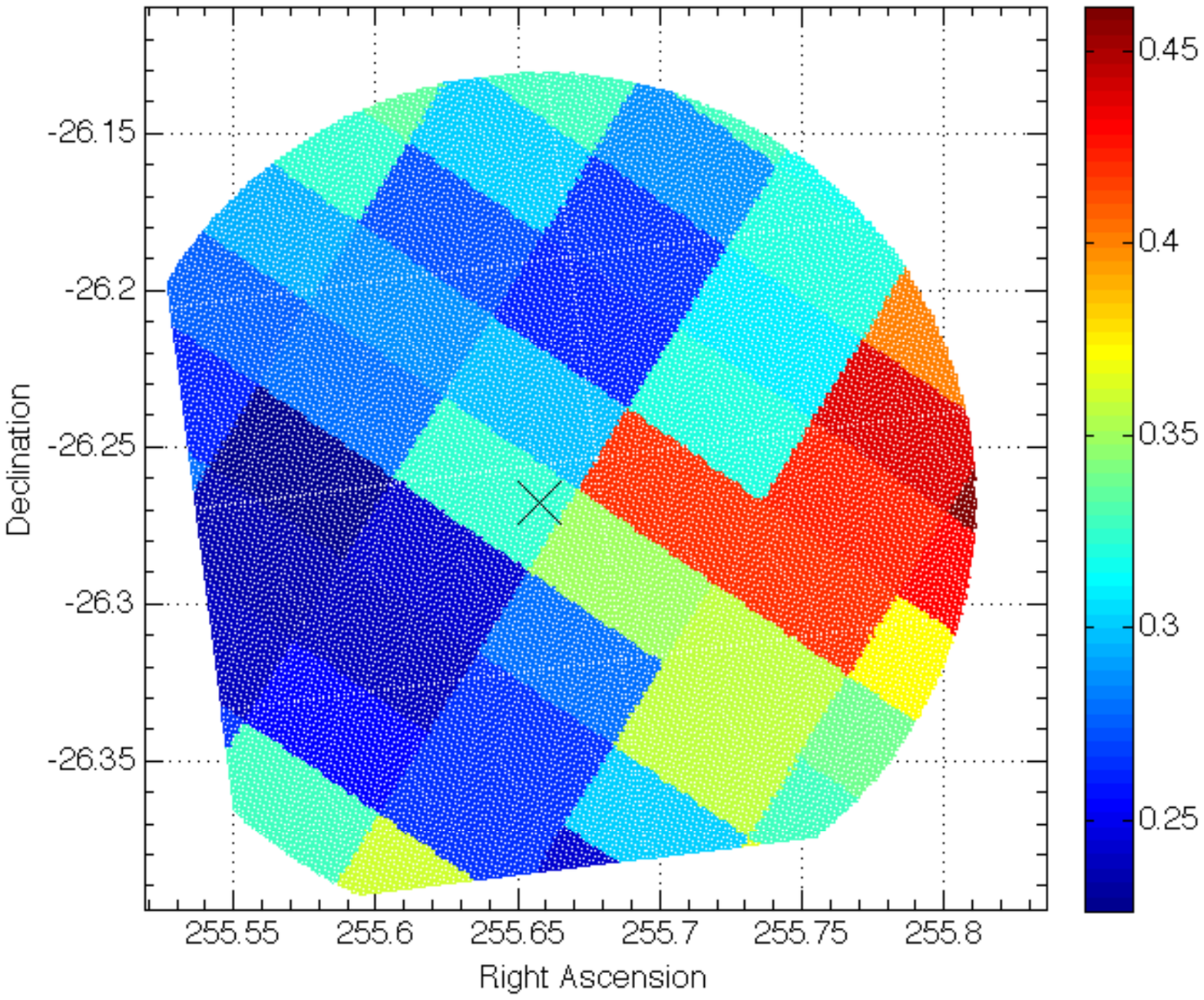}  &
\includegraphics[scale=0.21]{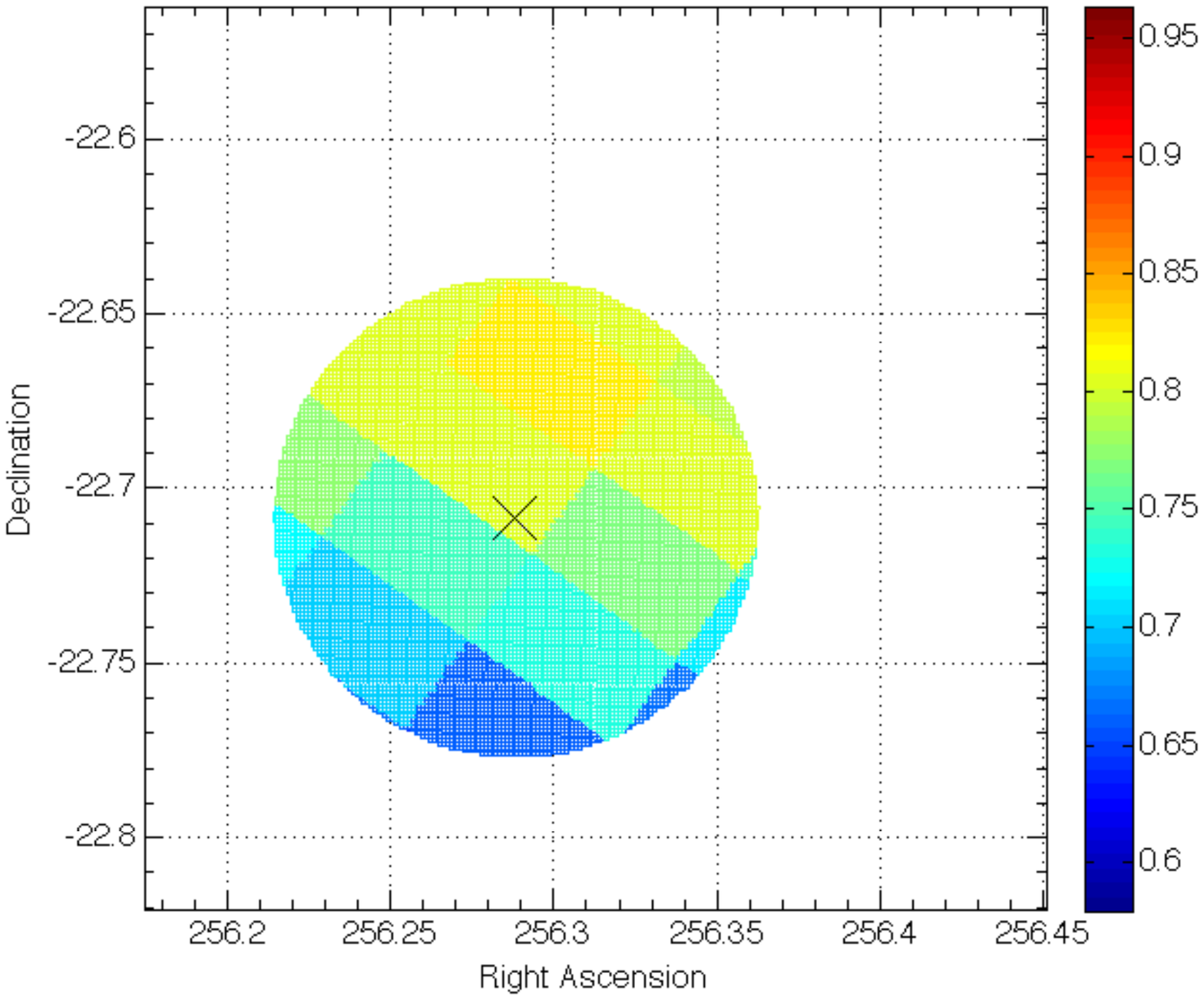}  &
\includegraphics[scale=0.21]{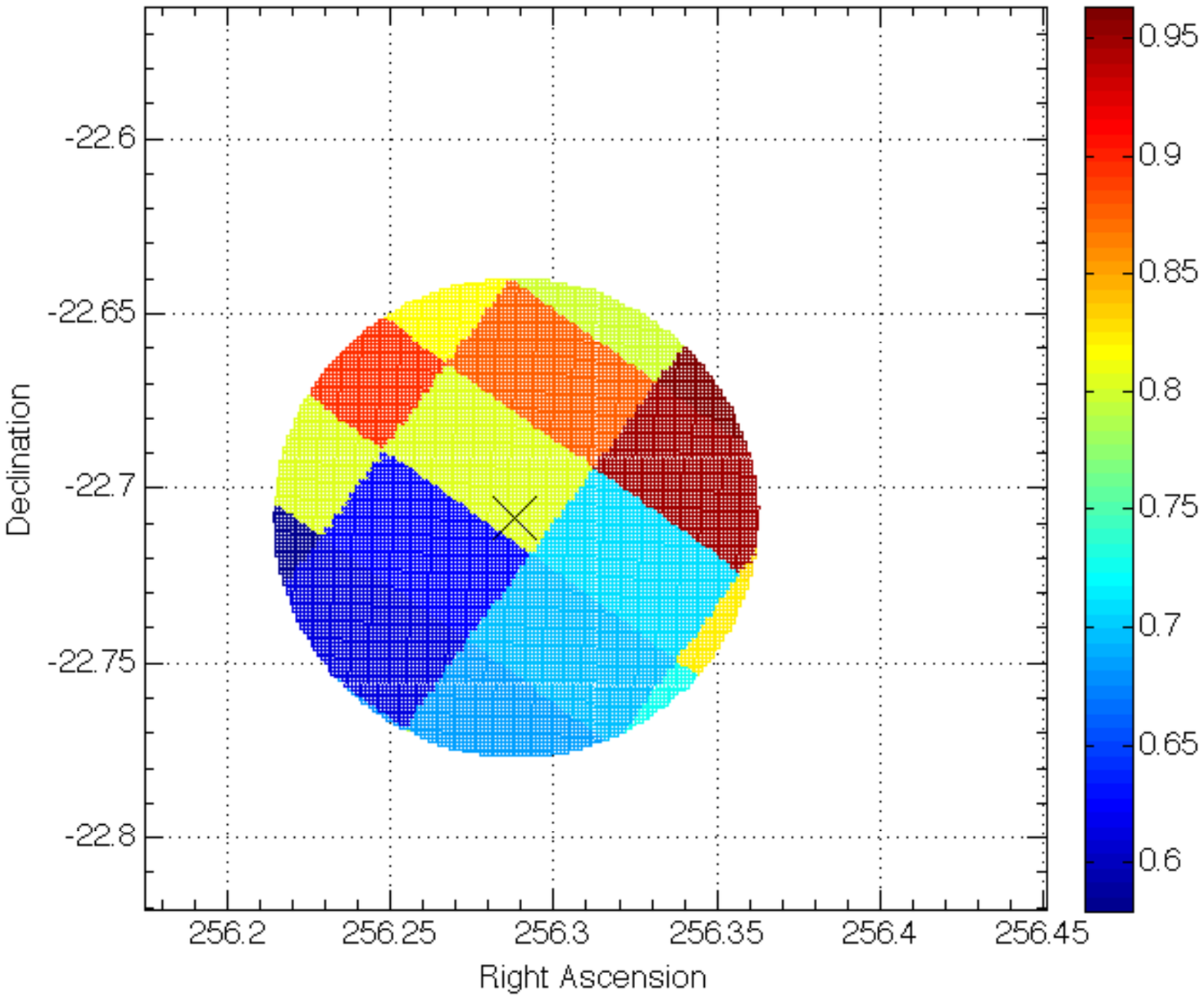} \\
NGC 6304 & & NGC 6333 & \\
\includegraphics[scale=0.21]{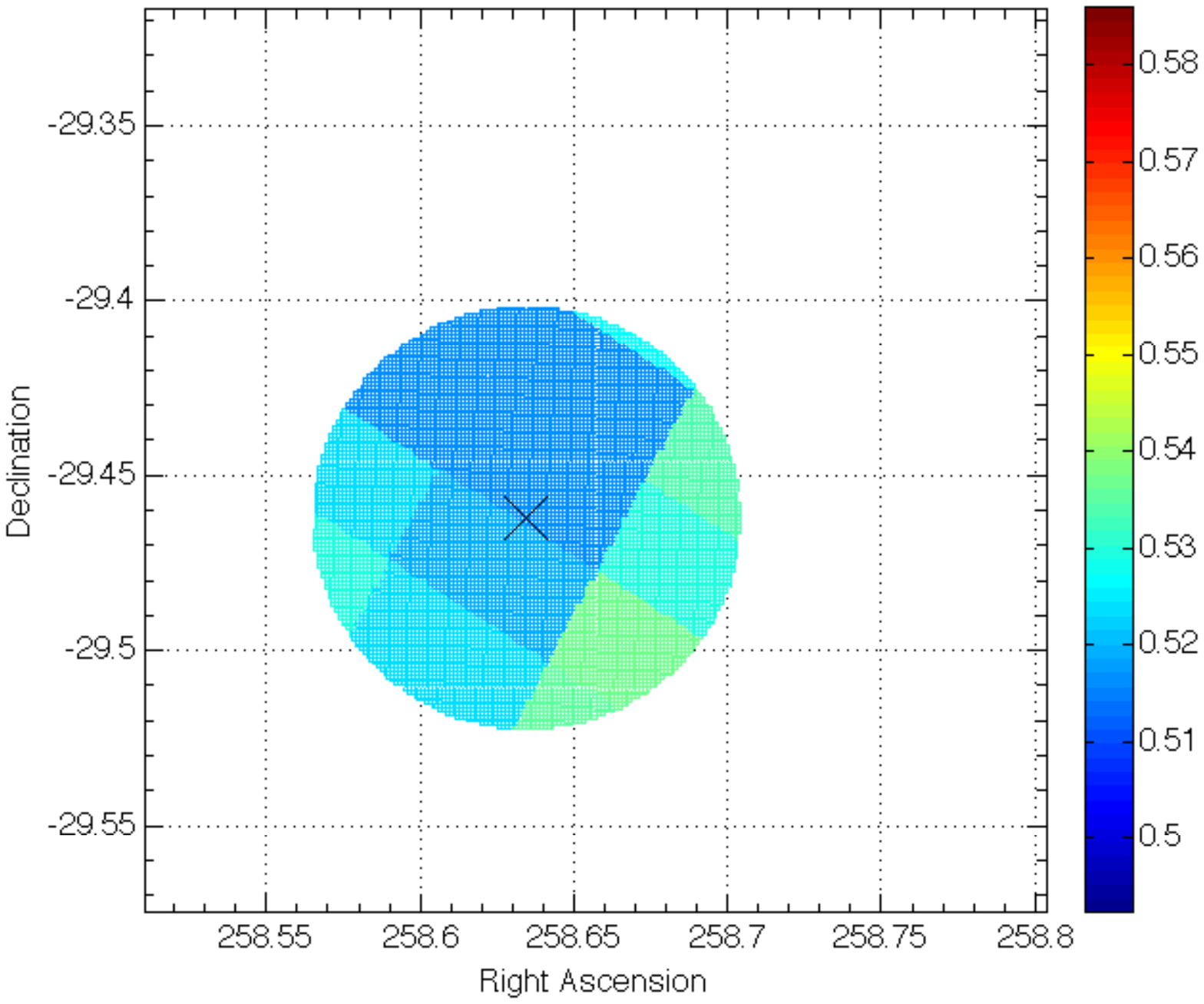}  &
\includegraphics[scale=0.21]{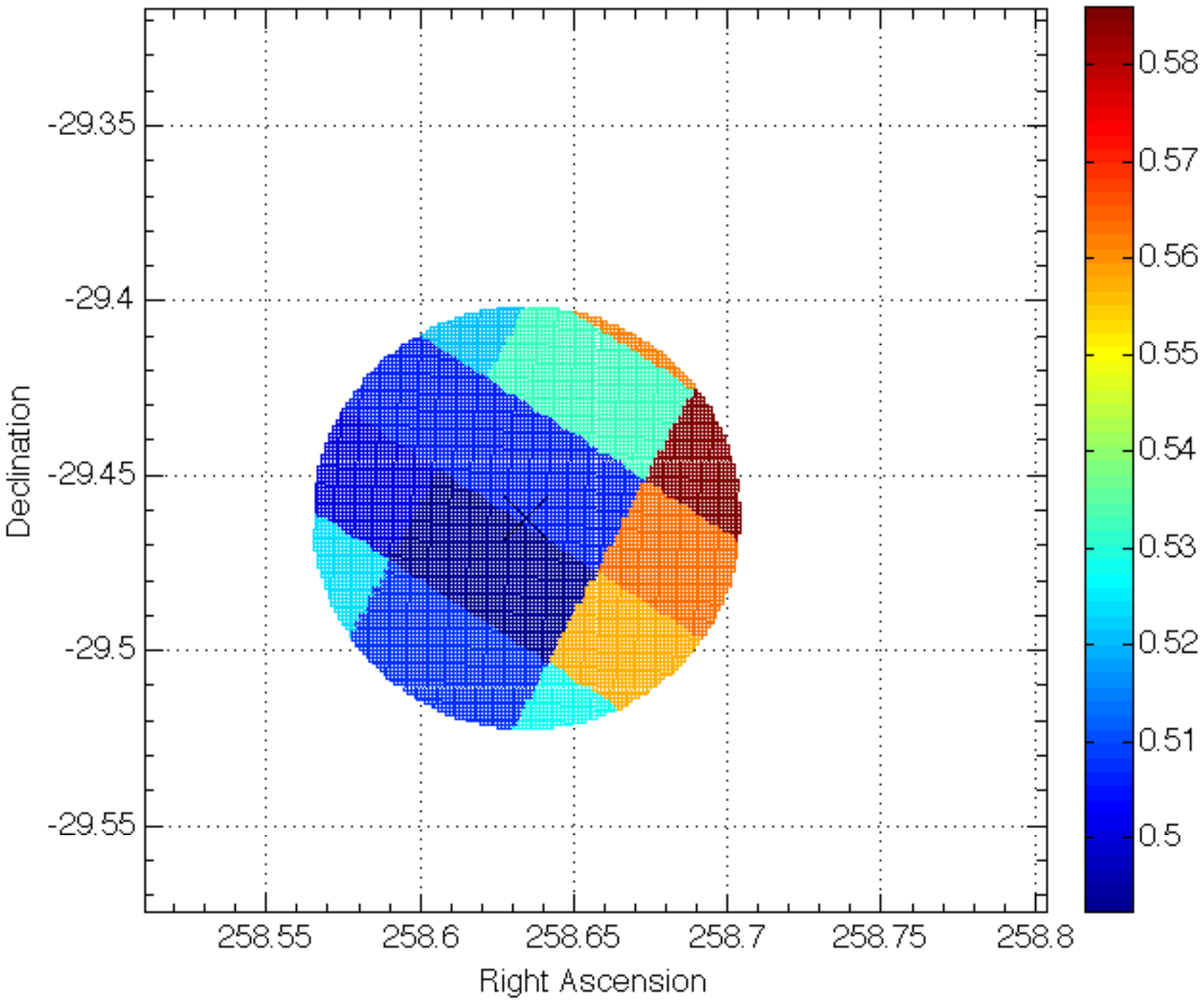}  &
\includegraphics[scale=0.21]{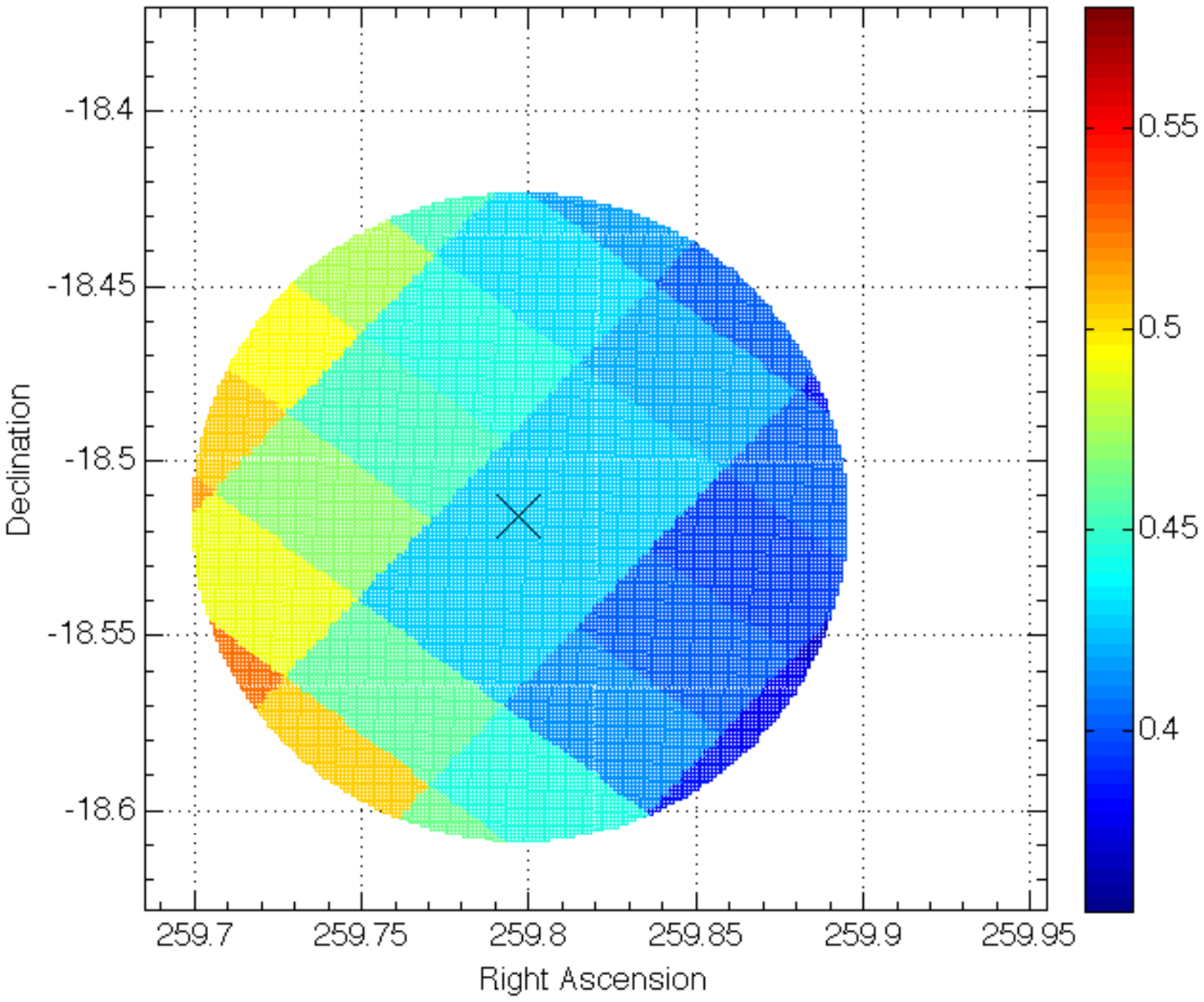}  &
\includegraphics[scale=0.21]{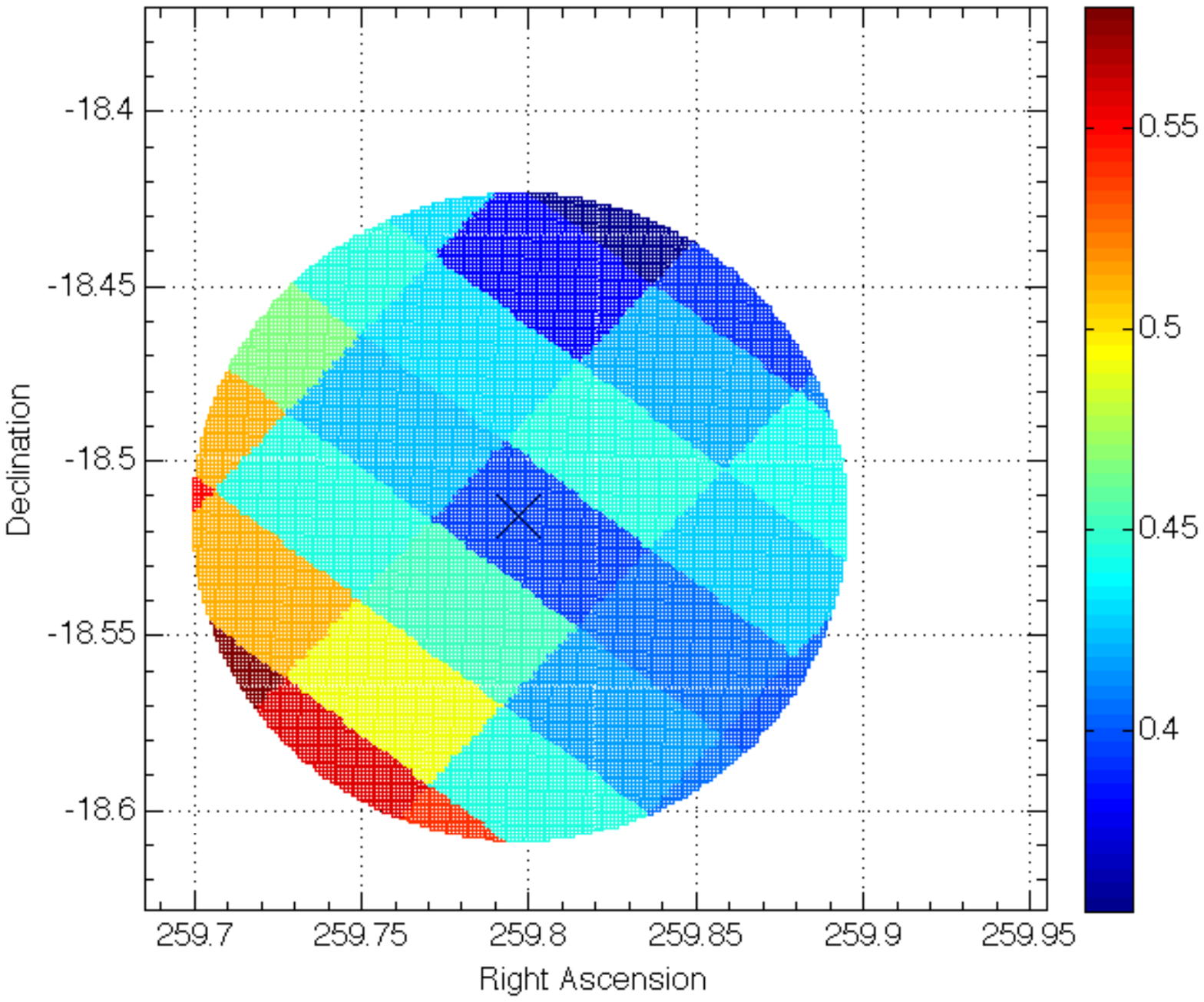} \\
NGC 6342 & & NGC 6352 & \\
\includegraphics[scale=0.21]{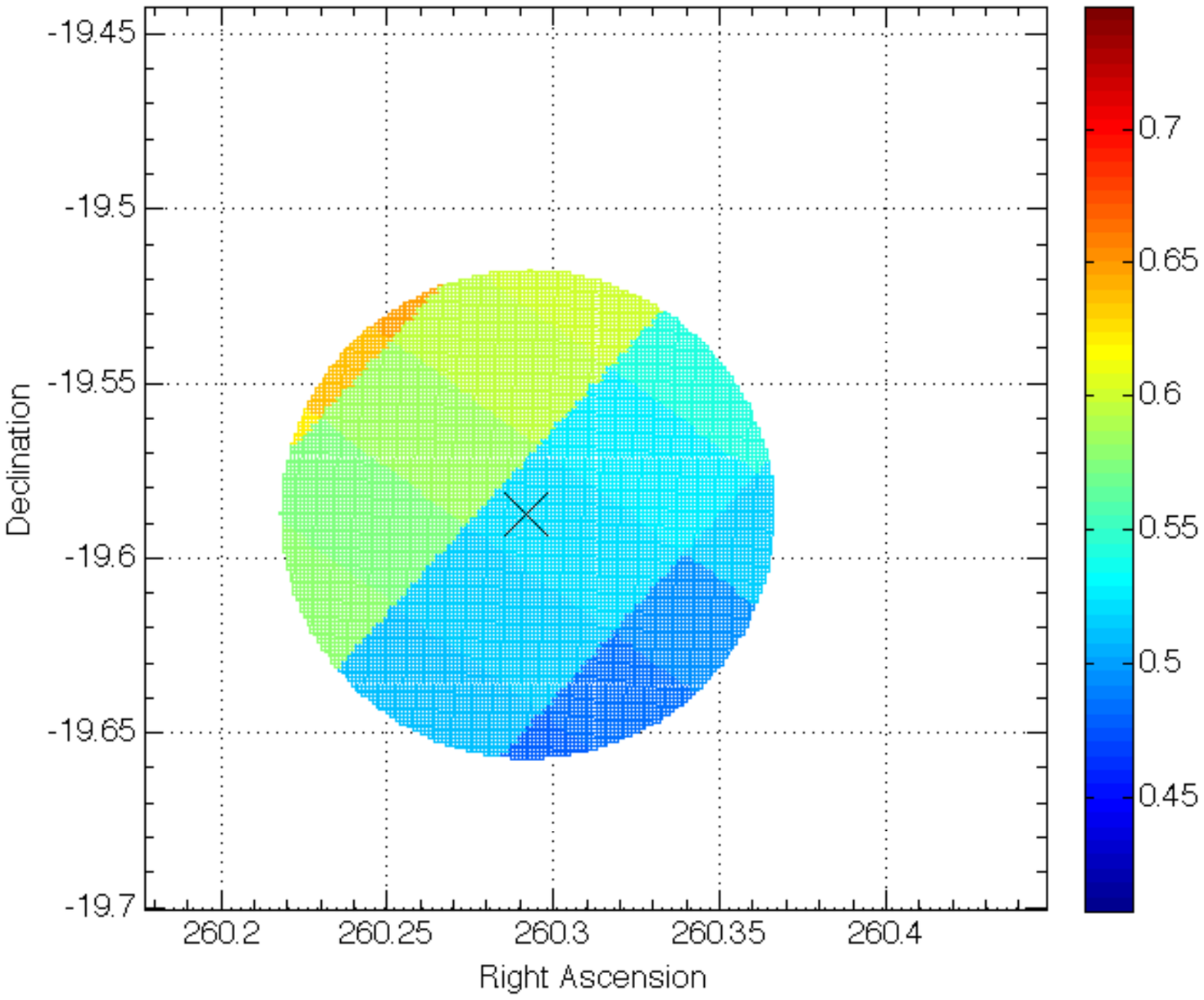}  &
\includegraphics[scale=0.21]{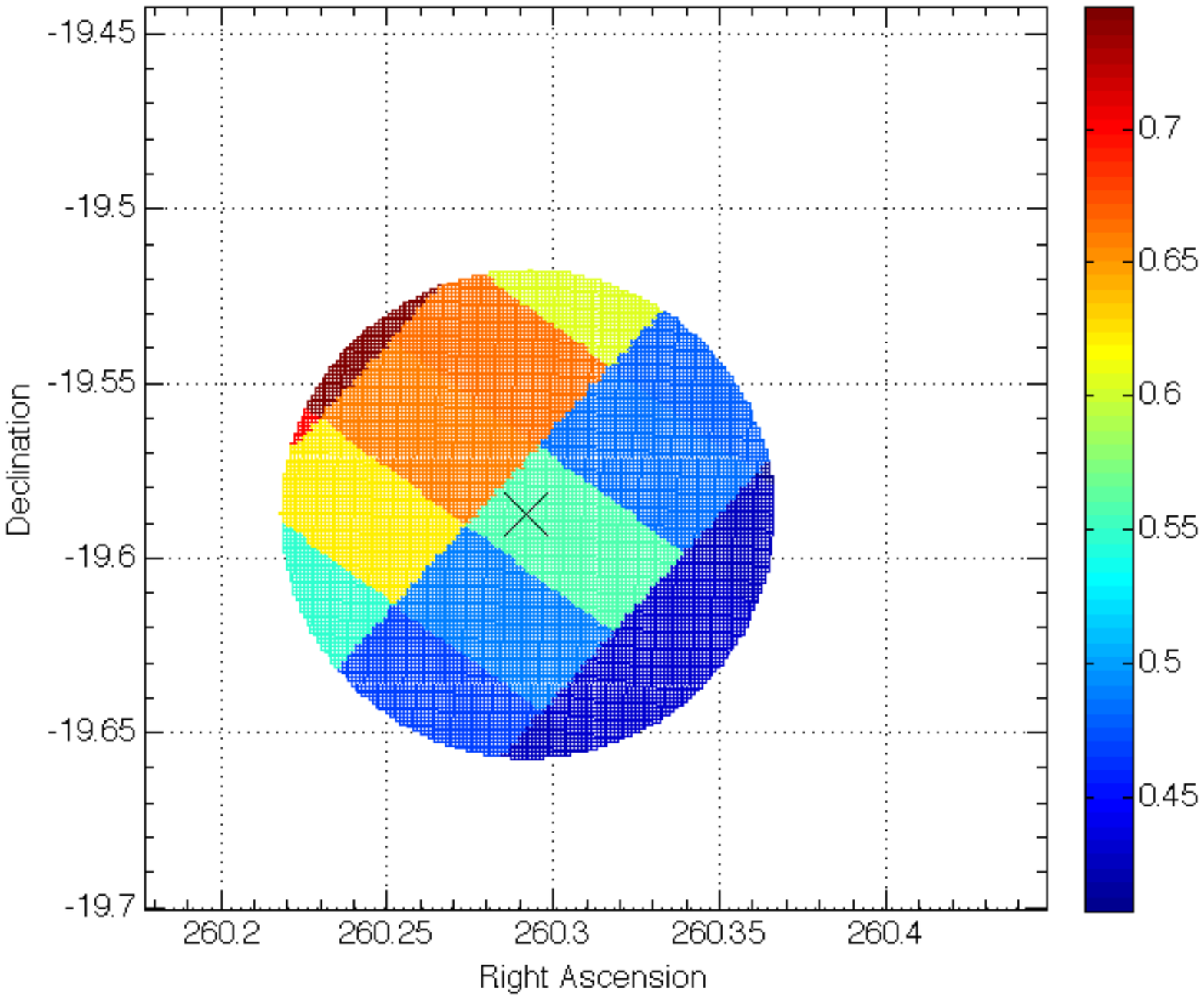}  &
\includegraphics[scale=0.21]{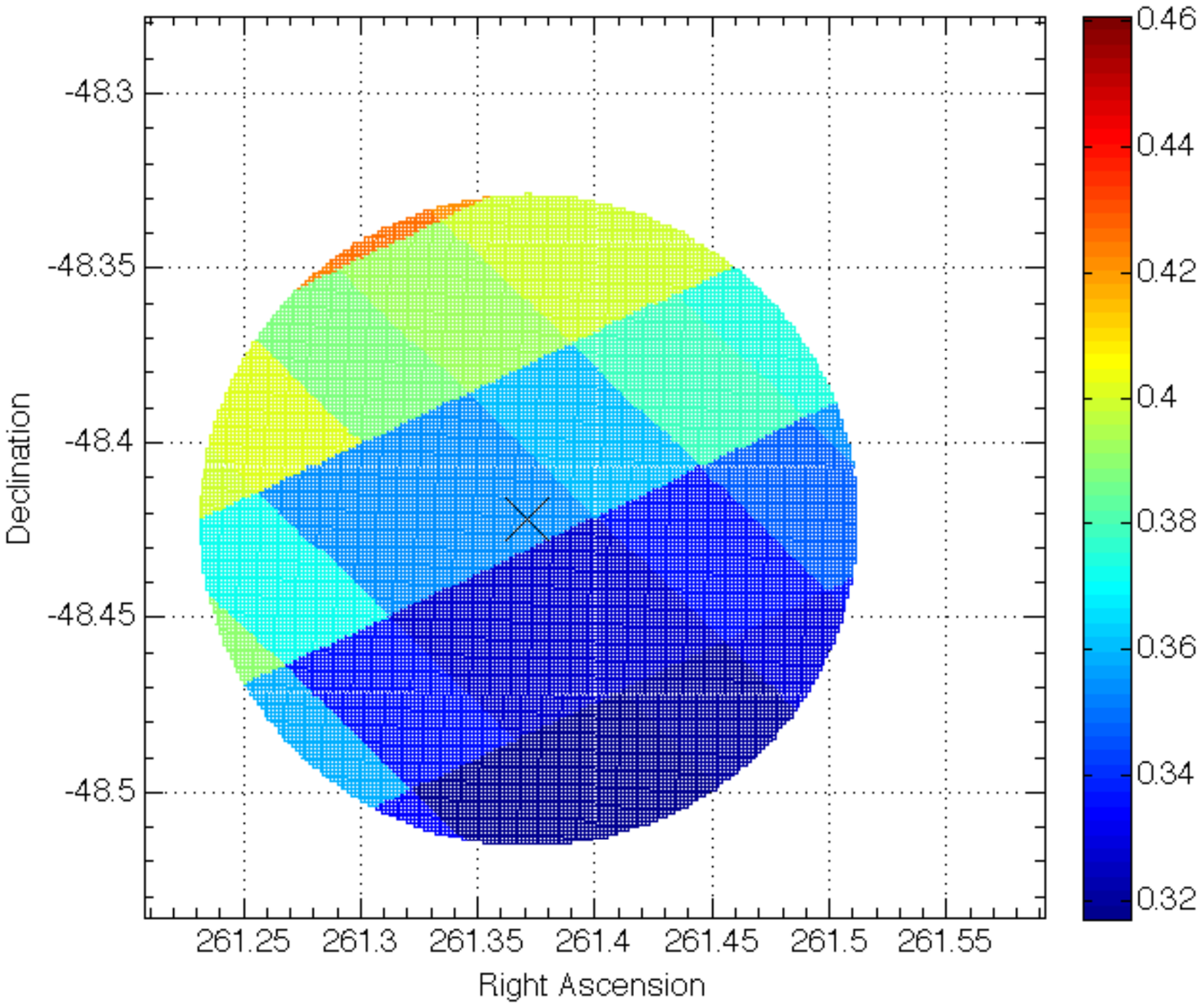}  &
\includegraphics[scale=0.21]{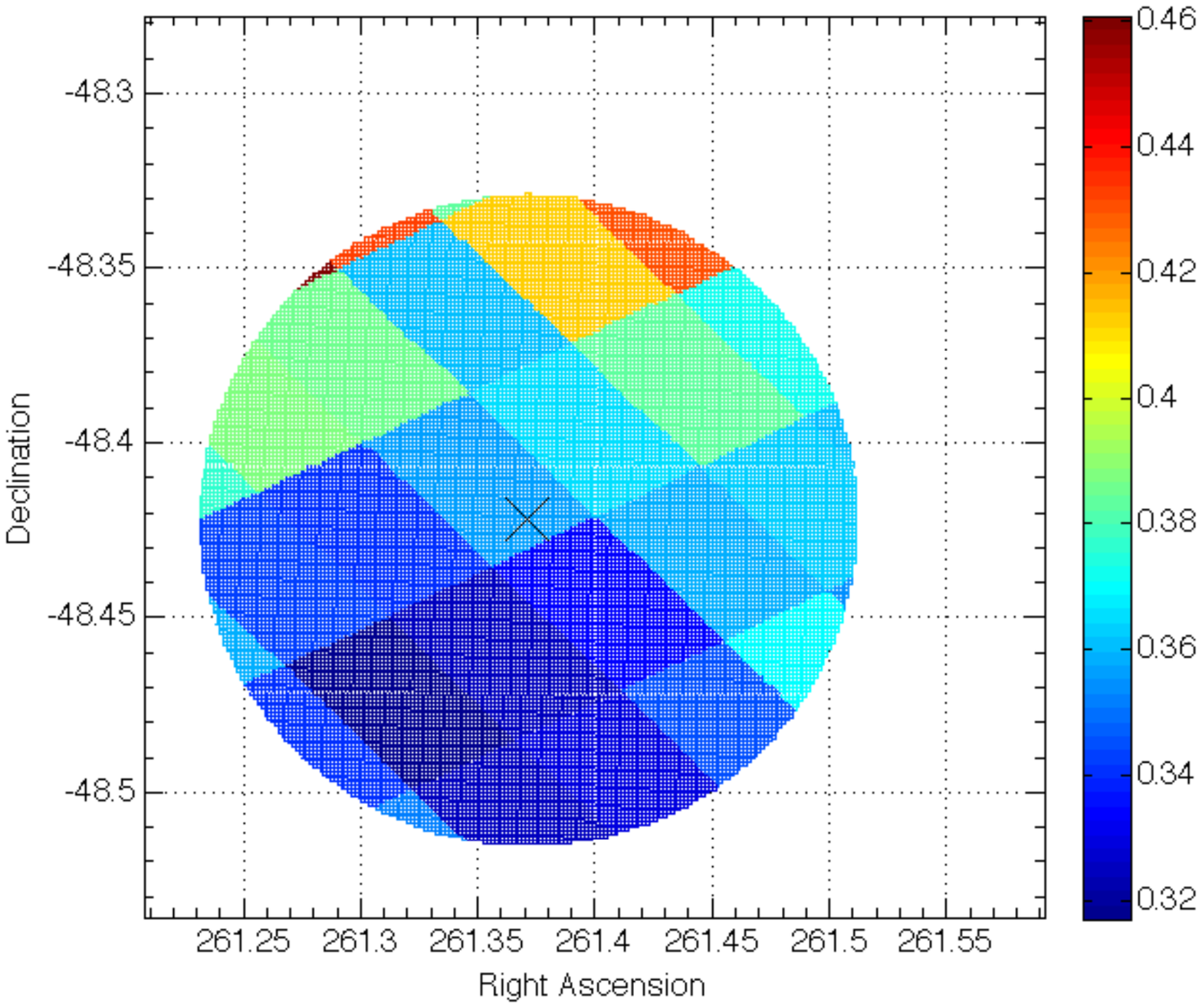} \\
\end{tabular}
\end{figure}

\begin{figure}[tp] \centering
\begin{tabular}{cccc}
NGC 6355 & & NGC 6397 & \\
\includegraphics[scale=0.21]{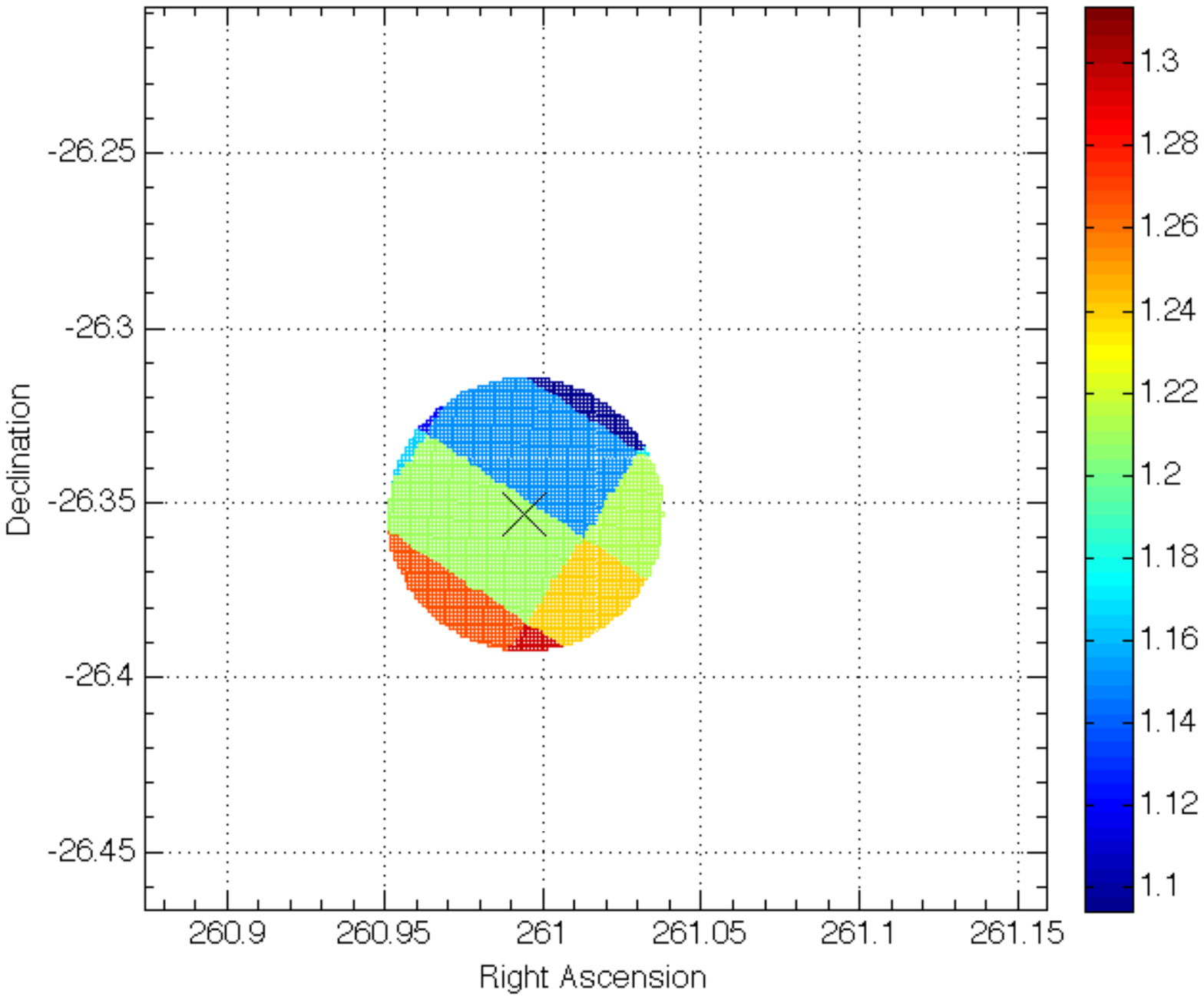}  &
\includegraphics[scale=0.21]{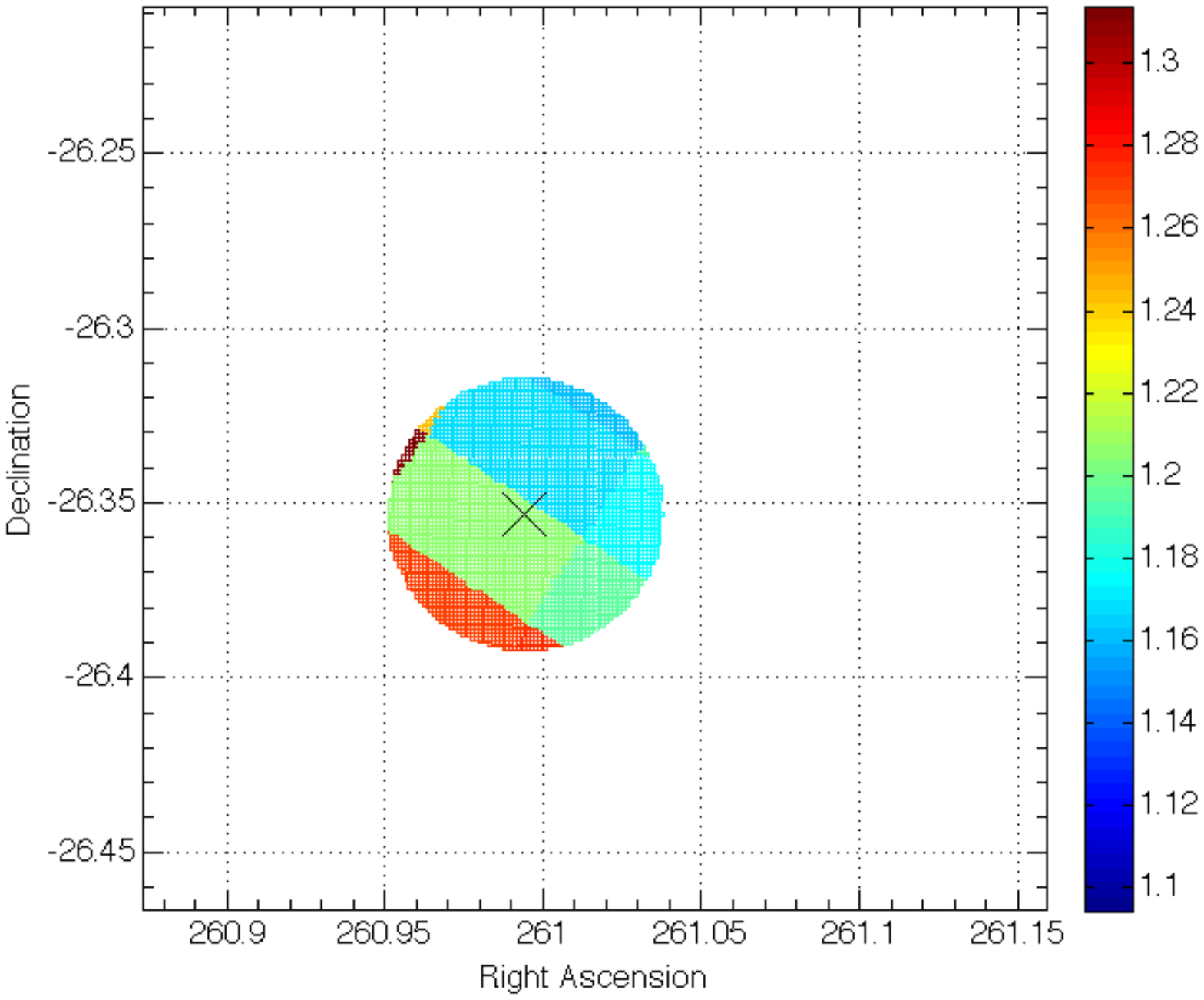}  &
\includegraphics[scale=0.21]{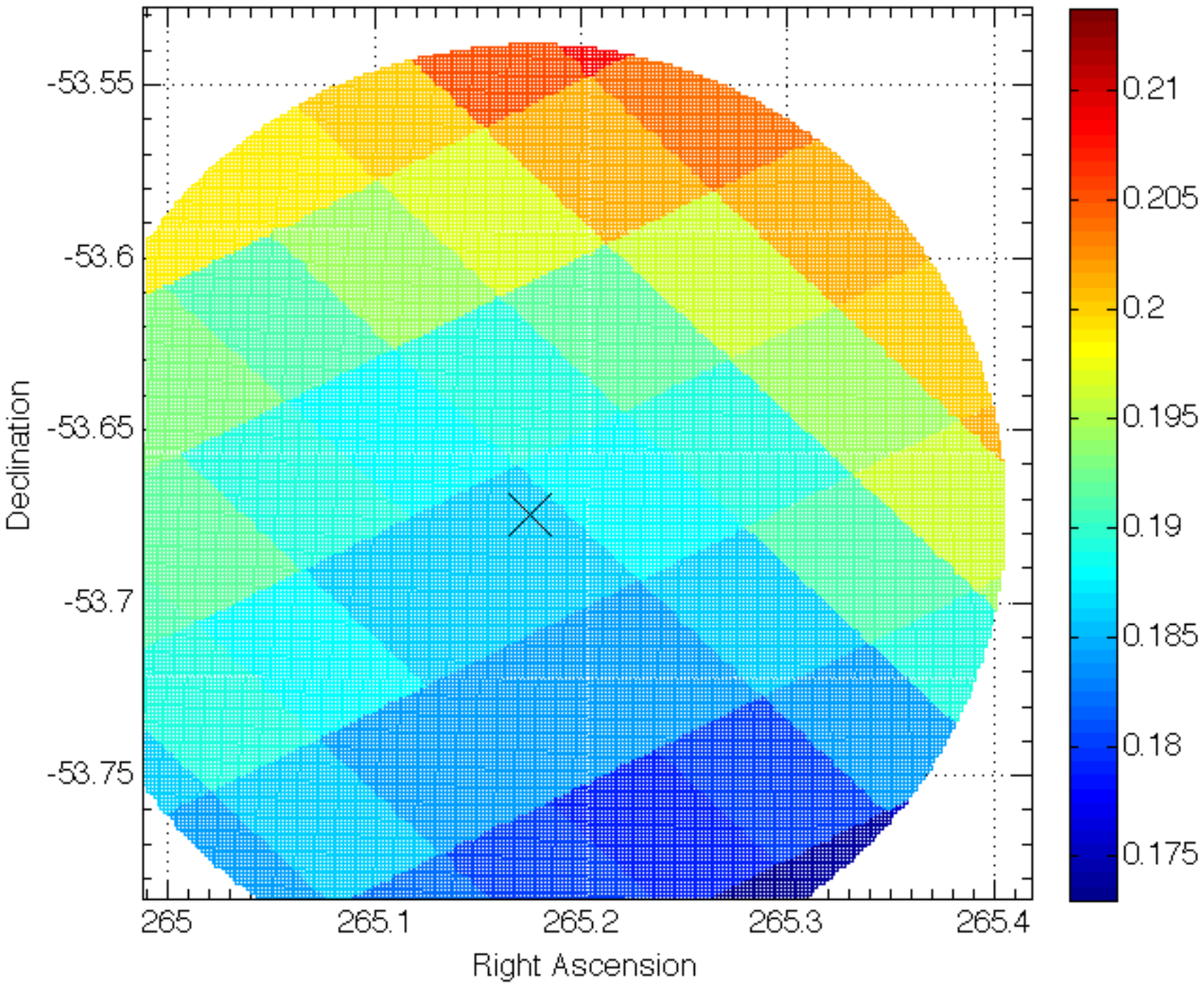}  &
\includegraphics[scale=0.21]{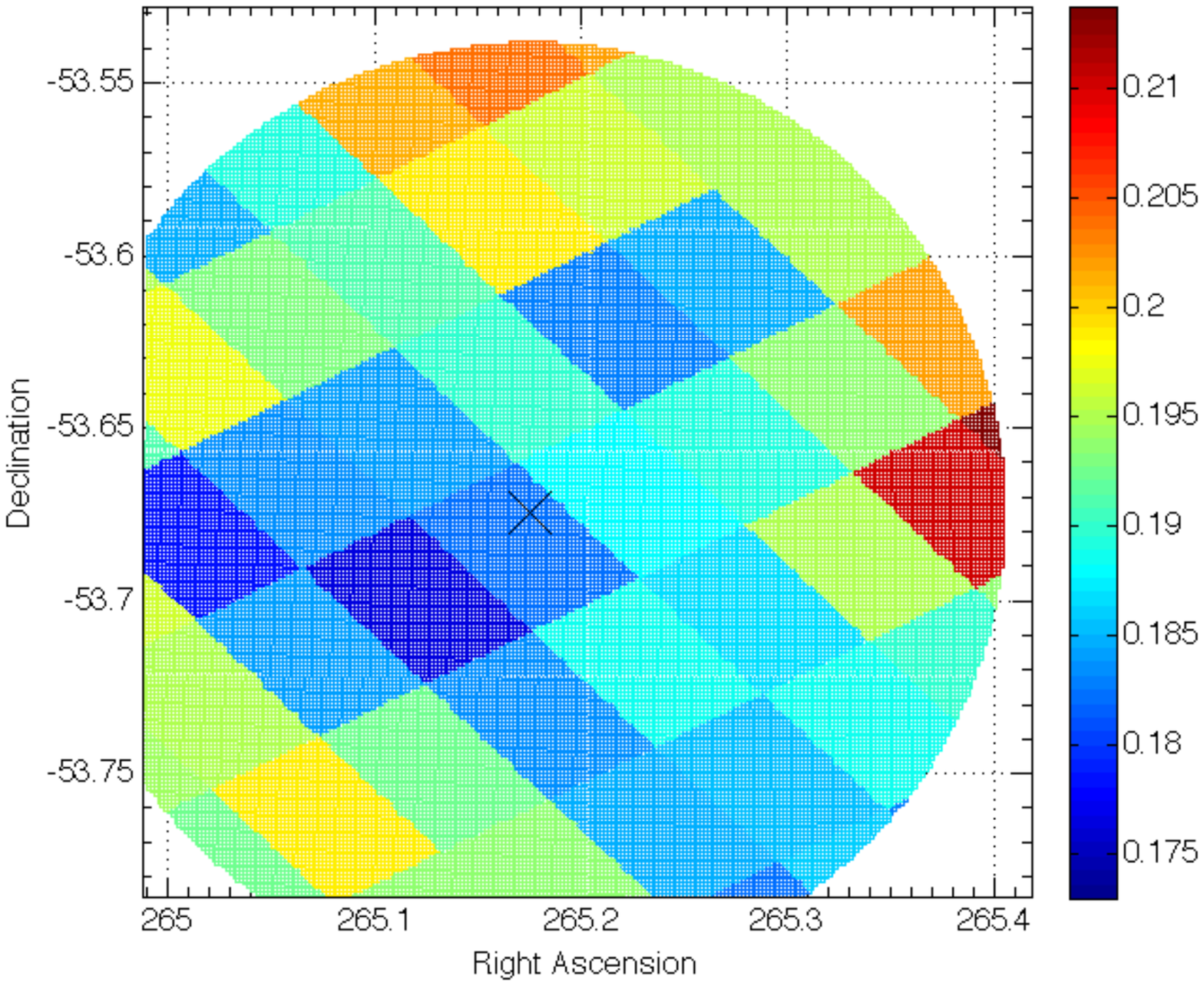} \\
NGC 6522 & & NGC 6541 & \\
\includegraphics[scale=0.21]{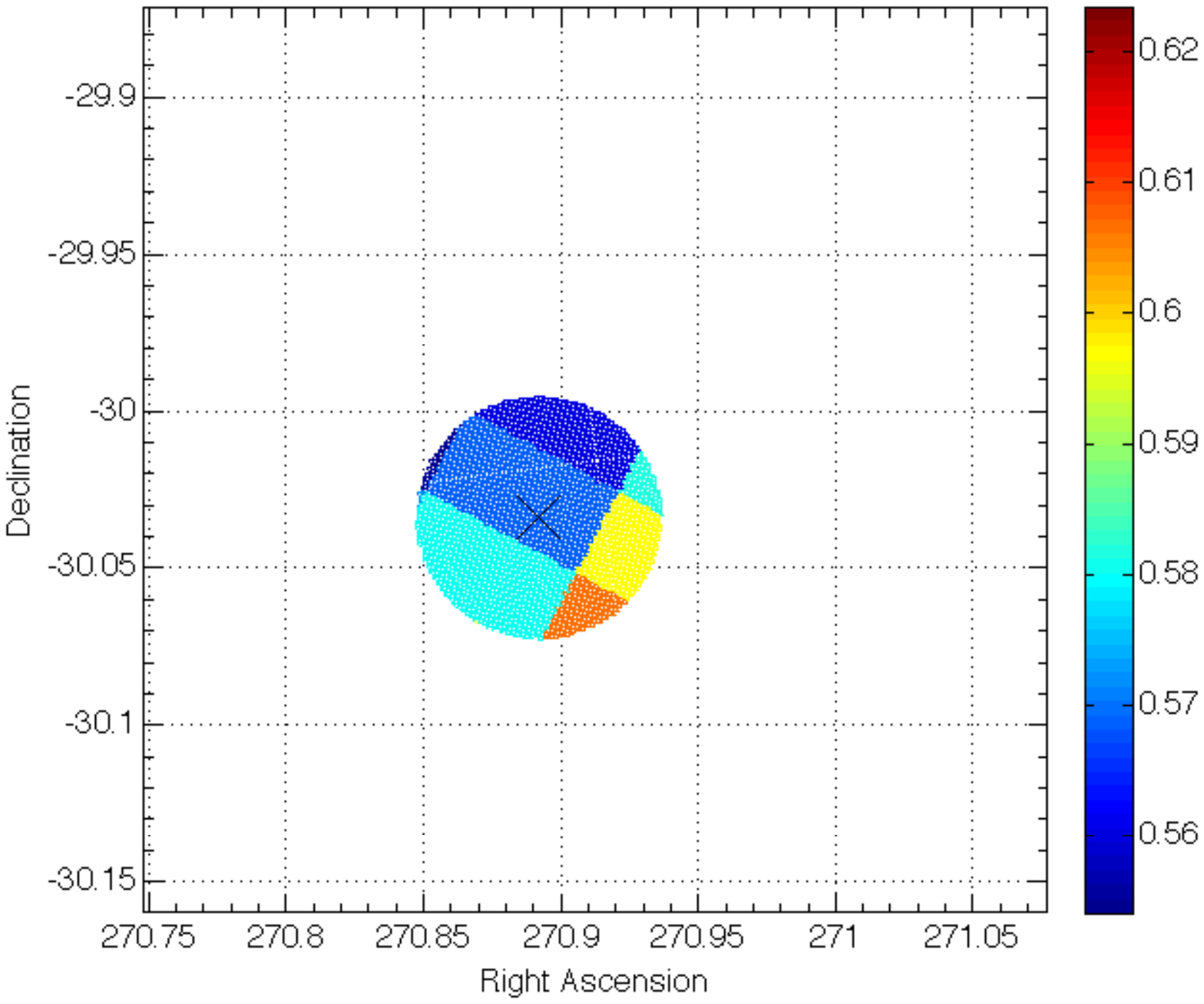}  &
\includegraphics[scale=0.21]{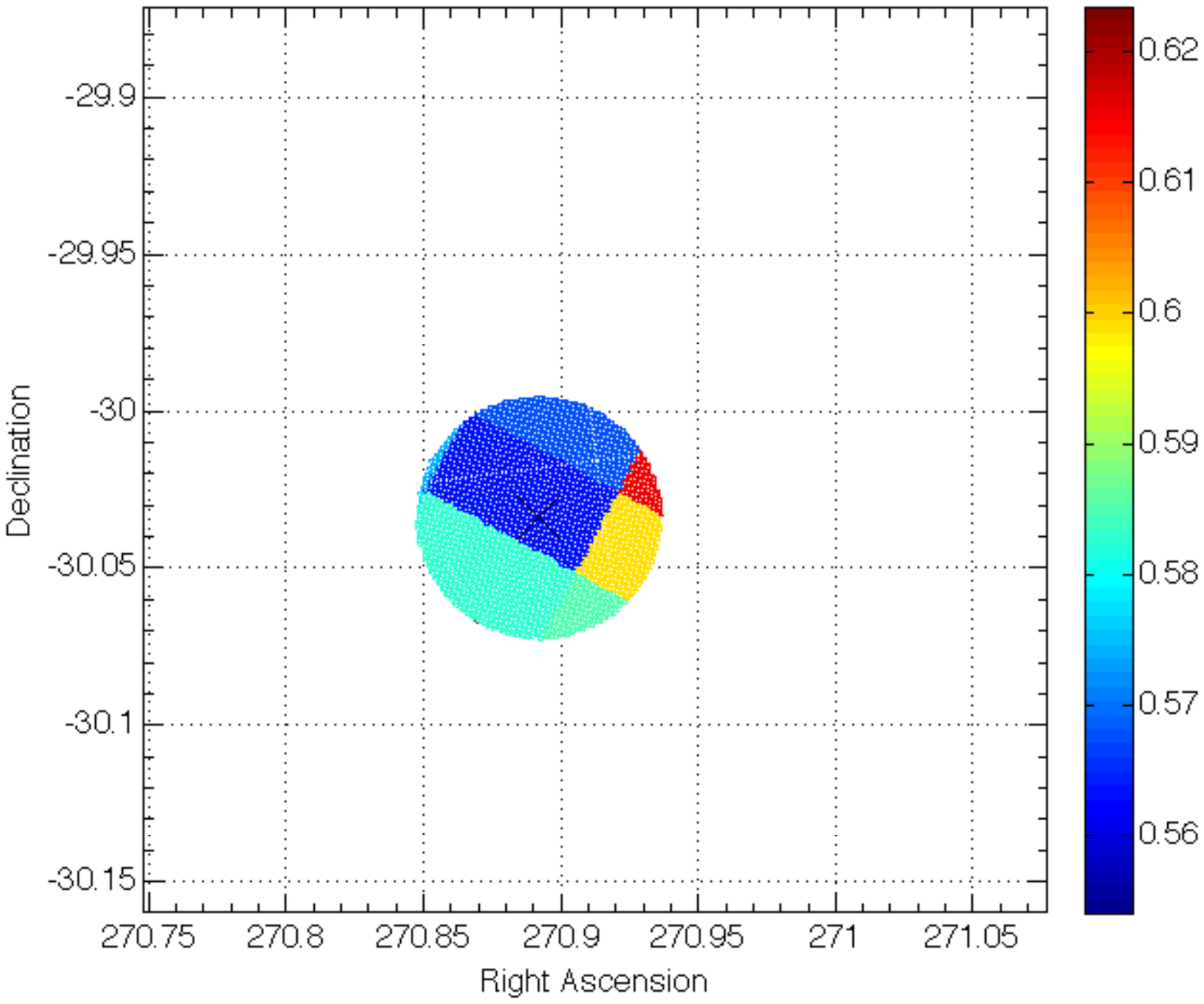}  &
\includegraphics[scale=0.21]{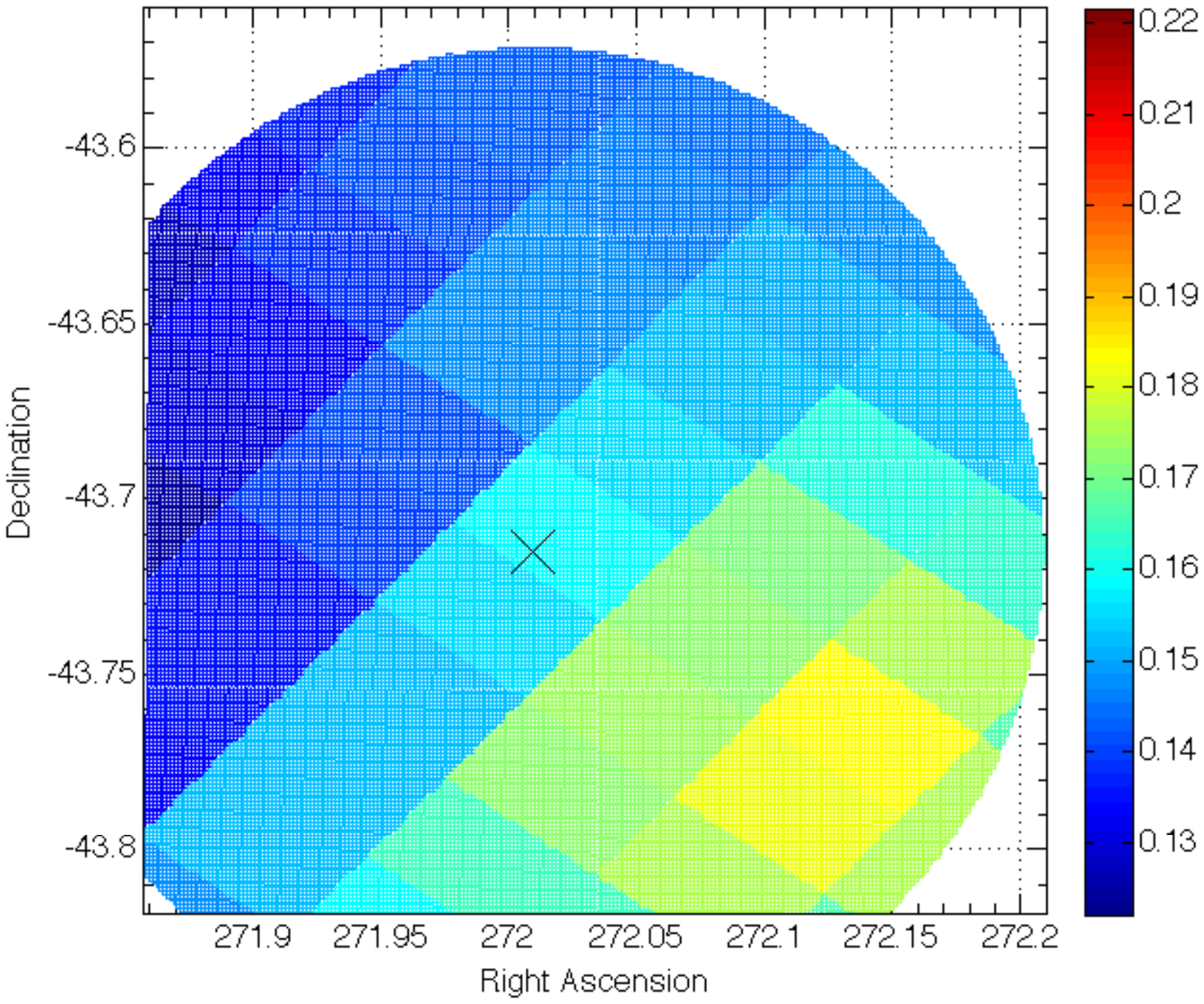}  &
\includegraphics[scale=0.21]{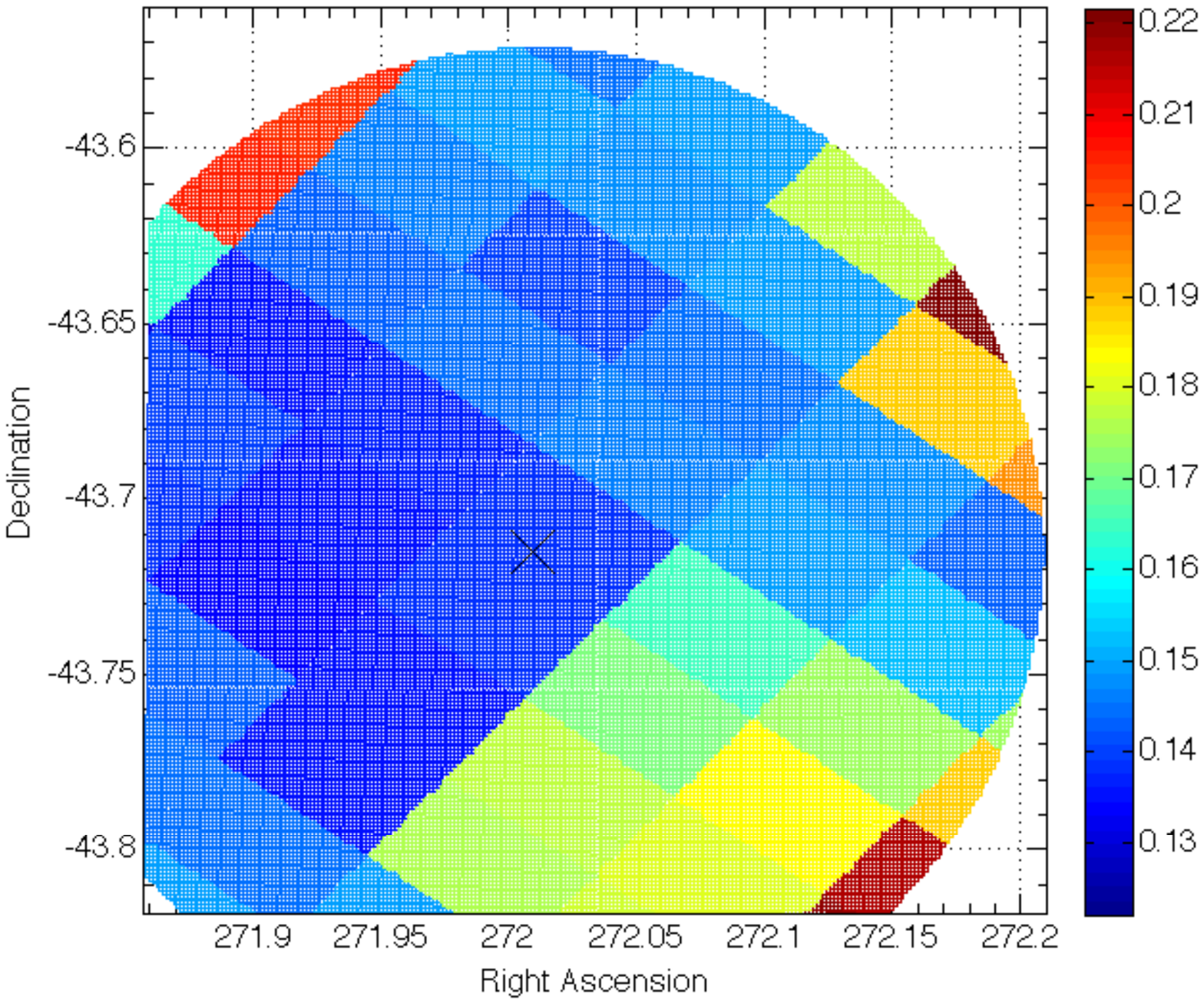} \\
NGC 6553 & & NGC 6558 & \\
\includegraphics[scale=0.21]{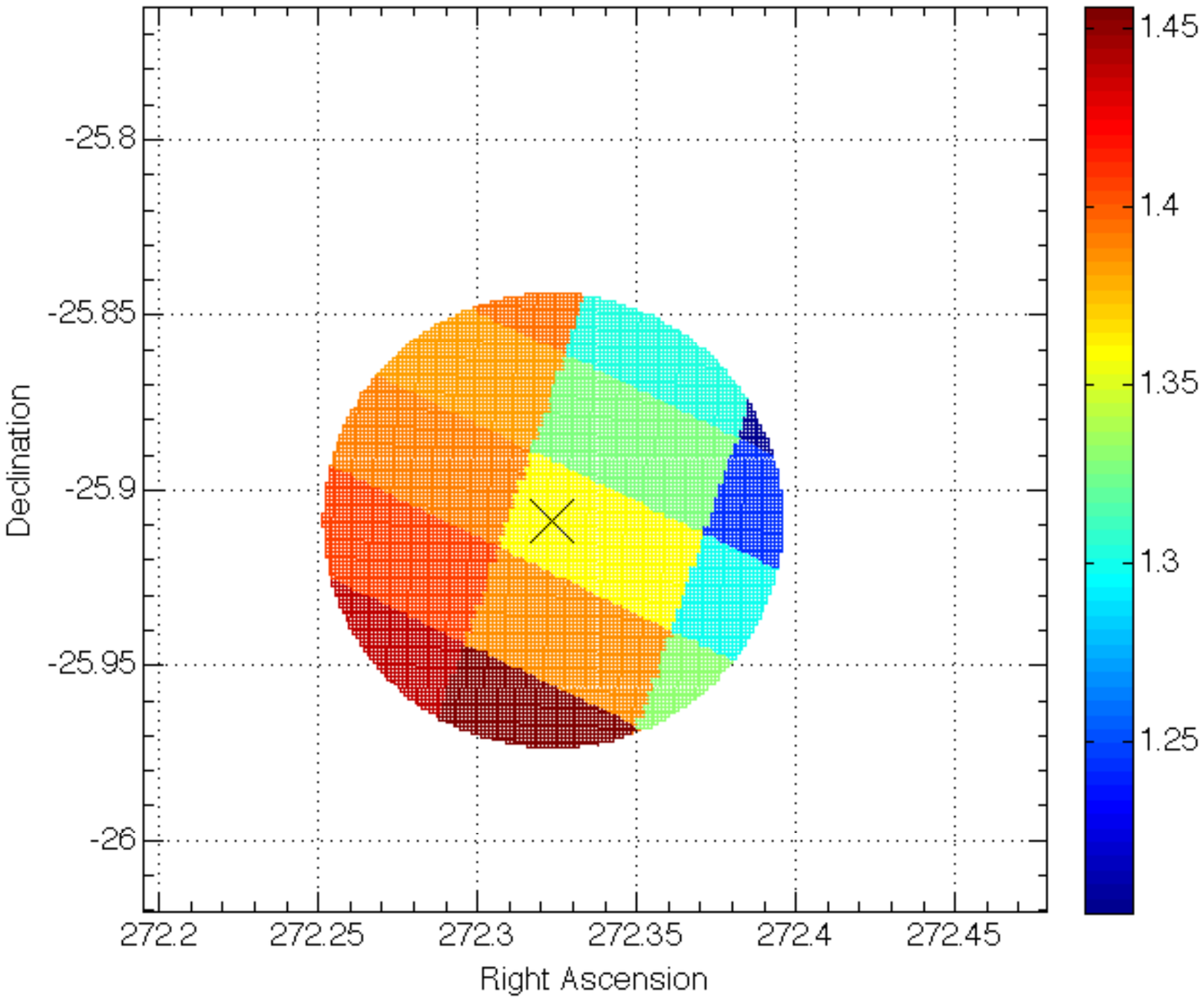}  &
\includegraphics[scale=0.21]{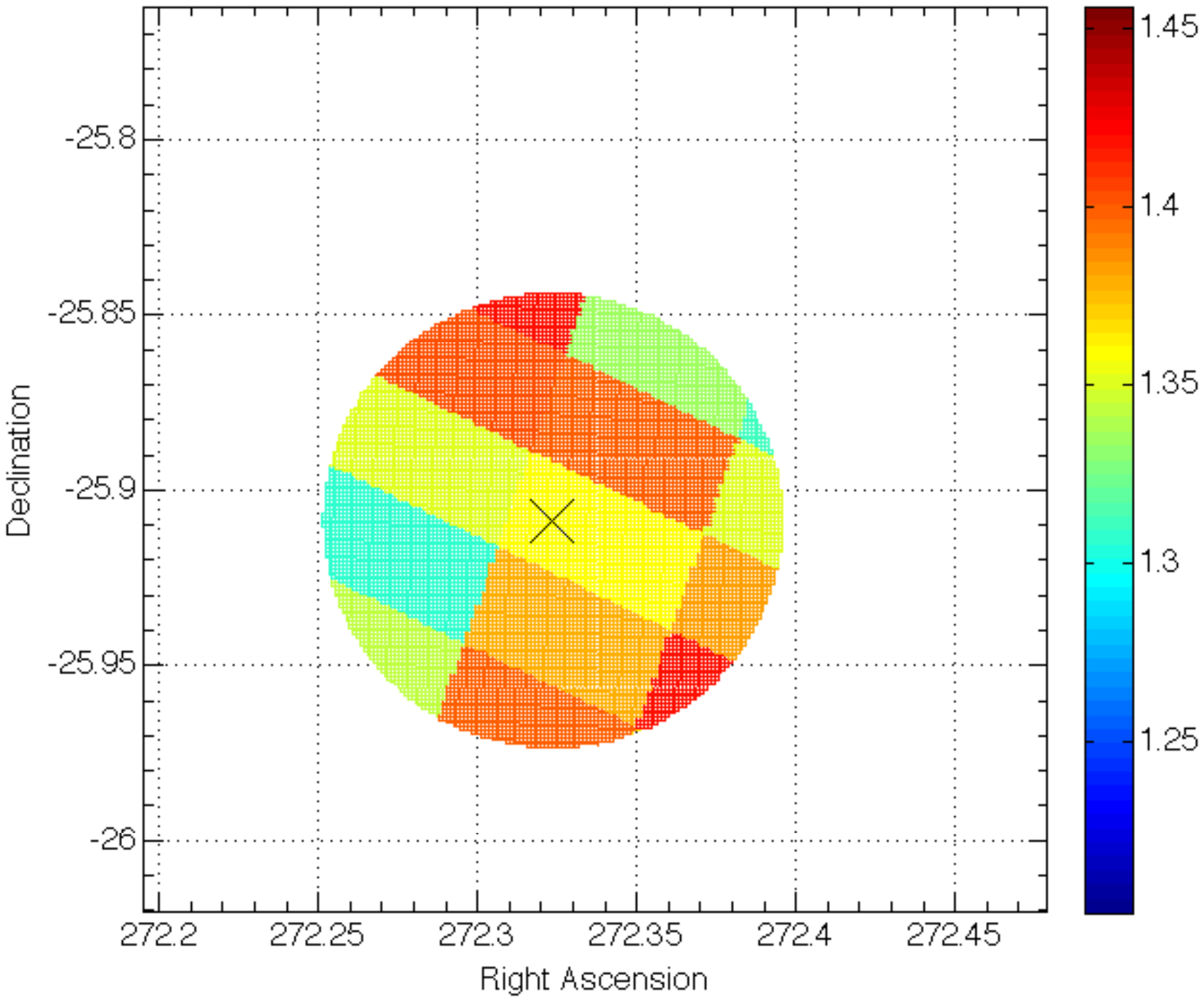}  &
\includegraphics[scale=0.21]{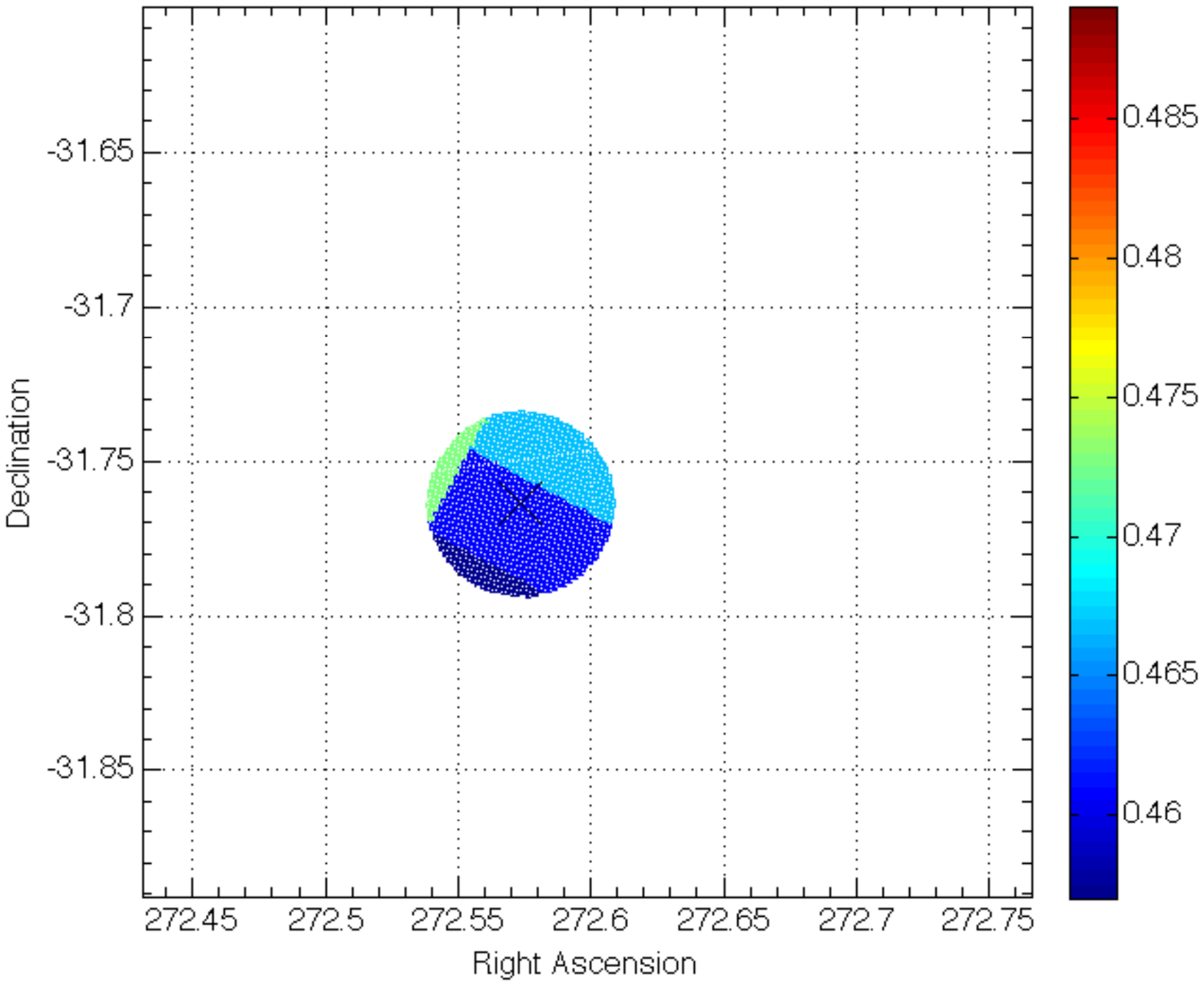}  &
\includegraphics[scale=0.21]{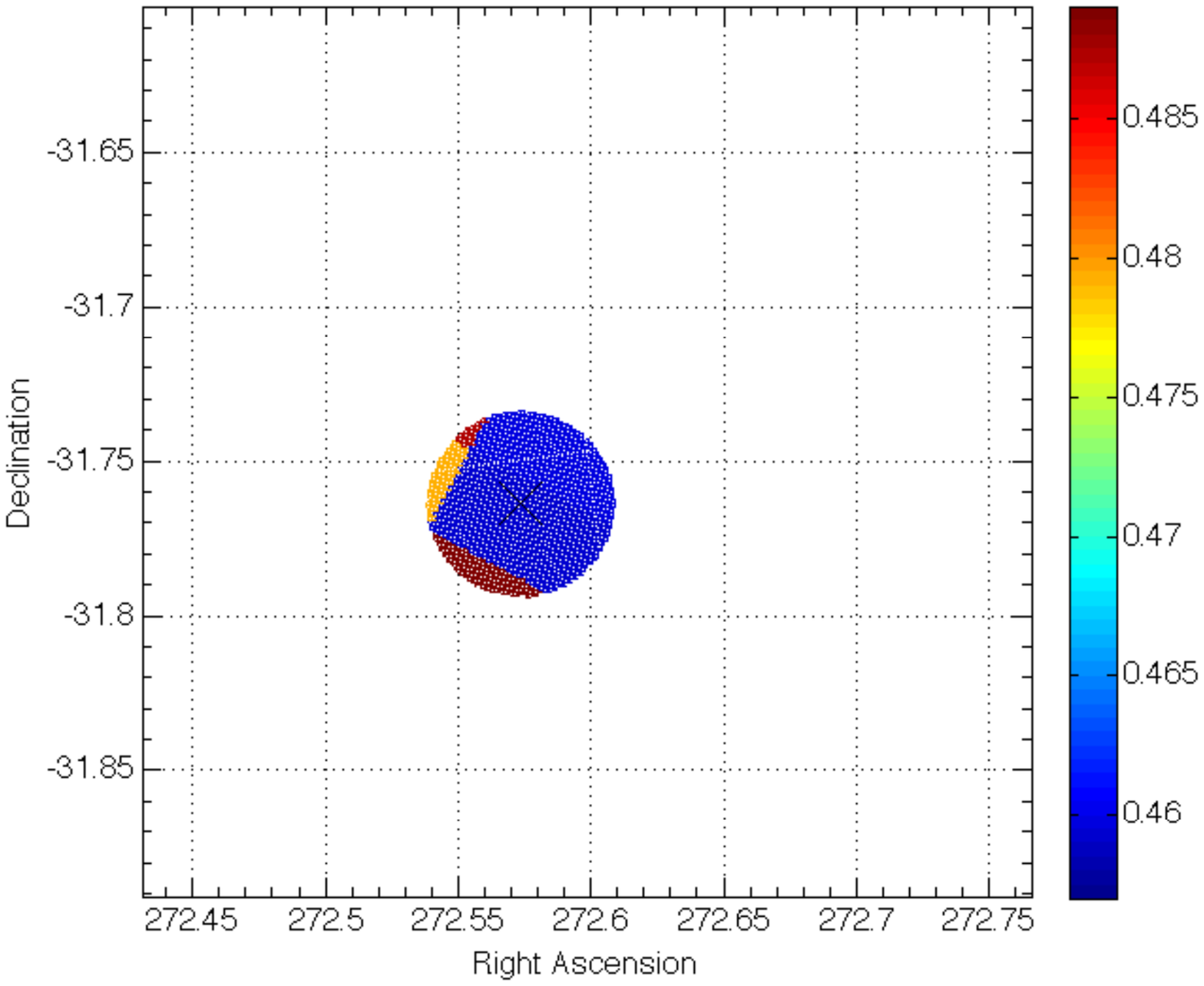} \\
\end{tabular}
\end{figure}

\begin{figure}[tp] \centering
\begin{tabular}{cccc}
NGC 6624 & & NGC 6626 & \\
\includegraphics[scale=0.21]{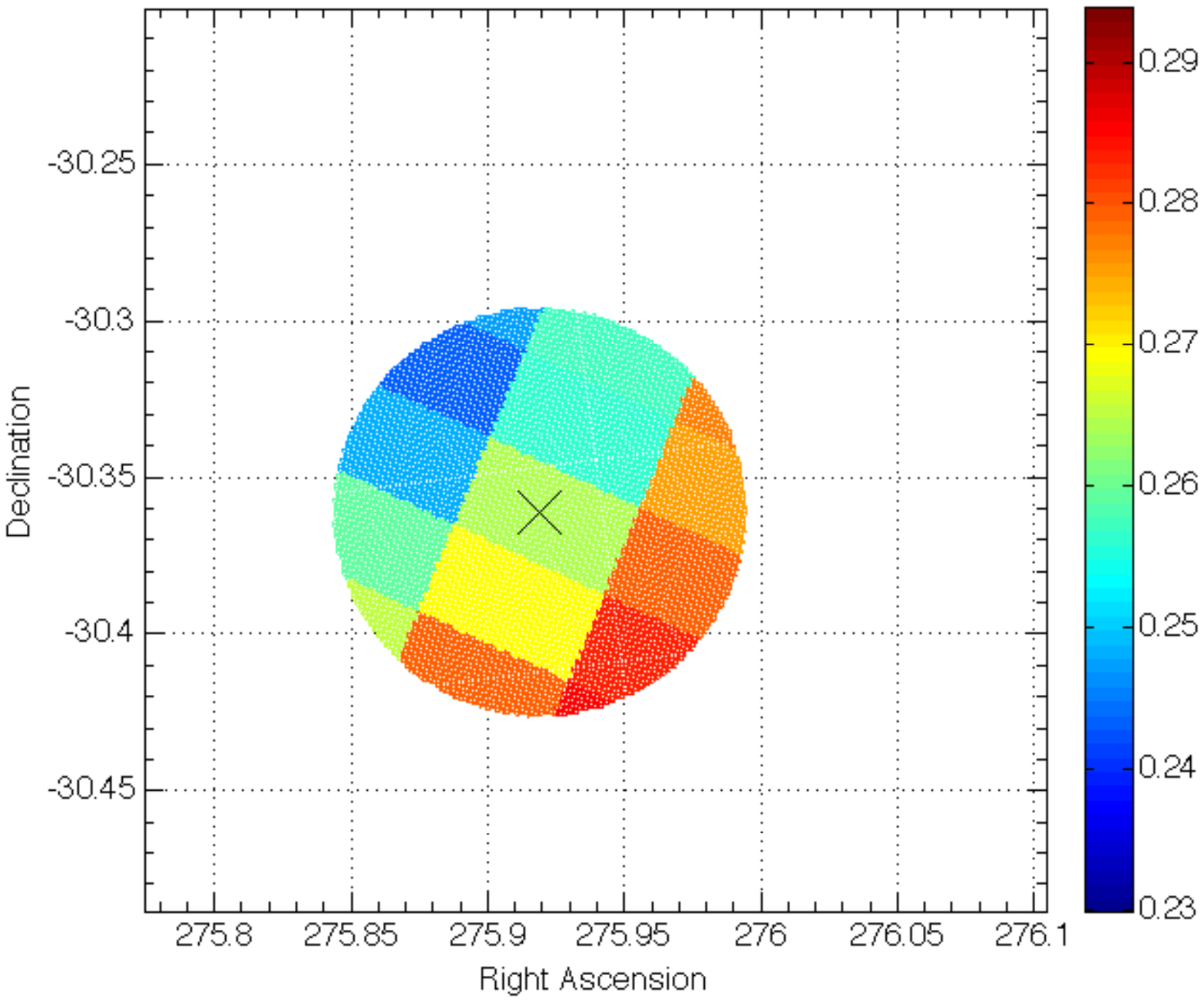}  &
\includegraphics[scale=0.21]{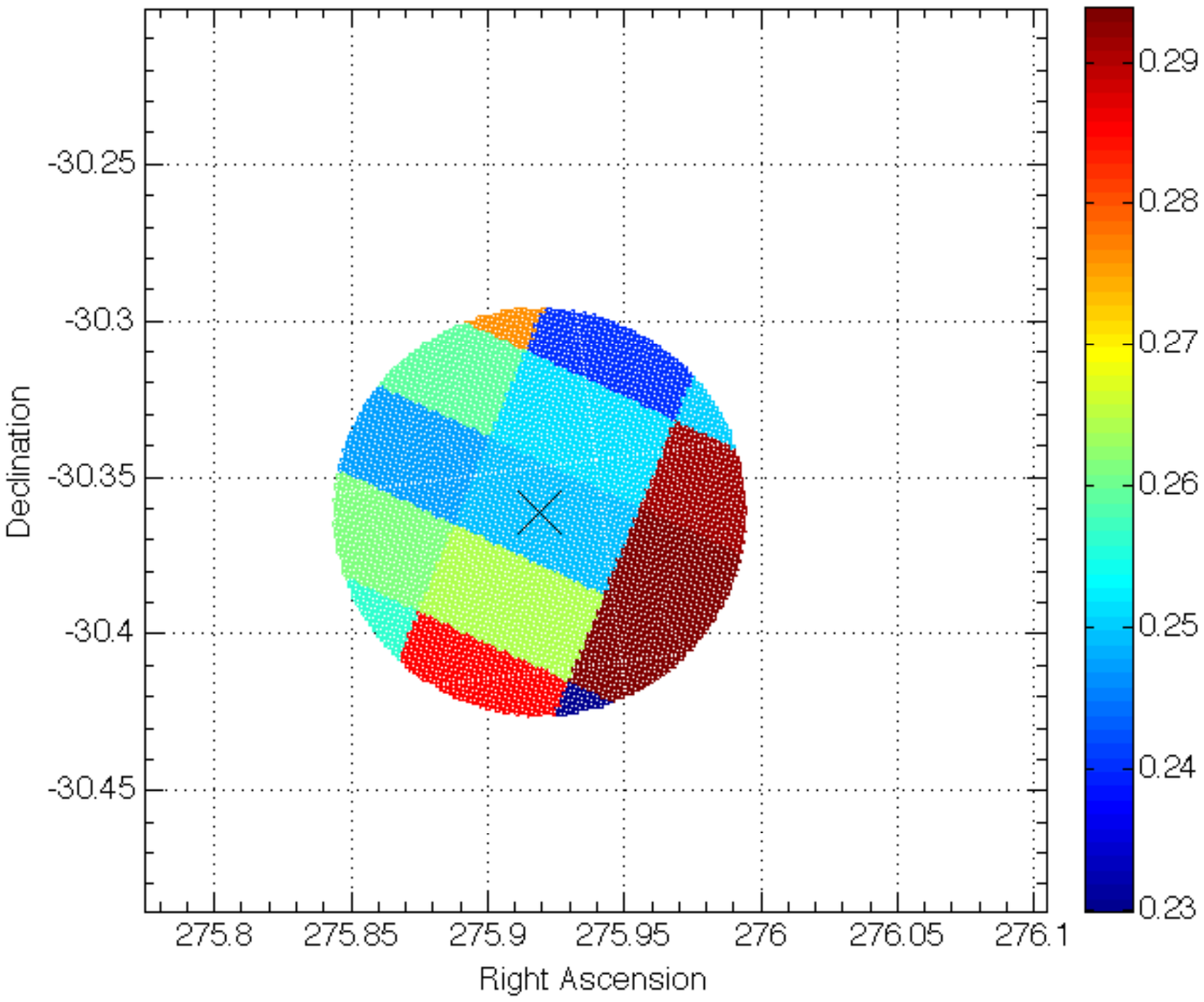}  &
\includegraphics[scale=0.21]{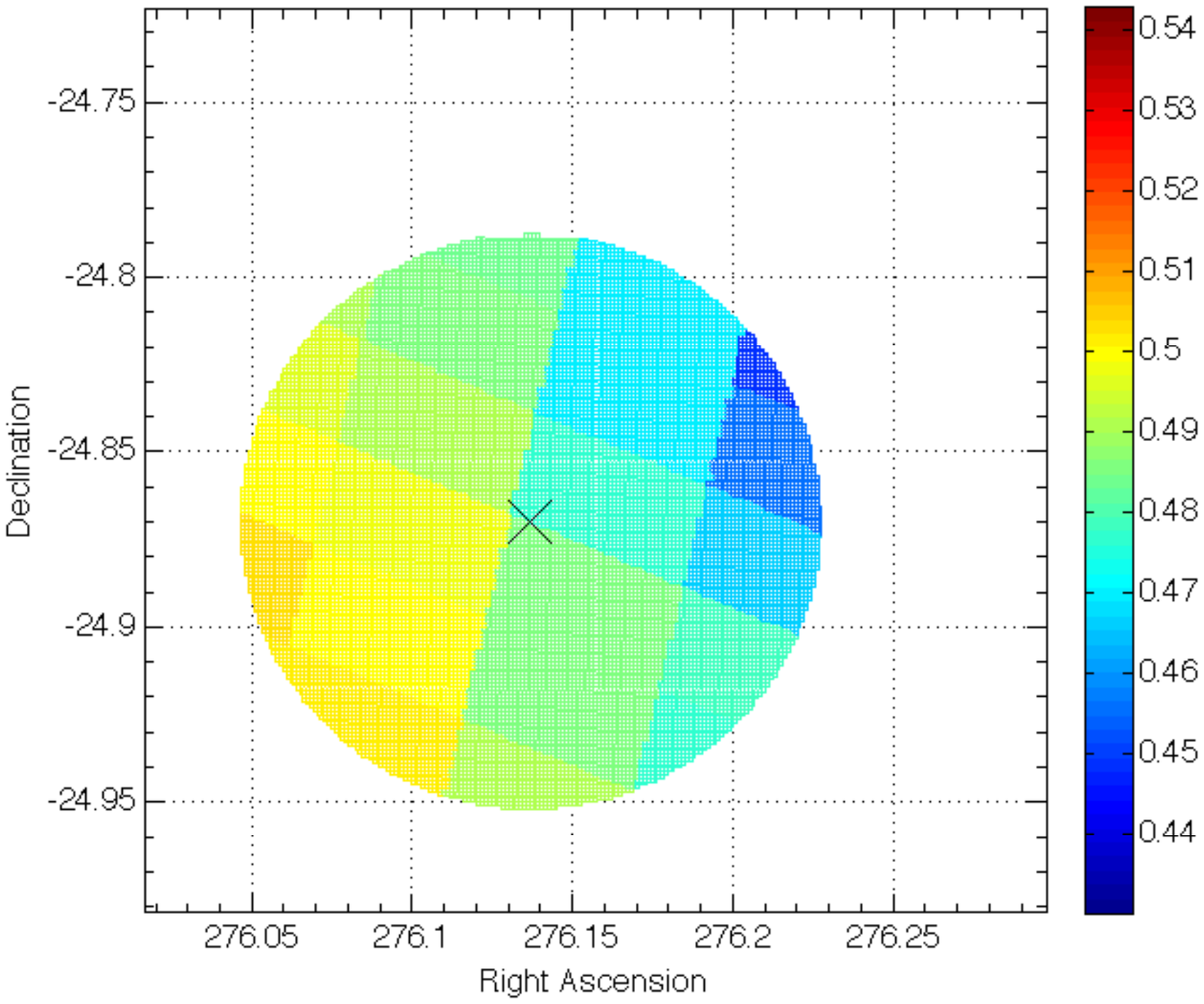}  &
\includegraphics[scale=0.21]{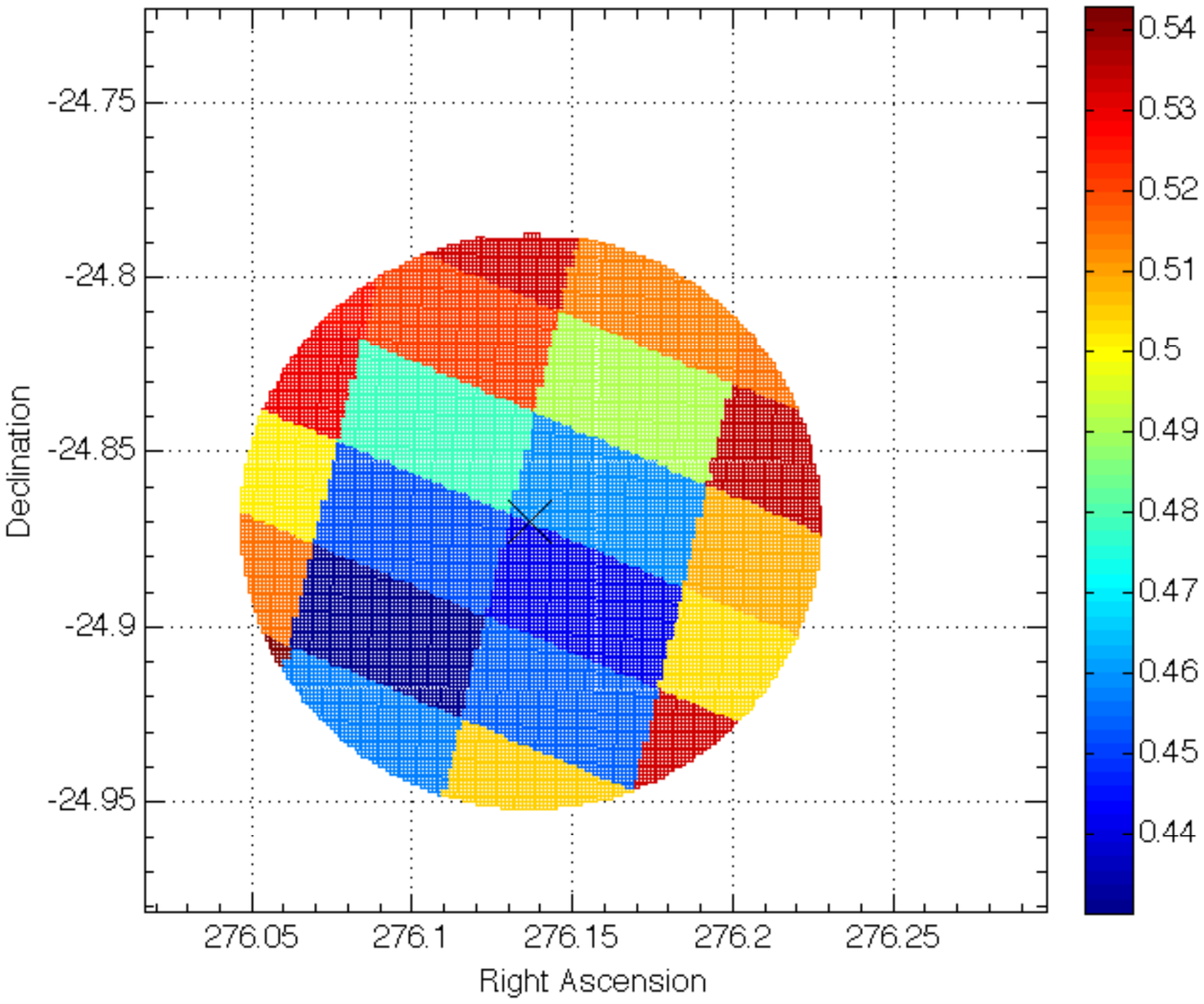} \\
NGC 6637 & & NGC 6642 & \\
\includegraphics[scale=0.21]{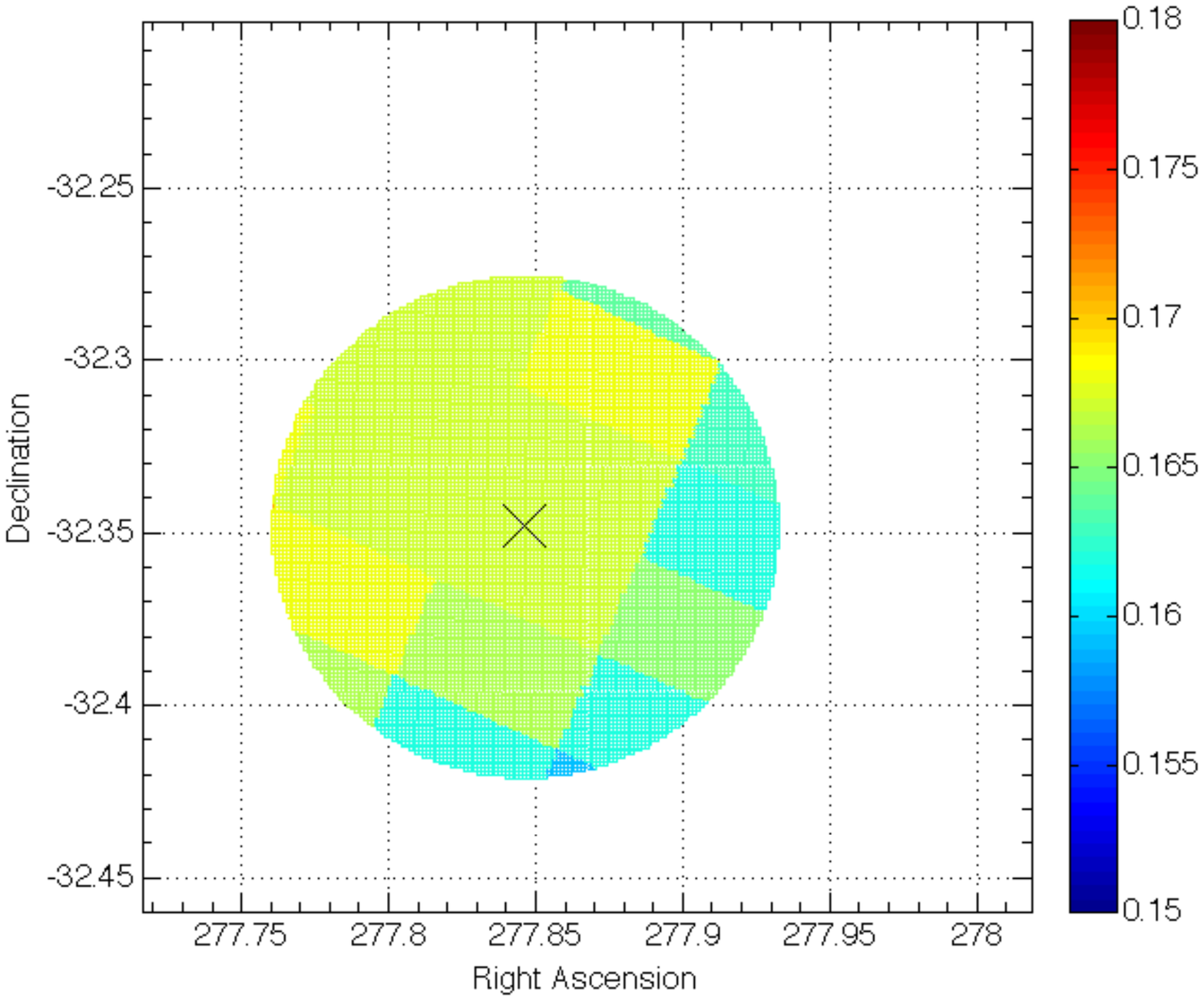}  &
\includegraphics[scale=0.21]{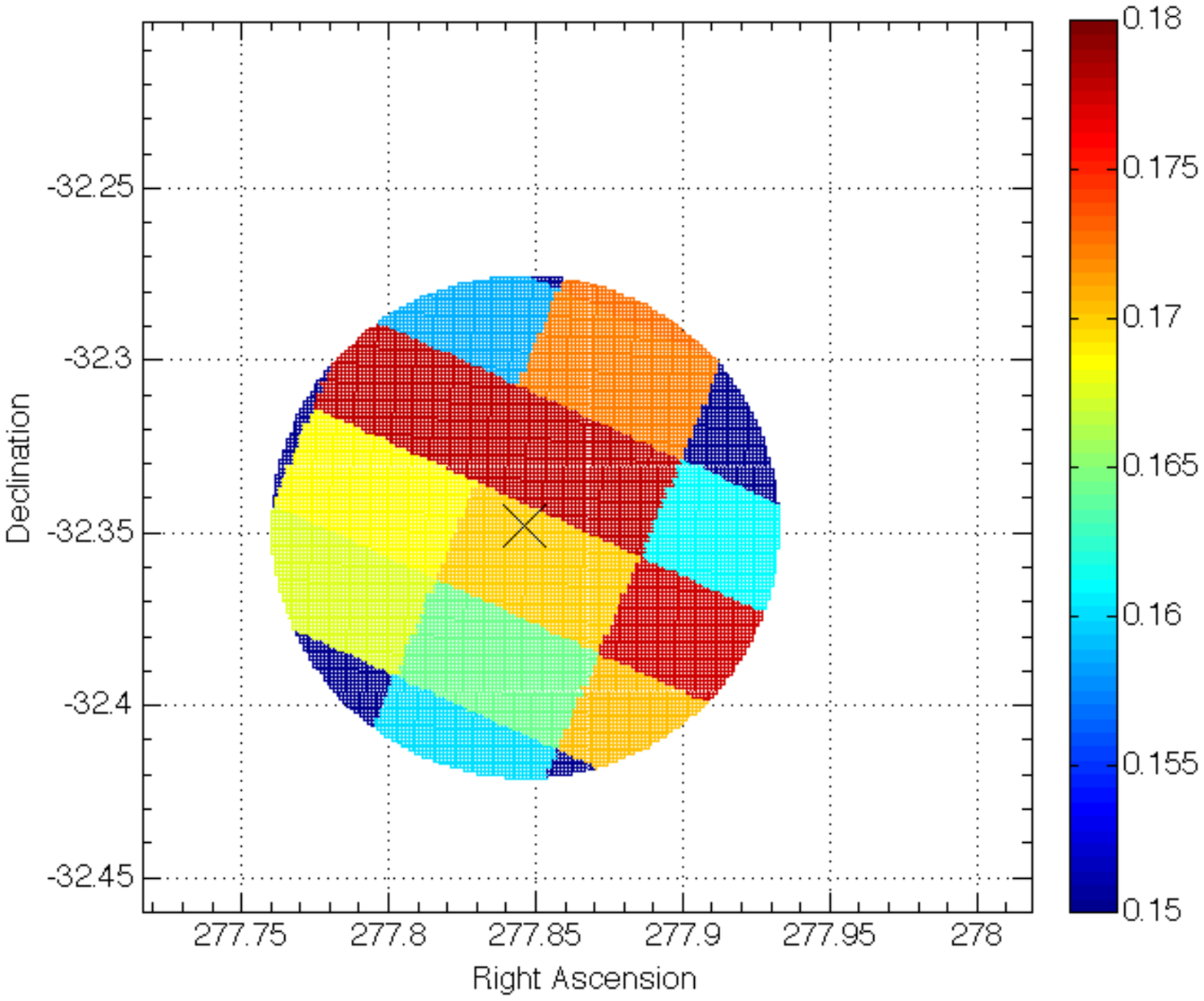}  &
\includegraphics[scale=0.21]{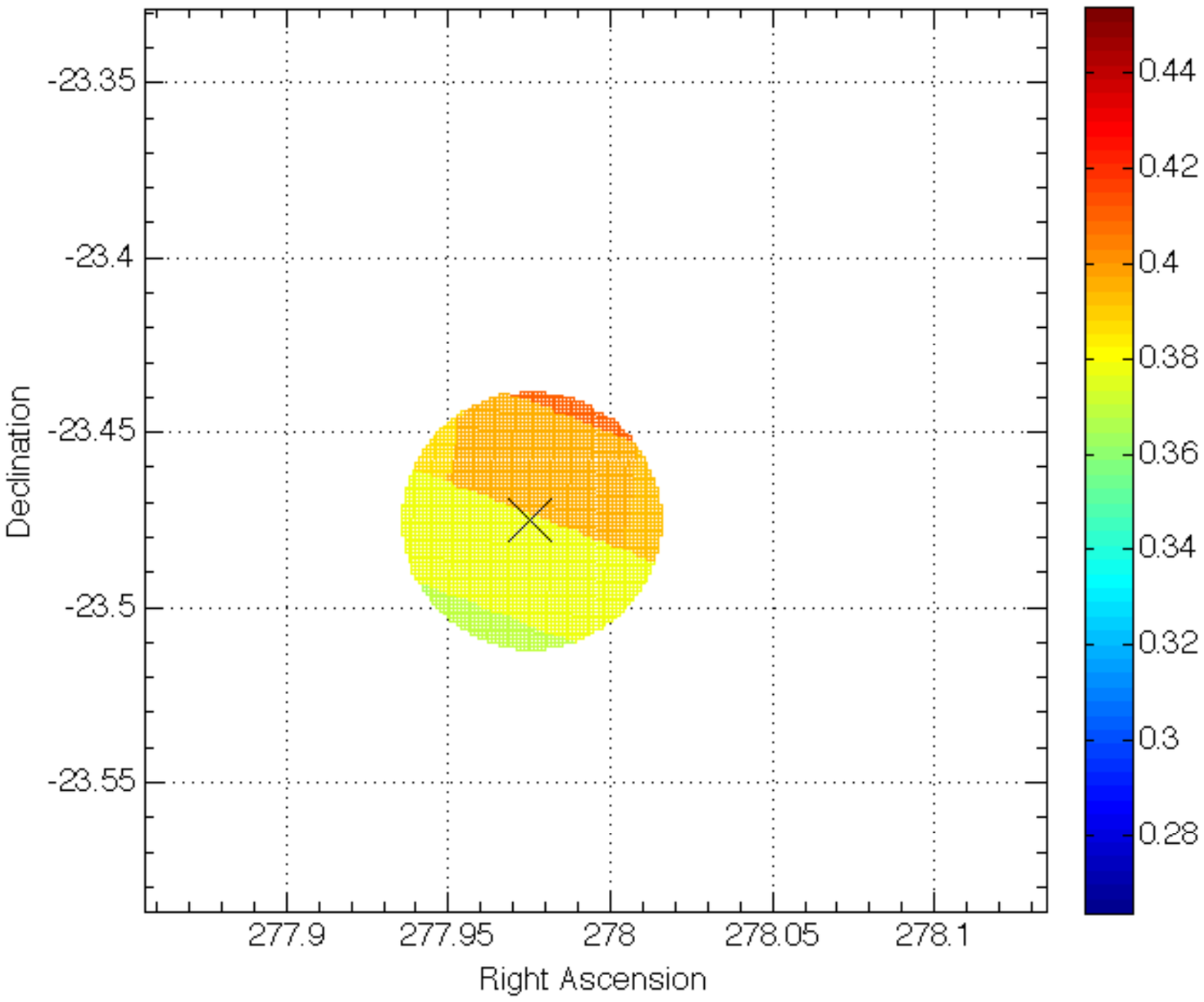}  &
\includegraphics[scale=0.21]{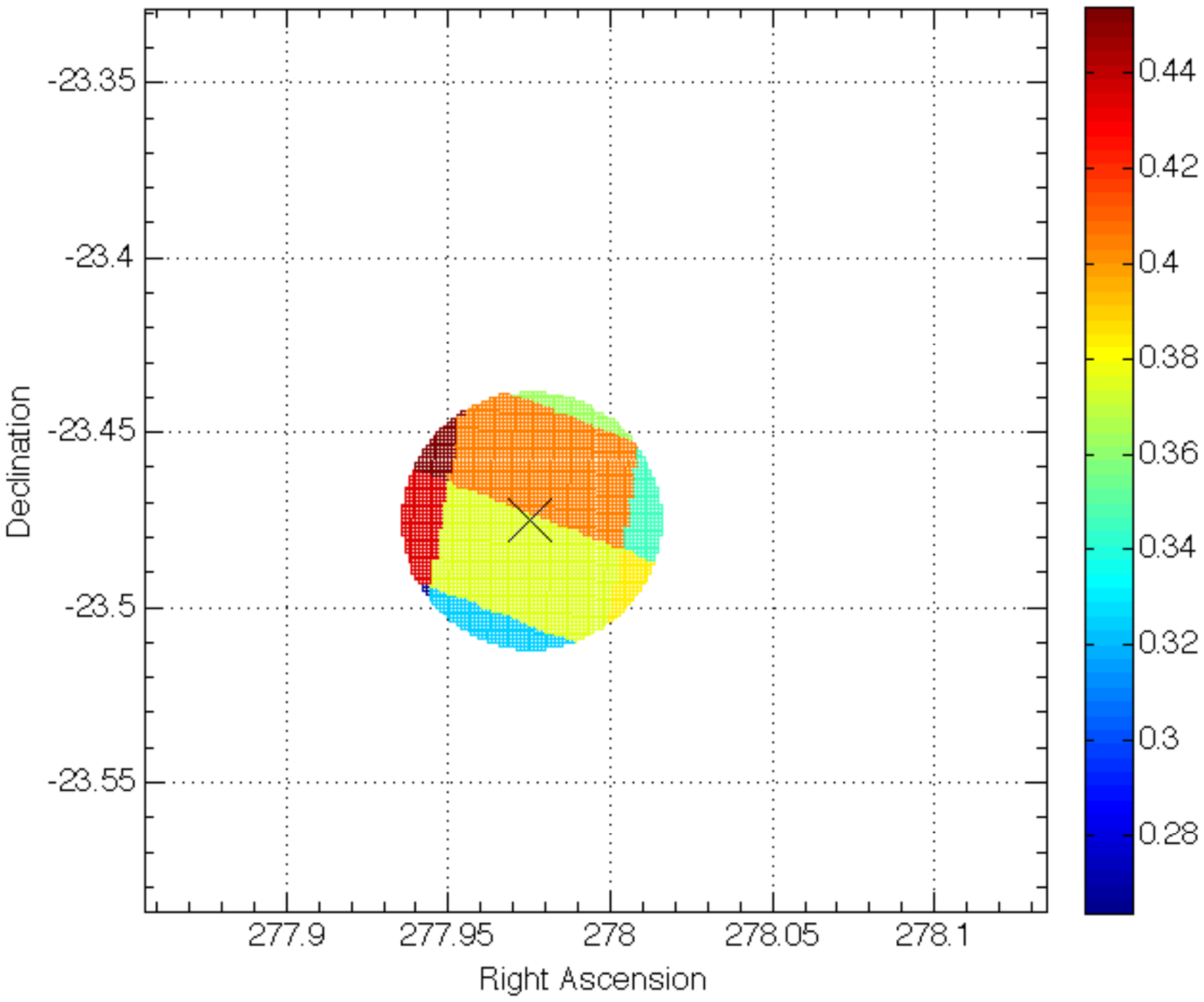} \\
NGC 6656 & & NGC 6681 & \\
\includegraphics[scale=0.21]{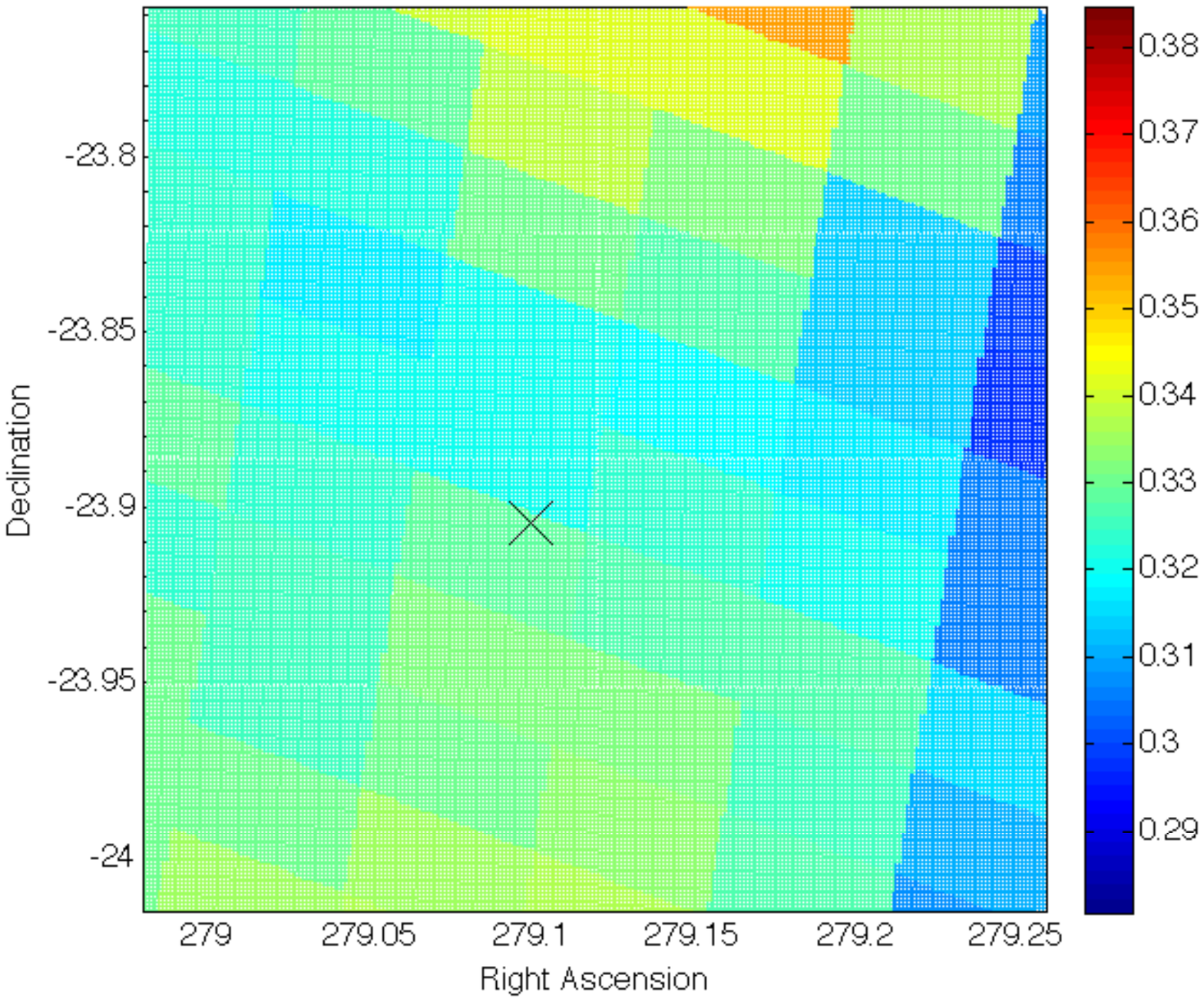}  &
\includegraphics[scale=0.21]{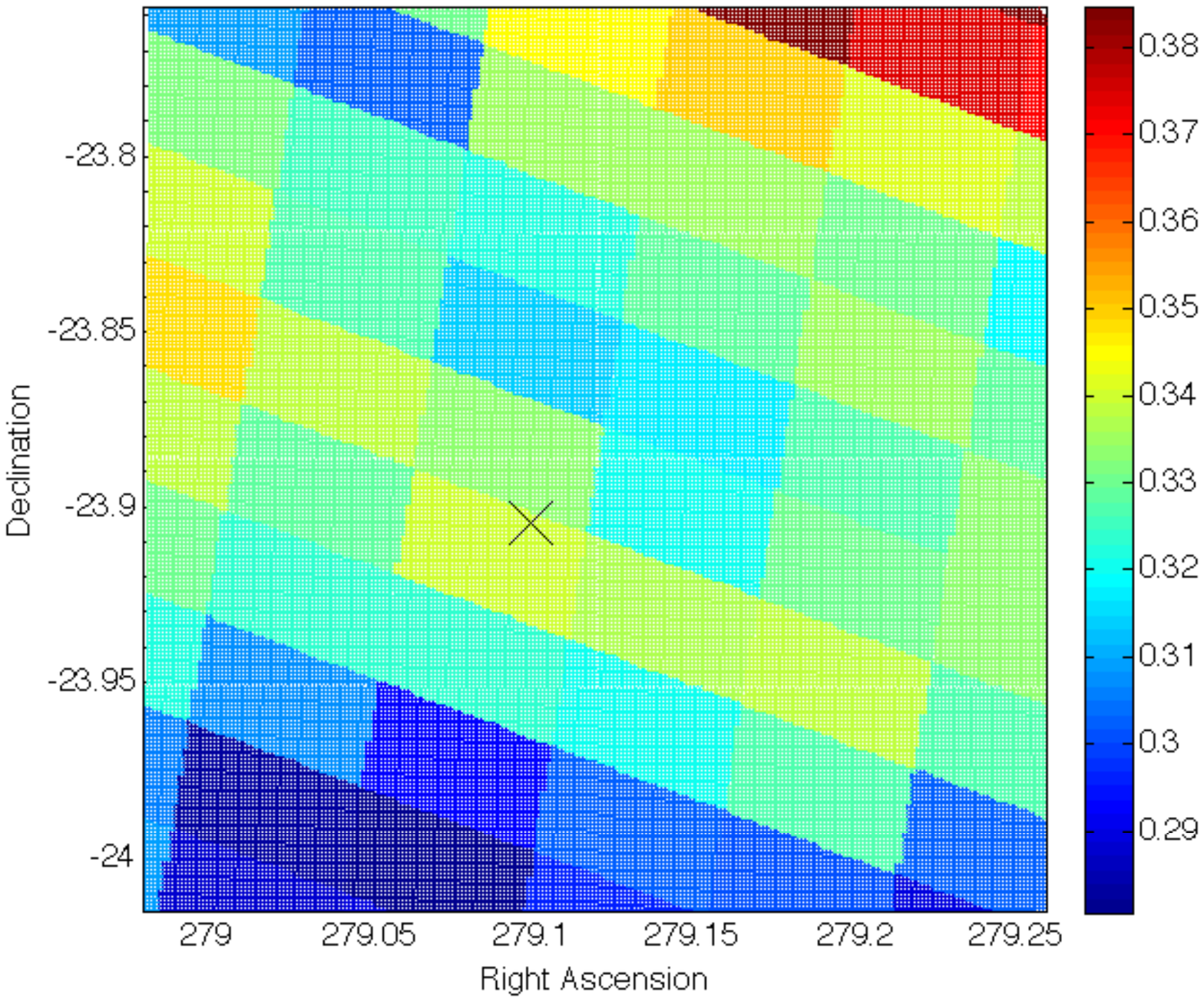}  &
\includegraphics[scale=0.21]{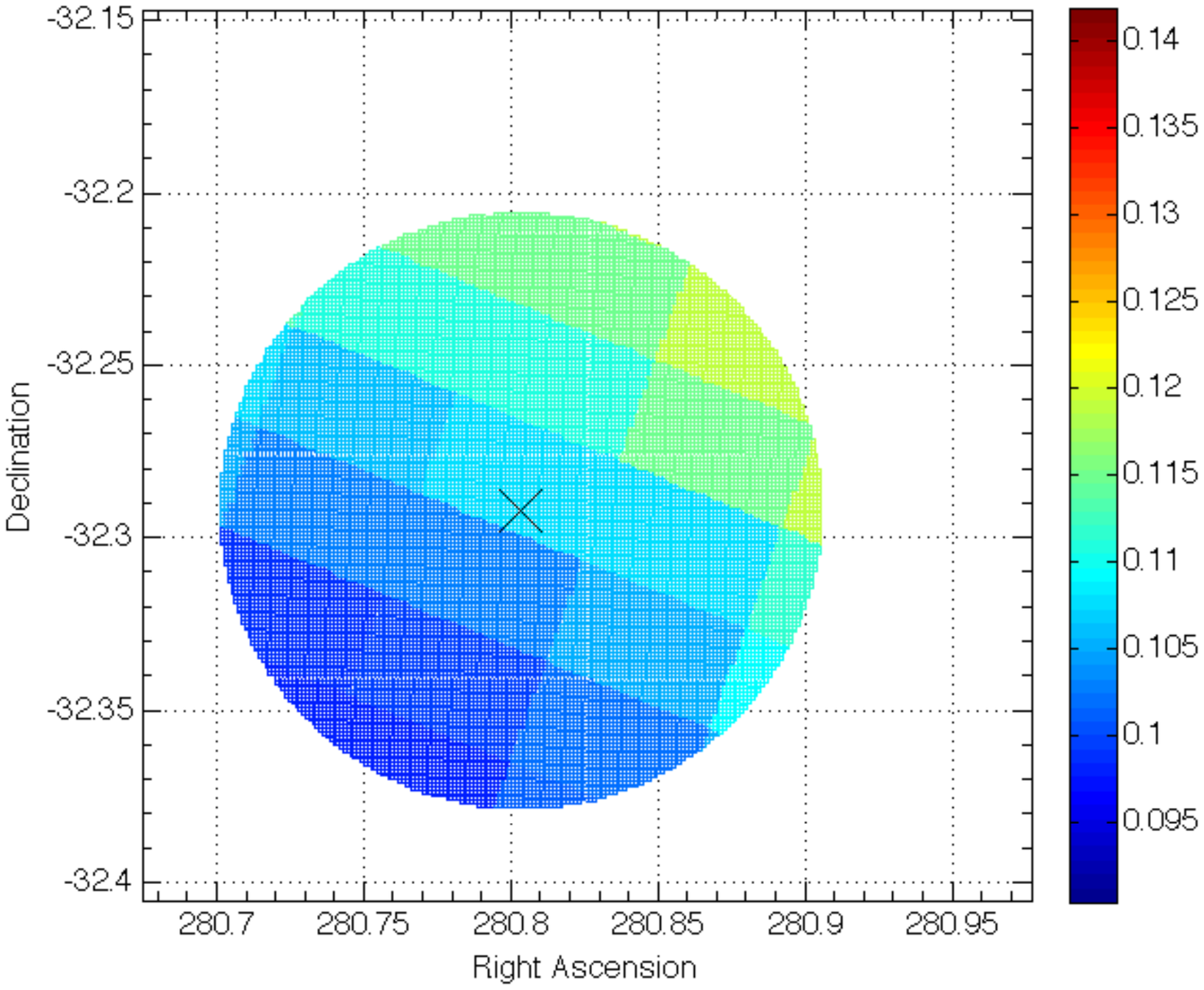}  &
\includegraphics[scale=0.21]{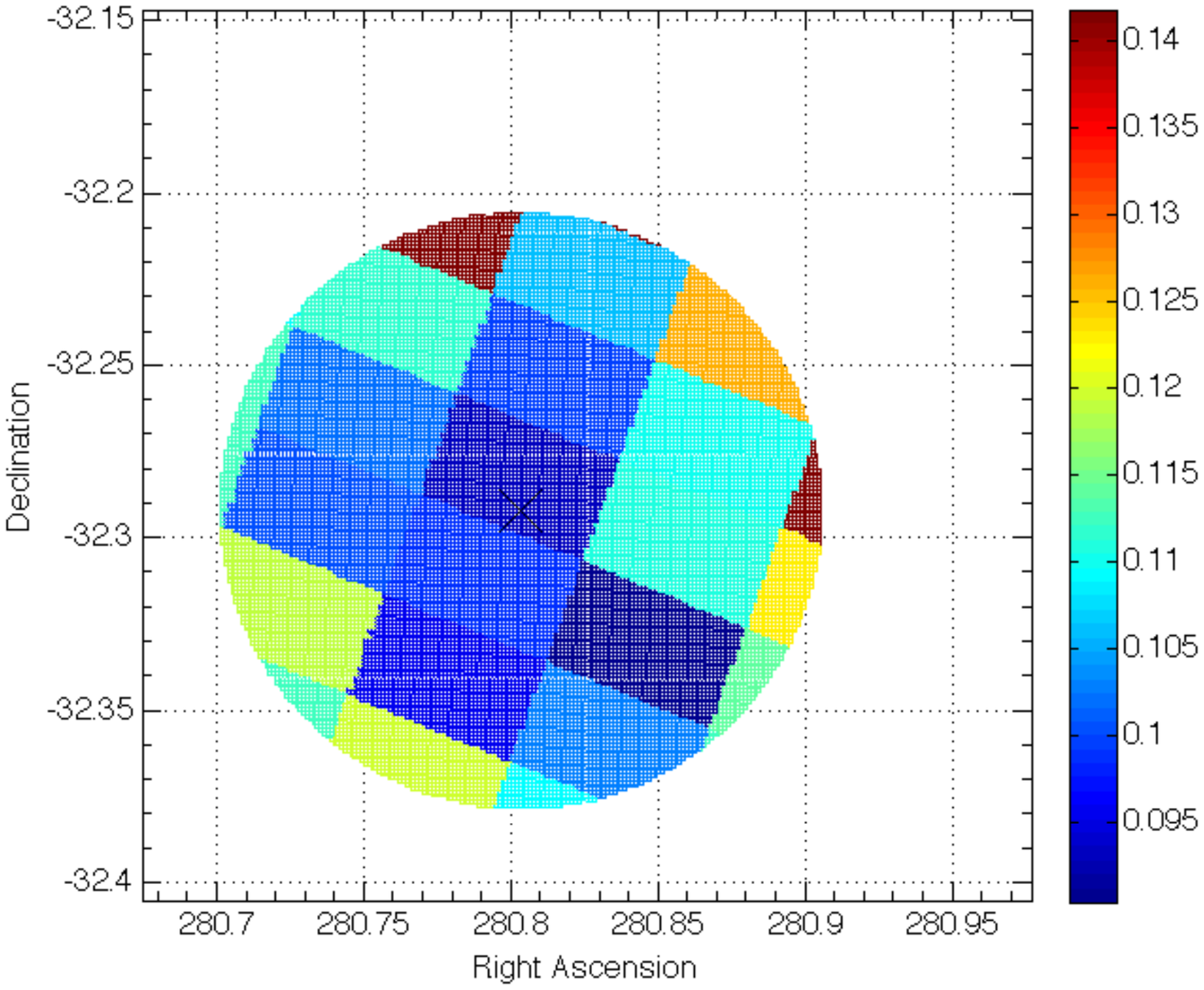} \\
NGC 6809 & & & \\
\includegraphics[scale=0.21]{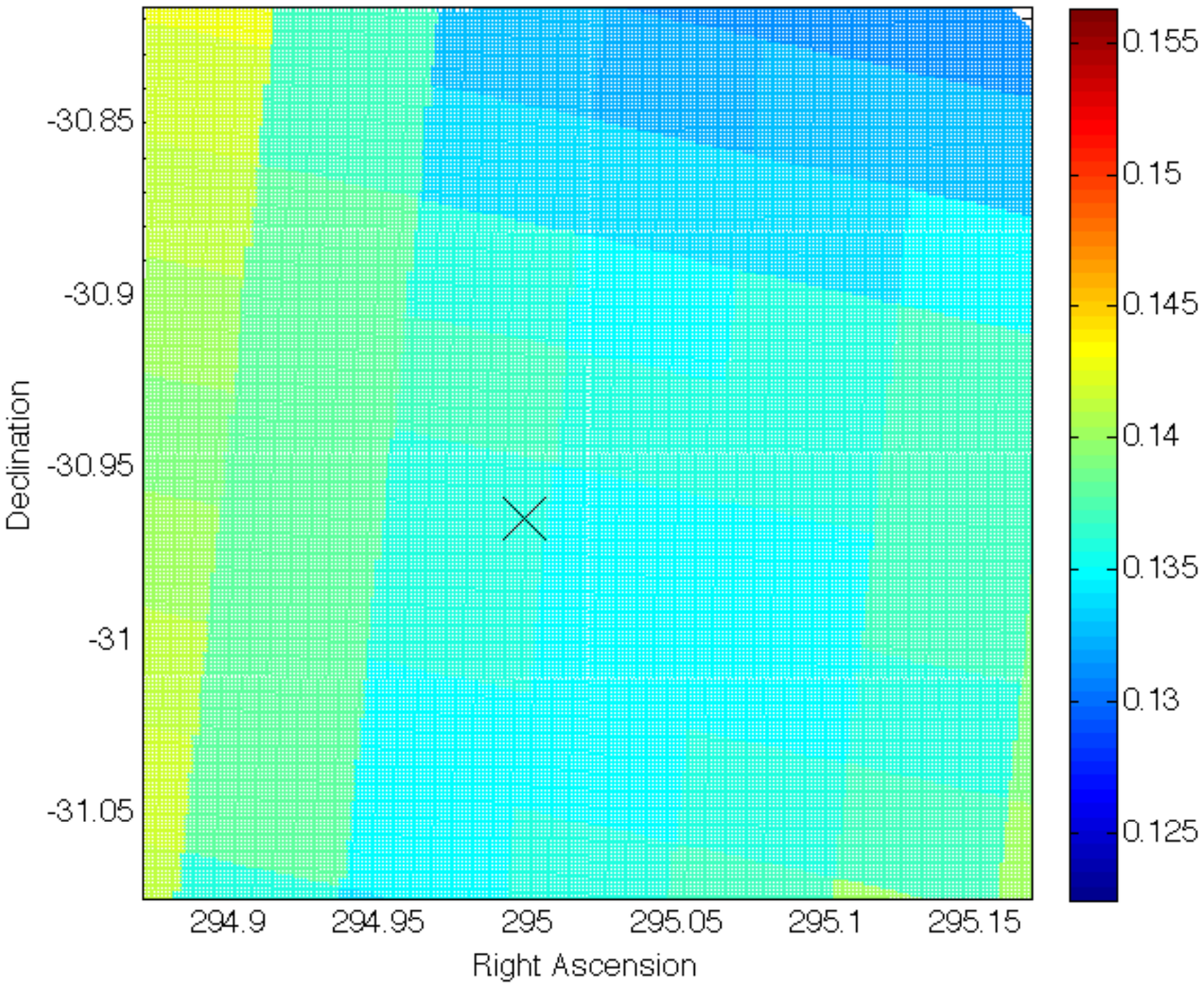}  &
\includegraphics[scale=0.21]{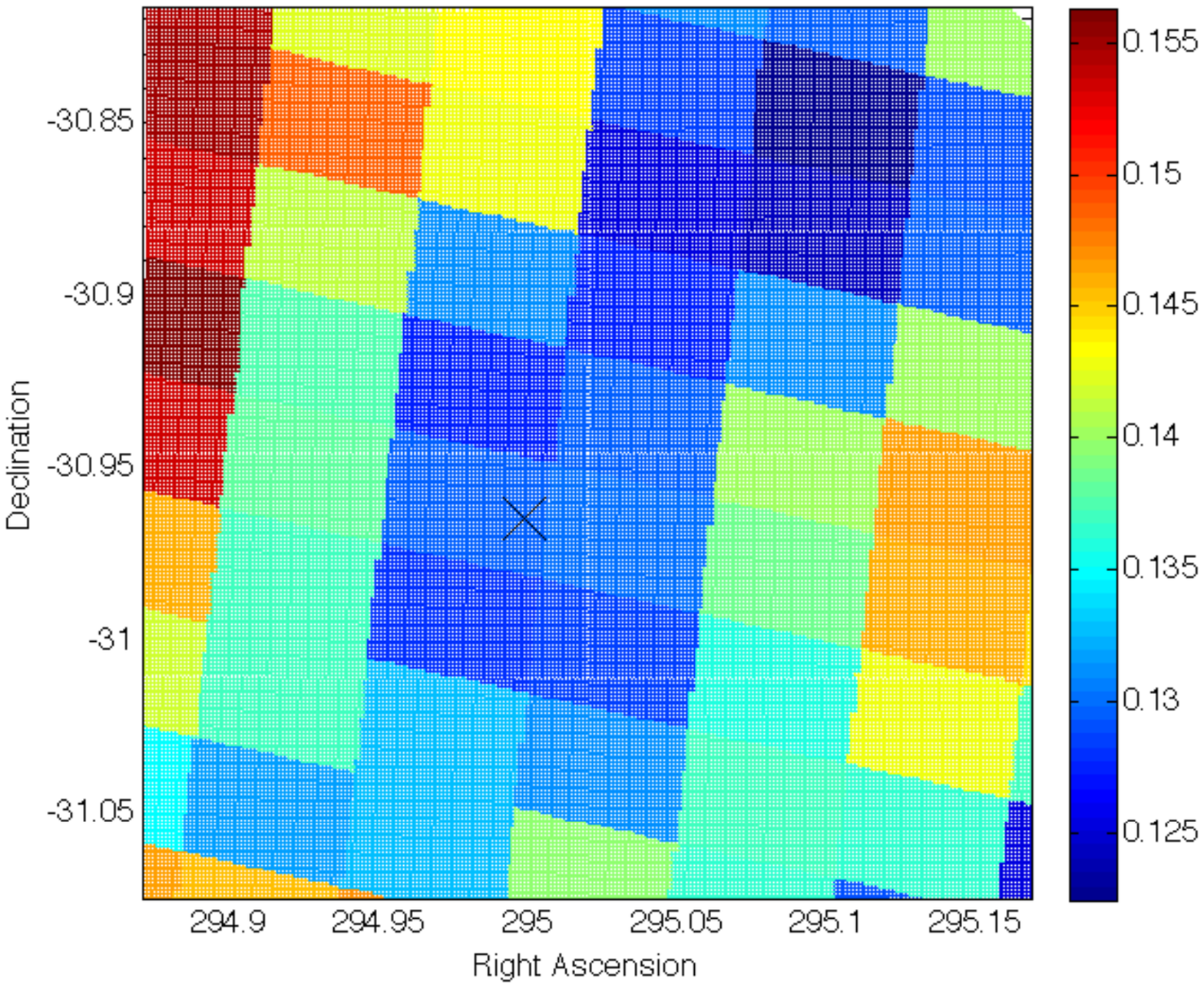}  &
&
\\
\end{tabular}
\end{figure}

\clearpage

\begin{figure}[tp] \centering
\begin{tabular}{cccc}
NGC 6121 & & NGC 6218 & \\
\includegraphics[scale=0.21]{ngc6121map1.pdf}  &
\includegraphics[scale=0.21]{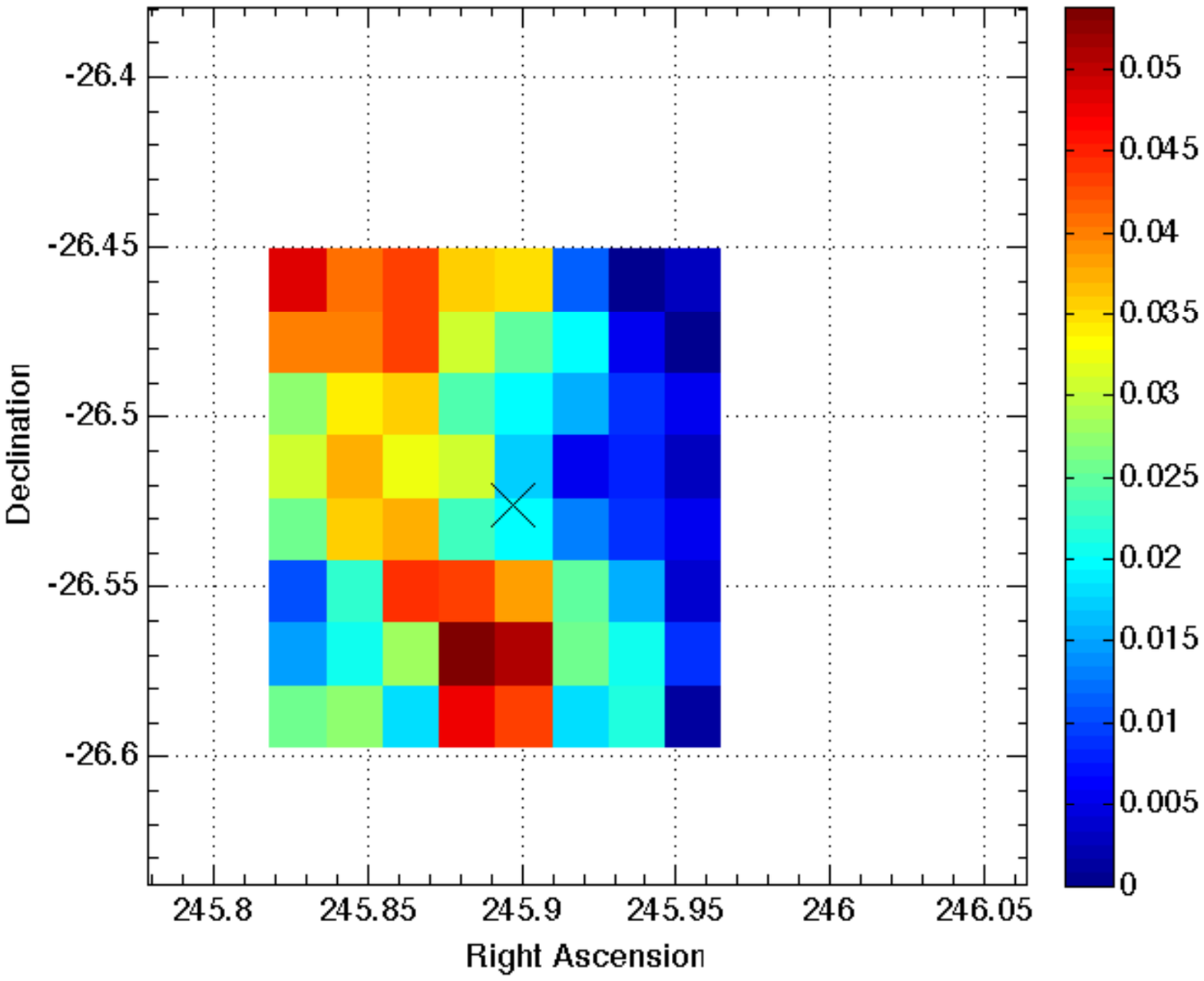}  &
\includegraphics[scale=0.21]{ngc6218map1.pdf}  &
\includegraphics[scale=0.21]{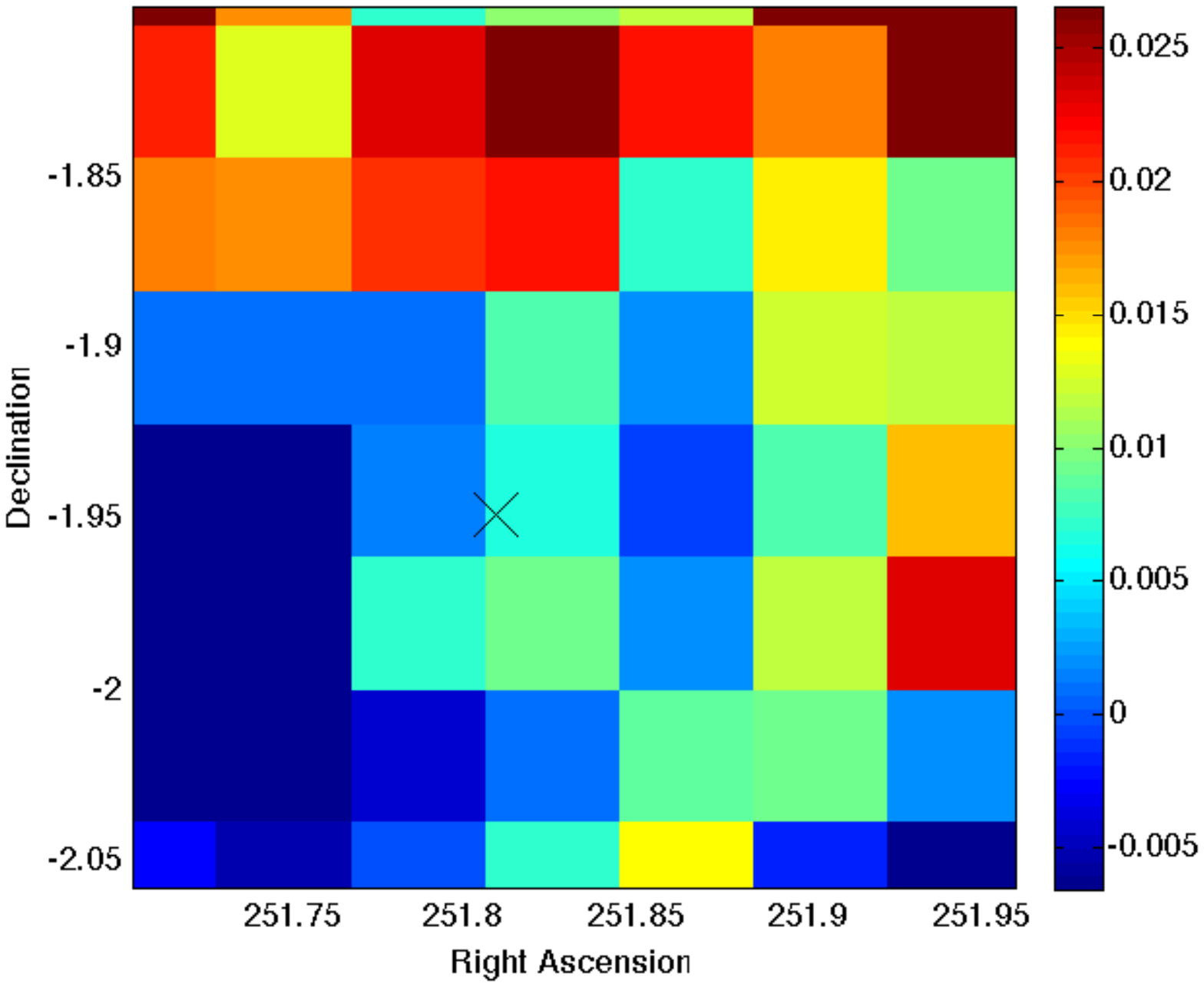} \\
NGC 6254 & & NGC 6266 & \\
\includegraphics[scale=0.21]{ngc6254map1.pdf}  &
\includegraphics[scale=0.21]{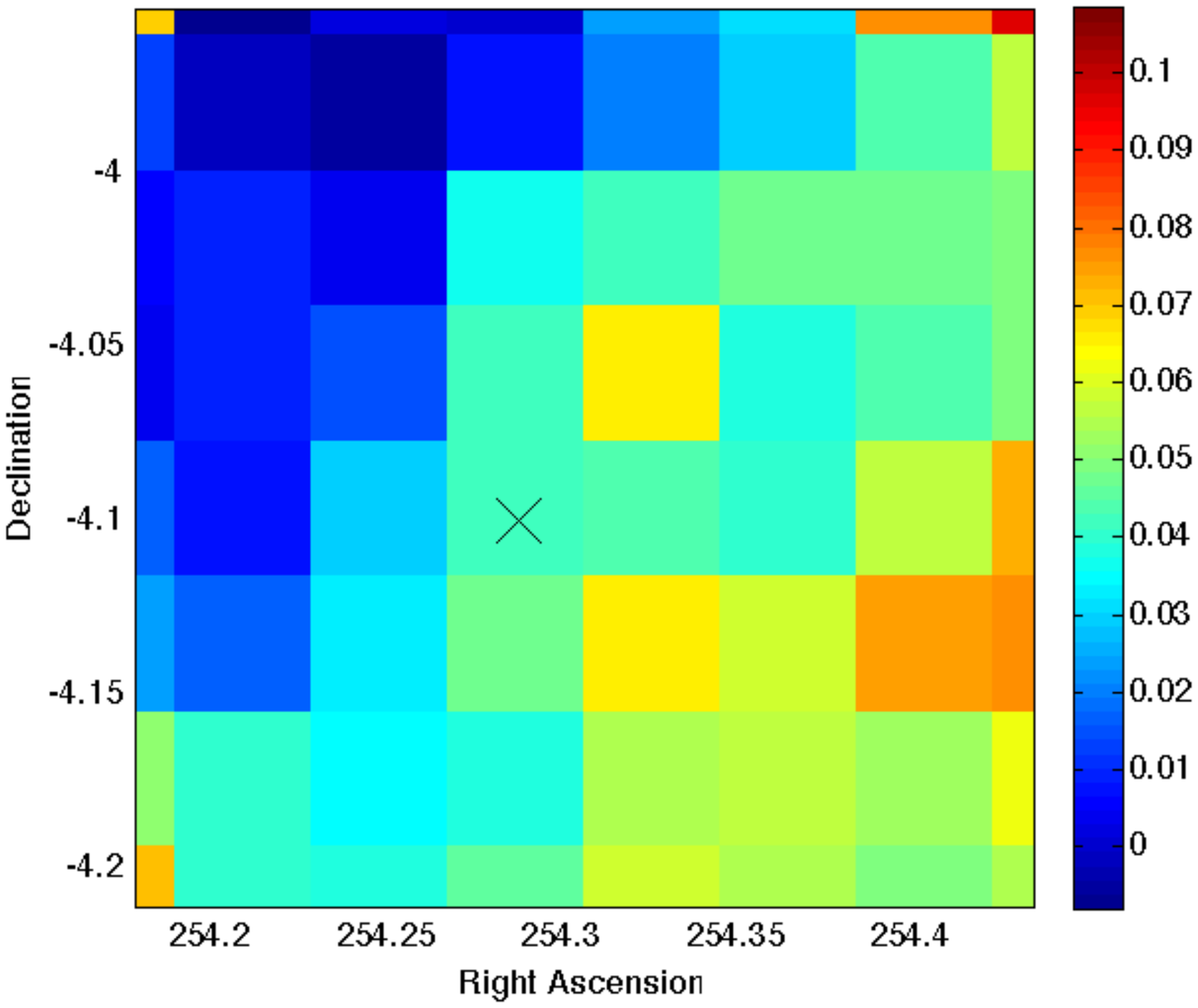}  &
\includegraphics[scale=0.21]{ngc6266map1.pdf}  &
\includegraphics[scale=0.21]{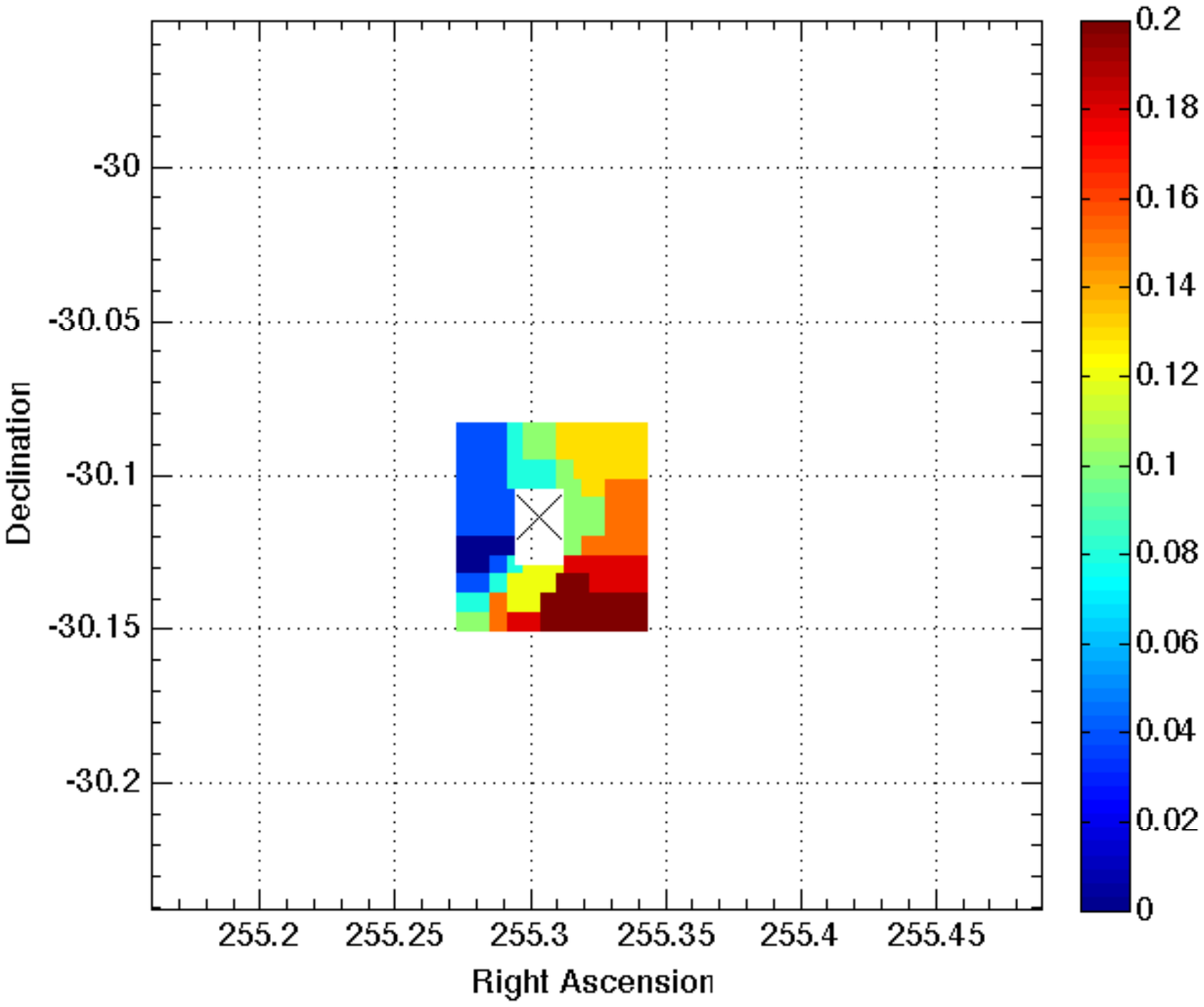} \\
NGC 6342 & & NGC 6553 & \\
\includegraphics[scale=0.21]{ngc6342map1.pdf}  &
\includegraphics[scale=0.21]{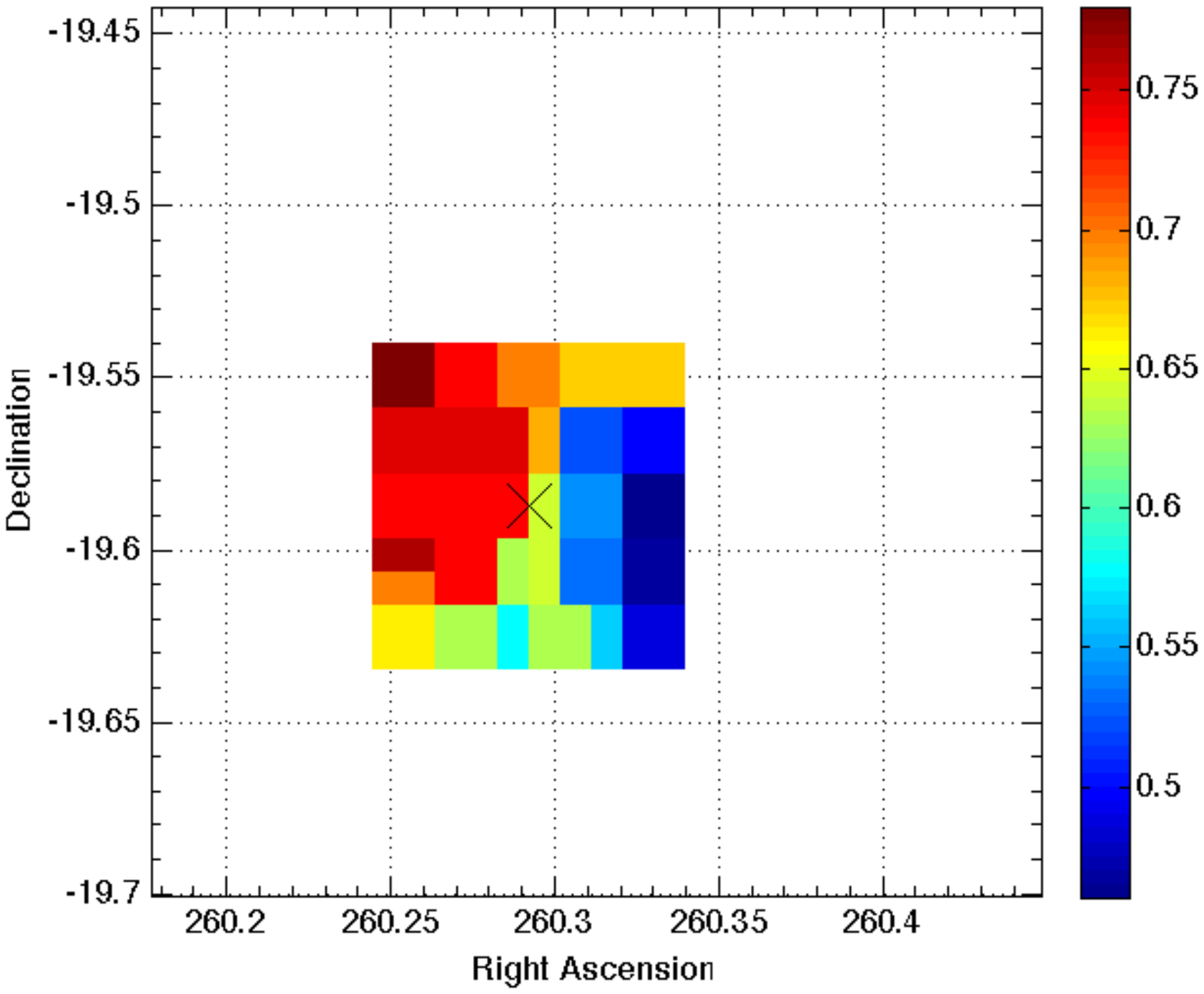}  &
\includegraphics[scale=0.21]{ngc6553map1.pdf}  &
\includegraphics[scale=0.21]{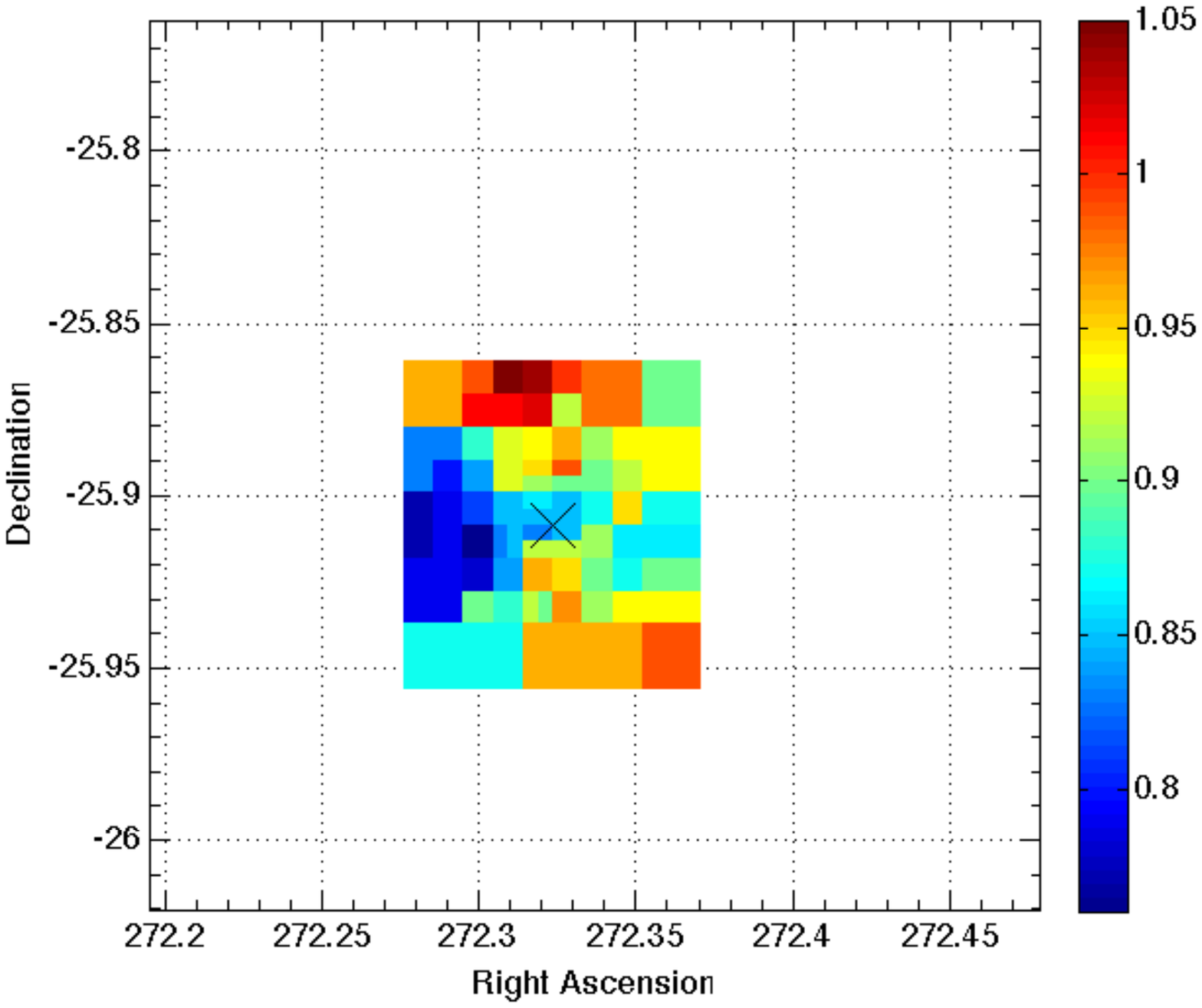} \\
\end{tabular}
\caption{Comparison of the E(B-V) extinction maps provided by our
  technique (left) with the extinction maps (right) provided by
  \citet{mo02} for NGC 6121, by \citet{vo02} for NGC 6218 and NGC 6254,
  by \citet{ge04} for NGC 6266, and by \citet{he99} for NGC 6342 and
  NGC 6553. All the maps from the literature show relative
  differential extinctions with respect to the extinction zero point
  of a fiducial region, except for NGC 6342 and NGC 6553, that show
  absolute extinctions. Notice that the extinction zero points between
  our maps and those from the literature can differ. }
\label{figmapcomp}
\end{figure}

\begin{figure}
\plottwo{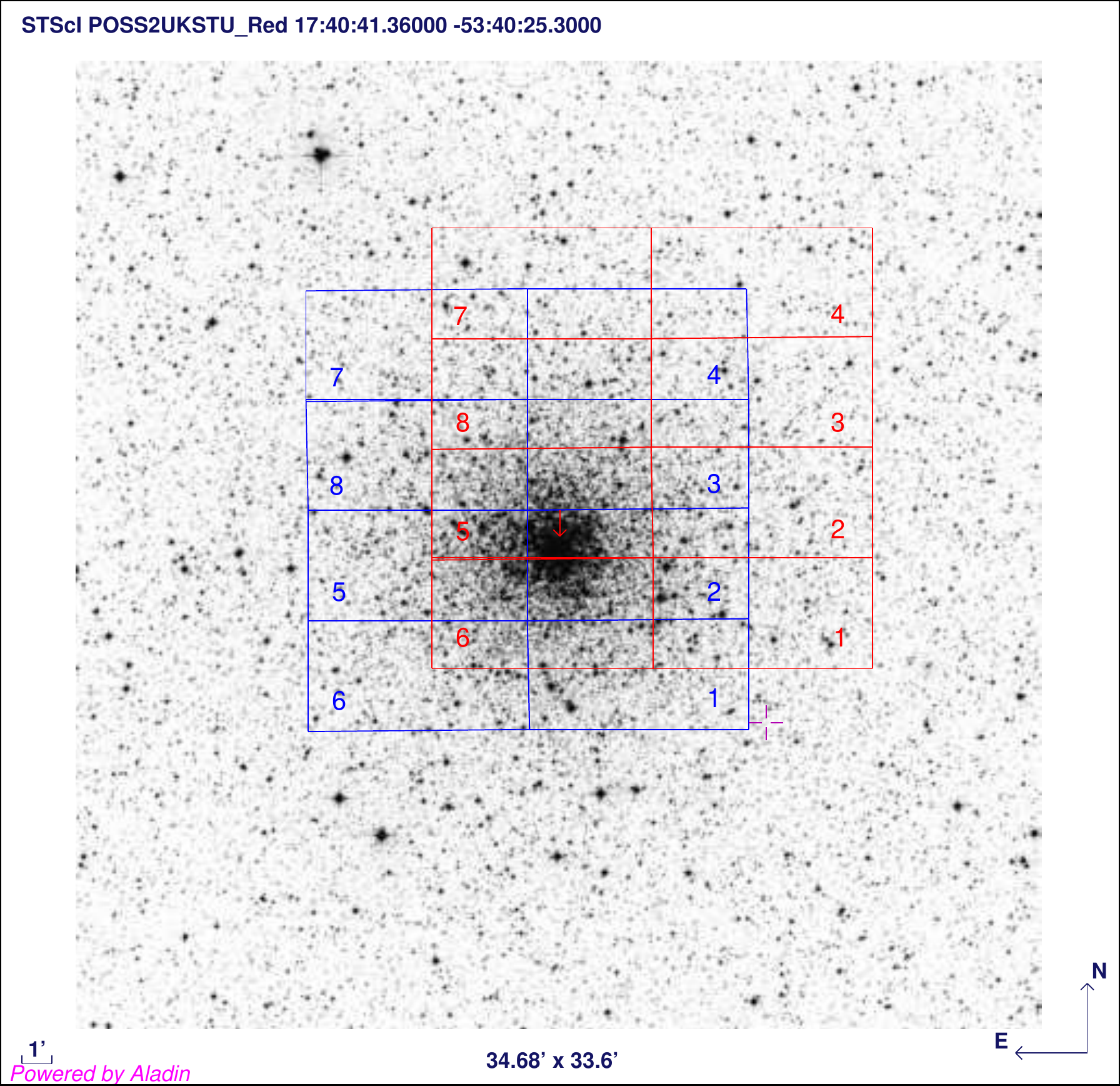}{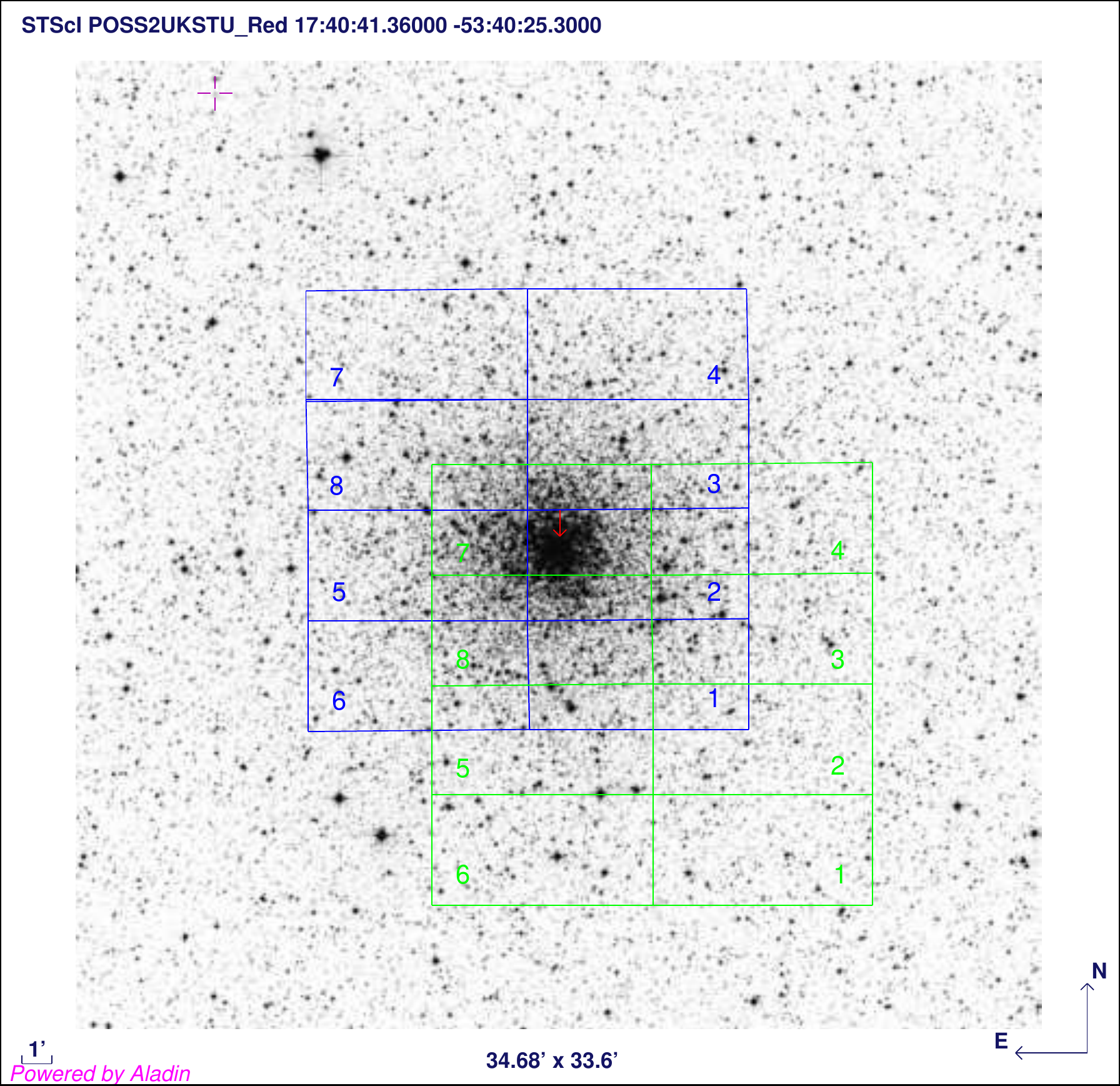}
\caption{Pointings used to make the instrumental calibration, bringing
  the photometry from the different chips to the chip2 system. We
  chose these pointings because they allowed us to observe stars in
  common between chip2 and all the other chips in just 3 pointings
  (chips 1, 5 and 6 have stars in common with chip 2 in the left
  configuration and chips 3, 4, 7, and 8 have stars in common with
  chip2 in the right configuration). We used stars from the cluster
  NGC6397, one of the clusters in our sample, to do this
  calibration. The blue pointing was the same as the one originally
  used in our observing run. That way we were able to compare the
  photometries of our original observing run and the calibrating run
  for all the chips in the camera and take care of any effect produced
  by a change in the sensitivity of the chips (see text).}
\label{figchipcal}
\end{figure}

\begin{figure}
%\epsscale{0.85}
\plotone{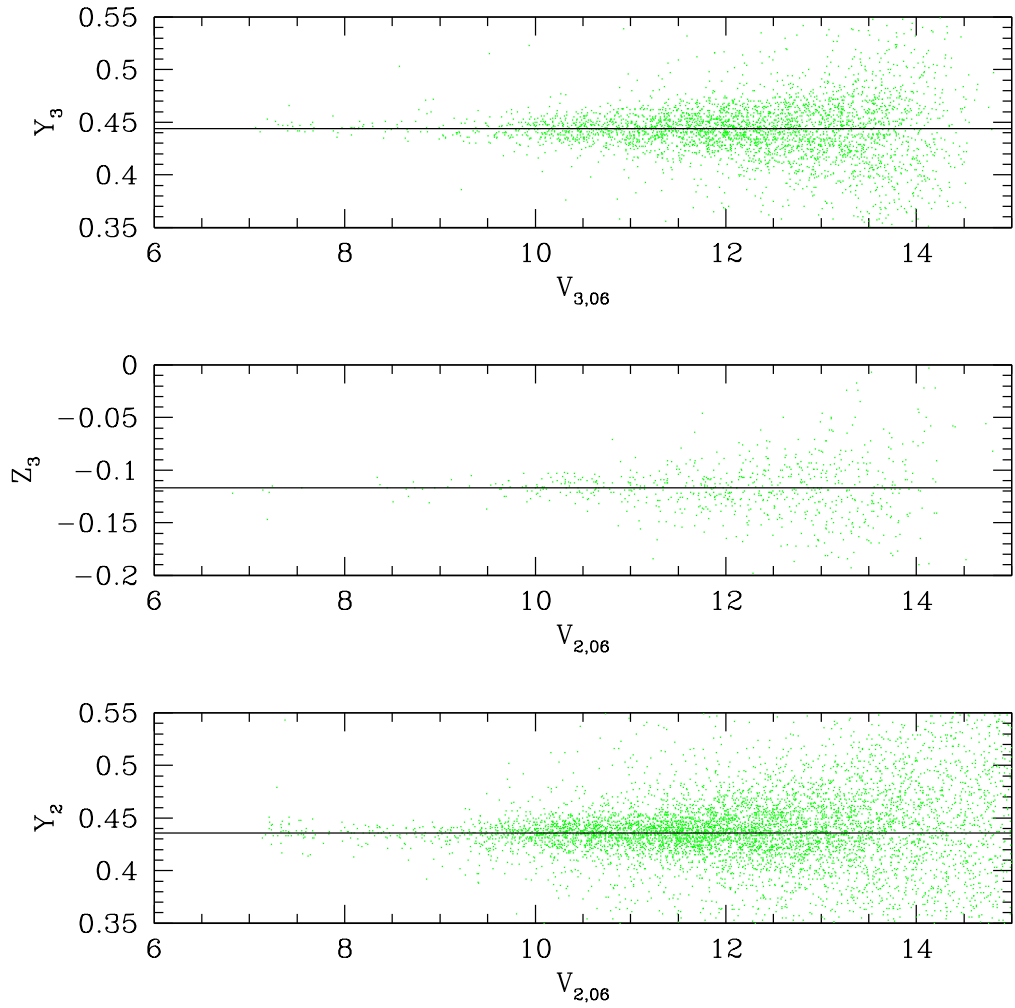}
\caption{Calculation of the zero points to bring the photometry to the
  chip2 system. In this figure we have plotted, as an example of the
  technique, the different zero points that we have to obtain to move
  the $V$ photometry from the chip3 to the chip2 reference system (see
  equations in the text). The solid lines show the final average
  values adopted.}
\label{figzp}
\end{figure}

\begin{figure}
\epsscale{1.25}
\plotone{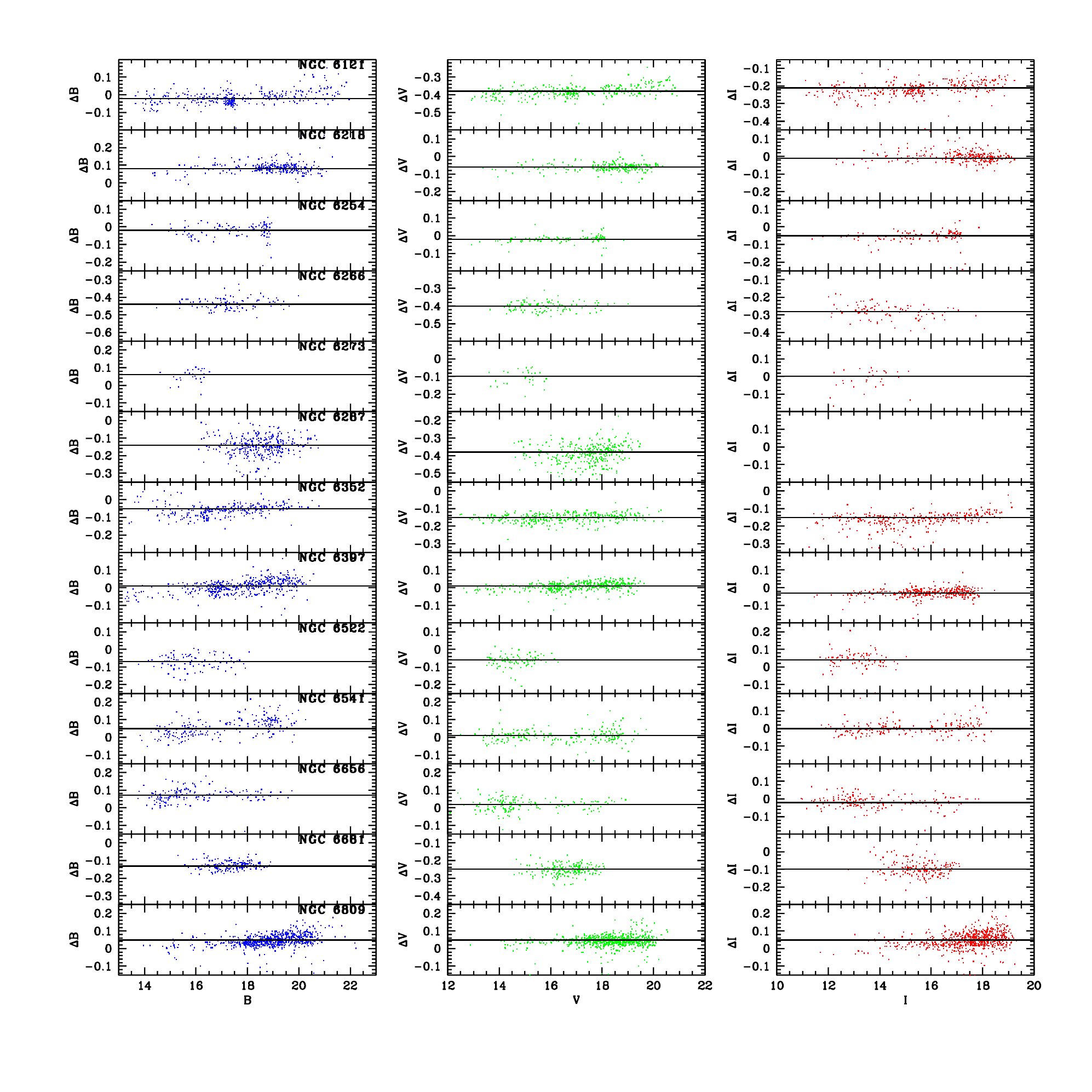}
\caption{Comparison of our raw ground-based photometry with the
  \citet{st00} calibrating stars for the clusters and filters available. We
  plot the magnitudes of the stars in our photometry for the different
  filters versus Stetson's values minus ours. The
  lines show where the average offsets lie.}
\label{figstetmag}
\end{figure}

\begin{figure}
%\epsscale{0.95}
\plotone{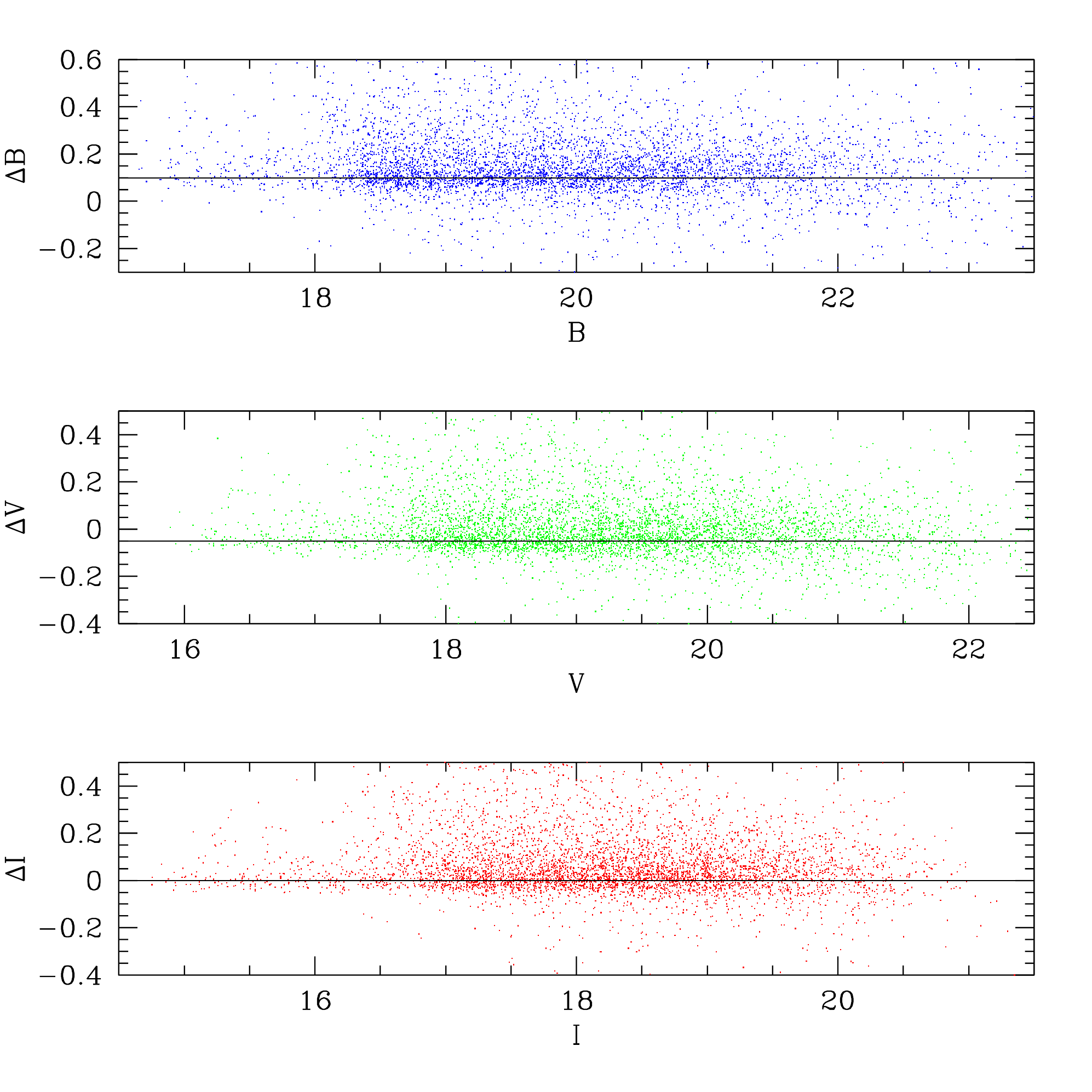}
\caption{Comparison of our raw ground-based photometry with the {\it
    HST} photometry. In this figure we show the steps to calculate the
  offsets in NGC 6218, as an example of the technique. We plot the
  magnitudes obtained from our ground-based observations versus the
  {\it HST} photometry minus our ground-based photometry.  We can
  observe the existence of the spread mentioned and explained in the
  text. The average offsets, shown by the lines, are calculated using
  a clipped weighted average (see text).}
\label{fighstmag}
\end{figure}  

\begin{figure}
\plotone{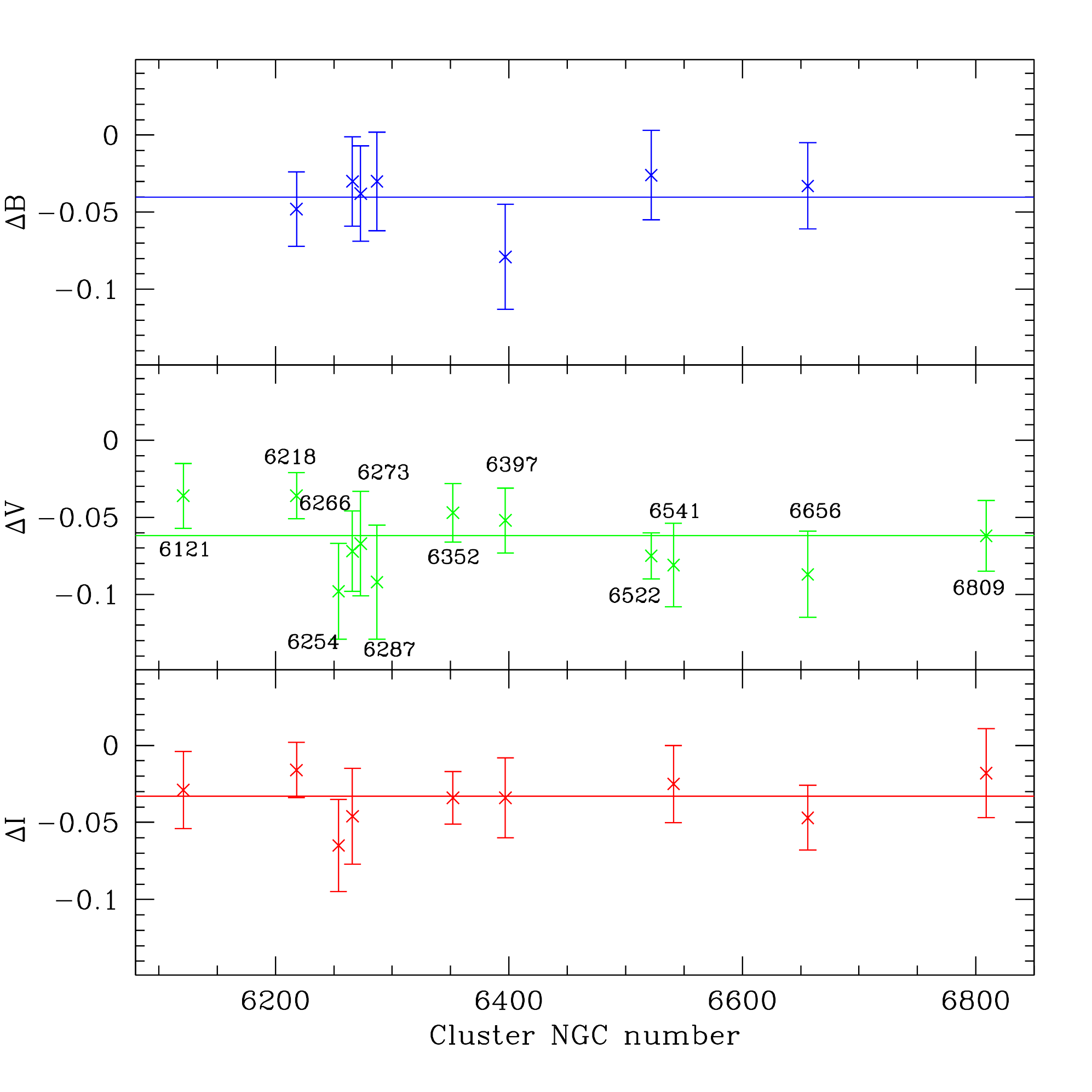}
\caption{Comparison of our photometry corrected with the available
  {\it HST} offsets, with the \citet{st00} calibrating stars. We plot the
  average differences of Stetson's values minus ours, for every
  cluster and filter. The lines show the weighted average offsets for
  every filter.}
\label{fighststet}
\end{figure}

\end{document}